\documentclass[11pt,a4paper]{article}

\usepackage{jheppub}
\usepackage{amsmath}
\usepackage[most]{tcolorbox}
\usepackage{dsfont}
\usepackage{ulem}
\usepackage{natbib}
\usepackage{xcolor}
\usepackage[hang,flushmargin]{footmisc}
\usepackage{tikz-cd}
\usepackage{enumitem}
\usepackage{amsfonts,amssymb, amscd,amsmath,latexsym,amsbsy}

\usepackage{subcaption}
\usepackage{bbm}

\usepackage{stmaryrd}
\usepackage{todonotes}
\usepackage{stackrel}
\usepackage{amsthm}

\usepackage{graphicx}

\usepackage{txfonts}
\usepackage{float}
\usepackage{nicematrix}

\usepackage{titlesec}
\usepackage{ulem}
\usepackage{color}
\usepackage[font=small,labelfont=bf]{caption}
\usepackage[flushmargin]{footmisc}
\usepackage{bigfoot}
\DeclareNewFootnote[para]{A}
\usepackage{tabularx}
\DeclareMathOperator{\sech}{sech}

\usepackage{addfont}
\addfont{OT1}{rsfs10}{\rsfs}

\usepackage{mathrsfs}
\usepackage{float}
\usepackage{stmaryrd}
\usepackage{todonotes}
\usepackage{cancel}
\usepackage{stackrel}
\usepackage{graphicx}
\usepackage{mathtools}
\usepackage{nicematrix}
\usepackage{bm}
\usepackage{dsfont}

\DeclareMathOperator{\Tr}{Tr}
\DeclareMathOperator{\tr}{tr}

\newcommand{\id}{\mathds{1}}

\newcommand{\bra}[1]{{\left< {#1} \right|}}
\newcommand{\ket}[1]{{\left| {#1} \right>}}

\title{\boldmath Ising 100: review of solutions}

\author{
Oğuz Alp Ağırbaş,
Anıl Ata,
Eren Demirci,
Ilmar Gahramanov,
Tuğba Hırlı,\\
R. Semih Kanber,
Ahmet Berk Kavruk,
Mustafa Mullahasanoğlu,
Zehra Özcan,\\
Cansu Özdemir,
Irmak Özgüç,
Sinan Ulaş Öztürk,
Uveys Turhan,\\
Ali Mert Yetkin,
Yunus Emre Yıldırım, and
Reyhan Yumuşak
}

\affiliation{Department of Physics, Bogazici University, 34342 Bebek, Istanbul, Türkiye}

\emailAdd{oguz.agirbas@std.bogazici.edu.tr}\emailAdd{anil.ata@std.bogazici.edu.tr}\emailAdd{erendemirci2002@hotmail.com}\emailAdd{ilmar.gahramanov@bogazici.edu.tr}\emailAdd{tugba3hirli@gmail.com}\emailAdd{rsmhknbr@gmail.com}\emailAdd{ahmetb.kavruk@gmail.com}\emailAdd{mustafa.mullahasanoglu@std.bogazici.edu.tr}\emailAdd{zehra.ozcan@std.bogazici.edu.tr}\emailAdd{cansu.ozdemir@mail.utoronto.ca}\emailAdd{irmak.ozguc@std.bogazici.edu.tr}\emailAdd{ulasoztrk@gmail.com}\emailAdd{uveys.turhan@std.bogazici.edu.tr}\emailAdd{ali.yetkin@std.bogazici.edu.tr}\emailAdd{yunus.yildirim1@std.bogazici.edu.tr}\emailAdd{reyhan.yumusak@std.bogazici.edu.tr}

\abstract{We present several known solutions to the two-dimensional Ising model. This review originated from the “Ising 100” seminar series held at Boğaziçi University, Istanbul, in 2024.}

\begin{document}

	\maketitle

\section{Introduction}

The Ising model is the simplest and most commonly used model for studying collective phenomena in statistical mechanics. It was first proposed and studied in the works of Lenz and Ising \cite{lenz1920beitrag}. In 1925, Ising found a solution for the one-dimensional chain, although this solution did not exhibit a phase transition \cite{Ising,Ising:1925em}. In 1944, Lars Onsager \cite{PhysRev.65.117} published an exact solution for the two-dimensional case in the absence of an external field and demonstrated the possibility of a phase transition in two-dimensional systems. As one of the earliest exactly solvable models, it laid the foundation for the development of other solvable models in statistical physics. At present, these models play a significant role in both theoretical and mathematical physics.

In this review article, we present several known solutions to the two-dimensional Ising model. In the appendix, we also include two lesser-known solutions to the one-dimensional Ising model.

\section{Combinatorial Approach \textit{(Eren Demirci)}}
\label{chapter:2}

After Onsager and Kaufman's \cite{PhysRev.65.117,PhysRev.76.1232,PhysRev.76.1244} exact solution of the two-dimensional Ising model, Kac and Ward \cite{PhysRev.88.1332} proposed a new approach based on combinatorics to facilitate calculations. Although this method was initially highly approximate, in this article, we first present the approximate form derived from their combinatorial approach, before discussing the corrections introduced by Potts and Ward \cite{Potts1955TheCM}. With these corrections, the combinatorial solution provides a viable alternative to the algebraic solutions. Additionally, we also review Potts' solution for the triangular lattice \cite{Potts1955CombinatorialSO}, which follows the same method and is included at the end of the article.

\subsection{Introduction}

To use as a starting point, Waerden's \cite{van_der_Waerden1941} square net method partition function is easy to obtain
\begin{equation}
    Z = (\cosh H)^h (\cosh H')^v \sum g(l,k)x^l y^k 
\end{equation}
where
\ $H=J/kT$, $H'=J'/kT$, $x=\tanh H$, $y=\tan H'$.
$g(l,k)$ is functional for closed polygons, $l$ and $k$ are correspondingly horizontal links and vertical links. Also, $h= |l|$, $v= |k|$.
Kaufman has shown that the exact formula for a two-dimensional $n \times m$ lattice partition function can be written as \cite{PhysRev.76.1232}

\begin{align*}
    Z=(2 \sinh 2 H)^{m n / 2}
\left\{\sum_{\sigma} \exp \left[m / 2\left( \sigma_2 \gamma_{2} + \sigma_4 \gamma_{4} + \cdots\right)\right]\right.\left.+\sum_{\sigma} \exp \left[m / 2\left( \sigma_1 \gamma_{1} + \sigma_3 \gamma_{3} + \cdots\right)\right]\right\} 
\end{align*}
where $\sigma_n = \pm 1$. Let's use the transfer matrix $M_n$ to arrive at the formula

\begin{equation*}
    \begin{pmatrix}
    e^{\frac{m}{2}\gamma_{2}} & e^{-\frac{m}{2}\gamma_{2}} \\
    e^{-\frac{m}{2}\gamma_{2}} & e^{\frac{m}{2}\gamma_{2}} 
\end{pmatrix}\begin{pmatrix}
    e^{\frac{m}{2}\gamma_{4}} & e^{-\frac{m}{2}\gamma_{4}} \\
    e^{-\frac{m}{2}\gamma_{4}} & e^{\frac{m}{2}\gamma_{4}} 
\end{pmatrix}\dots
\begin{pmatrix}
    e^{\frac{m}{2}\gamma_{2j}} & e^{-\frac{m}{2}\gamma_{2j}} \\
    e^{-\frac{m}{2}\gamma_{2j}} & e^{\frac{m}{2}\gamma_{2j}} 
\end{pmatrix}
\end{equation*}
Diagonalizing $M_1$ as
\begin{align*}
    \begin{pmatrix}
    -1 & 1 \\
    1 & 1 
\end{pmatrix}
\begin{pmatrix}
    e^{\frac{m}{2}\gamma_{2}}-e^{-\frac{m}{2}\gamma_{2}} & 0 \\
    0 & e^{\frac{m}{2}\gamma_{2}}+e^{-\frac{m}{2}\gamma_{2}}
\end{pmatrix}
\begin{pmatrix}
    \frac{-1}{2} & \frac{1}{2} \\
    \frac{1}{2} & \frac{1}{2} 
\end{pmatrix}
\end{align*}
Noticing that $M_n= S A_n S^{-1}$, the product of $M_i$ can be expressed as below
\begin{align*}
M_1M_2\dots M_n &= SA_1S^{-1}SA_2S^{-1}\dots SA_nS^{-1}\\
Tr(S(\prod_{r=1}^{n}A_{2r})S^{-1})&=Tr(S^{-1}S(\prod_{r=1}^{n}A_{2r}))\\
&=\prod_{r=1}^{n}\cosh{x_{2r}}+\prod_{r=1}^{n}\sinh{x_{2r}}
\end{align*}

\begin{align*}
    Z=\frac{1}{2}(2 \sinh 2 H)^{\frac{1}{2}mn}\left\{\prod_{r=1}^{n}\left(2 \cosh \frac{m}{2} \gamma_{2r}\right)
+\prod_{r=1}^{n}\left(2 \sinh \frac{m}{2} \gamma_{2r}\right)
+\prod_{r=1}^{n}\left(2 \cosh \frac{m}{2} \gamma_{2 r-1}\right)
+\prod_{r=1}^{n}\left(2 \sinh \frac{m}{2} \gamma_{2 r-1}\right)\right\}
\end{align*}
From the hyperbolic law of cosines
\begin{equation}
    \cosh \gamma_{j} = \cosh 2H^* \cosh 2H' - \sinh 2H^* \sin 2H' \cos (\frac{\pi j}{n})
\end{equation}

\begin{figure}[h]
\centering
\includegraphics[width=0.25\linewidth]{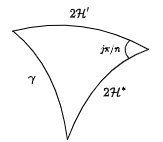}
\caption{\label{fig:hyper} Hyperbolic triangle whose edges are $2H'$, $2H^*$ and $\gamma$}
\end{figure}
where $ H^* $ represents the dual of $H$, it follows that $ e^{-2H} = \tanh H^* $.

For simplicity, let us define $y=\tanh H$ and $x=\sinh H$ and express the terms involving $H^*$ in terms of $x$ and $y$.
\begin{align*}
\cosh 2 H^* &= \frac{\cosh ^2 H^* + \sinh ^2 H^*}{\cosh ^2 H^* - \sinh ^2 H^*} 
=\frac{\operatorname{coth} H^* + \tanh H^*}{\operatorname{coth} H^* -\tanh H^*}\nonumber\\
&=\frac{e^{2 H}+e^{-2 H}}{e^{2 H}-e^{-2 H}}
=\frac{\cosh 2 H}{\sinh 2 H}\nonumber\\
&=\frac{\left(1+x^2\right)}{2 x} \\
\\
\cos 2 H &=\frac{\cosh ^2 H+\sinh ^2 H}{\cosh ^2 H-\sinh ^2 H}
=\frac{1+\tanh ^2 H}{1-\tanh ^2 H}\\
&=\frac{1+y^2}{1-y^2}    
\end{align*}
From this equation, it is easily can be seen that
\begin{align*}
    \cosh \gamma_{j}=\frac{\left(1+x^{2}\right)\left(1+y^{2}\right)}{2 x\left(1-y^{2}\right)}-\frac{\left(1-x^{2}\right)}{2 x} \frac{2 y}{1-y^{2}} \cos \frac{\pi j}{n}
\end{align*}
Let's look at $\gamma_{j}$ when $n$ is large enough
\begin{align*}
    \text{as} \: n\to\infty,\: \frac{j\pi}{n} \to 0, \:\text{hence:} \: \cos{(\frac{j\pi}{n})}\to 1
\end{align*}
As can be seen in Figure \ref{fig:hyper}, $\sinh\gamma_n\to 0$. When $\cos{(\frac{j\pi}{n})}\to 1$, the value of $j$ will have a negligible effect. Then $\cosh \gamma_{2j}=\cosh \gamma_{2j-1}$.

It must be noted that boundary effects, such as $j=n$ or some $j$ which is not too small than $n$, are ignored even when there are some values of $j$ are not inconsequential. Thus
\begin{align*}
    Z \sim\left(2\sinh 2 H\right)^{\frac{1}{2} m n} \prod_{r=1}^{n}\left(2 \cosh \frac{m}{2} \gamma_{2 r}\right)
\end{align*}
can be transformed with the aid of
\begin{equation}
\cosh l \theta=2^{l-1} \prod_{S=0}^{l-1}\left(\cosh \theta-\cos \frac{(2 S+1) \pi}{2 l}\right)    \label{eq:cos}
\end{equation}
into
\begin{align*}
    Z \sim(2 \sinh 2 H)^{l n} \prod_{r=1}^{n} \prod_{S=0}^{l-1}\left(\cosh \gamma_{2 r}-\cos \frac{(2 S+1) \pi}{2 l}\right) .
\end{align*}
We can extract $ \frac{1}{x} $ and $\frac{1}{(1-y^2)}$ separately, and $ l= \frac{m}{2}$ and assume $m$ is even
\begin{align*}
     Z \sim(2\cosh H)^{h}\left(2\cosh H^{\prime}\right)^{v} \prod_{r=1}^{n} \prod_{S=0}^{l-1}\left\{\left(1+x^{2}\right)\left(1+y^{2}\right)\right. \\
 \left.-2 y\left(1-x^{2}\right) \cos \frac{2 \pi r}{n}-2 x\left(1-y^{2}\right) \cos \frac{(2 S+1) \pi}{2 l}\right\}
\end{align*}
Assuming $2S+1\sim 2S$ in large scales we can see that
\begin{align*}
\sum g(l, k) x^{l} y^{k} & \sim \prod_{r=1}^{n} \prod_{S=1}^{\frac{m}{2}}\left\{\left(1+x^{2}\right)\left(1+y^{2}\right)\right. \\
& \left.-2 y\left(1-x^{2}\right) \cos \frac{2 \pi r}{n}-2 x\left(1-y^{2}\right) \cos \frac{2 \pi S}{m}\right\}
\end{align*}

\subsection{The Method}

From the Leibniz definition of the matrix determinant
\begin{equation*}
\det(A) = \sum_{\tau\in S_n}\text{sign}(\tau) \prod_{i = 1}^n a_{i, \, \tau(i)}
\end{equation*}
So, we know that we can reach every possible permutational path from the $n \times n$ matrix's determinant. To be more precise, we can achieve all closed polygons but the ones we achieve oriented. We can solve this problem with a little trick. Let's split the lattices into two and say that clockwise ones are on top and vice versa.
\begin{align*}
    \begin{vmatrix}
    1 & A_{1,2} & A_{1,3} & \ldots & A_{1,n} \\
    A_{2,1} & 1 & A_{2,3} & \ldots & A_{2,n} \\
    A_{3,1} & A_{3,2} & 1 & \ldots & A_{3,n} \\
    \ldots & \ldots & \ldots & \ldots & \ldots \\
    A_{n,1} & A_{n,2} & \ldots & A_{n, n-1} & 1
\end{vmatrix}
\end{align*}
Since all paths must include at least two points, $A_{rr} $ type elements have been taken as equal to $1$.
The last error originates from a plane divided by oriented cycles. We were studying on an $n \times m$ plane but since we split the plane into two parts we obtained cycles passing through between two planes. Let's say we count $n\times 2m$ model. Thus, the last version of the formula is
\begin{align*}
\sum g(l, k) x^{l} y^{k} &\sim \prod_{r=1}^{n} \prod_{S=1}^{m}\left\{\left(1+x^{2}\right)\left(1+y^{2}\right)\right.\\
&\left.-2 y\left(1-x^{2}\right) \cos \frac{2 \pi r}{n}-2 x\left(1-y^{2}\right) \cos \frac{2 \pi S}{m}\right\}    
\end{align*}
We know that the formula for the roots of the unity
$$\exp\left(\frac{2k\pi i}{n}\right)=\cos\frac{2k\pi}{n}+i\sin\frac{2k\pi}{n},\qquad k=1,\dots, n-1, n.$$
So we can rewrite this equation like when $\chi$'s are the $n^{th}$ root of unity and $\zeta$'s are the $m^{th}$ root of unity
\begin{equation} \label{sumus}
\begin{split}
\sum g(l, k) x^{l} y^{k} & \sim \prod_{\chi , \zeta}\left\{\left(1+x^{2}\right)\left(1+y^{2}\right)\right. \\
& \left.-y\left(1-x^{2}\right)(\chi + \chi ^{-1} )-x\left(1-y^{2}\right)(\zeta + \zeta ^{-1} )\right\} 
\end{split}
\end{equation}

\subsection{Construction of the Determinant}

Let us define the matrix element $a$, when $i$,$j$ indicates the points where $a$ stands on and $X$,$Y$ the direction where the path heads from this point. We do not have non-neighboring terms we connected. Thus we set $ a_{(i,j)XY}=0 $ if $(i,j)XY$ is improper. For example $a_{(1,3)XY}$ is zero. Even paths between neighboring points might be improper. For example, $a_{(1,2)XY}$ is also zero when $X$ is anything but the direction $R$.\\
$a_{(i,j)XY}=0 $ if $(i,j)XY$ is improper.\\
$a_{(i,j)XY}=x $ if $X= R$ or $L$ \\ 
$a_{(i,j)XY}=y $ if $X= U$ or $D$ 

According to the direction of rotation, we multiply $x$, $y$ with
\begin{align*}
     \alpha = e^{\frac{1}{4}i\pi}, \bar{\alpha} = e^{-\frac{1}{4}i\pi},  \alpha^0 = 1 
\end{align*}
With $\frac{1}{2}\pi$ indicating counter-clockwise rotation and $-\frac{1}{2}\pi$ indicating clockwise rotation. For example $A_{(i,j)RU} = \alpha x$, $A_{(i,j)UR} = \bar{\alpha} y$, $A_{(i,j)UU} = y$.

A cycle of length $C$ has $ (-1)^{C-1} $ as the multiplier and since we are dealing with cycles of even length the sign will be always negative and the sign of permutation or $r$ cycles will be $(-1)^r$. So the determinant elements will be in $(-1)^{\frac{\Omega}{2\pi}}(-1)x^ly^k $ form where $l$ and $k$ represents horizontal and vertical lines and $\Omega$ is the total angle through one turn. Due to the total angle, which can be either $2\pi$ or $-2\pi$, the sign of $x^ly^k$ is always positive.

\pagebreak
Finally, we can start constructing our determinant for the matrix corresponding to our $n \times m$ model which consists of $n\times m$ entries on each side. Every entry consists of four subentries, the directions $R$, $L$, $U$, and $D$. Thus we will have an $(4mn)\times(4mn) $ matrix, which is quite big, but most of its entries are zero. Let's look at an illustration of the matrix corresponding to our $n \times m$ model

\begin{figure}[h]
\centering
\includegraphics[width=1\linewidth]{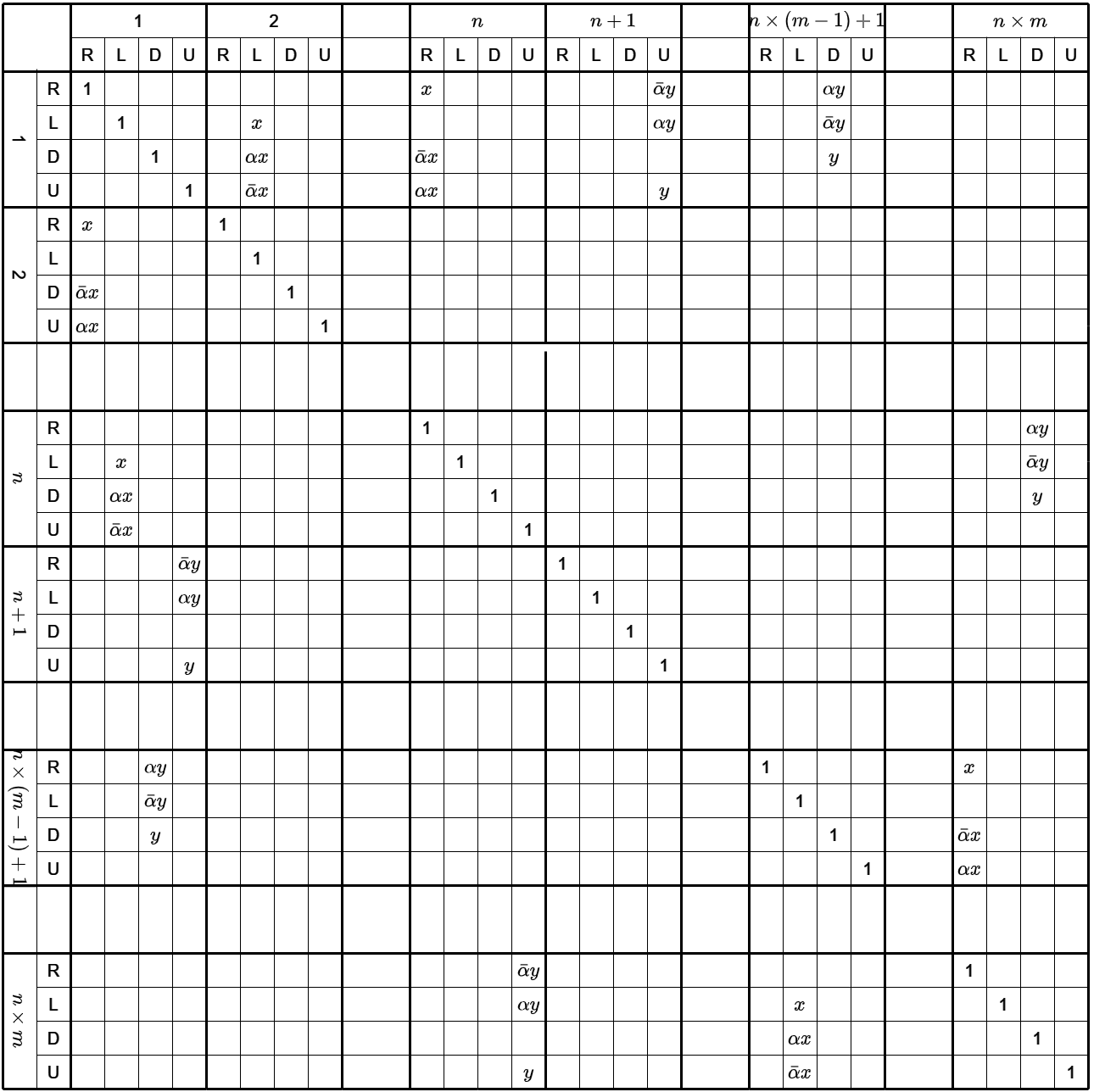}
\caption{\label{fig:revize}The illustration of $n\times m$ model.}
\end{figure}

\pagebreak
It must be noted that only one element from one column and one row is allowed. As the last problem concerning our method of approach, only two types of cycles are considered. Since all cycles can be reduced to one of these two types, this turns out to be not an issue. Solving the problem in terms of these two cycles will solve the entire problem.

\begin{figure}[h]
\centering
\includegraphics[width=0.25\linewidth]{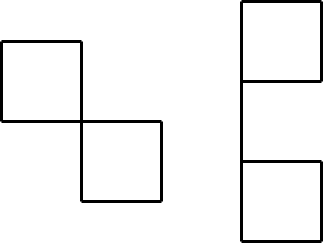}
\caption{\label{fig:inappropriate}The left cycle will be referred to as $(v)$ and the right one as $(f)$.}
\end{figure}
We can consider a $3\times 4$ model and determine its problematic cycles. One example of $(v)$ is

$$A_{(1,2)RD}A_{(2,5)DD}A_{(5,8)DL}A_{(8,9)LU}A_{(9,6)UL}A_{(6,5)LL}A_{(5,4)LU}A_{(4,1)UR}$$ 

\begin{figure}[h]
\centering
\includegraphics[width=0.2\linewidth]{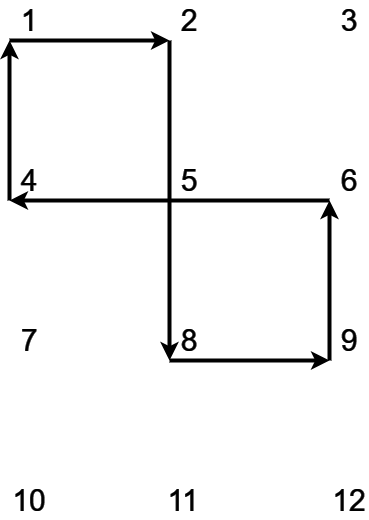}
\caption{\label{fig:inappropriate1} $\Omega$ for this one is 0.}
\end{figure}

Another is
$$A_{(1,2)RD}A_{(2,5)DR}A_{(5,6)RD}A_{(6,9)DL}A_{(9,8)LU}A_{(8,5)UL}A_{5,4)LU}A_{(4,1)UR}$$

\begin{figure}[h]
\centering
\includegraphics[width=0.2\linewidth]{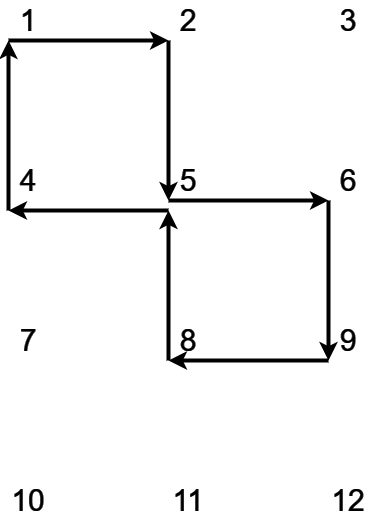}
\caption{\label{fig: inappropriate21} $\Omega$ for this one is $2\pi $.}
\end{figure}

\pagebreak
It is useful to remind ourselves that we multiply each cycle with $(-1)^{\frac{\Omega}{2\pi}}$. Consequently, Figure \ref{fig:inappropriate1} has a (+) sign while Figure \ref{fig: inappropriate21} has a (-) sign. Thus they cancel each other out. Then let's look at one example of (f): 
\begin{align*}
    A_{(1,2)RD}A_{(2,5)DL}A_{(5,4)LD}A_{(4,7)DR}A_{(7,8)RD}A_{(8,11)DL}A_{11,10)LU}A_{(10,7)UU}A_{7,4)UU}A_{(4,1)UR}
\end{align*}

\begin{figure}[h]
\centering
\includegraphics[width=0.2\linewidth]{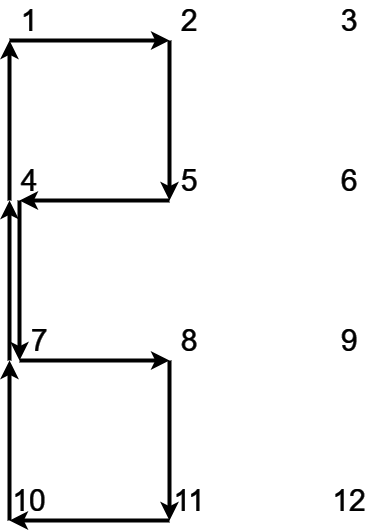}
\caption{\label{fig: inappropriate2}$\Omega$ for this one is $2\pi $.}
\end{figure}
Another one is

$$A_{(1,2)RD}A_{(2,5)DL}A_{(5,4)LD}A_{(4,7)DD}A_{(7,10)DR}A_{(10,11)RU}A_{(11,8)UL}A_{(8,7)LU}A_{7,4)UU}A_{(4,1)UR}$$

\begin{figure}[H]
\centering
\includegraphics[width=0.2\linewidth]{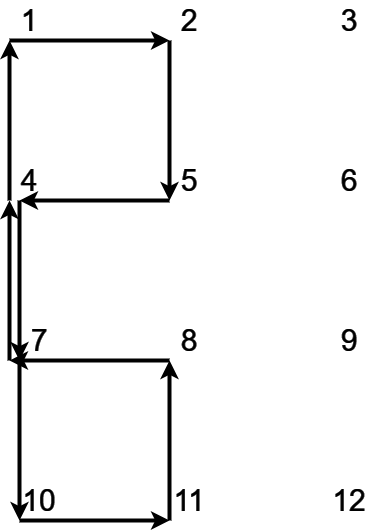}
\caption{\label{fig: inappropriate22}$\Omega$ for this one is 0.}
\end{figure}
We might notice that due to their $\Omega$'s, they cancel each other out. Then there will be no problematic cycles that remain. We do not have to make any corrections to solve this problem since it has not occurred.

\begin{figure}[H]
\centering
\includegraphics[width=0.25\linewidth]{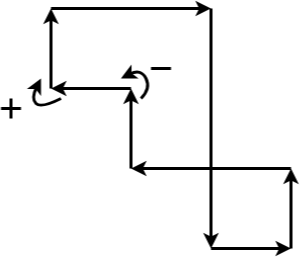}
\caption{\label{fig:inappropriate3} This cycle is also problematic.}
\end{figure}
As can be seen, every turn, plus or minus, brings its inverse part to close the cycle and they cancel each other out. This leads us to the realization that every problematic cycle will be a combination of $(v)$ and $(f)$. Thus, all of them will cancel each other out and there will be no inappropriate cycles that remain.
In our torus model, there are three types of loops similar to $R, R, R, R\dots R$: horizontal loops, vertical loops, and combinations of vertical and horizontal loops. They will be neglected for now, however we will account for them in the exact solution.

\begin{figure}[h]
\centering
\includegraphics[width=0.25\linewidth]{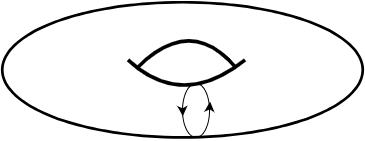}
\caption{\label{fig:torus} This is an example of vertical loops.}
\end{figure}

\subsection{Deriving the Determinant}
The determinant is too big to easily calculate by the method of minors but we can find the corresponding eigenvalues of the matrix and multiply them. Before attempting to construct the eigenvectors, notice that if an eigenvalue can be expressed as a function of $n \in \mathbb N$, $\lambda_n = f(n)$, it would be much easier to calculate the determinant of the matrix.

First let's define $b$ and $l$ as the coordinates of a point and instead of numbering in order, name them by their coordinates

\begin{figure}[h]
\centering
\includegraphics[width=0.9\linewidth]{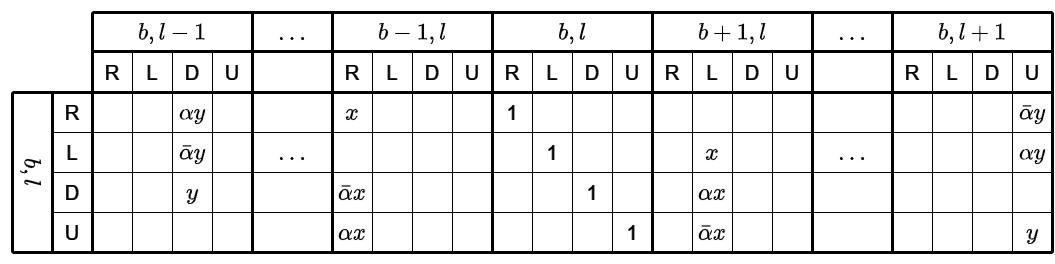}
\caption{\label{fig:row} The row block of a matrix for a point with $b,l$ coordinates.}
\end{figure}
As can be seen, of course, column blocks can change order, for every row block we achieve the same result with different coefficients. Say that $a_{b,l,d}$ is the element of an eigenvector which corresponds to the point $b,l$ and the direction of $d$. Then the equation for the eigenvalue will be as below.

\begin{align*}
    a_{(b-1,l,R)}x+ a_{(b,l,R)}+ a_{(b,l-1,D)}\alpha y+ a_{(b,l+1,U)}\bar{\alpha} y = \lambda a_{(b,l,R)}
\end{align*}
Applying this equation to the other directions results in:
\begin{align*}
    a_{(b,l-1,D)}y+ a_{(b,l,D)}+ a_{(b+1,l,L)}\alpha x+ a_{(b-1,l,R)}\bar{\alpha} x = \lambda a_{(b,l,D)}
\end{align*}
Since both $b$ and $l$ are dummy indices, we conclude that every equation of the same direction is of equivalent form. For example

\begin{align*}
    a_{(m-1,n,R)}x+ a_{(m,n,R)}+ a_{(m,n-1,D)}\alpha y+ a_{(m,n+1,U)}\bar{\alpha} y &= \lambda a_{(m,n,R)}\\
    a_{(o-1,p,R)}x+ a_{(o,p,R)}+ a_{(o,p-1,D)}\alpha y+ a_{(o,p+1,U)}\bar{\alpha} y &= \lambda a_{(o,p,R)}\\
    a_{(e-1,d,R)}x+ a_{(e,d,R)}+ a_{(e,d-1,D)}\alpha y+ a_{(e,d+1,U)}\bar{\alpha} y &= \lambda a_{(e,d,R)}\\
\end{align*}
Let's define $a_{(b,l,d)}$ as $r_{(b,l)}, l_{(b,l)}, d_{(b,l)}, u_{(b,l)}$ where $r$ corresponds to the right direction, $u$ to up, $d$ to down, $l$ to left. Then the equation for the right direction turns into
$$r_{(b-1,l)}x+ r_{(b,l)}+ d_{(b,l-1)}\alpha y+ u_{(b,l+1)}\bar{\alpha} y = \lambda r_{(b,l)} $$

To separate the position indices from the directed function corresponding to the point, let's assume that the position can be encoded by multiplication with a variable that corresponds to that index. Defining 2 variables: $\chi$ and $\zeta$, we can see that the equations become the same, and the places of the points do not matter for the equations. Since we are on a torus and the coordinates of points are just dummies, this is the most natural-looking form
$$r\chi^{b-1}\zeta^lx+ r\chi^{b}\zeta^l+ d\chi^{b}\zeta^{l-1}\alpha y+ u\chi^{b}\zeta^{l+1}\bar{\alpha} y = \lambda r\chi^{b}\zeta^l $$
Then
$$r\chi^{-1}x+ r+ d\zeta^{-1}\alpha y+ u\zeta\bar{\alpha} y = \lambda r $$
As can be seen, this equation is independent of $b$ or $l$, the coordinates of the point. Now to find the values of $\chi$ and $\zeta$ we may study the boundary where the equations are not $m$ and $n$ independent. At the point $(1,1)$ there are two boundary connections due to us working on the surface of a torus and at $(2,1)$ there is only one boundary connection which is one of the connections of $(1,1)$. Thus they are good cases to study

\begin{figure}[h]
\centering
\includegraphics[width=0.5\linewidth]{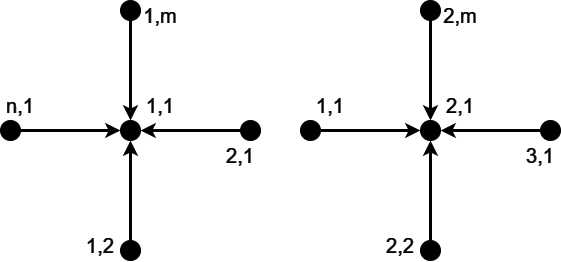}
\caption{\label{fig:dots} Direction illustration for $(1,1)$ and $(2,1)$ points.}
\end{figure}

\begin{align*}
    \zeta\chi^{n}rx+ \zeta\chi r + \zeta^{m}\chi d\alpha y+ \zeta^{2}\chi u\bar{\alpha} y &= \lambda \zeta\chi r\\
    \zeta\chi rx+ \zeta\chi^{2} r + \zeta^{m}\chi^{2} d\alpha y+ \zeta^{2}\chi^{2} u\bar{\alpha} y &= \lambda \zeta\chi^{2}r
\end{align*}
Ignoring the trivial case $\chi = \zeta = 0$, the above expression can be more concisely stated as below.

\begin{align*}
    \chi^{n-1}rx+ \zeta^{m-1} d\alpha y+ \zeta u\bar{\alpha} y &= (\lambda-1) r\\
    \chi^{-1} r x+ \zeta^{m-1} d \bar{\alpha} y+ \zeta u\bar{\alpha}  y &= (\lambda-1) r\\
\end{align*}
Notice that $\chi^{n-1} = \chi^{-1} $. Thus $\chi^n = 1$. Then $\chi$ is one of the $n^{th}$ roots of unity. Similarly, $\zeta$ is one of the $m^{th}$ roots of unity. Utilizing these facts, we may begin our construction of the eigenvectors. 

Then let us use the above method to tackle the other three equations corresponding to the other three directions: $l$, $d$, $u$.

\bigskip
Now we know that since $\chi$ and $\zeta$ of the point will cancel out extras there will be exactly the same equations for each row block.
Finally, equations corresponding to each of the four directions may be written as below

\begin{align*}
(1+x\chi^{-1})r + 0l +\alpha y \zeta^{-1} d + \bar{\alpha}y\zeta u &= \lambda r\\  
0r + (1+x\chi )l +\bar{\alpha} y\zeta^{-1} d + \alpha \zeta u &= \lambda l\\  
\alpha x \chi^{-1} r + \alpha x\chi l + (1+y\zeta^{-1})d + 0u &= \lambda d\\  
\bar{\alpha} x\chi^{-1} r + \bar{\alpha} x\chi l + 0d + (1+y \zeta) u &= \lambda u   
\end{align*}
As we have shown earlier every root of unity of $n^{th}$ (or $m^{th}$) degree fits in equation. Then we have $n\times m$ eigenvalues and we have to multiply all of them to calculate the determinant.

\begin{align*}
\prod_{j} \lambda_{j}(\chi, \zeta)&= \prod_{\chi, \zeta}\left|\begin{array}{cccc}
1+x \chi^{-1} & 0 & \alpha y \zeta^{-1} & \bar{\alpha} y \zeta \\
0 & 1+x \chi & \bar{\alpha} y \zeta^{-1} & \alpha y \zeta \\
\bar{\alpha} x \chi^{-1} & \alpha x \chi & 1+y \zeta^{-1} & 0 \\
\alpha x  \chi^{-1} & \bar{\alpha} x \chi & 0 & 1+y \zeta
\end{array}\right| \\
&= \prod_{\chi, \zeta} \left(1+x^{2}\right)\left(1+y^{2}\right)-y\left(1-x^{2}\right)\left(\chi+\chi^{-1}\right) - x\left(1-y^{2}\right)\left(\zeta+\zeta^{-1}\right) .    
\end{align*}
This is exactly the same as the Equation \eqref{sumus}, as it must have been. Thus this method is appropriate to calculate the partition function of the fishnet lattice method without corrections, an approximate solution.

\subsection{Formulating the Determinant}
To formulate the determinant let's split our original matrix into $E + M$, where $E$ is the unit matrix. Repeating blocks of M can be listed as

$$
A'(-1)=
\begin{bmatrix}
    0 & 0 & 0 & 0 \\
    0 & 1 & 0 & 0 \\
    0 & \alpha & 0 & 0 \\
    0 & \bar{\alpha} & 0 & 0
\end{bmatrix},
A(-1)=
\begin{bmatrix}
    0 & 0 & 0 & \alpha \\
    0 & 0 & 0 & \bar{\alpha} \\
    0 & 0 & 0 & 0 \\
    0 & 0 & 0 & 1  
\end{bmatrix}$$

$$
A(1)=
\begin{bmatrix}
    0 & 0 & \alpha & 0 \\
    0 & 0 & \bar{\alpha} & 0\\
    0 & 0 & 1 & 0 \\
    0 & 0 & 0& 0 
\end{bmatrix}
,
A'(1)=
\begin{bmatrix}
    1 & 0 & 0 & 0 \\
    0 & 0 & 0 & 0\\
    \bar{\alpha} & 0 & 0 & 0 \\
    \alpha & 0 & 0 & 0 
\end{bmatrix}
$$
We must define a permutation matrix R:
$$\begin{bmatrix}
    0 & 1 & 0 & \dots & 0 \\
    0 & 0 & 1 & \dots & 0 \\
    \dots & \dots & \dots & \dots & \dots \\
    0 & 0 & \dots &  0 & 1 \\
    1 & 0 & \dots & 0 & 0 
\end{bmatrix}$$
Our matrix can be expressed as

\begin{figure}[h]
\centering
\includegraphics[width=1\linewidth]{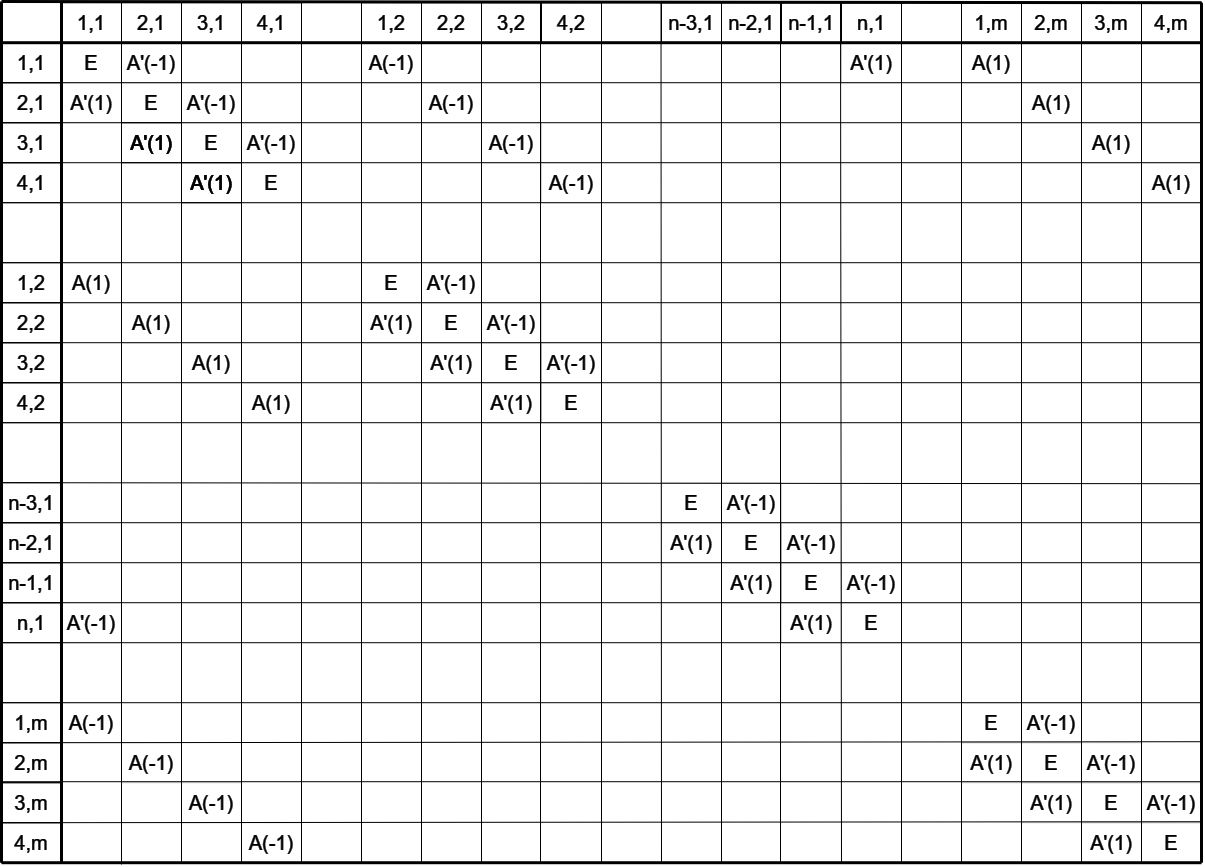}
\caption{\label{fig:compmatrix} Every $A$ matrix inserted in this $n\times m$ model.}
\end{figure}
It is easy to figure out the form of the matrix $M$.
\begin{align*}
M = x&(E_m\otimes R_n \otimes A(1) + E_m\otimes R^T_n \otimes A(-1))\\
&+y(R_m\otimes E_n \otimes A'(1) + R^T_m\otimes E_n\otimes A'(-1))    
\end{align*}
where $\otimes$ sign indicates direct product and $R^T$ is transpose of $R$.
We are left with
$$Z_{mn}^2 \sim P_1(H, H') =(2\cosh H \cosh H^*)^{2mn}|E+M|$$

\subsection{Obtaining the Exact Partition Function}
We must remember that at the beginning of this paper, we assume a plane going to infinity. But in the matter of fact, the partition function is calculated for finite models. So we must get rid of the error parts so we have a finite exact model at the end. We had used $H$ and $H'$ to count closed polygons but when we set them as $-H$ and $-H'$, every polygon stays the same except the loops on the torus, the ones we have to deal with. They have $(-1)^{n-1}$ as their multiplier but when we change their sign the multiplier will be $(-1)^{n-1}. (-1)^n= -1$. Then we can set three more $P$, $P_2$ changes all except one row of vertical interactions, and $P_3$ changes all except one column of horizontal interactions. $P_4$ changes all except one row of vertical interactions and one column of horizontal interactions.
Then:
\begin{equation}
 Z_{mn} = 1/2 \Bigl[ (P_1(-H,-H'))^{\frac{1}{2}}+(P_2(-H,-H'))^{\frac{1}{2}}+(P_3(-H,-H'))^{\frac{1}{2}}+(P_4(-H,-H'))^{\frac{1}{2}} \Bigr]  \label{eq:Zmn}
\end{equation}
We know that from previous pages when $l=m$
\begin{align*}
P_1(H,H')=& (2\sinh 2 H)^{l n} \prod_{r=1}^{n} \prod_{S=0}^{l-1}\left(\cosh \gamma_{2 r}-\cos \frac{(2 S+1) \pi}{2 l}\right)\\
=& (2\sinh 2 H)^{m n} \prod_{r=1}^{n} \prod_{S=1}^{m}\left( \cosh 2H^* \cosh 2H' - \sinh 2H^* \sin 2H' \cos \frac{2r\pi }{n}-\cos \frac{2 S \pi}{m}\right)\\
=& 2^{2m n} \prod_{r=1}^{n} \prod_{S=1}^{m}\left(\cosh 2H \cosh 2H + \sinh 2H' \cos \frac{2r \pi}{n}+\sinh 2H \cos \frac{2 S \pi}{m} \right)\\    
\end{align*}
Then it is easy to conclude that
\begin{align*}
P_1(-H,-H') =& 2^{2m n} \prod_{r=1}^{n} \prod_{S=1}^{m}\left(\cosh 2H \cosh 2H - \sinh 2H' \cos \frac{2r \pi}{n}-\sinh 2H \cos \frac{2 S \pi}{m}\right)\\
=&(2\sinh 2 H)^{m n} \prod_{r=1}^{n} \prod_{S=1}^{m}\left(\cosh 2H^* \cosh 2H' - \sinh 2H^* \sin 2H' \cos \frac{2r\pi }{n}-\cos \frac{2 S \pi}{m}\right)\\
=&(2\sinh 2 H)^{m n} \prod_{r=0}^{n} \prod_{S=0}^{m}\left(\cosh \gamma_{2r} -\cos \frac{2 S \pi}{m}\right) 
\end{align*}
With the aid of the trigonometric identities
$$2^m \prod_{k=1}^{m} \{ \cosh \gamma - \cos (2k \pi/m) \} = 4 \sinh^2{(\frac{1}{2} m \gamma)}  $$
We can reach
$$ P_1(-H,-H')=(2\sinh 2 H)^{m n} \prod_{r=0}^{n} 4 \sinh^2 (\frac{1}{2} m\gamma_{2r})$$
To find other$ P$'s we must define a matrix that aims to make one column or one row of interactions minus sign. To do that, we must change the sign of one of the elements in the permutation matrix. So we can change the permutation matrix $R$ as $T$
$$\begin{bmatrix}
    0 & 1 & 0 & \dots & 0 \\
    0 & 0 & 1 & \dots & 0\\
    \dots & \dots & \dots & \dots & \dots \\
    0 & 0 & \dots &  0 & 1 \\
    -1 & 0 & \dots & 0 & 0 
\end{bmatrix}$$
Then
\begin{align*}
    P_2 &= x(T^t_m\otimes E_n \otimes A(1) + T_m\otimes E_n \otimes A(-1))
    +y(E_m\otimes R^t_n \otimes A'(1) + E_m\otimes R_n\otimes A'(-1))\\
    P_3 &= x(R^t_m\otimes E_n \otimes A(1) + R_m\otimes E_n \otimes A(-1))
    +y(E_m\otimes T^t_n \otimes A'(1) + E_m\otimes T_n\otimes A'(-1))\\
    P_4 &= x(T^t_m\otimes E_n \otimes A(1) + T_m\otimes E_n \otimes A(-1))
    +y(E_m\otimes T^t_n \otimes A'(1) + E_m\otimes T_n\otimes A'(-1))
\end{align*}

\pagebreak
Then let's evaluate $P_2$
\begin{align*}
    P_2(-H,-H')=2^{2m n} \prod_{r=1}^{n} \prod_{S=1}^{m}\left(\cosh 2H \cosh 2H - \sinh 2H' \cos \frac{(2r-1) \pi}{n}-\sinh 2H \cos \frac{2 S \pi}{m}\right)
\end{align*}
Then from \ref{eq:cos} we can conclude that
\begin{align*}
    P_2(-H,-H')=(2\sinh 2H)^{m n} \prod_{r=1}^{n} ( 4\cosh^2 \frac{m}{2}\gamma_{2l})
\end{align*}
Then via similar computations, we can see that the other ones are
\begin{align*}
    P_3 &= (2\sinh 2 H)^{m n} \prod_{r=0}^{n} 4 \sinh^2 (\frac{1}{2} m\gamma_{2r-1})\\
    P_4 &= (2\sinh 2 H)^{m n} \prod_{r=0}^{n} 4 \cosh^2 (\frac{1}{2} m\gamma_{2r-1})
\end{align*}
From \ref{eq:Zmn} we can say that
\begin{align*}
    Z_{m,n}= \frac{1}{2}|2sinh 2 H| ^{\frac{mn}{2}}\left( \prod_{r=0}^{n} 2 \cosh (\frac{1}{2} m\gamma_{2r}) +\prod_{r=0}^{n} 2\sinh (\frac{1}{2} m\gamma_{2r})+ \prod_{r=0}^{n} 2\cosh (\frac{1}{2} m\gamma_{2r-1}) +\prod_{r=0}^{n} 2 \sinh (\frac{1}{2} m\gamma_{2r-1}) \right)
\end{align*}
So this is exactly the formula in the beginning that Kaufman has reached.

\subsection{Obtaining the Exact Partition Function of Triangular Model}
Now, we apply the same procedure to the triangular model and see if we can get the exact formula. Let's see what our model looks like

\begin{figure}[H]
\centering
\includegraphics[width=0.25\linewidth]{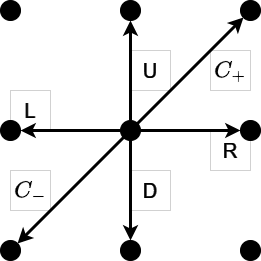}
\caption{\label{fig:tridots} Each point has six possibilities in this model.}
\end{figure}
As can be seen in Figure \ref{fig:tridots} now we have two more directions which are two crosses $C_+$ and $C_-$. These new directions obligate us to define a new variable $J_3$, then $H_3 = J_3/kT$, and $z$ which is equal to $\tanh H_3$ also new rotating factors
$$ \alpha = e^{\frac{i\pi}{6}}, \bar{\alpha} = e^{-\frac{i\pi}{6}}, \alpha^0 = 1 $$

\begin{figure}[H]
\centering
\includegraphics[width=0.6\linewidth]{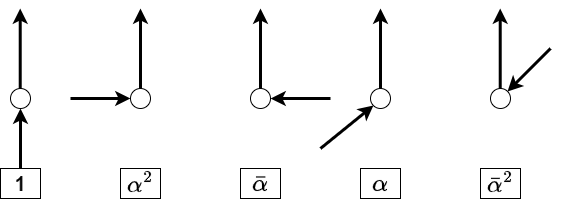}
\caption{\label{fig:angles} Possible turns are illustrated.}
\end{figure}
And the new matrix will be $6mn\times 6mn$. In light of this information let's construct a row block for a matrix of the $n\times m$ model

\begin{figure}[H]
\centering
\includegraphics[width=1.0\linewidth]{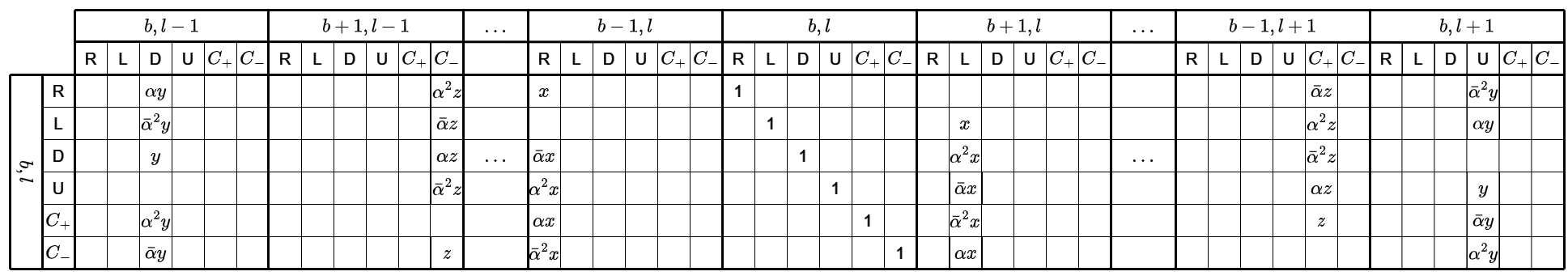}
\caption{\label{trirow} $b,l$ row block in a matrix.}
\end{figure}
We can formulate it easily by merely changing our previous formulation matrices a little bit. The $A$ matrices can be seen in the Figure \ref{trirow}

$$
A(1)=
\begin{bmatrix}
    1 & 0 & 0 & 0 & 0 & 0 \\
    0 & 0 & 0 & 0 & 0 & 0\\
    \bar{\alpha} & 0 & 0 & 0 & 0 & 0 \\
    \alpha^2 & 0 & 0 & 0 & 0 & 0  \\
    \alpha & 0 & 0 & 0 & 0 & 0 \\
    \bar{\alpha}^2 & 0 & 0 & 0 & 0 & 0
\end{bmatrix}
, \quad
A(-1)=
\begin{bmatrix}
    0 & 0 & 0 & 0 & 0 & 0 \\
    0 & 1 & 0 & 0 & 0 & 0\\
    0 & \alpha^2 & 0 & 0 & 0 & 0 \\
    0 & \bar{\alpha} & 0& 0 & 0 & 0  \\
    0 & \bar{\alpha}^2 & 0 & 0 & 0 & 0 \\
    0 & \alpha & 0 & 0 & 0 & 0 
\end{bmatrix}
$$

$$
A'(1)=
\begin{bmatrix}
    0 & 0 & \alpha & 0 & 0 & 0 \\
    0 & 0 & \bar{\alpha}^2 & 0 & 0 & 0\\
    0 & 0 & 1 & 0 & 0 & 0 \\
    0 & 0 & 0 & 0 & 0 & 0  \\
    0 & 0 & \alpha^2 & 0 & 0 & 0 \\
    0 & 0 & \bar{\alpha} & 0 & 0 & 0 
\end{bmatrix},
\quad
A'(-1)=
\begin{bmatrix}
    0 & 0 & 0 & \bar{\alpha}^2& 0 & 0 \\
    0 & 0 & 0 & \alpha & 0 & 0\\
    0 & 0 & 0 & 0 & 0 & 0 \\
    0 & 0 & 0 & 1 & 0 & 0  \\
    0 & 0 & 0 & \bar{\alpha} & 0 & 0 \\
    0 & 0 & 0 & \alpha^2 & 0 & 0 
\end{bmatrix}
$$

$$
A''(1)=
\begin{bmatrix}
    0 & 0 & 0 & 0 & \bar{\alpha} & 0 \\
    0 & 0 & 0 & 0 & \alpha^2 & 0\\
    0 & 0 & 0 & 0 & \bar{\alpha}^2 & 0 \\
    0 & 0 & 0 & 0 & \alpha & 0  \\
    0 & 0 & 0 & 0 & 1 & 0 \\
    0 & 0 & 0 & 0 & 0 & 0 
\end{bmatrix},\quad
A''(-1)=
\begin{bmatrix}
    0 & 0 & 0 & 0 & 0 & \alpha^2 \\
    0 & 0 & 0 & 0 & 0 & \bar{\alpha}\\
    0 & 0 & 0 & 0 & 0 & \alpha \\
    0 & 0 & 0 & 0 & 0 & \bar{\alpha}^2  \\
    0 & 0 & 0 & 0 & 0 & 0 \\
    0 & 0 & 0 & 0 & 0 & 1 
\end{bmatrix}
$$
So we can construct a formula for the matrix:
\begin{align*}
A= x&(E_m\times R^t_n \times A(1) + E_m\times R_n \times A(-1))\\
+ y&(R_m\times E_n \times A'(1) + R^t_m\times E_n \times A'(-1))\\
+ z&(R_m\times R^t_n \times A''(1) + R^t_m\times R_n \times A''(-1))
\end{align*}
Like before, we have to derive our eigenvectors with $r$, $l$, $d$, $u$ and $\chi^b$, $\zeta^l$ where $b$, $l$ are correspondingly horizontal and vertical coordinates of a point also $\chi$ and $\zeta$ are correspondingly $m^{th}$ and $n^{th}$ roots of unity. But we have additionally $c_+$, $c_-$. They stand for $C_+$ and $C_-$ row in every row block.\\
So now we can find the determinant of the $M$ matrix which is equal to $|A + E|$. To find eigenvectors we write down the equations of a row block

\begin{align*}
(1+x\chi^{-1})r + 0l +\alpha y \zeta^{-1} d + \bar{\alpha}^2 y\zeta u +\bar{\alpha} z \zeta\chi^{-1} c_+ + \alpha^2 z\zeta^{-1}\chi c_- &= \lambda r\\  
0r + (1+x\chi )l +\bar{\alpha}^2 y\zeta^{-1} d + \alpha \zeta u + \alpha^2 z \zeta\chi^{-1} c_+ + \bar{\alpha} z\zeta^{-1}\chi c_- &= \lambda l\\  
\bar{\alpha} x \chi^{-1} r + \alpha^2 x\chi l + (1+y\zeta^{-1})d + 0u +\bar{\alpha}^2 z \zeta\chi^{-1} c_+ + \alpha z \zeta^{-1}\chi c_- &= \lambda d\\  
\alpha^2 x\chi^{-1} r + \bar{\alpha} x\chi l + 0d + (1+y \zeta) u +\alpha z \zeta\chi^{-1} c_+ + \bar{\alpha}^2 z \zeta^{-1}\chi c_- &= \lambda u \\
\alpha x \chi^{-1} r + \bar{\alpha}^2 x\chi l + \alpha^2 y\zeta^{-1}d + \bar{\alpha} y\zeta u + (1 + z \zeta\chi^{-1}) c_+ + 0c_- &= \lambda c_+\\ 
\bar{\alpha}^2 x \chi^{-1} r + \alpha x\chi l + y\zeta^{-1}d + \alpha^2 y\zeta u + 0c_+ + (1 + z \zeta^{-1}\chi) c_- &= \lambda c_-
\end{align*}
Then it is easy to write the determinant of $M$

\begin{align*}
\prod_{j} \lambda_{j}(\chi, \zeta)&= \prod_{\chi, \zeta}\left|\begin{array}{cccccc}
(1 + x\chi^{-1})& 0& \alpha y \zeta^{-1}& \bar{\alpha}^2 y \zeta& \bar{\alpha} z \zeta\chi^{-1}& \alpha^2 z\zeta^{-1}\chi \\
0& (1+x\chi )& \bar{\alpha}^2 y\zeta^{-1}& \alpha \zeta& \alpha^2 z \zeta\chi^{-1}& \bar{\alpha} z\zeta^{-1}\chi \\
\bar{\alpha} x \chi^{-1}& \alpha^2 x\chi& (1+y\zeta^{-1})& 0&\bar{\alpha}^2 z \zeta\chi^{-1} & \alpha z \zeta^{-1}\chi \\
\alpha^2 x\chi^{-1}& \bar{\alpha} x\chi& 0& (1+y \zeta)& \alpha z \zeta\chi^{-1}& \bar{\alpha}^2 z \zeta^{-1}\chi\\
\alpha x \chi^{-1}& \bar{\alpha}^2 x\chi& \alpha^2 y\zeta^{-1}& \bar{\alpha} y\zeta& (1 + z \zeta\chi^{-1})& 0 \\
\bar{\alpha}^2 x \chi^{-1}& \alpha x\chi& y\zeta^{-1}& \alpha^2 y\zeta& 0& (1 + z\zeta^{-1}\chi) \\
\end{array}\right| \\
&= \prod_{\chi, \zeta} \left(1+x^{2}\right)\left(1+y^{2}\right)\left(1+z^{2}\right)+ 8xyz -y^2\left(1-x^{2}\right)\left(1-z^{2}\right) \left(\chi+\chi^{-1}\right)\\
 &- x^2\left(1-y^{2}\right)\left(1-z^{2}\right) \left(\zeta+\zeta^{-1}\right)- z^2\left(1-x^{2}\right)\left(1-y^{2}\right) \left(\chi \zeta +\chi^{-1}\zeta^{-1}\right) .    
\end{align*}
Let's convert $\chi$ and $\zeta$ into cosines and expand $x$, $y$, $z$

\begin{align*}
\prod_{k}^m \prod_{l}^n 
 \cosh2H &\cosh 2H' \cosh 2H_3 + \sinh 2H \sinh 2H' \sinh 2H_3 \\
 &- \sinh 2H \left(\cos \frac{2k\pi}{m}\right)
 - \sinh 2H' \left(\cos \frac{2l\pi}{n}\right)- \sinh 2H_3 \left(\cos \frac{2k\pi}{m} \cos \frac{2l\pi}{n}\right)    
\end{align*}
For now, we do not want to stay in a finite matrix, but to turn $Z_{mn}^2$ to $Z$ when $m$ and $n$ go to infinity (which is the thermodynamic limit). Then as
$$\lim_{m,n \to \infty}(Z_{mn})^{\frac{1}{2mn}} = Z$$
We can rewrite our equation like
\begin{align*}
\ln(Z) = \lim_{m,n \to \infty} \frac{1}{2mn} \sum_{k}^m \sum_{l}^n &\ln \Biggl[
 \cosh2H \cosh 2H' \cosh 2H_3\\
 &+ \sinh 2H \sinh 2H' \sinh 2H_3
 - \sinh 2H \left(\cos \frac{2k\pi}{m}\right)\\
 &- \sinh 2H' \left(\cos \frac{2l\pi}{n}\right)- \sinh 2H_3 \left(\cos \frac{2k\pi}{m} \cos \frac{2l\pi}{n}\right) \Biggr]
\end{align*}
As we know the definition of the Riemann Integral formula
$$\int\limits_a^b f(x)dx =
\lim_{n\to\infty}\left[\frac{b-a}{n}\sum_{k=1}^n f\big(a+k\tfrac{b-a}{n}\big)\right]$$
With the aid of this formula, our partition function takes the form as
\begin{align*}
\ln(Z) =\frac{1}{8\pi^2} & \int_0^{2\pi} \int \{\cosh2H \ln \cosh 2H' \cosh 2H_3\\
 &+ \sinh 2H \sinh 2H' \sinh 2H_3
 - \sinh 2H \left(\cos w_1 \right)\\
 &- \sinh 2H' \left(\cos w_2 \right)- \sinh 2H_3 \left(\cos (w_1+ w_2)\right) \}dw_1 dw_2
\end{align*}
Which is the same as the algebraic solution of the exact partition function.

In conclusion, the combinatorial method introduced by Kac and Ward for solving Ising models provides a valuable framework, which was later corrected to find the exact results in the work of Potts and Ward. While initially appearing to be a rough approximation, especially when applied to infinite lattices (which do not exactly correspond to real physical systems), this approach seems to be effective for large-scale systems. By considering sufficiently large lattices, we can achieve the exact solutions for both triangular and rectangular models, with only minor corrections required to account for small errors. Also, it must be considered that one can reach the hexagonal solution from the triangular solution. Thus, this method offers a practical and efficient means of solving Ising models.

\section{Kasteleyn's Dimer Method and Pfaffian Technique \textit{(Ahmet Berk Kavruk and Irmak Özgüç)}}
\label{chapter:3}

\def\revdots{\mathinner{\mkern1mu\raise1pt\vbox{\kern7pt\hbox{.}}\mkern2mu\raise4pt\hbox{.}\mkern2mu\raise7pt\hbox{.}\mkern1mu}}

    We reproduce Onsager's result for the Helmholtz free energy of the two-dimensional Ising model by employing Kasteleyn's methodology on dimer arrangements \cite{kast1961, kast1963} and the comprehensive treatment by McCoy and Wu \cite{mccoywu}. We review the necessary graph theoretical concepts and discuss how the Pfaffian method is used to solve the perfect matching problem for dimers on square lattices with cylindrical and toroidal boundary conditions. By following established methodologies, we illustrate the equivalence between the dimer model and the Ising model partition function. This provides a combinatorial perspective on Onsager's result and highlights the profound connections between graph theory and statistical mechanics.

\subsection{Graphs, Perfect Matchings and Hafnians} \label{Graphs, Perfect Matchings and Hafnians}

Before going on with the solution, we present a few definitions relating to graph theory that will be needed as we go on. We start with the definition of a \textit{graph}, which is defined as an ordered triple
 $G = (V,E, \epsilon)$ where
 
\begin{itemize}
 \item the \textit{vertex set} $V = V(G)$ is a finite, nonempty set of $G$
 \item the \textit{edge set} $E = E(G)$ is a finite set of $G$
 \item and the \textit{endpoint function} $\epsilon : E \rightarrow \mathcal{P} (V)$ is a function such that, for all $\epsilon \in E$, $\epsilon (e)$ is either a one-element subset of $V$ or a two-element subset of $V$.
\end{itemize} 

\begin{figure}[ht]
    \centering
    \includegraphics[scale=0.7]{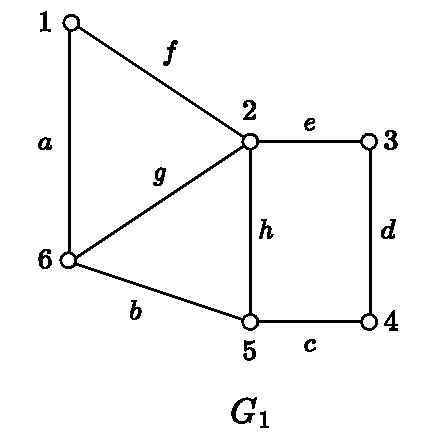}
    \caption{The graph $G_1$ with vertex set $V(G) = \{1, 2, 3, 4, 5, 6\}$, edge set $E(G) = \{a, b, c, d, e, f, g, h\}$ and endpoint function $\epsilon$.}
    \label{fig:graph_example}
\end{figure}

The edge $e$ is called a \textit{loop at vertex} $v$ if $\epsilon (e) = \{v\}$ and an \textit{edge from v to w} if $\epsilon(e) = \{ v, w\}$. The vertexes $v$ and $w$ are then called the \textit{endpoints} of  $e$
\cite{combinatorics}. One example of a graph is given in Figure \ref{fig:graph_example}.

\begin{figure}[h!]
    \centering
    \includegraphics[scale=0.7]{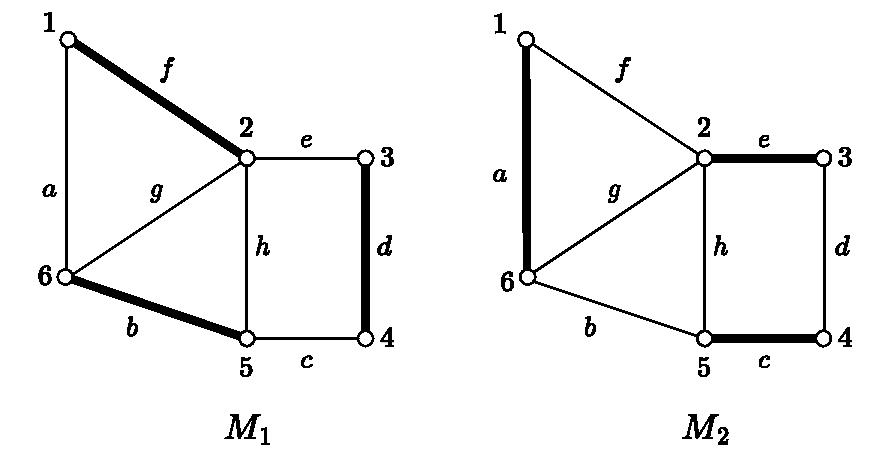}
    \caption{The matchings $M_1$ = \{\{1, 2\}, \{3, 4\} \{5, 6\}\} and $M_2$ = \{\{1, 6\}, \{2, 3\} \{4, 5\}\} }
    \label{fig:Perfect_Matching}
\end{figure}

In graph theory, a graph where every vertex is connected to precisely one of its nearest neighbors is called a \textit{perfectly matched} graph. A single graph can have multiple perfect matchings, as demonstrated in Figure \ref{fig:Perfect_Matching}. A fundamental question that arises is: "How many perfect matchings does a given graph possess?" This combinatorial problem, known as the "perfect matching problem," has significant ties to statistical mechanics under the name "dimer problem." While the perfect matching problem can be generalized to all graphs, this work will focus primarily on quadratic lattices.

 \begin{figure}[h!]
    \centering
    \includegraphics[scale=0.8]{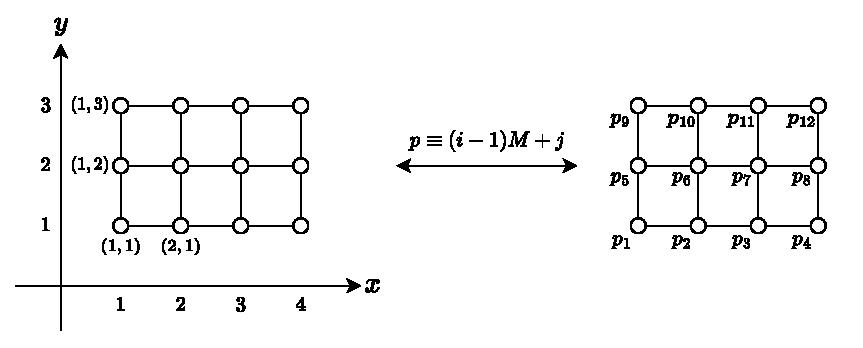} \caption{The graph $G(4,3)$ of $G(M,N)$ with $V=\{1, 2, \dots, (M \times N)\}$ and with $E = E_x \times E_y$, where 
    $E_x = \{ \{L, L+1\} : L \not\equiv 0 \ \ \text{(mod $M$)}\}$ $E_y = \{ \{L, L+M\} : 1 \leq L \leq M(N-1)\}.$}
    \label{fig:Quad_Lattice}
\end{figure}

\subsubsection{The Hafnian} \label{The Hafnian}
We could use the \textbf{Hafnian} when counting the number of perfect matchings in a graph. The Hafnian of a $2n \times 2n$ adjacency matrix is defined by 

\begin{equation}
        \bm{haf}A = \frac{1}{n!2^n}\sum_{\sigma \in S_{2n}} \prod_{j=1}^n A_{\sigma{(2j-1)}, \sigma{(2j)}},
\end{equation}
where $S_{2n}$ is the symmetric group on $[2n] = \{1, 2, \dots, 2n \}$, or equivalently 

\begin{equation*}
        \bm{haf}A = \sum_{M \in \mu } \prod_{(u,v) \in M} A_{u,v},
\end{equation*}
where $\mu$ is the set of all perfect matchings on the graph. Unfortunately, there is no effective way for us to calculate the Hafnian when the dimensions are large as the time complexity is $O(n^3 2^{n/2})$ \cite{bjorklund2019}. In the upcoming sections, however, we will see that we can do the calculation \textit{exactly}, also with an effective way of introducing the \textbf{Pfaffian}.


\subsection{Pfaffian, Transition Cycles and All That} \label{Pfaffian, Transition Cycles and All That}

To provide a foundation for later discussions, we introduce the quadratic lattice with free boundary conditions and show that \textbf{the Pfaffian} can be used to solve the "dimer problem". 

\subsubsection{On Lattices with Free Boundary Conditions} \label{On Lattices with Free Boundary Conditions}

Consider a quadratic $(M \times N)$ lattice to which one can attach dimers. By observation, one can see that at least $N$ or $M$ should be even to completely cover the lattice with dimers. We will take $M$ to be even for the rest of this chapter. We label the lattice sites as $1 \le i \le M$ for rows and $1 \le j \le N$ for columns, or by a single index $p$ where 

\begin{equation}\label{indexp}
    (i, j) \leftrightarrow p = (j-1)M + 1
\end{equation}

A dimer configuration on such lattice is denoted as 

\begin{equation}
    C = |p_1, p_2|p_3, p_4| \cdots |p_{(M \times N)-1}, p_{(M \times N)}|
\end{equation}

\begin{figure}[ht]
    \centering
    \includegraphics[scale=0.6]{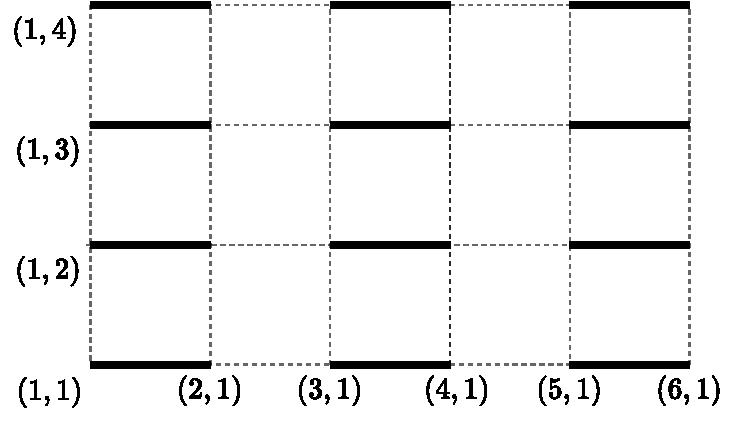}
    \caption{The configuration $C_0 = |1,2|3,4|\cdots|(M \times N)-1, (M \times N)|$.}
    \label{fig:C0_Configuration}
\end{figure}

However, this description for a dimer configuration is not unique. For example, the configuration $C_0$ in Figure \ref{fig:C0_Configuration} can also be described as $|2;1|3;4|\cdots$. To make the description unique, the following restrictions on the lattice points are added 

\begin{subequations}\label{restrictions}
\begin{equation}\label{restrict1}
    p_1 < p_2, \ p_3 < p_4,  \cdots, \ p_{(M \times N)-1} < p_{(M \times N)}
\end{equation}
\vspace{-4mm}
\begin{equation}\label{restrict2}
    p_1 < p_3 < p_5, \cdots, < p_{(M \times N)-1}. 
\end{equation}
\end{subequations}

\begin{figure}[h]
    \centering
    \includegraphics[scale=0.80]{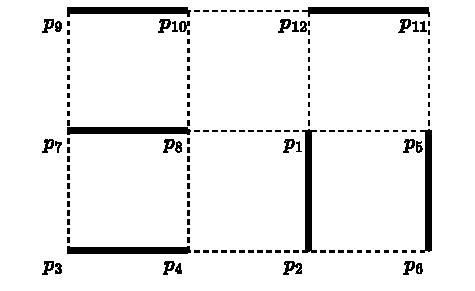}
    \caption{Both the conditions $  p_1 < p_2, p_3 < p_4, \cdots, p_{(M \times N)-1} < p_{(M \times N)}$ and $p_1 < p_3 < p_5, \cdots, < p_{(M \times N)-1}$ do not hold.}
    \label{fig:5no6no}
\end{figure}

We provide the following example to make these restrictions more clear. Consider the configuration given in 
Figure \ref{fig:5no6no}. The restrictions \eqref{restrictions} do not apply here.

\begin{figure}[H]
    \centering
    \includegraphics[scale=0.80]{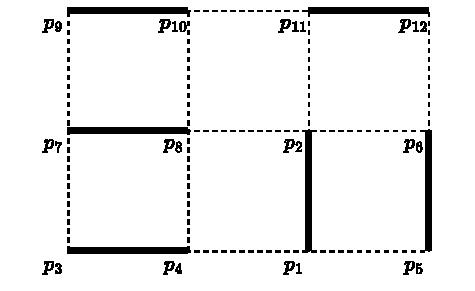}
    \caption{The condition $  p_1 < p_2, p_3 < p_4, \cdots, p_{(M \times N)-1} < p_{(M \times N)}$ holds but $p_1 < p_3 < p_5, \cdots, < p_{(M \times N)-1}$ does not hold.}
    \label{fig:5yes6no}
\end{figure}

A restriction that seems obvious is to require that in a dimer made up of $p_i$ and $p_{i+1}$, $p_{i+1}$ should stand in the position of the "larger" $p$ defined in \eqref{indexp}, thus we add the restriction \eqref{restrict1}. When it is applied, the configuration looks as in Figure \ref{fig:5yes6no}.

\begin{figure}[H]
    \centering
    \includegraphics[scale=0.80]{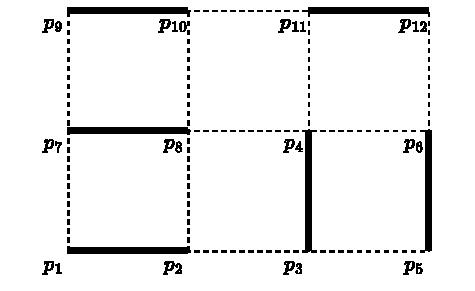}
    \caption{Both the conditions $  p_1 < p_2, p_3 < p_4, \cdots, p_{(M \times N)-1} < p_{(M \times N)}$ and $p_1 < p_3 < p_5, \cdots, < p_{(M \times N)-1}$ hold.}
    \label{fig:5yes6yes}
\end{figure}

Now that the "monomers" in a dimer are ordered, we go on to order the dimers as in \eqref{restrict2}. When both restrictions are applied, we get the configuration in Figure \ref{fig:5yes6yes}.

For an $(M \times N)$ lattice, we have a corresponding \textbf{adjacency matrix}  $A(p_{2j-1},p_{2j})$ with its non-zero elements denoted as

\begin{equation}
    A(p_{2j-1},p_{2j}) = z_i
\end{equation}

where $z_i$ refers to a class of bonds; in our case, $z_1$ refers to horizontal bonds, and $z_2$ refers to vertical bonds of dimers. The configuration generating function can be written as 

\begin{equation}\label{genfunc}
    Z^F_{M, N} = \sum_P A(p_1, p_2) A(p_3, p_4) \cdots A(p_{(M \times N)-1}, p_{(M \times N)}) 
\end{equation}
    
which is the Hafnian we defined previously. However, as mentioned before, there is no efficient way of calculating a Hafnian for large $M$ and $N$. Therefore, we introduce a new object that can provide an efficient way of calculating the generation function.

\subsubsection{The Pfaffian} \label{The Pfaffian}

    The Pfaffian of an $2n \times 2n$ adjacency matrix is defined as \cite{combinatorics}
    
\begin{equation*}
        \bm{Pf}A = \sum_{P \in S_{2n}} \delta_P \prod_{j=1}^n A(p_{2j-1}, p_{2j}),
\end{equation*}    
where $S_{2n}$ is the symmetric group on $[2n] = \{1, 2, \dots, 2n \}.$ For all $p_{2j-1}$, $p_{2j}$ with $1 \le j \le n$ we have the conditions
    
\begin{subequations}\label{conditions}
 \begin{equation}
      p_1 < p_2, \; p_3 < p_4, \; \cdots \; p_{2n-1} < p_{2n}
  \end{equation}
\vspace{-4mm}
  \begin{equation}
      p_1 < p_3 < \cdots < p_{2n-1}.
\end{equation}   
\end{subequations}

An important property of the Pfaffian for antisymmetric matrices is 

\begin{equation}
    \sqrt{\textbf{det}A} = \pm \textbf{Pf}A
\end{equation}
  
Since conditions \eqref{conditions} are exactly the same as the restrictions \eqref{restrictions}, it is easy to see that the Hafnian \eqref{genfunc} differs from a Pfaffian just by a factor of $\delta_P$. Say we select $A(p_{2j-1},p_{2j})$ as

\begin{equation}
    A(p_{2j-1},p_{2j})= S(p_{2j-1},p_{2j}) z_i
\end{equation}

with $|S(p_{2j-1},p_{2j})| = 1$. If there are such $S(p_{2j-1},p_{2j})$ that cancel the $\delta_p$ factor in the Pfaffian, the configuration generating function can be calculated using a Pfaffian. In other words, we want to show that there are $S(p_{2j-1},p_{2j})$ such that

\begin{equation}\label{z and pf}
    Z^F_{M, N} = \bm{Pf} A_F
\end{equation}

where $A_F$ is the following $(M \times N) \times (M \times N)$ matrix with non-zero elements

\begin{subequations}\label{free matrix elements}
 \begin{equation}
    A(i,j; i+1, j) = -A(i+1, j; i, j) = z_1, \; \; \; \; \;
    1 \le i \le M, \ 1 \le j \le N-1
\end{equation}
\vspace{-4mm}
\begin{equation}
    A(i,j; i, j+1) = -A(i, j+1; i, j) = (-1)^i z_2, \; \; \; \; \;
    1 \le i \le M-1, \ 1 \le j \le N
\end{equation}   
\end{subequations}

\begin{figure}[ht]
    \centering
    \includegraphics[scale=0.9]{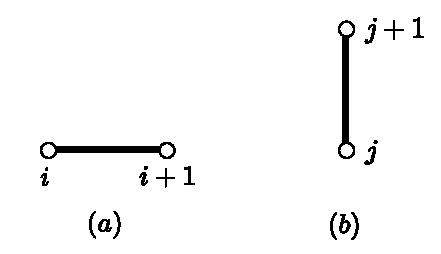}
    \label{fig:adjacencymatrixterms}
    \caption{\textit{(a)}, the bond corresponding to the matrix element $A(i,j; i+1, j) = -A(i+1, j; i, j) = z_1$ , and \textit{(b)}, the bond corresponding to the matrix element $A(i,j; i, j+1) = -A(i, j+1; i, j) = (-1)^i z_2$.}
\end{figure}

To show that such $S(p_{2j-1},p_{2j})$ exist, it is sufficient to show that $S(p_{2j-1},p_{2j})$ can be chosen as

\begin{equation}\label{choosing s}
    \delta_{p^{(1)}} S(p^{(1)}_{1},p^{(1)}_{2}) \cdots S(p^{(1)}_{MN-1},p^{(1)}_{MN}) = \delta_{p^{(2)}} S(p^{(2)}_{1},p^{(2)}_{2}) \cdots S(p^{(2)}_{MN-1},p^{(2)}_{MN})
\end{equation}

where $p^{(1)}$ and $p^{(2)}$ are any two permutations. This way, any two permutations in the Pfaffian will have the same sign, and thus \eqref{z and pf} will be satisfied. To make things simpler, we consider any two permutations $\bar{p}^{(1)}$ and $\bar{p}^{(2)}$ that do not satisfy \eqref{restrictions}. \eqref{choosing s} still holds for these permutations.

\begin{equation}
    \delta_{\bar{p}^{(1)}} S(\bar{p}^{(1)}_{1},\bar{p}^{(1)}_{2}) \cdots S(\bar{p}^{(1)}_{MN-1},\bar{p}^{(1)}_{MN}) = \delta_{\bar{p}^{(2)}} S(\bar{p}^{(2)}_{1},\bar{p}^{(2)}_{2}) \cdots S(\bar{p}^{(2)}_{MN-1},\bar{p}^{(2)}_{MN})
\end{equation}

To further our discussion, we introduce three new definitions: \textit{transition graphs}, \textit{transition cycles}, and \textit{double bonds} \cite{kast1963}.

\begin{figure}[h]
    \centering
    \includegraphics[scale=0.45]{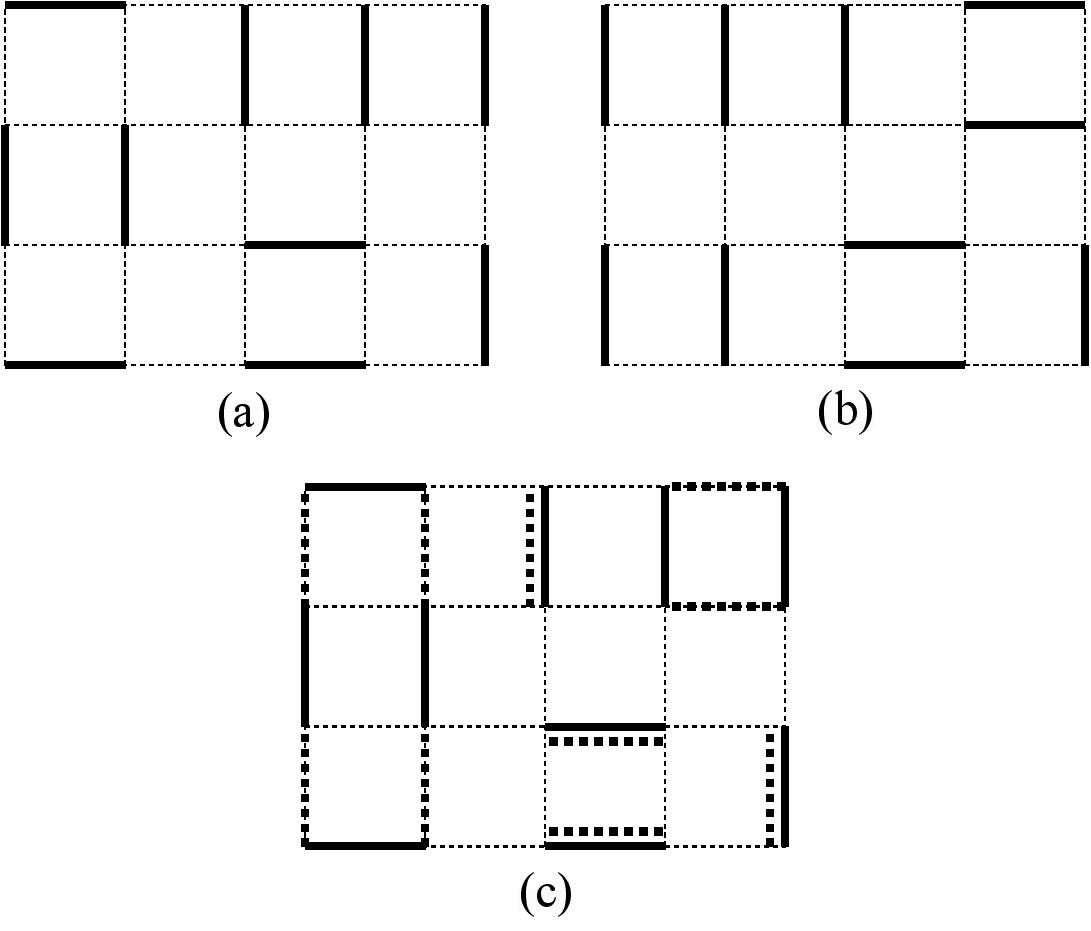}
    \caption{\textit{(a)} The configuration $C_1$. \textit{(b)} The configuration. $C_2$ \textit{(c)} The transition graph of $C_1$ and $C_2$. The closed polygons in the transition graph are called \textit{transition cycles}.}
    \label{fig:Transition Graph}
\end{figure}

Consider any two configurations $C_1$ and $C_2$. When the dimer bonds of both configurations are drawn on the same lattice, the figure obtained is called a \textit{transition graph}. As seen in Figure \ref{fig:Transition Graph}, transition graphs consist of two elements

 \begin{itemize}
        \item \textbf{Double bonds}: Two points connected by the bonds of both configurations.
        \item \textbf{Transition cycles}: Closed polygons with an even number of sides alternating between bonds of $C_1$ and $C_2$. 
    \end{itemize}   

We will first look at transition graphs that differ from each other by one transition cycle. 

\begin{figure}[H]
    \centering
    \includegraphics[scale=0.7]{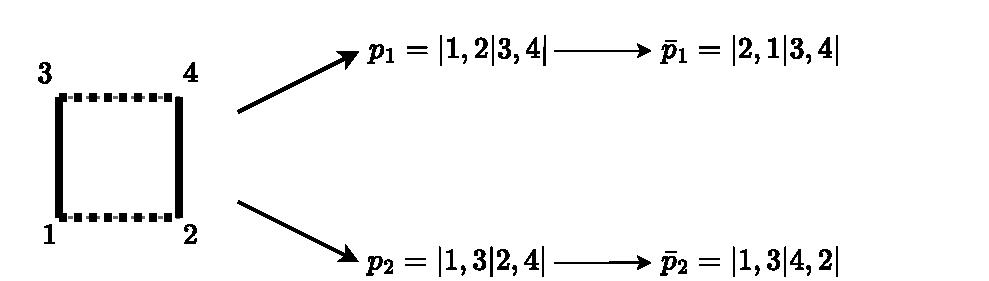}
    \caption{A $2\times2$ transition cycle and the permutations $p_1$, $\bar{p}_1$, $p_2$ and $\bar{p}_2$}
    \label{fig:2x2 transition graph}
\end{figure}

Consider two permutations $p_1$ and $p_2$ that obey our restrictions. These permutations have at least one corresponding permutation $\bar{p}_1$ and $\bar{p}_2$ that do not obey the restrictions. For example, take the transition cycle in Figure \ref{fig:2x2 transition graph}.

\begin{figure}[H]
    \centering
    \includegraphics[scale=0.6]{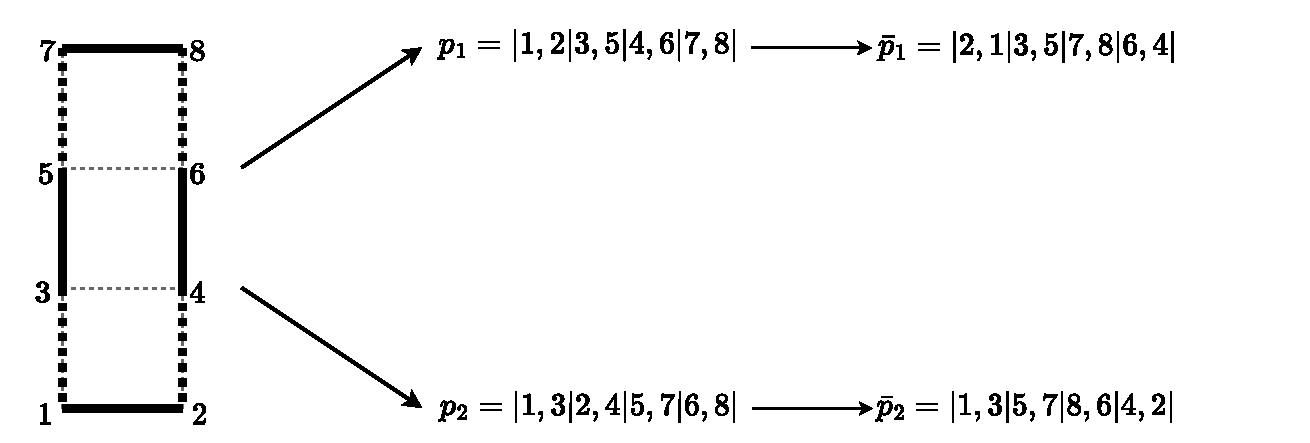}
    \caption{A $2\times4$ transition cycle and the permutations $p_1$, $\bar{p}_1$, $p_2$ and $\bar{p}_2$}
    \label{fig:2x4 transition graph}
\end{figure}

Here, the permutation $p_1$ is  $|1,2|3,4|$ and $p_2$ is $|1,3|2,4|$. The corresponding permutations $\bar{p}_1$ and $\bar{p}_2$ are chosen as the permutations in which the sites are ordered clockwise. So, $\bar{p}_1 = |2,1|3,4|$ and $\bar{p}_2 = |1,3|4,2|$. Let's look at another example, the transition cycle in Figure \ref{fig:2x4 transition graph}.

Here, we have $p_1$ as $|1,2|3,5|4,6|7,8|$ and $p_2$ as $|1,3|2,4|5,7|6,8|$. The corresponding permutations are $|2,1|3,5|7,8|6,4|$ and $|1,3|5,7|8,6|4,2|$ for $\bar{p}_1$ and $\bar{p}_2$ respectively.

In general; when we have permutations $p_1$ and $p_2$ and corresponding permutations $\bar{p}_1$ and $\bar{p}_2$, in which the sites are ordered clockwise, if 

\begin{equation}
    \bar{p}^{(1)} = |\bar{p}^{(1)}_1,\bar{p}^{(1)}_2| \bar{p}^{(1)}_3,\bar{p}^{(1)}_4|\cdots |\bar{p}^{(1)}_{(M \times N)-1},\bar{p}^{(1)}_{(M \times N)}|
\end{equation}

then

\begin{equation}
    \bar{p}^{(2)} = |\bar{p}^{(1)}_2,\bar{p}^{(1)}_3| \bar{p}^{(1)}_4,\bar{p}^{(1)}_5|\cdots |\bar{p}^{(1)}_{(M \times N)},\bar{p}^{(1)}_{1}|.
\end{equation}

If there is only one transition cycle, we have

\begin{equation}
    \delta_{\bar{p}^{(1)}} = - \delta_{\bar{p}^{(2)}}
\end{equation}

and for $t$ transition cycles

\begin{equation}
    \delta_{\bar{p}^{(1)}} = (-1)^t \delta_{\bar{p}^{(2)}}
\end{equation}

Therefore, any two permutations $p_1$ and $p_2$ will have the same sign if for each transition cycle

\begin{equation}
    S(\bar{p}_1^{(1)}, \bar{p}_2^{(1)}) S(\bar{p}_3^{(1)}, \bar{p}_4^{(1)}) \cdots S(\bar{p}_{2N-1}^{(1)}, \bar{p}_{2N}^{(1)}) =
    -S(\bar{p}_2^{(1)}, \bar{p}_3^{(1)}) S(\bar{p}_4^{(1)}, \bar{p}_5^{(1)}) \cdots S(\bar{p}_{2N}^{(1)}, \bar{p}_{1}^{(1)}) 
\end{equation}

or, in other words,

\begin{equation}\label{product of factors}
    \prod_{k=1}^{2N} S(\bar{p}_{k}^{(1)}, \bar{p}_{k+1}^{(1)}) = -1
\end{equation}

where $\bar{p}^{(1)}_{2N+1} \equiv \bar{p}^{(1)}_1$. 

From \eqref{product of factors}, we see that there are an odd number of factors $S(p_k, p_{k+1})$ with a minus sign. But how are the factors $S(p_k, p_{k+1})$ interpreted? An obvious interpretation is by drawing an arrow from the site $p_1$ to $p_2$ if $S(p_k, p_{k+1}) = 1$ and vice versa. 

\begin{figure}[ht]
    \centering
    \includegraphics[scale=0.9]{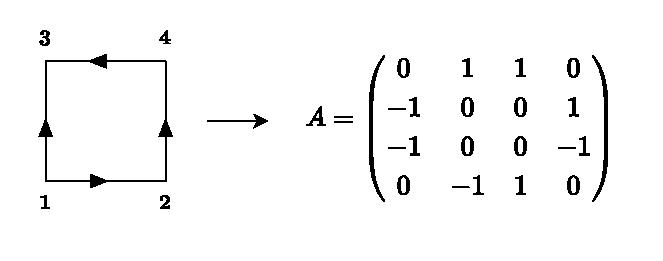}
    \caption{An oriented $2 \times 2$ lattice and its adjacency matrix. The factors $S(p_1, p_2)$ are interpreted by drawing an arrow from $p_1$ to $p_2$ if $S(p_1, p_2) = +1$ and vice versa.}
    \label{fig:adjacencymatrix2x2}
\end{figure}

Let's verify this with an example. Consider the oriented lattice and its adjacency matrix given in Figure \ref{fig:adjacencymatrix2x2}. 

Now let's check for this adjacency matrix that if Equation \eqref{choosing s} holds, all terms in the Pfaffian have the same sign.

\begin{equation*}
    \prod_{k=1}^{2N} S(\bar{p}_{k}^{(1)}, \bar{p}_{k+1}^{(1)}) = S(\bar{p}^{(1)}_1, \bar{p}^{(1)}_2)S(\bar{p}^{(1)}_2, \bar{p}^{(1)}_3)S(\bar{p}^{(1)}_3, \bar{p}^{(1)}_4)S(\bar{p}^{(1)}_4, \bar{p}^{(1)}_5)
\end{equation*}

The permutation was $\bar{p}^{(1)} = |2,1|3,4|$. Then, the product equals to 

\begin{equation*}
    S(2,1)S(1,3)S(3,4)S(4,2) = (-1) \times (1) \times (-1) \times (-1) = -1
\end{equation*}

Let's see if the terms in the Pfaffian have the same sign. The definition of the Pfaffian for any matrix was

\begin{equation*}
        \bm{Pf}A = \sum_{\sigma \in S_{2n}} \delta_p \prod_{j=1}^n A_{\sigma{(2j-1)}, \sigma{(2j)}},
\end{equation*} 

so the terms for our adjacency matrix $A$ are

\begin{equation*}
    \delta_p A_{12}A_{34} = 1 \times 1 \times (-1) = -1
\end{equation*}
\vspace{-5mm}
\begin{equation*}
    \delta_p A_{13}A_{24} = (-1)\times 1 \times 1 = -1
\end{equation*}
\vspace{-5mm}
\begin{equation*}
    \delta_p A_{14}A_{23} = (-1) \times 0 \times 0 = 0
\end{equation*}

Clearly, all terms of the Pfaffian have the same sign. To give a geometric interpretation, we introduce a new definition:

\textbf{Definition:} \textit{Orientation Parity of a Transition Cycle.} If the number of arrows pointing in either direction as we traverse the transition cycle is odd, the orientation parity of the transition cycle is said to be $(-1)$ and vice versa.

We previously noted that there are an odd number of factors $S(p_k, p_{k+1})$ with a minus sign, which is another way of saying that the orientation parity of a transition cycle is odd. Thus the following theorem is confirmed

\textbf{Theorem A}: If the orientation parity of every transition cycle in the lattice corresponding to the adjacency matrix $A_F$ is odd, then all terms in $\pm \bm{Pf}A_F$ will have the same sign, i.e. $\pm \bm{Pf}A_F = Z^F_{M, N} $

Although \textit{Theorem A} is confirmed, we still need to show that the orientation parity of every transition cycle in the lattice is odd. Let's try to obtain a characterization of transition cycles. We start by stating the following theorem: 

\textbf{Theorem B:} On a planar lattice, every transition cycle contains an even number of sites.

\begin{figure}[ht]
    \centering
    \includegraphics[scale=0.6]{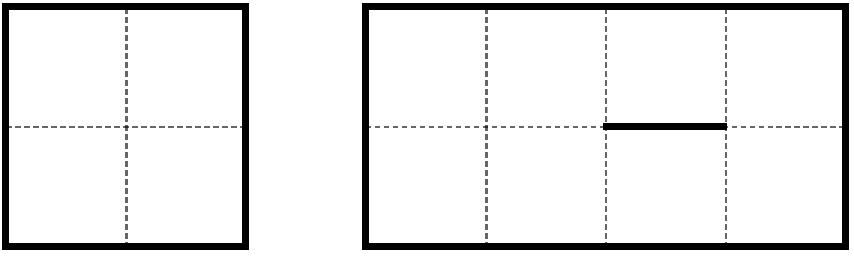}
    \caption{Polygons that are not transition cycles.}
    \label{fig:theoremb}
\end{figure}

\textit{\textbf{Proof:}} The interior of the transition cycle needs to be filled with transition cycles or double bonds, which require an even number of sites. If there are an odd number of sites inside the transition cycle, that means that there is a site that is not involved in a dimer bond, which goes against the definition of perfect matching. Thus, there can only be an even number of sites inside a transition cycle.

To use this property of transition cycles, we introduce the \textit{elementary polygon}:

\textbf{Definition:} \textit{Elementary Polygon.} A closed square with 4 sites and 4 sides is called an elementary polygon. 

A polygon with an odd number of arrows pointing in the clockwise direction is called \textit{clockwise odd} and vice versa. With this definition, we move on to the next theorem

\textbf{Theorem C:}
On a planar lattice, the arrows can always be oriented in a way that every elementary polygon is clockwise odd.

\begin{figure}[ht]
    \centering
    \includegraphics[scale=0.6]{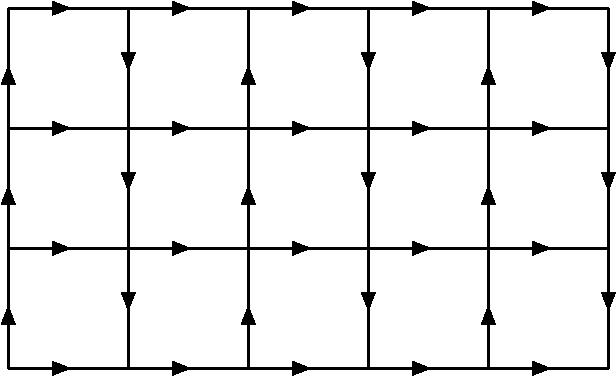}
    \caption{An oriented lattice in which all the elementary polygons are clockwise odd}
    \label{fig:theoremc}
\end{figure}

Instead of proving this theorem, we simply state that it is sufficient and much more efficient to verify \textit{Theorem C} directly on the lattice of interest as can be seen in Figure \ref{fig:theoremc}. Thus, \textit{Theorem A} is proven for all elementary polygons. To complete the proof for all transition cycles, we need to prove

\textbf{Theorem D:}
On an oriented planar lattice; if every elementary polygon is clockwise odd, then the orientation parity of any transition cycle is odd if the transition cycle encloses an even number of sites.

\begin{figure}[H]
    \centering
    \includegraphics[scale=0.4]{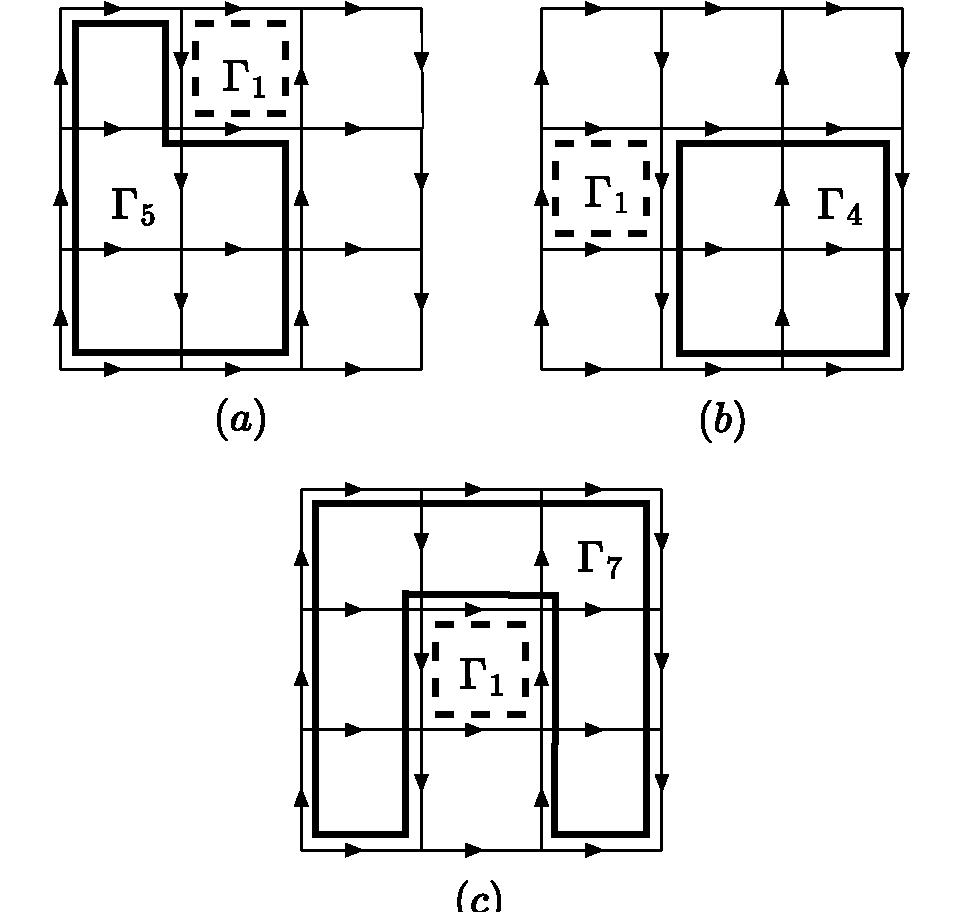}
    \caption{Examples for Theorem D: $a + a' - c$ must be odd if $p + c - 1$ is even }
    \label{fig:theoremdnew}
\end{figure}

\textit{\textbf{Proof:}} We start the proof by remarking that any polygon on the lattice can be thought of as being made up of a number of elementary polygons. Now, let $\Gamma_n$ be a polygon made up of $n$ elementary polygons. Say $\Gamma_n$ surrounds $p$ lattice points (even by assumption), contains $a$ clockwise arrows (odd by assumption), and has $c$ arrows common with the elementary polygon $\Gamma_1$. If $\Gamma_1$ contains $a'$ clockwise arrows (odd by \textit{Theorem C}), then $\Gamma_{n+1}$ contains $a + a' - c$ clockwise arrows and $p+c-1$ points. Since $\Gamma_{n+1}$ is a transition cycle, $p+c-1$ is even, which means that $c$ must be odd. Then, $a+a'-c$ is odd. Therefore, \textit{Theorem D} follows by induction.

We have now proven \textit{Theorem A}, which states that

\begin{equation}\label{eqn:paf_det}
    Z^F_{M, N} = \pm \bm{Pf} A_F
\end{equation}

Since all the configurations must have the same sign, we can see that $Z^F_{M, N} = +\bm{Pf} A_F$ by checking the configuration $C_0$.

\subsection{Evaluation of the Pfaffians} \label{Evaluation of the Pfaffians}

    Now we know how the elements of the adjacency matrix will be:

    \begin{equation}
        \begin{split}
            D(i, j; i+1, j) &= +z_1 \qquad \quad \ \ \text{for} \ 1 \leq i \leq m - 1, \ 1 \leq j \leq n \\
            D(i, j; i, j+1) &= (-1)^i z_2 \qquad \text{for} \ 1 \leq i \leq m, \ 1 \leq j \leq n - 1 \\
            D(i, j; i', j') &= 0 \qquad \qquad \qquad \quad \qquad \text{otherwise}
        \end{split}
    \end{equation}

    \begin{figure}[H]
        \centering
        \includegraphics[scale=0.8]{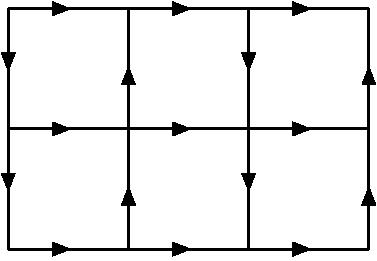}
        \caption{The oriented graph for $G(4,3)$}
        \label{fig:4-3graph}
    \end{figure}

Let's look at an example. If we take $M = 4$ and $N= 3$, we get the matrix for graph $G(4, 3)$ seen in Figure \ref{fig:4-3graph} with $z_1, z_2 = 1$ as

    \begin{equation}
        D =
        \begin{bmatrix}
             &  1  &     &     &  -1  &     &     &     &     &     &     &     \\
          -1 &     &  1  &     &     &  1 &     &     &     &     &     &     \\
             &  -1 &     &  1  &     &     &  -1  &     &     &     &     &     \\
             &     &  -1 &     &     &     &     &  1 &     &     &     &     \\
          1 &     &     &     &     &  1  &     &     &  -1  &     &     &     \\
             &  -1  &     &     &  -1 &     &  1  &     &     & 1  &     &     \\
             &     &  1 &     &     &  -1 &     &  1  &     &     &  -1  &     \\
             &     &     &  -1  &     &     & -1  &     &     &     &     &  1 \\
             &     &     &     &  1 &     &     &     &     &  1  &     &     \\
             &     &     &     &     &  -1  &     &     &  -1 &     &  1  &     \\
             &     &     &     &     &     &  1 &     &     &  -1 &     &  1  \\
             &     &     &     &     &     &     &  -1  &     &     &  -1 &    \\ 
        \end{bmatrix}
    \end{equation}

Now that we have the picture of the matrix, we only need to compute the determinant. To achieve this, we must first play with the matrix to make our job easier. We will do this with the well-known \textbf{Kronecker Product}.

\textbf{Definition.} (Kronecker Product) The \textbf{Kronecker Product} of the matrix $A \in \mathbb{M}^{p,q}$ with the matrix $B \in \mathbb{M}^{r,s}$ is defined as

\begin{equation*}
    A \otimes B = 
    \begin{bmatrix}
        a_{11} B & a_{12} B & \cdots & a_{1q} B\\
        a_{21} B & a_{22} B & \cdots & a_{2q} B\\
        \vdots  & \vdots  & \ddots & \vdots  \\
        a_{p1} B & a_{p2} B & \cdots & a_{pq} B
    \end{bmatrix}
\end{equation*}

    Kronecker product has the following properties where $A, B, C, D$ are matrices and $I_m$ is the $m \times m$ identity matrix.

    \begin{align*}
        I_m \otimes I_n &= I_{m\times n} \\
        (A \otimes B)(C \otimes D) &= (AC) \otimes (BD) \\
        (A \otimes B)^{-1} &= A^{-1} \otimes B^{-1} \\
        (A \otimes B)^T &= A^T \otimes B^T
    \end{align*}

   Notice that the dimension of $I_{m \times n}$ is $(m \times n) \times (m \times n)$. To use the Kronecker Product we introduced, we define the matrices $Q_k$ and $F_k$

    \begin{equation*}
        Q_k =
        \begin{bmatrix}
        0 & 1 & 0 & \cdots & 0 & 0 \\
        -1 & 0 & 1 &  \cdots & 0 & 0 \\
        0 & -1 & 0 &  \cdots & 0 & 0 \\
        \vdots  &  \vdots & \vdots & \vdots & \vdots & \vdots  \\
        0 & 0 & 0 & \cdots & 0 & 1 \\
        0 & 0 & 0 & \cdots & -1 & 0
        \end{bmatrix}
        , \hspace{4mm} F_k =
        \begin{bmatrix}
        -1 & 0 & 0 & \cdots & 0 & 0 \\
        0 & 1 & 0 & \cdots & 0 & 0 \\
        0 & 0 & -1 & \cdots & 0 & 0 \\
        \vdots  & \vdots  & \vdots & \vdots & \vdots & \vdots \\
        0 & 0 & 0 & \cdots & -1 & 0\\
        0 & 0 & 0 & \cdots & 0 & 1
        \end{bmatrix}
    \end{equation*}

    Using $Q_k$ and $F_k$, we can rewrite the matrix $D$ as

    \begin{equation}
        D = z_1 (I_N \otimes Q_M) + z_2 (Q_N \otimes F_M)
    \end{equation}

    For $M = 4$ and $N=3$, the picture of the matrix comes as the following:

    \begin{figure}[h]
        \centering
        \includegraphics[scale=0.60]{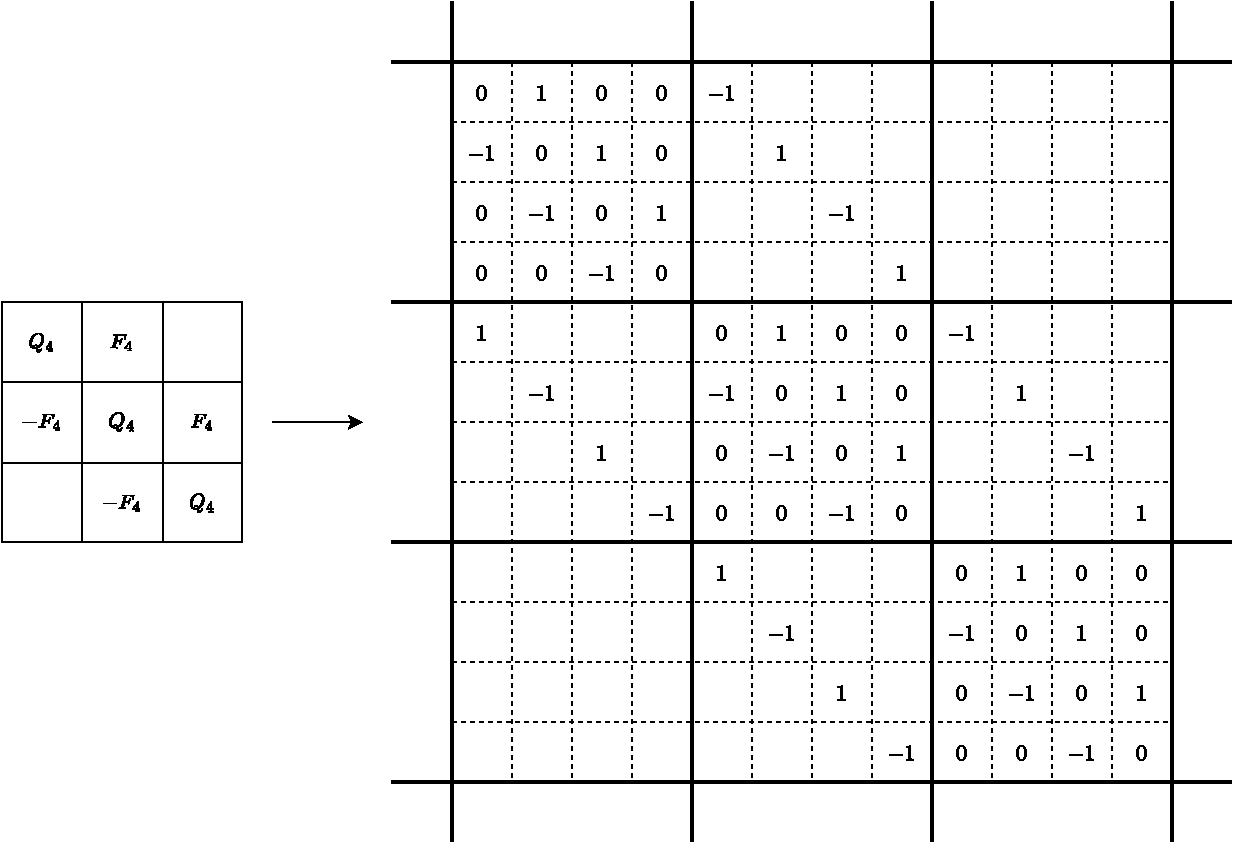}
        \label{fig:matrixpicture}
    \end{figure}

Turning back to the general $(M \times N) \times (M \times N)$ matrix at hand, we notice that the matrix $Q_k$ appears at both $z_1$ and $z_2$ side of the summation. So, we try to diagonalize it. Consider the eigenvalue equation $Qx = \lambda x$, in component form
\vspace{-2mm}
\begin{align*}
    x_2 &= \lambda x_1 \\
    -x_{k-1} + x_{k+1} &= \lambda x_k \\
    -x_{N-1} &= \lambda x_N
\end{align*}

It is the same as the difference equation 

\begin{equation}
    -v_{x-1} + v_{x+1} = \lambda x_k
\end{equation}

with the boundary conditions

\begin{equation}
    x_0 = x_{N+1} = 0
\end{equation}

   In the end, we get our eigenvalues $\lambda_k$ and our similarity transformation matrix $U_k$ for $Q_k$ as follows

    \begin{equation}
        U_k(l, l') = i^l \sin \left(\frac{\pi l l'}{k+1} \right), \qquad \qquad \lambda_k(l) = 2i \cos \left(\frac{\pi l}{k+1} \right).
    \end{equation}

If we let $U = U_N \otimes U_M$ operate on $D$, we get

\begin{equation}
            U^{-1} D U = z_1 (I_N \otimes \Tilde{Q}_M) +  z_2(\Tilde{Q}_N \otimes U^{-1}_M F_M U_M) 
\end{equation}

     With $\Tilde{D} = z_1 (I_N \otimes \Tilde{Q}_M) +  z_2(\Tilde{Q}_N \otimes U^{-1}_M F_M U_M)$, our diagonalization is nearly complete. Now focusing our attention to $F_MU_M = U_MV_M$, we see

    \begin{equation}
    \hspace{-2mm}
        \begin{bmatrix}
        -1 & 0 & 0 & \cdots & 0\\
        0 & 1  &  0 & \cdots & 0\\
        0 & 0  & -1  &  \cdots & 0\\
        \vdots  & \vdots   & \vdots & \vdots & \vdots\\
        0 & 0 & 0 & \cdots   &1
        \end{bmatrix}
        \begin{bmatrix}
        u_{11} & u_{12} & u_{13} & \cdots & u_{1M} \\
        u_{21} & u_{22} & u_{23} & \cdots & u_{2M} \\
        u_{31} & u_{32} & u_{33} & \cdots & u_{3M} \\
        \vdots & \vdots & \vdots & \vdots & \vdots \\
        u_{M1} & u_{M2} & u_{M3} & \cdots & u_{MM}
        \end{bmatrix}    
        =
        \begin{bmatrix}
        -u_{11} & -u_{12} & -u_{13} & \cdots & -u_{1M} \\
        u_{21} & u_{22} & u_{23} & \cdots & u_{2M} \\
        -u_{31} & -u_{32} & -u_{33} & \cdots & -u_{3M} \\
        \vdots & \vdots & \vdots & \vdots & \vdots \\
        u_{M1} & u_{M2} & u_{M3} & \cdots & u_{MM}
        \end{bmatrix}
    \end{equation}

    We can deduce that all of the entries have the following form

    \begin{equation}
        (UV)_{kj} = (-1)^k  i^k \sin \left(\frac{\pi kj}{M+1} \right) = \cos{(\pi k)} i^k \sin \left(\frac{\pi kj}{M+1} \right)
    \end{equation}

    Turning this into a sum of angles equation, we get
    
    \begin{align}
            i^k\sin\left(\frac{\pi k(M+1) - \pi kj}{M+1} \right) = i^k \left[ \sin(\pi k)\cos\left(\frac{\pi kj}{M+1} \right) \right.  
            \left. - \cos(\pi k) \sin\left(\frac{\pi kj}{M+1} \right) \right] 
    \end{align}

    now, since $\sin(\pi k) = 0$, in the end, we have

    \begin{gather}
        U_M(k, M+1-j) = i^k\sin\left(\frac{\pi k(M+1) - \pi kj}{M+1} \right) = - \cos(\pi k) \sin\left(\frac{\pi kj}{M+1} \right) \\ \label{therefore}
        \therefore -U_M(k, M+1-j) = (-1)^k U_M(k, j)
    \end{gather}

In the expression (\ref{therefore}), we can notice the change in entry numbers 
    
    \begin{align*}
        k &\longrightarrow k \\
        j &\longrightarrow M+1-j
    \end{align*}

    See that the row entry $k$ stays the same but the column entry $j$ does not. Then left-multiplying $U_m$ with $F_m$ is practically the same thing as right-multiplying the $U_m$ with a permutation matrix. Therefore we have $F_MU_M = U_MV_M$ with 

    \begin{equation}
        V_M = -
        \begin{bmatrix}
        0 & 0 & \cdots & 1 \\
        \vdots & \vdots & \vdots & \vdots \\
        0 & 1 & \cdots & 0 \\
        1 & 0 & \cdots & 0
        \end{bmatrix} 
    \end{equation}

    In the end, we get 

    \begin{equation}
        \Tilde{D} = z_1(I_N \otimes \Tilde{Q}_M) + z_2(\Tilde{Q}_N \otimes V_M)
    \end{equation}

    Let's see what $\Tilde{D}$ looks like when $M=4$ and $N=3$ while $z_1 = z_2 = 1$

    \begin{equation}
        \Tilde{D} = 
        \begin{bmatrix}
        \Tilde{Q}_4 + \lambda_3(1) V_4 & 0 & 0 \\
         0          & \Tilde{Q}_4 + \lambda_3(2) V_4 & 0 \\
         0          & 0 &  \Tilde{Q}_4 + \lambda_3(3) V_4  \\
        \end{bmatrix} 
    \end{equation}

    The resulting matrix $\Tilde{D}$ has diagonally aligned block matrices that look like the letter \textbf{X}, for example, the block with the bottom right entry $(xm, xm)$ is

\begin{equation}\label{rowcolumn_x}
    \begin{bmatrix}
    z_1 \lambda_M(1)&         &                                        &                                           &        &   -z_2 \lambda_N(x)   \\
                    & \ddots  &                                        &                                           &\revdots&                        \\
                    &         & z_1 \lambda_M\left( \frac{M}{2} \right)& -z_2 \lambda_N(x)                         &        &                       \\
                    &         & -z_2 \lambda_N(x)                       & z_1 \lambda_M\left( \frac{M}{2}+1 \right)&        &                       \\
                    & \revdots&                                        &                                           & \ddots &                        \\
    -z_2 \lambda_N(x)&        &                                        &                                           &        & z_1 \lambda_M(m)       \\
    \end{bmatrix}
\end{equation}

\begin{figure}[h!]
    \centering
    \includegraphics[scale=0.45]{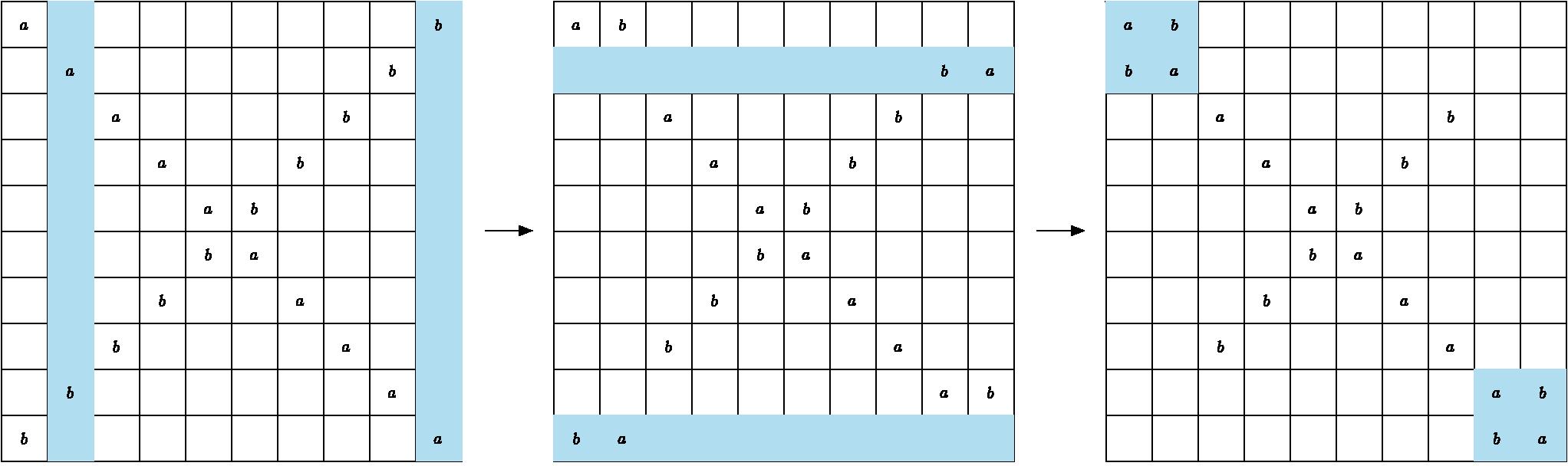}
    \caption{Row and column operations with the resulting matrix.}
    \label{fig:rowcolumn}
\end{figure}

We can further diagonalize the submatrix (\ref{rowcolumn_x}) with row and column operations seen in Figure \ref{fig:rowcolumn}. The signs will not be changed since we do two operations per sub-submatrix.

    At last, we can take the determinant of $\Tilde{D}$. 
    \vspace{-2mm}
    \begin{align*}
        \textbf{det}\Tilde{D} = \prod_{k=1}^{M/2} \prod_{j=1}^N \textbf{det} 
        \begin{bmatrix}
        z_1\lambda_M(k) & -z_2\lambda_N(j) \\
        -z_2\lambda_N(j) & z_1 \lambda_M(M+1-k)  
        \end{bmatrix} 
    \end{align*}

    Looking at $\lambda_M(M+1-k)$
    \vspace{-2mm}
\begin{equation*}
        \lambda_M(M+1-k) = 2i \cos\left(\pi - \frac{k}{M+1} \right) = -2i\cos\left(\frac{k}{M+1}\right) = -\lambda_M(k)
\end{equation*}

    \begin{equation}
        \therefore 
        \textbf{det}\Tilde{D} = \prod_{k=1}^{M/2} \prod_{j=1}^N \textbf{det} 
        \begin{bmatrix}
        z_1\lambda_M(k) & -z_2\lambda_N(j) \\
        -z_2\lambda_N(j) & -z_1 \lambda_M(k)  
        \end{bmatrix} 
    \end{equation}

Continuing 
    \begin{align} 
        \textbf{det}\Tilde{D} &= \prod_{k=1}^{M/2} \prod_{j=1}^N \textbf{det} 
        \begin{bmatrix}
        z_1\lambda_M(k) & -z_2\lambda_N(j) \\
        -z_2\lambda_N(j) & -z_1 \lambda_M(k)  
        \end{bmatrix} 
        \\
        &= \prod_{k=1}^{M/2} \prod_{j=1}^N \left( 4z_1^2 \cos^2 \left(\frac{\pi k}{M+1}\right) + 4z_2^2 \cos^2 \left(\frac{\pi j}{M+1}\right)\right) \\
        &= 4^{MN/2} \prod_{k=1}^{M/2} \prod_{j=1}^N \left( z_1^2 \cos^2 \left(\frac{\pi k}{M+1}\right) + z_2^2 \cos^2 \left(\frac{\pi j}{M+1}\right)\right)
    \end{align}
    
We have thus found $\textbf{det}\Tilde{D} = \textbf{det}D$. We can now easily apply this to the quadratic planar lattice, recalling
Equation (\ref{eqn:paf_det}) and

\begin{equation}
    \lim_{\substack{M\to\infty \\ N\to\infty}}(Z_{M,N})^{\frac{1}{MN}} \approx Z^{QL}
\end{equation}

\noindent If we plug everything, take the natural logarithm of it and use the fact that $\ln \left( \prod a_i \right) = \sum \ln(a_i)$ we get

\begin{equation*}
    \ln{Z^{QL}} = \lim_{\substack{M\to\infty \\ N\to\infty}} \sum_{k=1}^{\frac{1}{2}M} \sum_{l=1}^{N} \frac{1}{MN}
    \left(
    \ln{\left[ 4z_1^2\cos^2 \left(\frac{k\pi}{M+1} \right) + 4z_2^2 \cos^2 \left(\frac{l\pi}{N+1} \right)\right]^\frac{1}{2}}      
    \right) 
\end{equation*}

\noindent Now using the sum of angles theorem for cosine and arranging the boundaries

\begin{equation*}
    \ln{Z^{QL}} = \lim_{\substack{M\to\infty \\ N\to\infty}} \sum_{k=1}^{M} \sum_{l=1}^{N} \frac{1}{MN} 
    \left(
    \ln{2 \left[ z_1^2\left(\cos \left(\frac{2k\pi}{M+1} \right)+1\right) + z_2^2 \left(\cos \left(\frac{2l\pi}{N+1} \right)+1\right)\right]} 
    \right)      
\end{equation*}

\noindent Substituting $w_1 = \frac{2\pi}{M+1}, \ w_2 = \frac{2\pi}{N+1}$, changing the boundaries as necessary, and taking the exponential, we finally get $Z^{QL}$ as

\begin{align}
    \Aboxed{Z^{QL} &= \exp\left[ (2\pi)^{-2} \int_0^\pi dw_1 \int_0^\pi dw_2 \ln 2(z_1^2 + z_2^2 + z_1^2\cos(w_1) + z_2^2\cos(w_2)\right]}
\end{align}

\subsection{Towards the Ising Model: Boundary Conditions} \label{Towards the Ising Model: Boundary Conditions}

Before we apply the dimer problem to the Ising model, we need to introduce boundary conditions. In this section, we consider two types of boundary conditions: cylindrical and toroidal

\subsubsection{On Lattices with Cylindrical Boundary Conditions} \label{On Lattices with Cylindrical Boundary Conditions}

The first boundary conditions we look at are cylindrical boundary conditions. Previously, to ensure that $(M \times N)$ would be even, we took $M$ to be even. We have two different cases for the value of $N$:

\begin{enumerate}
 \item Both $M$ and $N$ are even. 
 \item $N$ is odd and $M$ is, by definition, even. In this case, there are two distinct subcases:
\end{enumerate}

\begin{figure}[h]
    \centering
    \includegraphics[scale = 0.6]{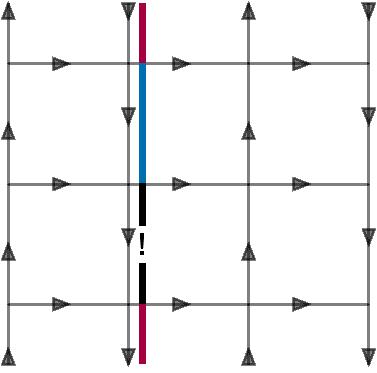}
    \caption{The oriented graph for $G(4, 3)$}
    \label{fig:noloop}
\end{figure}

\textit{(2.a)} Cyclic boundary conditions are applied in the vertical direction. In this case, there are no transition cycles looping the cylinder since there are an odd number of bonds, but transition cycles have an even number of bonds by definition. An example can be seen in Figure \ref{fig:noloop}.

Then, $A_{c, v}$ is then defined by
\vspace{-1mm}

\begin{subequations}
\begin{equation}
    A(i,j; i+1, j) = -A(i+1, j; i, j) = z_1, \; \; \; \; \;
    1 \le i \le M, \  1 \le j \le N-1
\end{equation}
\begin{equation}
    A(i,j; i, j+1) = -A(i, j+1; i, j) = (-1)^i z_2, \; \; \; \; \;
    1 \le i \le M-1, \ 1 \le j \le N
\end{equation}    
\end{subequations}

and an extra element coming from the boundary condition

\begin{equation}
    A(i, N; i, 1) = -A(N, i; 1, i) =(-1)^i z_2.
\end{equation}

This case is no different from the lattice with free boundary conditions in terms of types of transition cycles, and thus the "if" statement of \textit{Theorem A} is satisfied. Then, for dimer configuration with odd $M$ and even $N$ and cyclic boundary conditions in the vertical direction we have

\begin{equation}
    Z^{c, v}_{M, N} = \bm{Pf} A_{c, v}.
\end{equation}

\textit{(2.b)} Cyclic boundary conditions are applied in the horizontal direction. In this case, there is a new type of transition cycle: transition cycles that loop around the cylinder. This requires that we need to prove

\begin{equation}
     Z^c_{M, N} = \bm{Pf} A_c
\end{equation}

again for the cases $(1)$ and $(2.b)$, where $A_c$ is defined as
\vspace{-1mm}

\begin{subequations}
   \begin{equation}
    A(i,j; i+1, j) = -A(i+1, j; i, j) = z_1, \; \; \; \; \;
    1 \le i \le M, 1 \le j \le N-1
\end{equation}
\begin{equation}
    A(i,j; i, j+1) = -A(i, j+1; i, j) = (-1)^i z_2, \; \; \; \; \;
    1 \le i \le M-1, 1 \le j \le N
\end{equation} 
\end{subequations}

and

\vspace{-2mm}
\begin{equation}
    A(M, j; 1, j) = -A(1, j; M, j) = -z_1.
\end{equation}

As mentioned before, there are two types of transition cycles for these cases:

\begin{itemize}
    \item Cycles that do not loop around the cylinder
    \item Cycles that loop around the cylinder exactly one time. We will call these \textit{class 2 transition cycles}.
\end{itemize}

\begin{figure}[ht]
    \centering
    \includegraphics[scale=0.5]{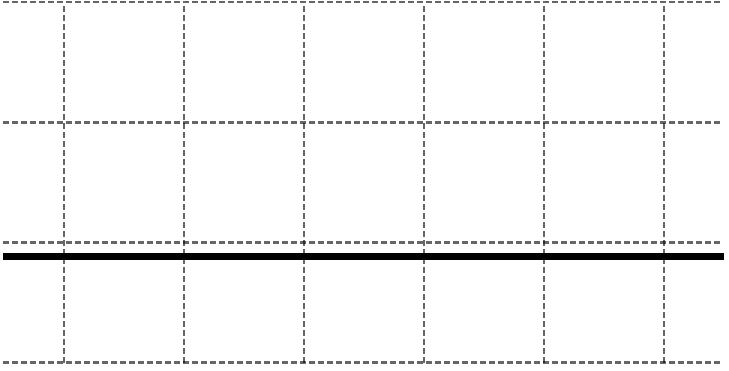}
    \caption{An elementary class 2 transition cycle on a cylinder}
    \label{fig:class2transcycle}
\end{figure}

Let's look at class 2 transition cycles since we know how to deal with the cycles that do not loop around the cylinder. One type of class 2 transition cycle has no vertical bonds. We call these class 2 transition cycles \textit{elementary class 2 transition cycles}. There are $N$ of these transition cycles in total, each differing from each other by one vertical translation. An example can be seen in Figure \ref{fig:class2transcycle}.

We need to include the terms corresponding to class 2 transition cycles in the Pfaffian. We have two choices as to how we define these terms:

\begin{equation}\label{cylinder choice 1}
    A(1, i; M, i) = -A(M, i; 1, i) = z_1
\end{equation}
\vspace{-1mm}
or 
\vspace{-1mm}
\begin{equation}\label{cylinder choice 2}
    A(1, i; M, i) = -A(M, i; 1, i) = -z_1.
\end{equation}

\begin{figure}[H]
\centering
\includegraphics[scale=0.5]{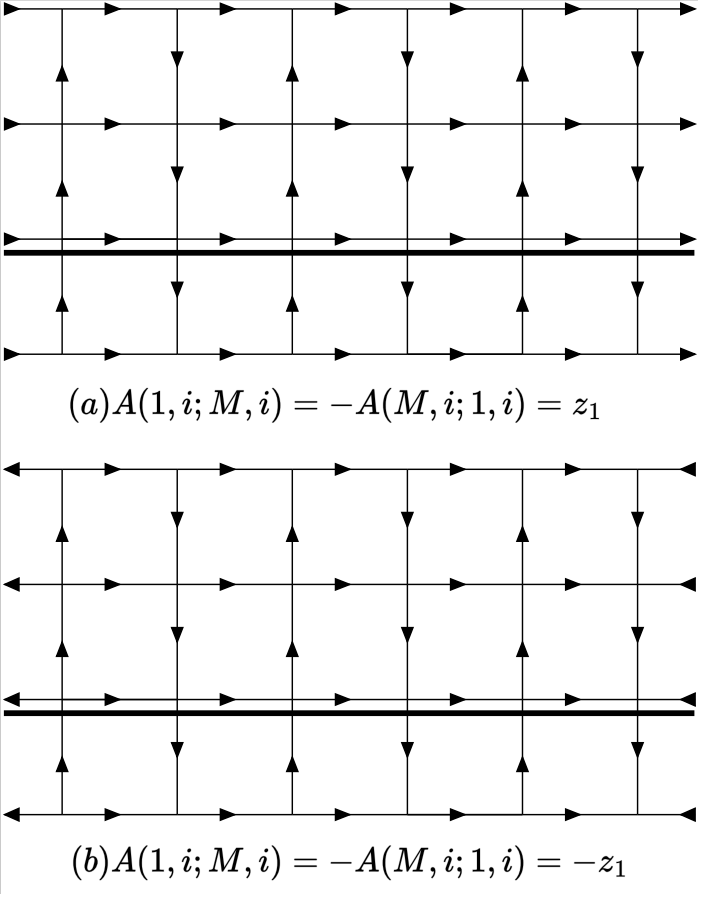}
\caption{\textit{(a)} An oriented lattice corresponding to \eqref{cylinder choice 1}. \textit{(b)} An oriented lattice corresponding to \eqref{cylinder choice 2}. For orientation parity to be odd, we choose the terms of the Pfaffian as \textit{(b)}.} 
\label{fig:orientedcylindricallattice}
\end{figure}

The orientation parity of all class 1 transition cycles is clearly odd for both of these choices. However, the orientation parity of all elementary class 2 transition cycles will only be odd for the choice \eqref{cylinder choice 2}, as seen in Figure \ref{fig:orientedcylindricallattice}.

Now that we know the orientation parity of every elementary class 2 transition cycle is odd, we just need to show that the orientation parity of all class 2 transition cycles is odd. Then, the conditions of \textit{Theorem A} will be satisfied, and $Z^c_{M, N} = \bm{Pf} A_c$. This leads to the following theorem:

\textbf{Theorem E:} On a square lattice with cylindrical boundary conditions, if the orientation parity of all elementary class 2 transition cycles is odd, then the orientation parity of all class 2 transition cycles are odd as well. 

\begin{figure}[ht]
    \centering
    \includegraphics[scale=0.7]{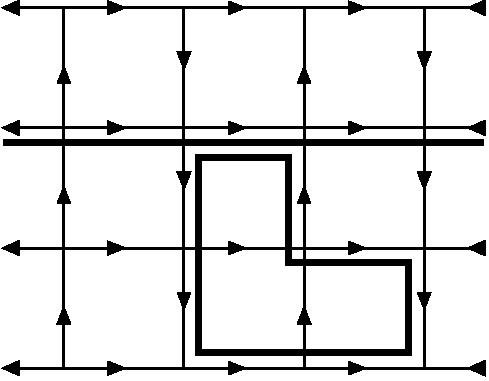}
    \caption{A class 2 transition cycle as the superposition of an elementary class 2 transition cycle and a polygon.}
    \label{fig:theoreme}
\end{figure}

Before starting the proof, we remark that if we divide the cylinder into two pieces by omitting the bonds and sites of any class 2 transition cycle, we are left with two pieces, each containing an even number of sites, to fill the pieces entirely with double bonds. Now, with a similar remark to that of \textit{Theorem D}, we say that an arbitrary class 2 transition cycle can be thought of as the superposition of an elementary class 2 transition cycle and a number of closed polygons, each having bonds in common with the elementary class 2 transition cycle and having the rest of the bonds only on one side of the elementary class 2 transition cycle, as seen in Figure \ref{fig:theoreme}.

\textit{\textbf{Proof:}} Let $\Gamma_n$ be the polygon having $a$ clockwise bonds (odd by \textit{Theorem D}), surrounding $p$ lattice points (even by \textit{Theorem B}), and having $c$ arrows in common with $\Gamma_e$, an elementary class 2 transition cycle. Say $\Gamma_e$ has $a'$ clockwise bonds
(odd by assumption) and contains $M$ points (even by assumption). Then, their superposition $\Gamma_{n+e}$ has $a + a' - c$ clockwise arrows and contains $M + p + c - 1$
points. $M + p + c - 1$ is even by the first remark, so $a + a' - c$ must be odd.

With the proof of \textit{Theorem E}, we have shown that 

$$
Z^c_{M, N} = \bm{Pf} A_c
$$

still holds for planar lattices with cylindrical boundary conditions which may or may not have transition cycles looping around the whole cylinder. 

\subsubsection{On Lattices with Toroidal Boundary Conditions} \label{On Lattices with Toroidal Boundary Conditions}

We now extend our discussion to planar lattices with toroidal boundary conditions. This introduces transition cycles that loop the torus $h$ times in the horizontal direction and $v$ times in the vertical direction.

Similar to the discussion in \textbf{Section \ref{The Pfaffian}}, for transition cycles with $h=v=0$ we can choose $A(i, j; i', j')$ as  

\begin{subequations}
   \begin{equation}
    A(i,j; i+1, j) = -A(i+1, j; i, j) = z_1, \; \; \; \; \;
    1 \le i \le M, 1 \le j \le N-1
\end{equation}
\begin{equation}
    A(i,j; i, j+1) = -A(i, j+1; i, j) = (-1)^i z_2, \; \; \; \; \;
    1 \le i \le M-1, 1 \le j \le N
\end{equation} 
\end{subequations}

to make the orientation parity of those transition cycles odd. For the terms corresponding to transition cycles with nonzero $h$ and $v$, we have four possible assignments: 

\begin{subequations}\label{toroidal choices}
\begin{equation}\label{toroidal choice 1}
        \begin{split}
            A_1(M, j; 1, j) = -A_1(1, j; M, j) = z_1  \\
            A_1(i, N; i, 1) = - A_1(i, 1; i, N) = (-1)^kz_2
        \end{split}
\end{equation}
\begin{equation}\label{toroidal choice 2}
        \begin{split}
            A_2(M, j; 1, j) = -A_2(1, j; M, j) = z_1  \\
            A_2(i, N; i, 1) = - A_2(i, 1; i, N) = -(-1)^kz_2
        \end{split}
\end{equation}
\begin{equation}\label{toroidal choice 3}
        \begin{split}
            A_3(M, j; 1, j) = -A_3(1, j; M, j) = -z_1  \\
            A_3(i, N; i, 1) = - A_3(i, 1; i, N) = (-1)^kz_2
        \end{split}
\end{equation}
\begin{equation}\label{toroidal choice 4}
        \begin{split}
            A_4(M, j; 1, j) = -A_4(1, j; M, j) = -z_1  \\
            A_4(i, N; i, 1) = - A_4(i, 1; i, N) = -(-1)^kz_2
        \end{split}
\end{equation}
\end{subequations}

\begin{figure}[H]
    \centering
    \includegraphics[scale = 0.4]{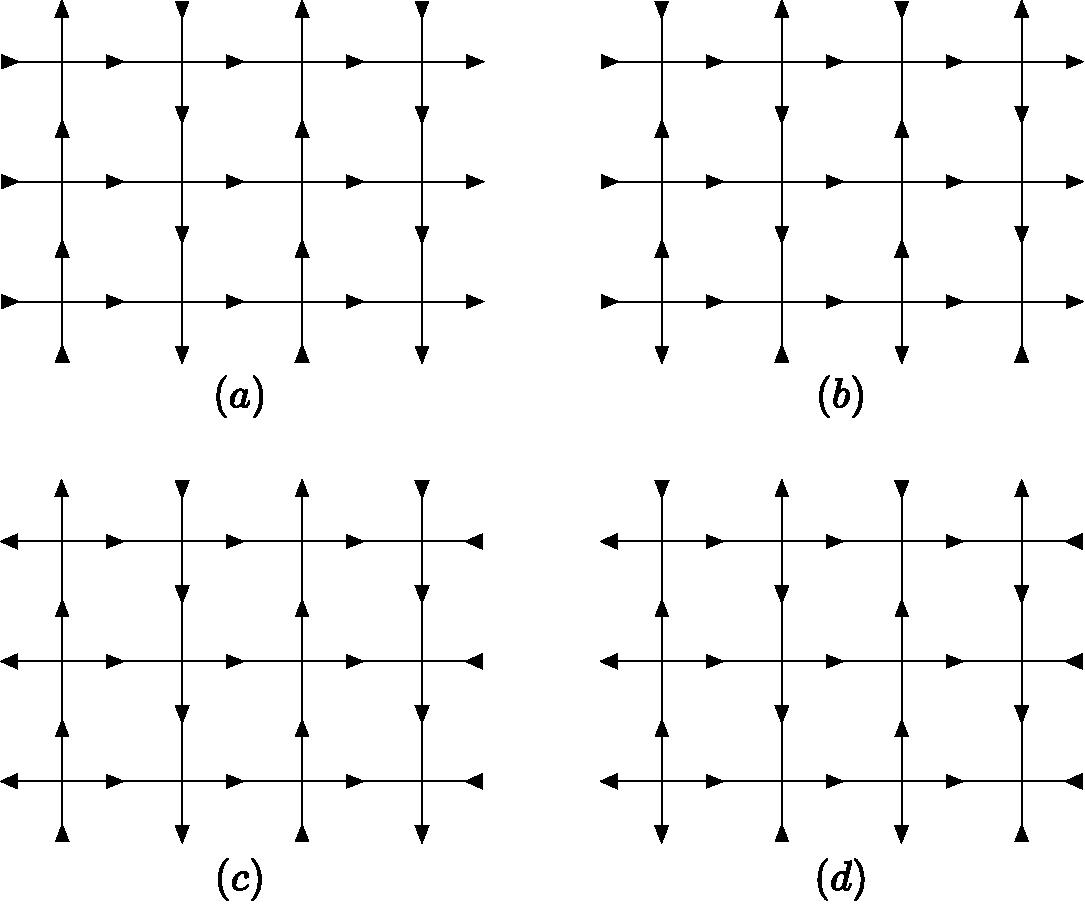}
    \caption{Oriented lattices corresponding $(a)$ to \eqref{toroidal choice 1}; $(b)$ to \eqref{toroidal choice 2}; $(c)$ to \eqref{toroidal choice 3}; $(d)$ to \eqref{toroidal choice 4}.}
    \label{fig:toroidallattices}
\end{figure}

Figure \ref{fig:toroidallattices} shows the oriented lattices corresponding to the possible assignments.
    
\begin{figure}[H]
    \centering
    \includegraphics[scale = 0.4]{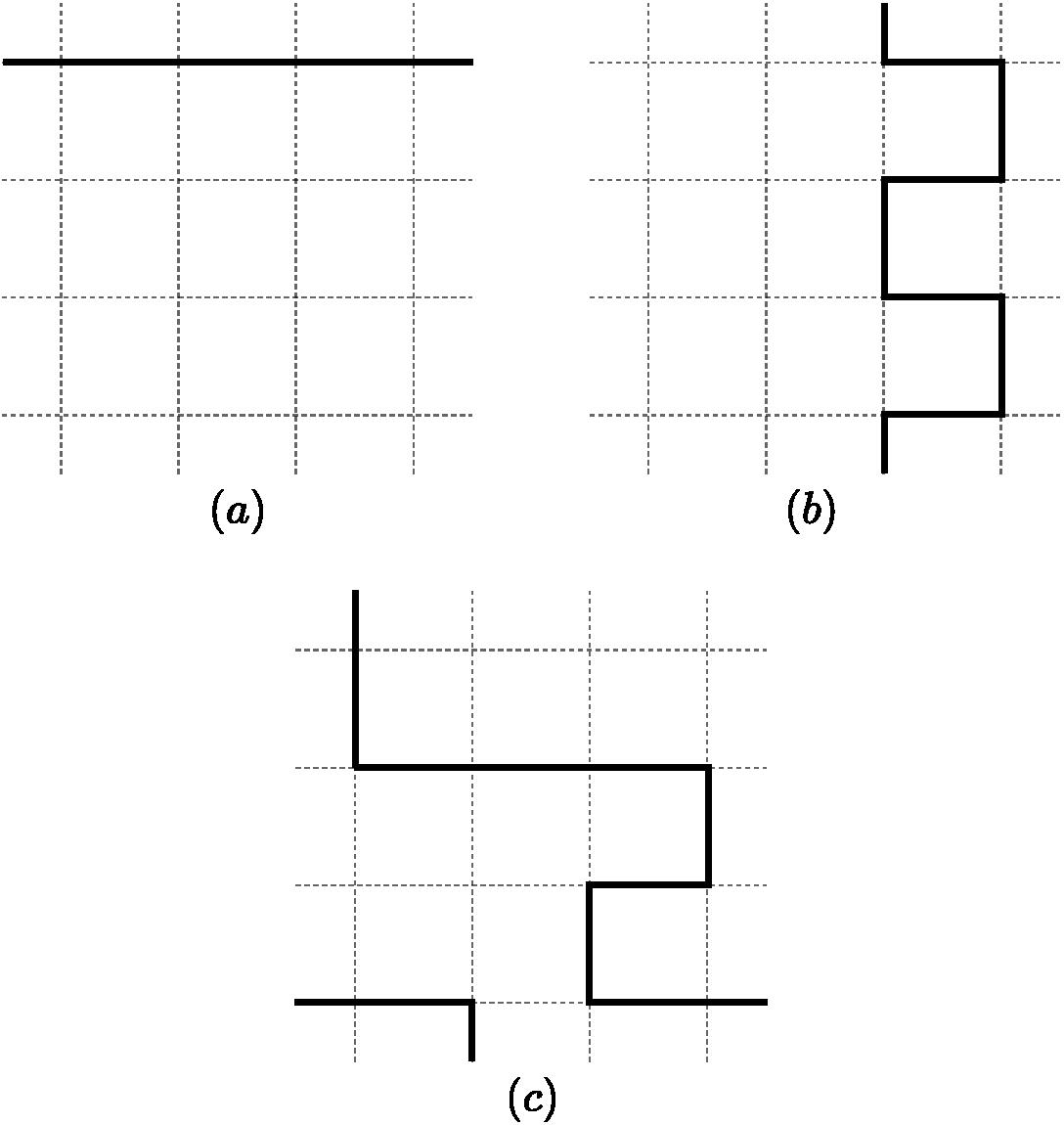}
    \caption{Three transition cycles that do not have odd orientation parity when the assignments \eqref{toroidal choices} are applied. $(a)$ has $v=0$, $h=1$; $(b)$ has $v=1$, $h=0$; $(c)$ has $v=h=1$.}
    \label{fig:3transitioncycles}
\end{figure}

In \textbf{Section \ref{On Lattices with Cylindrical Boundary Conditions}}, it was possible to choose one from the two choices that assured orientation parity of all transition cycles would be odd. Unfortunately, none of the assignments \eqref{toroidal choices} assures that all transition cycles will have odd orientation parity simultaneously for one choice of assignment. For example, consider the transition cycles given in Figure \ref{fig:3transitioncycles}.

The rest of this section is dedicated to showing that we can find a suitable linear configuration of the Pfaffians of the four matrices defined by \eqref{toroidal choices}, particularly

\begin{equation}
    Z^(t)_{M, N} = \frac{1}{2} \left[ -\bm{Pf}(A_1) +  \bm{Pf}(A_2) + \bm{Pf}(A_3) + \bm{Pf}(A_4) \right]
\end{equation}

Checking which assignment of the sign makes which transition cycle has odd orientation parity for all transition cycles is impossible. So, we need a method of finding the orientation parity of a given transition cycle. In \textbf{Section \ref{The Pfaffian}}, our leading argument for all terms in the Pfaffian to have the same sign was that any two terms in the Pfaffian should have the sign, which we proved. This is clearly replaceable with the statement that if the sign of any dimer configuration is the same as the sign of a fixed dimer configuration, then the corresponding terms in the Pfaffian will have the same sign. The most obvious choice for the fixed dimer configuration is the configuration $C_0$ defined in Figure \ref{fig:C0_Configuration}. With these considerations, we move on with the following definition:

\textbf{Definition:} \textit{$C_0$ Transition Cycle}. A transition cycle made up of the bonds of an arbitrary transition cycle $C$ and the configuration $C_0$ is called a \textit{$C_0$ transition cycle}.

\begin{figure}[H]
    \centering
    \includegraphics[scale = 0.6]{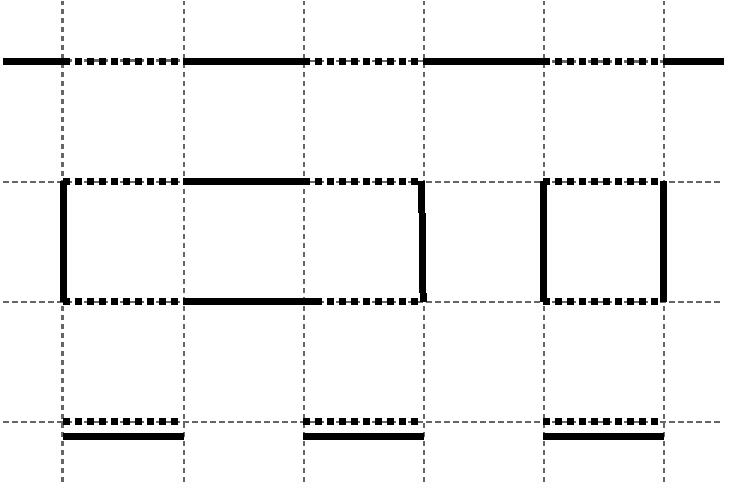}
    \caption{A transition graph with several $C_0$ transition cycles.}   \label{fig:c0transgraph}
\end{figure}

We can now restate \textit{Theorem A} as
     
\textbf{Theorem A':} If the orientation parity of every $C_0$ transition cycle is odd, then all terms in the Pfaffian will have the same sign.

This theorem makes our job a lot easier: now we just need a way of finding the orientation parity of every $C_0$ transition cycle instead of every transition cycle for the different assignments. For $v=h=0$, the matrices defined by \eqref{toroidal choices} give odd orientation parity for all $C_0$ transition cycles since they already satisfy the condition for \textit{Theorem A}. What about nonzero $v$ and $h$? Thankfully, the orientation parity of $C_0$ transition cycles in this case depends only on whether the values of $v$ and $h$ are even or odd. Consider the following four classes of configurations

\begin{itemize}
    \item $v = h = even$, denoted as $(e, e)$
    \item $v = odd$ and $h = even$, denoted as $(o, o)$
    \item $v = even$ and $h = odd$, denoted as $(e, o)$
    \item $v = h = odd$, denoted as $(o, o)$
\end{itemize}

It is actually impossible for the $v$ and $h$ values of a $C_0$ transition cycle to have a common divisor. Thus, for non-zero and even $v$, $h = 0$ and vice versa.

\begin{figure}[H]
    \centering
    \includegraphics[scale = 0.4]{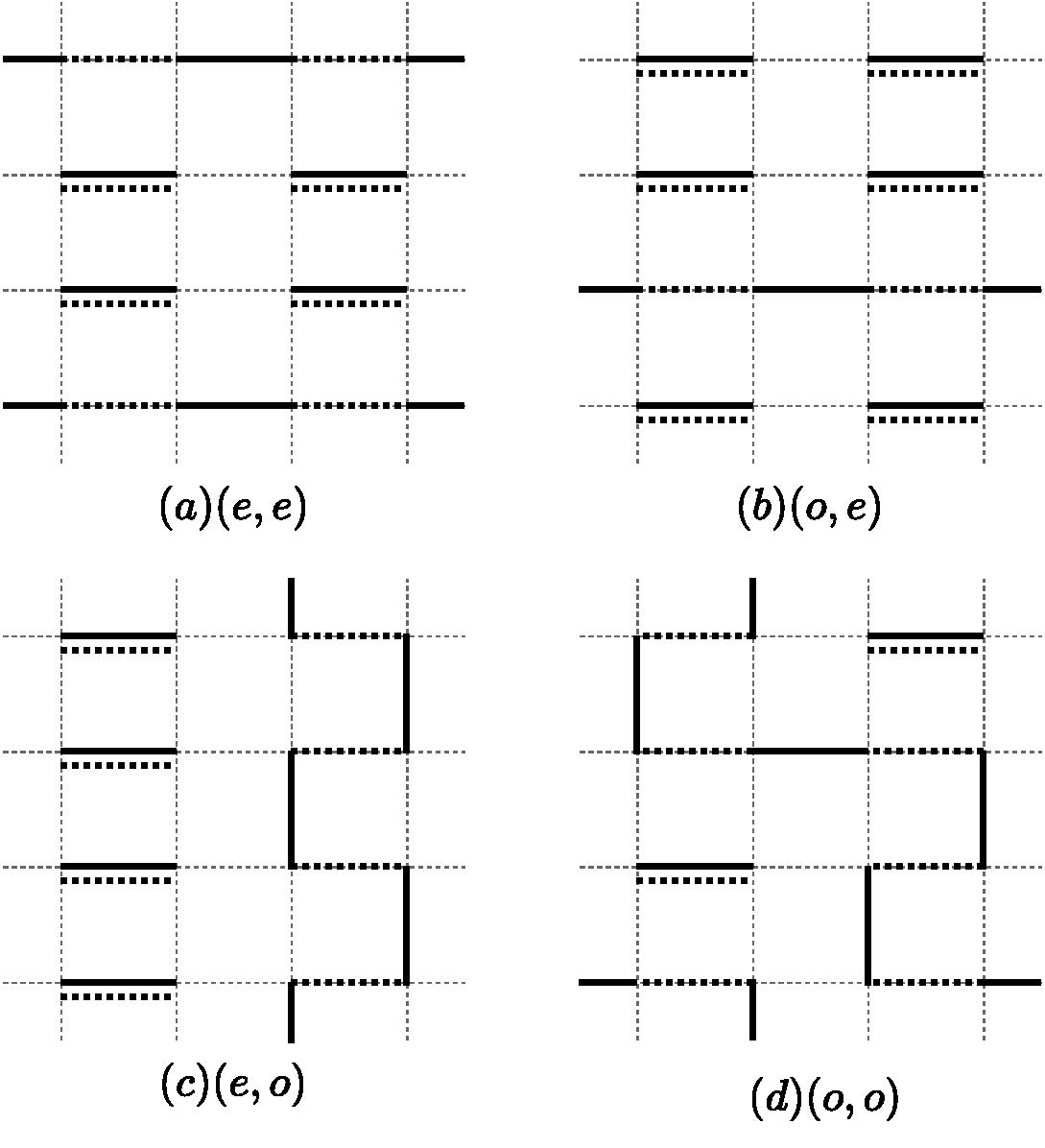}
    \caption{Examples of the four classes of configurations $(a)$ for $(e, e)$, $(b)$ for $(o, e)$, $(c)$ for $(e, o)$ and $(d)$ for $(o, o)$}   \label{fig:oddevenconfigs}
\end{figure}

Examples for these four classes are given in Figure \ref{fig:oddevenconfigs}.

As discussed previously, the Pfaffian will give the configuration generating function only if all terms in the Pfaffain have the same sign, i.e. if the orientation parity of all the transition cycles is odd. For the examples in Figure \ref{fig:oddevenconfigs}, we can see that \eqref{toroidal choice 1} gives odd orientation parity for $(a)$, \eqref{toroidal choice 2} for $(b)$, \eqref{toroidal choice 3} for $(c)$ , and \eqref{toroidal choice 4} for $(d)$. We can generalize this for all cases of $C_0$ transition cycles. From these arguments, the signs of $C_0$ transition cycles are summarized in the Pfaffian are given in the table below.

\begin{table}[h]
\centering
\begin{tabular}{|c|c|c|c|c|}
\hline
Configuration & $A_1$ & $A_2$ & $A_3$ & $A_4$ \\
\hline
$(e,e)$ & $+$ & $+$ & $+$ & $+$ \\
\hline
$(o,e)$ & $-$ & $-$ & $+$ & $+$ \\
\hline
$(e,o)$ & $-$ & $+$ & $-$ & $+$ \\
\hline
$(o,o)$ & $-$ & $+$ & $+$ & $-$ \\
\hline
\end{tabular}
\caption{Signs in which the $C_0$ transition cycles appear in the Pfaffian for the four cases}
\end{table}

From the table, we can see that the linear combination of the Pfaffians that will correctly give the configuration generating function is 

\begin{equation}
    Z(t)_{M, N} = \frac{1}{2} \left[ -\bm{Pf}(A_1) +  \bm{Pf}(A_2) + \bm{Pf}(A_3) + \bm{Pf}(A_4) \right]
\end{equation}

\subsection{The Two-Dimensional Ising Model} \label{The Two Dimensional Ising Model}

\subsubsection{The Partition Function} \label{The Partition Function}

    We impose toroidal boundary conditions on our lattice with no magnetization ($H=0$) and write the partition function as

\begin{equation}
    \begin{split}
        Z_{M,N}^{(t)} &= \sum_{\sigma = \pm 1}\exp\left[\beta E_1 \sum_{k=1}^{M} \sum_{j=1}^{N}  \sigma_{j,k}\sigma_{j, (k+1)} + \beta E_2 \sum_{k=1}^{M} \sum_{j=1}^{N}  \sigma_{j,k}\sigma_{(j+1), k} \right] \\
        &= \sum_{\sigma = \pm 1} \left[ \prod_{k=1}^{M} \prod_{j=1}^{N}  e^{\beta E_1 \sigma_{j,k}\sigma_{j, (k+1)}}\right]
                                 \left[ \prod_{k=1}^{M} \prod_{j=1}^{N}  e^{\beta E_2 \sigma_{j,k}\sigma_{(j+1), k}}\right]
    \end{split}
\end{equation}

Since $\sigma$ only takes the values $\pm 1$, we have

\begin{equation}
    e^{\beta E \sigma \sigma'} = \cosh{\beta E} + \sigma \sigma' \sinh{\beta E}
\end{equation}

Therefore we get the partition function as

\begin{equation}
        Z_{M,N}^{(t)} = (\cosh{\beta E_1}\cosh{\beta E_2})^{MN} \sum_{\sigma = \pm 1}\prod_{k=1}^{M} \prod_{j=1}^{N} (1 + z_1\sigma_{j,k}\sigma_{j, (k+1)}) (1 + z_2\sigma_{j,k}\sigma_{(j+1), k})      
\end{equation}

\noindent where $z_1 = \tanh{\beta E_1}$ and $z_2 = \tanh{\beta E_2}$. Now we expand the products over $j$ and $k$. It is evident that any term with a factor $\sigma_{j,k}$ or $\sigma_{j,k}^3$ will vanish. Since $\sigma_{j,k}^2 = \sigma_{j,k}^4 = 1$, we get

\begin{equation}
    \sum_{\sigma = \pm 1} 1 = 2^{MN}
\end{equation}

\noindent and in the end we find

\begin{equation}
    Z_{M, N}^{(t)} = (2\cosh{\beta E_1}\cosh{\beta E_2})^{MN}\sum_{p,q} N_{p,q}z_1^pz_2^q
\end{equation}

The summation term in our last equation is understood as the weighted sum over all closed polygons satisfying properties 

    \begin{itemize}
        \item The nearest neighbors cannot make more than one bond.
        \item An even number of bonds appear from every point.
        \item The figure contains $p$ horizontal and $q$ vertical bonds.
    \end{itemize}   
    
    where each bond of the polygon has a weight  $\tanh{\beta}$ which is the energy of the bond.

\begin{figure}[h!]
        \centering
        \includegraphics[scale = 0.4]{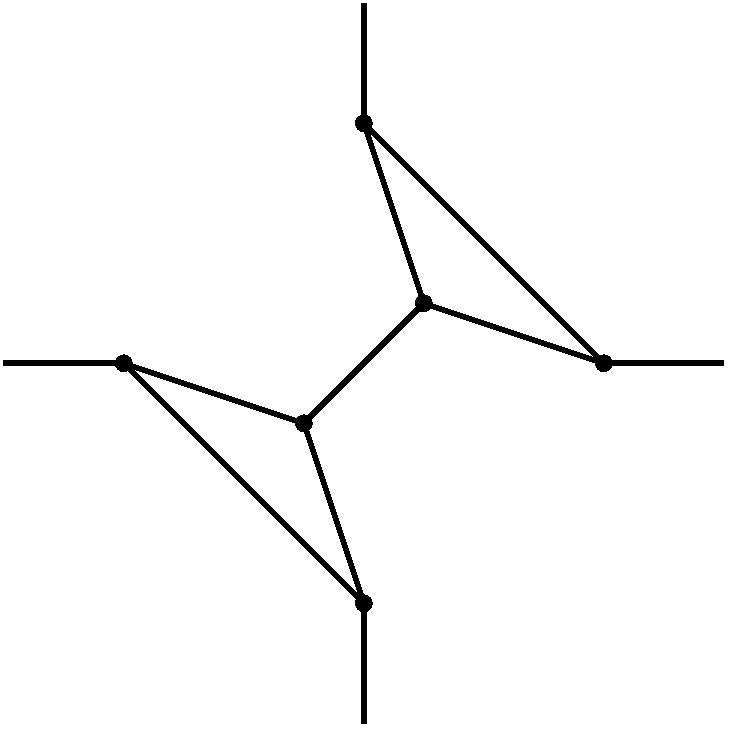}
        \caption{The cluster used to convert the Ising problem into a dimer problem.}
        \label{sixsitecluster}
    \end{figure}

    We can replace each site on the Ising lattice with such a cluster that the summation term becomes the generating function for closest-packed dimer configurations. One such cluster is given in Figure \ref{sixsitecluster}.

 The sites become clusters with six points inside, we name the points as \textit{R, L, U, D, 1} , and \textit{2}. We will be working with toroidal boundary conditions, so we will have $4$ different matrices from $4$ different orientations. One such matrix can be given by

    \begin{equation}
    \bar{A}(j, k; j, k) =
    \begin{bNiceMatrix}[first-row, first-col]
       & R & L & U & D  & 1 & 2  \\ 
    R  &  0&  0& -1& 0  & 0 & 1  \\
    L  & 0 &  0&  0& -1 & 1 & 0  \\
    U  & 1 & 0 &  0&  0 & 0 & -1 \\
    D  &  0&  1& 0 &  0 & -1&  0 \\
    1  &  0& -1&  0&  1 & 0 &  1 \\
    2  & -1&  0&  1&  0 &-1 & 0 
    \end{bNiceMatrix}
    \end{equation}
    \begin{equation}
    \bar{A}(j, k; j, k+1) = 
    \begin{bNiceMatrix}[first-row, first-col]
       & R & L & U & D  & 1 & 2  \\ 
    R  &  0&z_1& 0 & 0  & 0 & 0  \\
    L  & 0 &  0&  0& 0  & 0 & 0  \\
    U  & 0 & 0 &  0&  0 & 0 & 0  \\
    D  &  0&  0& 0 &  0 & 0 & 0  \\
    1  &  0& 0 &  0&  0 & 0 & 0  \\
    2  & 0 &  0&  0&  0 & 0 & 0 
    \end{bNiceMatrix}
    \end{equation}
    \begin{equation}
    \bar{A}(j, k; j+1, k) = 
    \begin{bNiceMatrix}[first-row, first-col]
       & R & L & U & D  & 1 & 2  \\ 
    R  &  0&  0& 0 & 0  & 0 & 0  \\
    L  & 0 &  0&  0& 0  & 0 & 0  \\
    U  & 0 & 0 &  0& z_2& 0 & 0  \\
    D  &  0&  0& 0 &  0 & 0 & 0  \\
    1  &  0& 0 &  0&  0 & 0 & 0  \\
    2  & 0 &  0&  0&  0 & 0 & 0 
    \end{bNiceMatrix}
    \end{equation}  
    with $\bar{A}(j, k; j, k+1) = -\bar{A}^T(j, k+1; j, k)$ and $\bar{A}(j, k; j+1, k) = -\bar{A}^T(j+1, k; j, k)$.

We can further reduce $\bar{A}$ to four separate $4(M \times N) \times 4(M \times N)$ matrices which we will be denoting with $A$ with the property

\begin{equation}
    \textbf{det}\bar{A} = \textbf{det}A.
\end{equation}

We also do this by row and column operations.[citation needed] We get $A$ as

\begin{equation}
A(j, k; j, k) =
\begin{bNiceMatrix}[first-row, first-col]
   & R & L & U & D  \\ 
R  &  0&  1& -1& -1 \\
L  & -1&  0&  1& -1 \\
U  &  1& -1&  0&  1 \\
D  &  1&  1& -1&  0 
\end{bNiceMatrix}
\end{equation}

\begin{equation}
A(j, k; j, k+1) = 
\begin{bNiceMatrix}[first-row, first-col]
   & R&  L&  U& D  \\ 
R  & 0&  z_1&  0& 0 \\
L  & 0&  0&  0& 0 \\
U  & 0&  0&  0& 0 \\
D  & 0&  0&  0& 0 
\end{bNiceMatrix}
\end{equation}

\begin{equation}
A(j, k; j+1, k) = 
\begin{bNiceMatrix}[first-row, first-col]
   & R&  L&  U& D  \\ 
R  & 0&  0&  0& 0 \\
L  & 0&  0&  0& 0 \\
U  & 0&  0&  0& z_2 \\
D  & 0&  0&  0& 0 
\end{bNiceMatrix}
\end{equation}

with $A(j, k; j, k+1) = -A^T(j, k+1; j, k)$ and $A(j, k; j+1, k) = -A^T(j+1, k; j, k)$.

Now that we have found our matrix, all that's left is to take the determinant. To do that, we again decompose our matrix with the help of \textbf{Kronecker Product}. Define the ${N \times N}$ matrix

\begin{equation}
H_N^{\pm} =
\begin{bmatrix}
0 & 1 & 0 & 0 & \cdots & 0 & 0 \\
0 & 0 & 1 & 0 & \cdots & 0 & 0 \\
0 & 0 & 0 & 1 & \cdots & 0 & 0 \\
\vdots  & \vdots  & \vdots & \vdots & \vdots & \vdots & \vdots  \\
0 & 0 & 0 & 0 & \cdots & 0 & 1 \\
\pm 1 & 0 & 0 & 0 & \cdots & 0 & 0
\end{bmatrix}
\end{equation}

The matrix $H_N$ is a special form of the matrices called \textbf{Toeplitz matrices} which have the eigenvalues $\lambda_n^+ = e^{2\pi i \frac{n}{N}}$ and $\lambda_n^- = e^{2\pi i \frac{n+1}{N}}$. The matrices $H_N^+$ and $H_N^-$ is also unitary, so they have the property

\begin{equation}
    H^{+}(H^{+})^T = H^{-}(H^{-})^T = 1
\end{equation}

    We can now decompose all of our matrices, albeit with a new and easier-to-read notation, as

    \begin{equation}
    \begin{split}
        A_{\pm \pm'} =&
        I_{N} \otimes I_{M} \otimes 
        \begin{bmatrix}
             0 &  1& -1& -1 \\
            -1 &  0&  1& -1 \\
             1 & -1&  0&  1 \\
             1 &  1& -1&  0 
        \end{bmatrix}
        \\
        &+ 
        I_{N} \otimes H^\pm_{M} \otimes 
        \begin{bmatrix}
             0 &  z_1&  0& 0 \\
             0 &    0&  0& 0 \\
             0 &    0&  0& 0 \\
             0 &    0&  0& 0 
        \end{bmatrix}
        + 
        I_{N} \otimes (H^\pm_{M})^T \otimes  
        \begin{bmatrix}
             0 &    0&  0& 0 \\
            -z_1&    0&  0& 0 \\
             0 &    0&  0& 0 \\
             0 &    0&  0& 0 
        \end{bmatrix}
        \\
        &+ 
        H^{\pm'}_{N} \otimes I_{M} \otimes 
        \begin{bmatrix}
             0 &    0&  0& 0 \\
             0 &    0&  0& 0 \\
             0 &    0&  0& z_2 \\
             0 &    0&  0& 0 
        \end{bmatrix}
        + 
        (H^{\pm'}_{N})^T \otimes I_{M} \otimes 
        \begin{bmatrix}
             0 &    0&  0& 0 \\
             0 &    0&  0& 0 \\
             0 &    0&  0& 0 \\
             0 &    0&  -z_2& 0 
        \end{bmatrix}       
    \end{split}
\end{equation}

Here $A_{\pm \pm'}$ correspond to $A_i$ where $i=1,\dots,4$. Let $U_n$ be the $(n \times n)$ matrix which diagonalizes $H_n$. Then we just define $U = U_N \otimes U_M \otimes I_4$ and let it do its work, getting a submatrix as

    \begin{equation*}
    A(\theta_1, \theta_2) =
    \begin{bmatrix}
        0                   &  1+z_1 e^{i\theta_1} &          -1         & -1                   \\
       -1-z_1 e^{-i\theta_1}&          0           &           1         & -1                   \\
        1                   &          -1          &           0         &  1+ z_2 e^{i\theta_2}\\
        1                   &          1           & -1-z_2 e^{-i\theta_2}&  0 
    \end{bmatrix}
\end{equation*}

where $\theta_i^+ = 2\pi n/L_i$ and $\theta_i^- = \pi (2n-1)/L_i$ for $i=1, 2$ with $L_1 = M$, $L_2 = N$; $n = 1,2, \dots, L_i$.

    Taking the determinant, we get 

    \begin{align*}
    \textbf{det}A_{\pm \pm'} = \prod_{\theta_1^\pm} \prod_{\theta_2^{\pm'}} 
                                [(1+z_1^2)(1+z_2^2) - 2z_1(1-z_2^2)\cos{\theta_1}
                                - 2z_2(1-z_1^2)\cos{\theta_2}].
    \end{align*}

where $\theta_i^+ = 2\pi n/L_i$ and $\theta_i^- = \pi (2n-1)/L_i$ for $i=1, 2$ with $L_1 = M$, $L_2 = N$; $n = 1,2, \dots, L_i$. Also $z_1 = \tanh{\beta E_1}$ and $z_2 = \tanh{\beta E_2}$. So we get $Z^{(t)}_{M,N}$ as 

\begin{equation}
    Z^{(t)}_{M,N} = (2\cosh{\beta E_1}\cosh{\beta E_2})^{MN}\frac{1}{2}\sum_i \pm \sqrt{\textbf{det}A_i}
\end{equation}

The signs of the determinants depend on the temperature, and it needs hard work to determine them. Luckily, the signs are irrelevant when we want to compute the free energy in the thermodynamic limit.

\begin{equation}
    -F\beta = \lim_{\substack{M\to\infty \\ N\to\infty}} \frac{1}{MN}\ln{Z^{(t)}_{M,N}}
\end{equation}

This is an even easier translation to the integral

\begin{equation}
    \begin{split}
        -F\beta = \ln(2\cosh{\beta E_1} \cosh{\beta E_2}) + \frac{1}{2}(2\pi)^{-2} &\int_0^{2\pi} dw_1 \int_0^{2\pi} dw_2 \ln[(1+z_1^2)(1+z_2^2) \\ 
        &- 2z_1(1-z_2^2)\cos{w_1} - 2z_2(1-z_1^2)\cos{w_2}] 
    \end{split}
\end{equation}

or simply

\begin{equation}
    \begin{split}
        -F\beta = \ln2 + \frac{1}{2}(2\pi)^{-2} \int_0^{2\pi} dw_1 &\int_0^{2\pi} dw_2 \ln[\cosh{2\beta E_1}\cosh{2\beta E_2} \\ 
        &- \sinh{2\beta E_1}\cos{w_1} - \sinh{2\beta E_2}\cos{w_2}].
    \end{split}
\end{equation}

Thus we have successfully reproduced Onsager's solution \cite{PhysRev.65.117} for the two-dimensional Ising model by reformulating its partition function through perfect matchings (dimers) on a planar lattice. Using Kasteleyn’s Pfaffian method, we efficiently counted these matchings. We demonstrated how proper edge orientations ensure consistent signs in the Pfaffian expansion. Adapting this method to various boundary conditions—free, cylindrical, and toroidal—we derived the Ising partition function as a linear combination of Pfaffians, highlighting the connection between graph theory and statistical mechanics in exactly solvable lattice models.

\section{Solution via Star-Triangle Relation \textit{(Tuğba Hırlı and R. Semih Kanber)}}
\label{chapter:4}

Many solutions of the Ising model consist of some combinatorial techniques or transfer matrices. In this solution, we will try to obtain an exact solution only by considering the local properties of the model, which is why Baxter and Enting believe that this is the simplest solution so far. In this solution, we will start with the eight-vertex model and by using the star-triangle relation, we will show that the nearest (or next nearest) neighbor correlation will be the same as the nearest-neighbor correlation on the square lattice under the assumption of the thermodynamic limit (see \cite{baxterarticle}). In the enlightenment of this idea, we will see that the correlation function is a function of two variables only. After simplifying the correlation function, we will obtain a linear relationship between the nearest and next nearest neighbor correlations of the honeycomb lattice.

\subsection{The Honeycomb Lattice}
\begin{figure}[h]
    \centering
\includegraphics[width=0.5\textwidth]{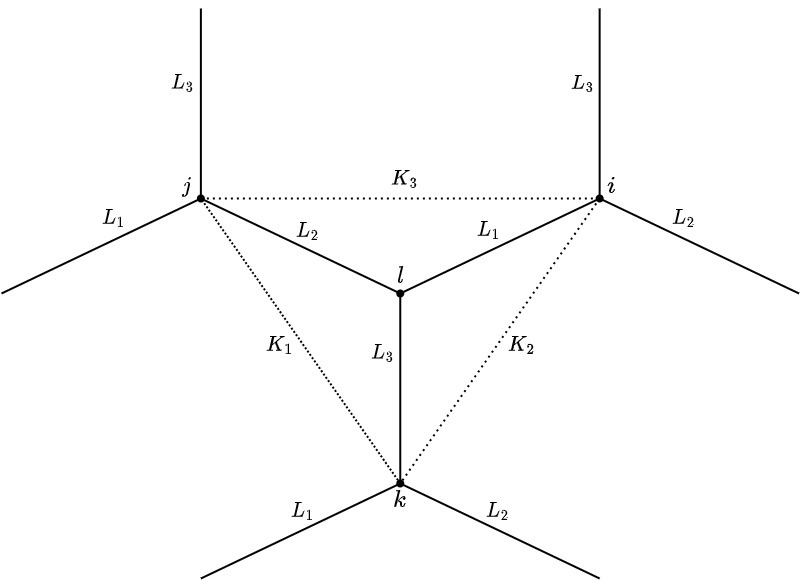}
    \caption{A star (shown with solid line) and triangle (shown with dotted line)
in a honeycomb lattice.}
    \label{figure 1}
\end{figure}
Consider an anisotropic honeycomb lattice with different interaction coefficients in each direction. Considering that spins ($\sigma_i$) in each site $i$ has interaction energy with its adjacent spins as -$ J'_r \sigma_i \sigma_l $, where $r= 1,2,3$ according to the direction of interaction, we can write the partition function for Figure \ref{figure 1} as

\begin{equation}\label{Honeycombpartition}
    Z^H_N{L} = \sum _{\sigma} \exp[L_1 \Sigma\sigma_i \sigma_l + L_2 \Sigma \sigma_j \sigma_l + L_3 \Sigma \sigma_k \sigma_l]
\end{equation}
where
\begin{equation} \label{l}
    L_r=J'_r/k_{B}T
\end{equation}
Now, let's define $\xi$ to be

$$ \xi(\sigma_i , \sigma_j , \sigma_k) = \sum _{\sigma_l} \Xi(\sigma_l | \sigma_i, \sigma_j, \sigma_k) $$
where $\Xi(\sigma_l | \sigma_i, \sigma_j, \sigma_k)= \exp[ \sigma_l (L_1 \sigma_i + L_2 \sigma_j + L_3 \sigma_k)] $.
\\
Considering that the sum in $\xi$ is only over $\sigma_l$ which can take values $\pm 1$, $\xi $ can be written as
 \begin{equation}\label{xi}
     \xi(\sigma_i ,\sigma_j , \sigma_k) = 2\cosh(L_1 \sigma_i + L_2 \sigma_j + L_3 \sigma_k) 
 \end{equation}
 Using \eqref{xi}, the partition function in \eqref{Honeycombpartition} takes the form

 \begin{equation}\label{honeycombwithxi}
     Z^H_N{L} =\sum_{\sigma} \prod_{i,j,k} \xi(\sigma_i , \sigma_j , \sigma_k)
 \end{equation}
Returning to Figure \ref{figure 1}, the partition function for the triangular lattice can be written using the interaction coefficients $K_r$. We can write the partition function for the triangular lattice as we did in \eqref{Honeycombpartition}.

\begin{equation}\label{triangularpartition}
    Z^T_N{K} = \sum _{\sigma} \exp[K_1 \Sigma\sigma_k \sigma_j + K_2 \Sigma \sigma_k \sigma_i + K_3 \Sigma \sigma_j \sigma_i]
\end{equation}
where
\begin{equation} \label{k}
    K_r = J_r/ k_{B}T
\end{equation}
\begin{figure}[h]
    \centering
    \includegraphics[width=0.4\textwidth]{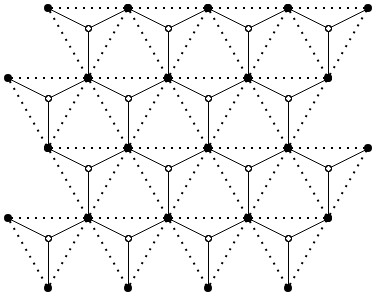}
    \caption{Honeycomb lattice}
    \label{baxterhoneycomblattice}
\end{figure}
Considering that the exponential part on the right-hand side of \eqref{triangularpartition} and the term within the $\cosh$ function \eqref{xi} have the same variables $(\sigma_i, \sigma_j, \sigma_k)$, each of which can take values $\pm 1 $, there must be a parameter $R$ such that
\begin{equation}\label{xirelation}
    \xi(\sigma_i, \sigma_j, \sigma_k) = R\exp(K_1 \sigma_k \sigma_j + K_2 \sigma_k \sigma_i + K_3 \sigma_i \sigma_j)
\end{equation}
Although our focus will be on the equation \eqref{xirelation}, after this point it will be easy to show the relationship between two partition functions for the honeycomb lattice model and the triangular lattice model. The number of spins that are considered is crucial. In the partition function of the triangular lattice, we only considered those that are visualized as solid circles in Figure \ref{baxterhoneycomblattice}, while for the honeycomb lattice, we considered all of them. Realizing that the number of solid circles is equal to the number of open circles, if we consider N particles for the triangular lattice then we should consider $2N$ particles for the honeycomb lattice. Then the relation between the two partition functions is

\begin{equation}\label{relationbetweenlattices}
     Z_{2N}^H {L} = R^N Z_N^T {K} 
\end{equation}
We will now focus on finding the relation between the interaction coefficients $K_i$ and $L_i$. Using the equation \eqref{xirelation} and substituting the values (+1, +1, +1) and (+1, +1, -1) for $(\sigma_i, \sigma_j, \sigma_k)$ respectively, we get the star-triangle relation
\begin{subequations}\label{sıralı}
    \begin{equation}
        2\cosh(L_1+L_2+L_3)= R\exp(K_1+ K_2+ K_3)
    \end{equation}
    \begin{equation}
        2\cosh(L_1+L_2-L_3)= R\exp(-K_1- K_2+ K_3)
    \end{equation}
\end{subequations}
Eliminating $K_3$ and $R$ between \eqref{sıralı}, we get the following relation 
\begin{equation}\label{star-triangle relation}
    \exp(2K_1+ 2K_2) = \frac{cosh(L_1+L_2+L_3)}{cosh(L_1+L_2-L_3)}
\end{equation}
By permuting the suffix $1,2,3$, two other equations can be obtained.
\medskip
\subsubsection{The Kagome Lattice}
Now we will introduce a new lattice, namely the Kagome Lattice. 
\begin{figure}[h]
    \centering
    \includegraphics[width=0.5\textwidth]{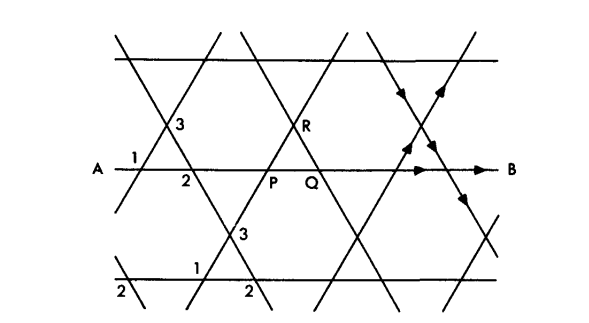}
    \caption{The Kagome Lattice}
    \label{kagome}
\end{figure}
We will put arrows on the edges, and allow only an even number of arrows pointing to each site. At each site $i$ we have eight possible configurations of arrows. Moreover, to each arrangement $j$, we assign an energy $\varepsilon_{ij}$ and the Boltzmann weight in the following manner

\begin{equation}
    \omega_{ij} = \exp{\left(\frac{\varepsilon_{ij}}{k_\beta T}\right)}
\end{equation}
\medskip
Then we can express our partition function
 \begin{equation}\label{partition1}
     Z = \sum_C \prod_i \omega_{i,j(i,C)}
 \end{equation}
where the sum is over all configurations $C$, the product is over all sites $i$, and $j(i,C)$ is the arrow arrangement at site $i$ for configuration $C$.
\\
Consider the arrow configurations on two types of interactions in the figure below. For $a_1$ in the figure, there are two different configurations; one of them consists of reversed arrows of the first one. In the zero-field case, We have the same weight for $w_{i1} $ and $w_{i2} $.
\begin{figure}[h]
    \centering
    \includegraphics[width=0.4\textwidth]{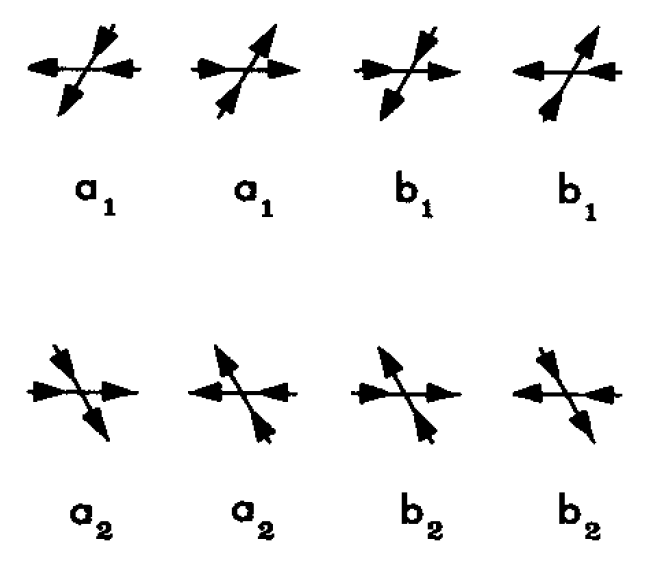}
    \caption{Arrow configurations}
    \label{arrows}
\end{figure}
\begin{figure}[h]
    \centering
    \includegraphics[width=0.8\textwidth]{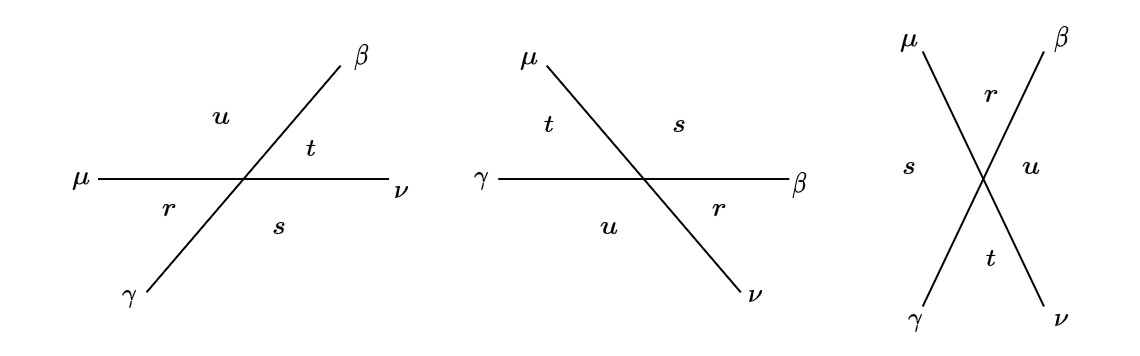}
    \caption{Three types of interaction are shown below}
    \label{interactions}
\end{figure}
\\
Here, let us define $w_i(\mu, \gamma | \beta, \nu)$ as the Boltzmann weight of the corresponding arrow configuration. The partition function can then be defined as

\begin{equation}
    Z = \sum_\alpha \prod_i w_i(\alpha_\mu, \alpha_\gamma | \alpha_\beta, \alpha_\nu)
\end{equation}
Where the sum is over all choices of $\alpha$ of the arrow spins, and the product is over all sites; for each site, the symbol $i$ denotes its type.
\\
Now let us look at two types of triangles with one of them pointing down and the other one pointing up.

\begin{figure}[h]
    \centering
    \includegraphics[width=0.8\textwidth]{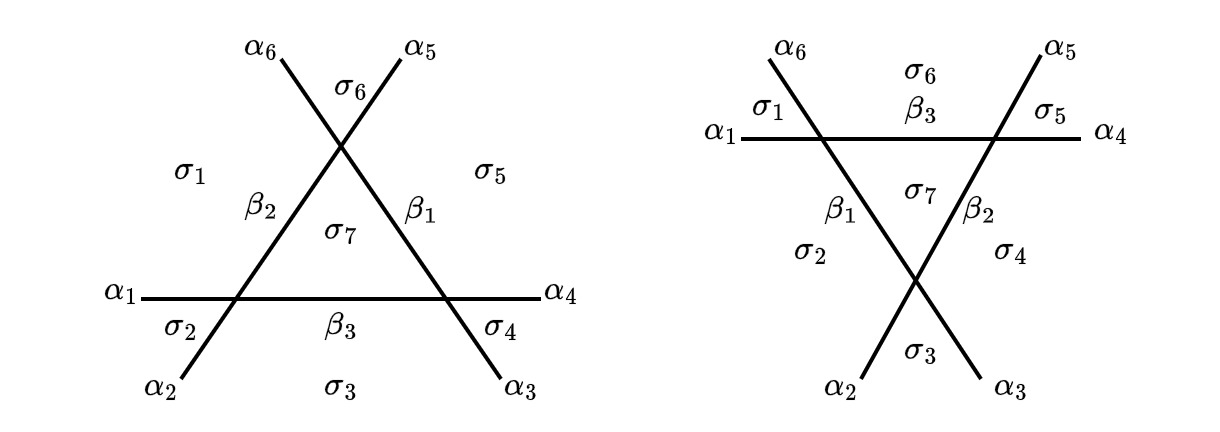}
    \caption{The two types of triangles in the Kagome Lattice}
\end{figure}
\begin{figure}[h] 
    \centering
    \includegraphics[width=0.8\textwidth]{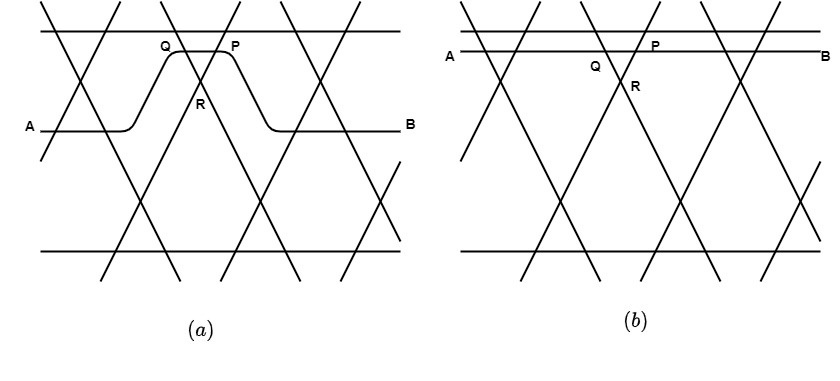}
    \caption{The line AB shifted above R, and following the same procedure the line AB shifted up a complete row.}
    \label{pqr}
\end{figure}
\begin{figure}[h] 
    \centering
    \includegraphics[width=1\textwidth]{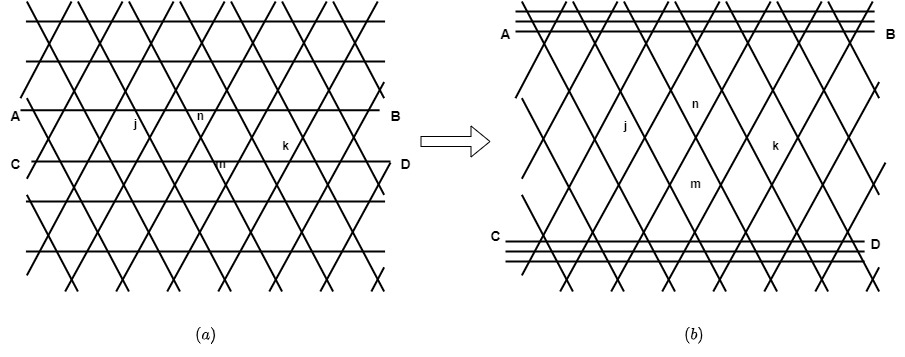}
    \caption{The correlations in the central row such as $\langle \alpha_j \alpha_k\rangle$ and $\langle \sigma_m \sigma_n \rangle $ }
    \label{lineab}
\end{figure}

Clearly, the weights of the pointing-up triangle and the pointing-down triangle are respectively

\begin{subequations}
    \begin{equation}
        w_1(\alpha_1, \alpha_2 | \beta_2,\beta_3) w_3(\alpha_6, \beta_2| \alpha_5, \beta_1) w_2(\beta_1, \beta_3 | \alpha_4, \alpha_3) 
    \end{equation}
    \begin{equation}
         w_2(\alpha_6, \alpha_1 | \beta_3,\beta_1) w_3(\beta_1, \alpha_2| \beta_2, \alpha_3) w_1(\beta_3, \beta_2 | \alpha_5, \alpha_4)  
    \end{equation}
\end{subequations}
We say that the Kagome Lattice is solvable if weights are the same for both types of triangle, i.e. if 
\begin{equation} \label{condition}
    \begin{split}   
    & 
    \sum_{\beta_1, \beta_2, \beta_3}  w_1(\alpha_1, \alpha_2 | \beta_2,\beta_3) w_3(\alpha_6, \beta_2| \alpha_5, \beta_1) w_2(\beta_1, \beta_3 | \alpha_4, \alpha_3)
    \\& = 
    \sum_{\beta_1, \beta_2, \beta_3} w_2(\alpha_6, \alpha_1 | \beta_3,\beta_1) w_3(\beta_1, \alpha_2| \beta_2, \alpha_3) w_1(\beta_3, \beta_2 | \alpha_5, \alpha_4)
\end{split}
\end{equation}
Let's consider any pointing-up triangle in the Kagome lattice, e.g. the triangle $PQR$ in Figure \ref{kagome}. The contribution of this triangle to the partition function that is summed over arrow spins on internal edges is the left-hand side of \eqref{condition}.
\\
By the solvability condition we stated above, nothing would change if we replace this contribution with the right-hand side of \eqref{condition}; thus, the partition function remains unchanged as the horizontal line $AB$ has been shifted above the site R, as in the Figure \ref{pqr}. The site P is still the intersection of AB with the SW-NE line, and still has the weight function $w_1$. The rest is similar. Furthermore, this procedure not only leaves the partition function $Z$ unchanged; but it also leaves any correlation unchanged.
\newpage
Repeating the same procedure for the horizontal line above $AB$, then for $AB$ itself results in Figure \ref{lineab}.
\\
This leaves the spins at sites ($i, j, m, n$) unaffected. 

\subsubsection{Back to The Honeycomb Lattice}

\begin{figure}[h]
    \centering
    \includegraphics[width=0.8\textwidth]{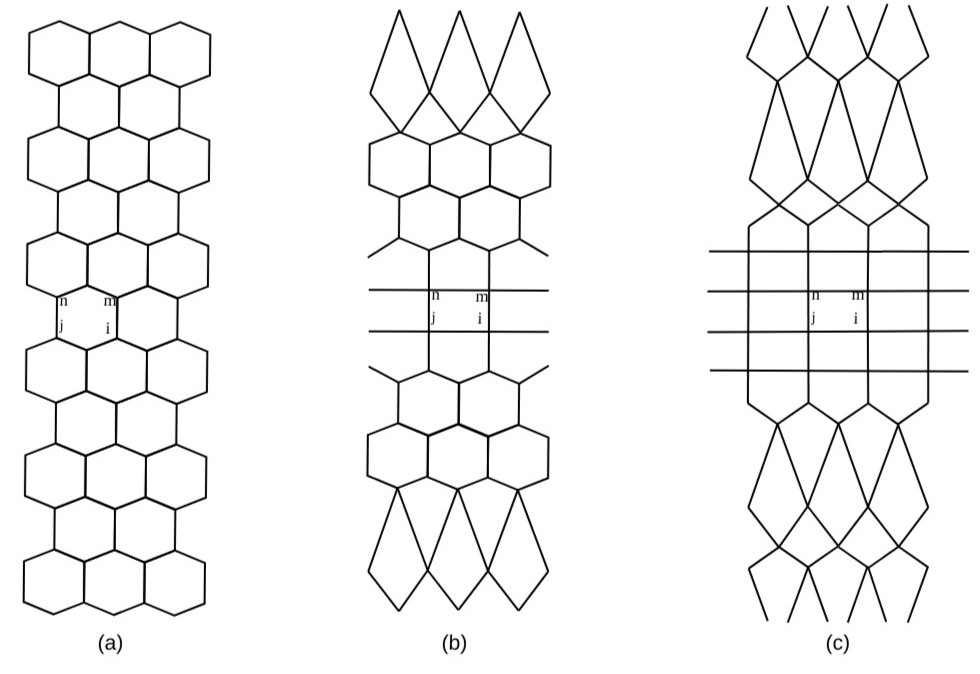}
    \caption{The effect of repeated star-to-triangle and triangle-to-star transformations on the honeycomb lattice}
    \label{shiftingeffect}
\end{figure}

The same transformation can be done in the honeycomb lattice, by converting the lattice consisting of the dotted lines in $(a)$ to $(c)$ as in Figure \ref{shiftingeffect}.
\\
Let $2M$ be the number of horizontal lines. Then after shifting, each region above and below $R$ contains $M$ horizontal lines. In the limit of large $M$, edges $j$ and $k$ lie deep inside the square lattice region. Therefore, the correlation $\langle \sigma_j  \sigma_m \rangle$ is that of the usual square lattice drawn diagonally.
The result is the lattice shown in Figure \ref{shiftingeffect}(c). It contains a central square lattice region, with interaction coefficient $L_3$ for all vertical edges, and $K_3$ for all horizontal edges. Since $L_1, L_2, K_1, K_2$ are all real, the boundary condition only introduces positive weights, and in the thermodynamic limit, it cannot affect even-spin correlations within the square region.
\\
The two-spin correlations between $\sigma_i, \sigma_j, \sigma_m, \sigma_n$ must therefore be those of the square lattice. In particular
\begin{equation}\label{corr1}
    \langle \sigma_i \,  \sigma_j \rangle = g(K_3, L_3)
\end{equation}
\begin{equation}\label{corr2}
    \langle \sigma_i \, \sigma_m \rangle = g(L_3, K_3)
\end{equation}
where $g(K,L)$ is the correlation function.
\\
Thus, we have reduced the first and second neighbor correlations of the honeycomb lattice from functions of three variables to two variables, namely, $K_3$ and $L_3$.

\subsection{Local relation between honeycomb first and second-neighbour correlations}

Let $P(\alpha, \beta, \gamma)$ be the probability of the spins $\sigma_i, \sigma_j, \sigma_k$ in the \ref{figure 1}; and $P(\delta | \alpha, \beta, \gamma) $ to be the probability of spin $\sigma_l$ to have the value $\delta$ while $\sigma_i, \sigma_j, \sigma_k$ have values $ \alpha , \beta, \gamma$ respectively.
\begin{equation}\label{probabilityrelation}
 P(\alpha, \beta, \gamma, \delta)  = P(\delta | \alpha, \beta, \gamma) P(\alpha, \beta, \gamma) 
 \end{equation}
For the $ P(\delta | \alpha, \beta, \gamma) $, with $\sigma_i$, $\sigma_j$, $\sigma_k$ fixed, the spin $\sigma_l$ will be isolated from the rest of the honeycomb lattice. So, it only sees the magnetic field caused by $\sigma_i$, $\sigma_j$, $\sigma_k$, which equals to $k_B T(L_1\alpha + L_2 \beta+ L_3 \gamma)$.
\pagebreak
 
$$ P(\delta | \alpha, \beta, \gamma) = \frac{\exp(\delta(L_1\alpha + L_2 \beta+ L_3 \gamma))}{Z} = \frac{\cosh(L_1\alpha + L_2 \beta+ L_3 \gamma) + \delta \sinh(L_1\alpha + L_2 \beta+ L_3 \gamma)}{2 \cosh(L_1\alpha + L_2 \beta+ L_3 \gamma)} $$

\begin{equation}
    = \frac{1}{2} (1+\delta \tanh(L_1\alpha + L_2 \beta+ L_3 \gamma))
\end{equation}

Since $\alpha$, $\beta$ and $\gamma$ can take the value $\pm 1$, we can write this probability as follows

\begin{equation}\label{probiwthw}
 P(\delta | \alpha, \beta, \gamma)=\frac{1}{2} ( 1 + \delta (w_1 \alpha + w_2 \beta +w_3 \gamma -w\alpha \beta \gamma))
\end{equation}
By multiplying by $\alpha \beta \gamma \delta$, $\alpha \delta$, $\beta \delta$,  $\gamma \delta$ and summing over $\alpha$, $\beta$, $\gamma$, $\delta$, we get 4 different equations for the 4 unknowns in \eqref{probiwthw}.

\begin{equation} \label{wmatrix}
    \begin{pmatrix}
    w \\ w_1 \\w_2 \\ w_3
\end{pmatrix}
= \frac{1}{4}
\begin{pmatrix}
-1 & 1 & 1 & -1 \\
1 & 1 & 1 & 1 \\
-1 & 1 & -1 & 1 \\
-1 & -1 & 1 & 1 
\end{pmatrix}
\begin{pmatrix}
    \tanh(L_1-L_2-L_3) \\ \tanh(L_1+L_2-L_3) \\ \tanh(L_1-L_2+L_3) \\ \tanh(L_1+L_2+L_3) 
\end{pmatrix}
\end{equation}
\medskip
If we solve for the $w$
\begin{equation}
        w= \frac{1}{4}(- \tanh(L_1-L_2-L_3)+\tanh(L_1+L_2-L_3)+ \tanh(L_1-L_2+L_3)-\tanh(L_1+L_2+L_3)) 
\end{equation}
$$
=\frac{1}{4} \sinh{2L_2}\left( \frac{1}{c_1 c_3}-\frac{1}{c_2c} \right ) = \frac{1}{4} \sinh{2L_2}\left (\frac{c_2c-c_1c_3}{cc_1c_2c_3} \right)
$$
where $c=\cosh(L_1+L_2+L_3)$ and $ c_i=\cosh(-L_i+L_j+L_k)$.
Using the properties of the hyperbolic trigonometric functions, we obtain the following

\begin{equation}\label{wfirst}
    w = \frac{1}{4} \frac{\sinh{2L_1} \sinh{2L_2}  \sinh{2L_3} }{cc_1c_2c_3}
\end{equation}
To get a simpler equation, we need to derive some relations using the star-triangle equation. So we will start using \eqref{sıralı}
\begin{equation} \label{e4k}
    \frac{cc_1}{c_2c_3} = \exp(4K_1)
\end{equation}
\begin{equation}
    \exp(4K_1) -1 = \sinh{2L_2}\sinh{2L_3}/c_2c_3
\end{equation}
$$
\exp(2K_1)[\exp(2K_1) - \exp(-2K_1)]= \sinh{2L_2}\sinh{2L_3}/c_2c_3
$$
\begin{equation}\label{sinh2k12l1}
    2\sinh{2K_1} = \frac{\sinh{2L_2}\sinh{2L_3}}{c_2 c_3 \sqrt{cc_1/c_2c_3}}
\end{equation}
$$\sinh{2K_1} \sinh{2L_1} = \frac{\sinh{2L_1}\sinh{2L_2}\sinh{2L_3}}{2\sqrt{cc_1c_2c_3}}$$
With the help of the above equations and \eqref{sıralı}, $R$ can be found to be
$$ R^2 = 4\sqrt{cc_1c_2c_3}   $$
\begin{equation}\label{R}
    R^2 = 2k \sinh{2L_1}\sinh{2L_2}\sinh{2L_3}
\end{equation}
As one can see, the right-hand side of the equation \eqref{sinh2k12l1} is symmetric on $L_1$, $L_2$, and $L_3$, so the permutation of the suffixes in the left-hand side will be equal to each other.
\begin{equation}\label{overk} 
    \sinh{2K_i}\sinh{2L_i} = \frac{\sinh{2L_1}\sinh{2L_2}\sinh{2L_3}}{2(cc_1c_2c_3)^{1/2}}
\end{equation}
Then we have
$$ \sinh{2K_1}\sinh{2L_1} = \sinh{2K_2}\sinh{2L_2} = \sinh{2K_3}\sinh{2L_3}    $$
Since we have a value that is independent of the suffixes, let's define $k$ as
\begin{equation}\label{kdefn}
    \sinh{2K_i}\sinh{2L_i} = k^{-1}
\end{equation}
Now we can return to our calculation for the probability equation. Using \eqref{overk} in \eqref{wfirst} we can write the equation for $w$ as follows
\begin{equation}\label{w}
w= \frac{1}{4}\frac{\sinh{2L_1} \sinh{2L_2}  \sinh{2L_3} }{cc_1c_2c_3}= \frac{1}{2} \frac{\sinh{2K_1} \sinh{2L_
1}}{\sqrt{cc_1c_2c_3}}= \frac{\sinh{2K_1} \sinh{2K_2}}{\sinh{2L_3}}
\end{equation}
For the rest of the unknowns in \eqref{wmatrix}, it will be sufficient to show the derivation of $w_1$, since the others can be calculated with the same procedure.
\begin{equation*}
        w_1= \frac{1}{4}( \tanh(L_1-L_2-L_3)+\tanh(L_1+L_2-L_3)+ \tanh(L_1-L_2+L_3)+\tanh(L_1+L_2+L_3))
\end{equation*}

$$
w_1 = \frac{\sinh{2L_1}}{4}\left(\frac{1}{cc_1} + \frac{1}{c_2 c_3}  \right) =  \frac{\sinh{2L_1}}{4} \frac{\cosh{2L_1} + \cosh{2L_2}\cosh{2L_3}}{cc_1c_2c_3}
$$
For simplicity, we multiply and divide with $\sinh{2K_1}$, so we can use \eqref{w}
$$
w_1= \frac{\sinh{2K_1} \sinh{2L_
1}}{2\sqrt{cc_1c_2c_3}} \frac{\cosh{2L_1} + \cosh{2L_2}\cosh{2L_3}}{2\sinh{2K_1} \sqrt{cc_1c_2c_3}} =w \frac{\cosh{2L_1} + \cosh{2L_2}\cosh{2L_3}}{2\sinh{2K_1} \sqrt{cc_1c_2c_3}} 
$$
\\
We derive the following using \eqref{overk} and \eqref{e4k}
\begin{equation}
w_1= w \frac{\cosh{2L_1} + \cosh{2L_2}\cosh{2L_3}}{\sinh{2L_2 \sinh{2L_3}}} = w \coth{2K_1}
\end{equation}
\\
When the simplification is done for the $w_2$ and $w_3$, we find the following equation (where $r =1, \, 2,\, 3$)
\begin{equation}\label{otherw}
    w_r=w \coth{2K_r}
\end{equation}
Now, if we use \eqref{probabilityrelation} in \eqref{corr2}, then multiply both sides by $ \gamma \delta $, we get
$$ \gamma \delta  P(\alpha, \beta, \gamma, \delta) = \frac{1}{2} ( 1 +  (w_1 \alpha \gamma + w_2 \beta \gamma +w_3 -w\alpha \beta )) P(\alpha, \beta, \gamma) $$
Summing over $\alpha$, $\beta$, $\gamma$, $\delta$, we get
 \begin{equation}\label{15}
     \langle \sigma_k \sigma_l \rangle = w_1 \langle \sigma_i \sigma_k \rangle + w_2 \langle \sigma_j \sigma_k \rangle +w_3 - w \langle \sigma_i \sigma_j \rangle
 \end{equation}
by using the equation for the correlation function
$$ \langle \sigma_m \sigma _n \rangle = Z^{-1}\sum _ {\sigma} \sigma_m \sigma_n \exp(L\Sigma \sigma_i \sigma_j)   $$
The equation \eqref{15} shows the linear relation between first and second neighbor correlations.

\subsection{Functional equation for the correlations}

Combining the results in \eqref{15}, \eqref{corr1}, and \eqref{corr2} we get

\begin{equation}\label{26}
    g(L_3, K_3) = w_1 g(K_2, L_2) + w_2 g(K_1, L_1) +w_3 -wg(K_3, L_3)
\end{equation}
By negating alternating rows and columns on a square we can derive a relation as below
\begin{equation}
    g(K,L) = -g(-K,L) = g(K,-L)=-g(-K,-L)
\end{equation}
Let's look at the spin correlations of $\sigma_i$ and $\sigma_j$ to derive this relationship.
$$
\langle \sigma_i \sigma_j \rangle = \sum _ {\sigma} \sigma_i \sigma_j \exp(K \Sigma \sigma_i \sigma_j + L \Sigma \sigma_i \sigma_{m})
$$
First, substitute $-K$ for $K$; and to make it easier, just sum over the spins $\sigma_i$ and $\sigma_j$ in both situations.
$$
g(K,L) = + \exp(K+L\sigma_m)- \exp(-K+L\sigma_m) - \exp(-K-L\sigma_m) + \exp(K-L\sigma_m)
$$
$$
g(-K,L)= + \exp(-K+L\sigma_m)- \exp(K+L\sigma_m) - \exp(K-L\sigma_m) + \exp(-K-L\sigma_m)
$$
Now it is easy to see that $g(K,L)= -g(-K,L)$
\\
To find the relation for $L$, let's calculate the correlation for $-L$ instead of $L$. In this calculation, we also need to show the sum over $\sigma_m$.
$$
g(K,L)= + \exp(K+L) + \exp(K-L)- \exp(-K+L) - \exp(-K-L) $$
$$- \exp(-K-L) - \exp(-K+L) + \exp(K-L) + \exp(K+L)
$$
It is obvious to see the equality written below
\begin{equation}
    g(K,L)= g(K,-L)
\end{equation}
\\
Instead of \eqref{corr1}, we can use a function $f(K,k)$, defined in terms of
\begin{equation}\label{ffunc}
    g(K,L) = \coth{2K} f(K,k)
\end{equation}
\\
Here we will call $K$ the argument, and $k$ the modulus of $f(K,k)$.
\\
Eliminating $L_1$ and $L_2$ between \eqref{sıralı}, we get

\begin{equation} \label{identity}
        \cosh{2K_1} \cosh{2K_2} \sinh{2K_3} + \sinh{2K_1} \sinh{2K_2} \cosh{2K_3}  = \sinh{2K_3} \cos{2L_3}
\end{equation}
Substituting \eqref{ffunc} into the equation \eqref{26}, using the \eqref{w} and \eqref{otherw}, we get
\begin{multline}   
    \coth{2L_3}f(L_3, k) = \frac{\cosh{2K_1}\cosh{2K_2}}{\sinh{2L_3}}f(K_2,k) + \frac{cosh{2K_2}\cosh{2K_1}}{\sinh{2L_3}}f(K_1,k)+ 
    \\
    \frac{\sinh{2K_1}\sinh{2K_2}\cosh{2K_3}}{k}f(K_3,k) - \frac{\sinh{2K_1}\sinh{2K_2} \cosh{2K_3}}{k}
\end{multline}

\begin{equation}
\begin{split}
    f(K_1,k)+&f(K_2,k)-f(K_3,k)\left ( \frac{\sinh{2L_3}\sinh{2K_1}\sinh{2K_2}\cosh{2K_3}}{\cosh{2K_1}\cosh{2K_2} k} \right) = \\&
\frac{\cosh{2L_3}}{\cosh{2K_1}\cosh{2K_2}} f(L_3,k) - \frac{\sinh{2L_3}\sinh{2K_1}\sinh{2K_2}\cosh{2K_3}}{k \,\cosh{2K_1}\cosh{2K_2}}
\end{split}
\end{equation}

If we use \eqref{identity} in the equation above we get

$$
    f(K_1,k)+f(K_2,k)+f(K_3,k) -1 = \frac{\cosh{2L_3}}{\cosh{2K_1}\cosh{2K_2}} \left (f(K_3,k)+f(L_3,k) - 1 \right)
$$
Although this is a neat equation, to be able to insert a variable that is symmetric in $K_1$, $K_2$ and $K_3$, we rewrite it as 
\begin{equation}\label{bsymmetric}
    k^{-1}b \sech{2K_1} \sech{2K_2} \sech{2K_3} = f(K_1,k)+ f(K_2,k)+ f(K_3,k) -1
\end{equation}
where $b$ is defined as

\begin{equation}\label{beq}
    b = \coth{2K_3} \coth{2L_3} (f(K_3,k) + f(L_3,k) -1)
\end{equation}
Eliminating $L_3$ from the equation \eqref{identity}, we can derive another equation for $k$. First, we take the square of this identity

$$
 \frac{\scalebox{0.7}{$\cosh^2{2K_1}\cosh^2{2K_2}\sinh^2{2K_3} + \sinh^2{2K_1}\sinh^2{2K_2}\cosh^2{2K_3}+ 2\cosh{2K_1}\cosh{2K_2}\cosh{2K_3}\sinh{2K_1}\sinh{2K_2}\sinh{2K_3} $} }{\scalebox{0.7}{$\sinh^2{2K_3}$}} = \cosh^2{2L_3}
$$
Since we need to eliminate $L_3$, we can write $1+ \frac{1}{k^2 \sinh^2{2K_3}}$ instead of $\cosh^2{2L_3}$, by using the equation for $k$ and properties of hyperbolic trigonometric functions. Then the equation for $k$ can be written as follows

$$
k^2= \left( \scalebox{0.65}{$\cosh^2{2K_1}\cosh^2{2K_2}\sinh^2{2K_3} + \sinh^2{2K_1}\sinh^2{2K_2}\cosh^2{2K_3}+ 2\cosh{2K_1}\cosh{2K_2}\cosh{2K_3}\sinh{2K_1}\sinh{2K_2}\sinh{2K_3} - \sinh^2{2K_3}$} \right)^{-1}  
$$
Let us call the denominator $A$, and look at each term separately at first.
\bigskip

$ 
\scalebox{0.8}{$1^{\text{st}} \,\, \text{term} \xrightarrow{ } \cosh^2{2K_1}\cosh^2{2K_2}\sinh^2{2K_3} = \cosh^2{2K_1}\cosh^2{2K_2}\cosh^2{2K_3} \tanh^2{2K_3}$}
$

$ \scalebox{0.8}{$2^{\text{nd}} \,\, \text{term} \xrightarrow{ } \sinh^2{2K_1}\sinh^2{2K_2}\cosh^2{2K_3} = \cosh^2{2K_1}\cosh^2{2K_2}\cosh^2{2K_3} \tanh^2{2K_1}\tanh^2{2K_2}$}    $

$
\scalebox{0.7}{$3^{\text{rd}} \,\, \text{term} \xrightarrow{ } 2\cosh{2K_1}\cosh{2K_2}\cosh{2K_3}\sinh{2K_1}\sinh{2K_2}\sinh{2K_3} = 2\cosh^2{2K_1}\cosh^2{2K_2}\cosh^2{2K_3}\tanh{2K_1}\tanh{2K_2}\tanh{2K_3}$}
$
\\
Then summing them all, $A$ can be written as
\begin{equation}
\begin{split}
    A= \cosh^2{2K_1}\cosh^2{2K_2}\cosh^2{2K_3}& ( \tanh^2{2K_3} + \tanh^2{2K_1}\tanh^2{2K_2} +\\& 2\tanh{2K_1}\tanh{2K_2}\tanh{2K_3}) - \sinh^2{2K_3}
\end{split}
\end{equation}
We derive the following using the properties of hyperbolic trigonometric functions

\begin{equation}
    A = \frac{\tanh^2{2K_3} + \tanh^2{2K_2} + 2\tanh{2K_1}\tanh{2K_2}\tanh{2K_3}  }{(1-\tanh^2{2K_1})(1-\tanh^2{2K_2}) (1-\tanh^2{2K_3})} - \frac{\tanh^2{2K_3}}{1- \tanh^2{2K_3}} 
\end{equation}

Denoting $v_r = \tanh{K_r}$ and inserting this into the equation above, we get
$$
A= \frac{16v_1v_2v_3(1+v_1^2)(1+v_2^2)(1+v_3^2) + 16v_1^2v_2^2(1+v_3^2)^2+4v_3^2(1+v_1^2)^2(1+v_2^2)^2}{(1-v_1^2)^2(1-v_2^2)^2(1-v_3^2)^2} - \frac{4v_3^2}{(1-v_3^2)^2}$$
Remembering that $k=\sqrt{\frac{1}{A}}$ and after simplifying $A$, we have

\begin{equation}
    k= \frac{(1-v_1^2)(1-v_2^2)(1-v_3^2)}{4 [(1+v_1v_2v_3)(v_1+ v_2v_3)(v_2+v_1v_3)(v_3+v_1v_2) ]^{1/2}}
\end{equation}

\subsection{Solution of the Functional Equation}

So far, we have derived some relations that will be essential for the exact solution of the honeycomb lattice. Now we will start the solution by defining some functions. From \eqref{kdefn} we can write $L_3$ as a function of $K_3$, and $k$. So with \eqref{beq} we can write $b $ as
\begin{equation}
    b = b(K_3,k)
\end{equation}
Even though in \eqref{beq}, $b(K_3, k)$ is written in terms of $K_3$ and $k$, the rest of the terms in \eqref{bsymmetric} are symmetric in $K_1$, $K_2$, $K_3$, then $b$ is a symmetric function of $K_1, K_2$, and $K_3$ i.e.
\begin{equation}\label{bKsym}
    b(K_2, k) = b(K_3, k)
\end{equation}
However, $K_3$ and $K_2$ are independent variables, which means that the function  $b$ cannot depend on $K_3$ and $K_2$ separately. Thus $b$ cannot depend on its argument. 
\begin{equation}\label{lastb}
    b(K,k) = b(k)
\end{equation}
Now we will differentiate \eqref{bsymmetric} along a line in $(K_1,K_2,K_3)$ space where $K_3$, $k$, and hence $L_3$, are fixed.
\\
The derivative of \eqref{bsymmetric} with respect to $K_1$ and $K_2$ is equal since \eqref{bsymmetric} is symmetric with respect to $K_1$ and $K_2$. Taking the derivative with respect to $K_1$

\begin{equation}
    k^{-1}b \left( -\frac{2 \sinh{2K_1}}{\cosh^2{2K_1}}\right) \sech{2K_2}\sech{2K_3}  = \frac{\partial f(K_1,k)}{\partial K_1} 
\end{equation}

Then we can use $\sinh{2K_1}\sinh{2L_1}$ instead of $k^{-1} $. We also need to multiply and divide the term includes $\tanh^2{2K_1}$ with $\sinh{2K_1}$.

\begin{equation}
    -2b\tanh^2{2K_1} -2b\tanh^2{2K_1}\frac{\cosh{2K_1} \sinh{2K_1} \sinh{2K_3}}{\sinh{2K_1}\cosh{2K_2}\cosh{2K_3}} = \coth{2L_1} \frac{\partial f(K_1,k)}{\partial K_1}
\end{equation}

Finally, we get
\begin{equation}
    -b\tanh{2K_1}\tanh{2K_2}\tanh{2K_3} = b\tanh^2{2K_1} + \frac{1}{2}\coth{2L_1} \frac{\partial f(K_1,k)}{\partial K_1}
\end{equation}
The left-hand side will be denoted as $a(K_1,k)$.

\begin{equation}
    a(K_1,k) = b\tanh^2{2K_1} + \frac{1}{2}\coth{2L_1}f'(K_1,k)
\end{equation}

Since we have stated that the derivative on $K_1$ and $K_2$ must be equal, i.e.
\begin{equation}\label{adefn}
    a(K_1,k) = a(K_2,k) 
\end{equation}
Just as \eqref{bKsym} implies \eqref{lastb}, so does \eqref{adefn} implies

\begin{equation}\label{lasta}
    a(K,k) = a(k)
\end{equation}
Then
\begin{equation}\label{aseconddefn}
    a(k) = b\tanh^2{2K_r} + \frac{1}{2}\coth{2L_r}f'(K_r,k)
\end{equation}
Now we can extract the derivative of $f(K,k)$ with respect to $K$ i.e. $f'(K,k)$ using\eqref{aseconddefn}
$$ f'(K,k) = 2 \frac{a(k) - b(k) \tanh^2{2K}}{\coth{2L}}  $$
We will write it as a function of $K$ and $k$ by using \eqref{kdefn}

$$ \coth^2{2L} = \frac{1+\sinh^2{2L}}{\sinh^2{2L}} = 1+ k^2\sinh^2{2K}  $$

\begin{equation}\label{fprime}
    f'(K,k) = 2\frac{a(k) - b(k)\tanh^2{2K} }{\sqrt{1+k^2\sinh^2{2K}}}
\end{equation}
We know that the correlation function $g(K,L)$ must be bounded, the \eqref{ffunc} results in $f(0,k) = 0$. Integrating \eqref{fprime} therefore gives, $0 \leqslant K \leq \mathcal{\infty}$.

\begin{equation}\label{ffuncab}
    f(K,k) = a(k) A(K,k) - b(k)B(K,k)
\end{equation}
where
\begin{subequations}\label{integrals}
    \begin{equation}
        A(K,k) = \int_0^{2K} \frac{dx}{\sqrt{1+k^2\sinh^2{2x}}}
    \end{equation}
    \begin{equation}
        B(K,k) = \int_0^{2K} \frac{\tanh^2{x}}{\sqrt{1+k^2\sinh^2{2x}}}
    \end{equation}
\end{subequations}
Hence for any given $k$, the function $f$ is a linear combination of the functions $A(K,k)$, and $B(K,k)$ which will be calculated in the next section using elliptic integrals.
\\
Knowing that the correlation function is bounded, $L_3 \xrightarrow{} \mathcal{\infty}$ as $K_3\xrightarrow{} 0$ for fixed $k$ from equations \eqref{beq} and \eqref{lastb}; and it follows that $f(\mathcal{\infty}, k) = 1$. Then we can deduce that $a(k)$, and $b(k)$ satisfy the linear relation
\begin{equation} \label{eq1}
    a(k)A(\mathcal{\infty},k) -b(k) B(\mathcal{\infty},k) = 1
\end{equation}

\subsection{Free Energy for Kagome Lattice}

Before we start to find the solution for $a(k)$, and $b(k)$, we need to use some thermodynamic equations in which we can see the relation between the internal energy and the correlation function. The free energy for the Kagome Lattice can be split into two parts as
\begin{equation}\label{free}
    F_{KG} = F_{SQ} + 2F_{FR}
\end{equation}
Here $F_{KG}$ is the free energy for the entire lattice, $F_{SQ}$ is the free energy of the central region, and $F_{FR}$ is the free energy of the upper or lower region. The free energy per site can be written as
\begin{equation}\label{totalfree}
    f_r = - \frac{1}{N} k_B T \ln{Z}
\end{equation}
We pointed out that our model has three types of interaction each of whom has N sites as shown in \eqref{interactions}. Hence in the central region, we have $N$ sites that have the interaction of type 3. In the upper and lower regions, there are $\frac{N}{2}$ sites with type 1 and type 2 interaction.
\\
For the Kagome Lattice, we had the arrow arrangements, but we can formulate it in terms of magnetic spins on faces. If we change one spin then two arrows would change direction. Hence \eqref{partition1} can be rewritten as
\begin{equation}\label{partititon2}
    Z = \frac{1}{2}\sum_{\sigma} \prod_i \omega_{i,j(i\sigma)}
\end{equation}
The partition function of our lattice is now
\begin{equation}
    Z = \frac{1}{2} Z_H (L_1,L_2,L_3) Z_T (K_1,K_2,K_3) 
\end{equation}
Putting \eqref{relationbetweenlattices} in this equation
\begin{equation}
    2Z = R^N Z_T^2 (K_1,K_2,K_3)
\end{equation}
Taking the logarithm of both sides
\begin{equation}
    \ln{2} + \ln{Z} = N \ln{R} + 2 \ln{Z_T}
\end{equation}
Substituting \eqref{totalfree} into this equation
\begin{equation}
    \ln{2} - \frac{N(f_1 + f_2 + f_3)}{k_B T} - N \ln{R} = 2 \ln{Z_T}
\end{equation}
Defining
\begin{equation}\label{psifr}
    \psi = \frac{f_r}{k_B T} = - \lim_{N \xrightarrow{} \infty} N^{-1} \ln{Z}
\end{equation}
Then we have
\begin{equation}
    \psi_T(L_1, L_2, L_3) = \frac{1}{2} [ \ln{R} + \psi_{SQ}(K_1, L_1) + \psi_{SQ}(K_2, L_2) + \psi_{SQ}(K_3, L_3) ]
\end{equation}
For the triangular lattice, we had $N$ sites, so for the honeycomb lattice we have 2$N$ sites. Hence the free energy per site must be divided by 2 for the honeycomb lattice, and using \eqref{relationbetweenlattices}
\begin{equation}
    \psi_H (K_1, K_2, K_3) = \frac{1}{4} [- \ln{R} + \psi_{SQ}(K_1, L_1) + \psi_{SQ}(K_2, L_2) + \psi_{SQ}(K_3, L_3) ]
\end{equation}
\begin{equation}
    -\psi_H + \frac{1}{2}\psi_T  = \frac{1}{2}\ln{R}
\end{equation}
So we have
\begin{equation}
    \psi_H (K_1, K_2, K_3) = \frac{1}{2}[-\ln{R} + \psi_T(L_1,L_2,L_3)  ]
\end{equation}
The correlation functions are derivatives of $\psi$'s.

\begin{equation}
    \frac{\partial \psi_H}{\partial K_3} =  \frac{1}{2} \left[- \frac{\partial \ln{R}}{\partial K_3} + \frac{\psi_T}{\partial L_1} \frac{\partial L_1}{\partial K_3} + \frac{\psi_T}{\partial L_2} \frac{\partial L_2}{\partial K_3} + \frac{\psi_T}{\partial L_3} \frac{\partial L_3}{\partial K_3} \right]
\end{equation}
\begin{equation}
        g(K_3,L_3) = \frac{1}{2} \left[ -\frac{\partial \ln{R}}{\partial K_3} + g(K_1,L_1) \frac{\partial L_1}{\partial K_3} + g(K_2,L_2) \frac{\partial L_2}{\partial K_3}  + g(K_3,L_3) \frac{\partial L_3}{\partial K_3}  \right]
\end{equation}
\\
So we have the set of equations
\begin{equation}
\begin{split}
    &\frac{\partial \ln{R}}{\partial K_3} = 2\omega_3 \,\,\,\,\,\,\,\,\,\,\,\,\,\,\,\,\,\,\,\,\,\,\,\,\,\,\,\,\,\,\,\,\,\,\, \frac{\partial L_1}{\partial K_3} = 2 \omega_2 \\ &\frac{\partial L_2}{\partial K_3} = 2\omega_1 \,\,\,\,\,\,\,\,\,\,\,\,\,\,\,\,\,\, \,\,\,\,\,\,\,\,\,\,\,\,\,\,\,\,\,\,\,\,\, \frac{\partial L_3}{\partial K_3} = -2\omega
\end{split}
\end{equation}

\subsection{Differential Equations for \texorpdfstring{$a(k)$}{} and \texorpdfstring{$b(k)$}{}}

To determine $a(k)$ and $b(k)$, we note that
\begin{equation}\label{corrwithpsi}
    g(K,L) = -\frac{\partial \psi(K,L)}{\partial K}
\end{equation}
Remember that $\psi(K, L)$ must be a symmetric function of $K$ and $L$. Differentiating \eqref{corrwithpsi} with respect to $L$, and using \eqref{ffunc}
$$
\left ( \frac{\partial g(K,L)}{\partial L} \right)_K = -\frac{\partial\psi (K,L)}{\partial L \partial K} = \left ( \frac{\partial g(L,K)}{\partial K} \right)_L
$$
$$
\coth{2K} \left( \frac{\partial f(K,k)}{\partial k} \right)_K \frac{\partial k}{\partial L} = \coth{2L} \left ( \frac{\partial f(L,k)}{\partial k} \right)_L \frac{\partial k}{\partial K}
$$

$$
2\cosh{2K} \cosh{2L} \left (\frac{\partial f(K,k)}{\partial k} \right)_K = 2\cosh{2K} \cosh{2L} \left( \frac{\partial f(L,k)}{\partial k} \right)_L
$$
So, we reach a symmetry relation on the function $f(K,k)$ as follows
\begin{equation} \label{sym}
    \left (\frac{\partial f(K,k)}{\partial k} \right)_K = \left( \frac{\partial f(L,k)}{\partial k} \right)_L
\end{equation}
The $f(L,k)$ on the right-hand side of equation \eqref{sym} can be written in terms of $f(K,k)$ by using \eqref{beq}, with the suffix 3 removed. Doing this, and using \eqref{kdefn}, \eqref{lastb}, and \eqref{fprime} the relation \eqref{sym}; one can find the derivative of $f(K,k)$ with respect to $k$.

\begin{equation}
    f(L,k) = b\tanh{2K}\tanh{2L} + 1 - f(K,k)
\end{equation}

$$
\frac{\partial f(K,k)}{\partial k} = b' \tanh{2K}\tanh{2L} + 2b\sech^2{2K} \tanh{2L} \frac{\partial K}{\partial k}-\frac{\partial f(K,k)}{\partial K} \frac{\partial K}{\partial k} 
-\frac{\partial f(K,k)}{\partial k}
$$

$$\frac{\partial K}{\partial k} = -\frac{\sinh^2{2K} \sinh{2L}}{2 \cosh{2K}}  $$
Substituting this into the equation above
$$
\frac{\partial f(K,k)}{\partial k} 
=\left( b'(k) \tanh{2K} - b(k)\frac{\sinh^2{2K} \sinh{2L} }{\cosh^3{2K}} + a(k) \frac{\sinh^2{2K} \sinh{2L}}{\cosh{2K}}   \right)\tanh{2L}
-\frac{\partial f(K,k)}{\partial k}
$$
Then multiply both sides with $k$ and use \eqref{kdefn}
$$ 2k \frac{\partial f(K,k)}{\partial k} =  \left( b'(k) k - \frac{b}{\cosh^2{2K}} + a(k) - b \tanh^2{2K}     \right) \tanh{2K} \tanh{2L} $$

$$ = \left( b'(k) k - b(k) + a(k)\right) \tanh{2K} \tanh{2L}    $$

\begin{equation}\label{derivativeoffwtrk}
    k\frac{\partial f(K,k)}{\partial k} = \frac{1}{2}\left( b'(k) k - b(k) + a(k)\right) C(K,k)
\end{equation}
where
\begin{equation}
    C(K,k) = \frac{\tanh{2K}}{\sqrt{1+ k^2 \sinh^2{2K}}}
\end{equation}
Defining $k'^2 = 1- k^2$, we have the following differential equations both of which can be verified by differentiating with respect to $K$

\begin{subequations}
    \begin{equation}\label{diff1}
        k\frac{\partial A(K,k)}{\partial k} = B(K,k) - A(K,k) + C(K,k)
    \end{equation}
    \begin{equation}
        k\frac{\partial }{\partial k} (k'^2 B(K,k)) = k'^2 B(K,k) - A(K,k) + C(K,k)
    \end{equation}
\end{subequations}
For \eqref{diff1}
\begin{equation}
    k \frac{\partial}{\partial k} \left( \frac{\partial A(K,k)}{\partial K}   \right) = \frac{\partial B}{\partial K} - \frac{\partial A}{\partial K} + \frac{\partial C}{\partial K}
\end{equation}

We are going to look $\frac{\partial C}{\partial K}$

$$
\frac{\partial C}{\partial K} = \frac{2\sech^2{2K}\sqrt{1+k^2\sinh^2{2K}}}{1+k^2\sinh^2{2K}}
-\frac{ \frac{1}{2}(1+k^2\sinh^2{2K})^{-\frac{1}{2}} 4k^2 \sinh{2K} \cosh{2K} \tanh{2K}}{1+k^2\sinh^2{2K}} 
$$

$$ \frac{\partial C}{\partial K} = \frac{2 + 2k^2 \sinh^2{2K}- 2k^2\sinh^2{2K}\cosh^2{2K} }{\cosh^2{2K}(1+k^2\sinh^2{2K})^{\frac{3}{2}}}  $$
\\
Then we can proceed to show that equation \eqref{diff1} holds
\begin{multline*}
    k \frac{\partial}{\partial k} \left(\sqrt{1+k^2\sinh^2{2K}} \right )^{-1}   = -\frac{k^2\sinh^2{2K}}{(1+k^2\sinh^2{2K})^{3/2}} =\\ -\frac{2(1+k^2\sinh^2{2K})}{\cosh^2{2K}(1+k^2\sinh^2{2K})^{3/2}} + \frac{2(1+k^2\sinh^2{2K}) - 2k^2\sinh^2{2K}\cosh^2{2K} }{\cosh^2{2K} (1+k^2\sinh^2{2K})^{3/2}}
\end{multline*}

Substituting the expression \eqref{ffuncab} for $f(K,k)$ into \eqref{derivativeoffwtrk}, and using \eqref{diff1} 
$$
k\left( a'(k)A(K,k) + a(k) \frac{\partial A(K,k)}{\partial k} - b'(k)B(K,k) - b(k)\frac{\partial B(K,k)}{\partial k} \right)
=\frac{1}{2}\left[ a(k) - b(k) + k b'(k) \right] C(K,k)
$$

\begin{equation}
    A(K,k)\left[ ka'(k) - a(k) + c(k) \right] +\left[ B(K,k) + \frac{C(K,k)}{2} \right] \left[ a(k) - kb'(k) - c(1+k^2) \right] = 0
\end{equation}
where;
\begin{equation} \label{c}
    c(k) = \frac{b(k)}{k'^2}
\end{equation}
For given $k$, we have that $A(K,k)$, $B(K,k)$, and $C(K,k)$ are linearly independent. This implies
\begin{subequations} \label{aveb}
    \begin{equation}
        ka'(k) = a(k) - c(k)
    \end{equation}
    \begin{equation}
        kb'(k) = a(k) - (1+k^2)c(k)
    \end{equation}
\end{subequations}
Now, we need to derive a differential equation for $c(k)$ using the equations above. So first we eliminate $a(k)$.
$$
a'(k)= kb'(k)+(1+k^2)c(k)
$$
$$
ka'(k)=k(kb'(k)+(1+k^2)c(k))' =(kb'(k)+(1+k^2)c(k)) -c(k)
$$

$$
kb''(k)+kc(k)+c'(k)+k^2c'(k)=0
$$
We can derive $b''(k)$ using \eqref{c}
$$
b''(k)=c''(k)(1-k^2)-4kc'(k)-2c(k)
$$
If we insert this into the equation to eliminate $b(k)$, we get the following
$$
k(1-k^2)c''(k)+(1-3k^2)c'(k)-kc(k)=0
$$
We can write the first two terms on the left-hand side as a total derivative as follows

\begin{equation} \label{diff}
    \frac{d}{dk} \left(k(1-k^2)\frac{dc(k)}{dk}  \right) -kc(k)=0
\end{equation}

\subsection{Final Determination of the Correlations}

We have derived equations for the all functions we need to calculate $f(K,k)$. To be able to calculate, we need to show two types of elliptic integrals.

\begin{equation}
    \int \frac{\mathrm{d}x}{\sqrt{(1-x^2)(1-k^2x^2)}} \hspace{18mm} \int \frac{\sqrt{1-k^2x^2}}{\sqrt{1-x^2}} \mathrm{d}x
\end{equation}

These integrals are called elliptic integrals of the first and second kind in the Legendre normal form respectively. The $k$ is called the modulus of these integrals and $k'=\sqrt{1-k^2}$ is called the complementary modulus. If we use a substitution such as $x=\sin \phi$ these integrals can be written as follows

\begin{subequations}
    
\begin{equation}
F(\phi ,k) = \int_{0}^{\phi} \frac{\mathrm{d} \phi}{\sqrt{1-k^2\sin^2\phi}} 
\end{equation}
\begin{equation}
E(\phi , k) = \int_{0}^{\phi} \sqrt{1-k^2\sin^2 \phi } \,\, \mathrm{d}\phi
\end{equation}
\end{subequations}
If we set the upper boundary to be $\phi=\pi/2$, then we will obtain complete elliptic integrals.
\begin{equation}
    \mathcal{K}(k) =  \int_{0}^{\pi /2} \frac{\mathrm{d} \phi}{\sqrt{1-k^2\sin^2\phi}} 
\end{equation}
\begin{equation}
    \mathcal{E} (k)= \int_{0}^{\pi /2} \sqrt{1-k^2\sin^2 \phi }\,\, \mathrm{d}\phi
\end{equation}
The series representation of $\mathcal{K}$ is shown below

\begin{equation}
    \mathcal{K} = \frac{\pi }{2 } \left ( 1+ \left(\frac{1}{2} \right)^2 k^2 + \left ( \frac{1 \cdot 3}{2 \cdot 4} \right)^2k^4 + \cdots + \left ( \frac{(2n-1)!!}{2^n n!} \right)^2 k^{2n}+ \cdots \right )  = \frac{\pi}{2} F \left (\frac{1}{2}, \frac{1}{2} ; 1; k^2  \right)
\end{equation}
\medskip
Now, we can start to calculate $f(K,k)$. 
In these calculations, we will use Gradshteyn and Ryzhik \cite{baxtergr} (hereafter referred to as GR). Since the differential equation in \eqref{diff} is singular at $k=1$, considering \eqref{k} and \eqref{l} we can divide our calculations for high-temperature and low-temperature cases.

\subsubsection{Low-Temperature Case: \texorpdfstring{$k<1$}{}}
$\mathcal{K}(k)$ and $\mathcal{K}(k')$ is the solution of the differential equations of the type \eqref{diff} (GR \S 8.124.1).
\begin{equation}
    c(k)=\lambda \mathcal{K}(k) + \mu \mathcal{K}(k')
\end{equation}
If we look at the behaviour of $c(k)$ as $k \xrightarrow{ } 0 $ we can see that $\mathcal{K}(k) \xrightarrow{ } \pi/2 $ and $\mathcal{K}(k') \xrightarrow{ } \infty $. Considering \eqref{ffunc}, we know that $|f(K,k)|<1$. The $\mu$ must be 0, otherwise $f(K,k)$ becomes infinity for fixed $K$.
\\
With the help of \eqref{aveb} we write the solutions for a and b as follows (GR \S 8.123.4)
\begin{equation}
    a(k)=\lambda \mathcal{E}(k) \,\,\,\,\,\,\,\,\,\,\,\,\,\,\,\,\,\,\,\,\,\,\,\,\,\,\,\,\,\,\,\,\, b(k)= \lambda k'^2 \mathcal{K}(k)
\end{equation}
For \eqref{integrals} we first need to change the variable as $\tan \alpha = \sinh{x}$. Then $A(K,k)$ turns out as follows

$$
A(K,k)=\int \frac{\mathrm{d}\alpha}{\sqrt{\cos^2 \alpha +k^2 \sin^2 \alpha}}
$$
Using $k'$ instead of $k$, we derive an integral similar to $\mathcal{K}(k')$. The only thing we need to consider is the boundaries of the integral. To interpret $A$ and $B$ as elliptic integrals, we need to choose the upper boundary as $\pi/2$. Then $K$ should be chosen as infinity considering the change of variables we did in the first step.
\begin{equation}
    A(\infty , k) = \int_{0}^{\pi/2} \frac{\mathrm{d} \alpha}{\sqrt{1-k'^2\sin^2 \alpha}} = \mathcal{K}(k')
\end{equation}
Doing the same procedure for $B(K,k)$ gives

$$
B(\infty,k) = \int_{0}^{\pi/2} \frac{\sin^2 \alpha \mathrm{d}\alpha}{\sqrt{\cos^2 {\alpha} + k^2\sin^2 {\alpha}}} = \int_{0}^{\pi/2} \frac{\sin^2 \alpha \mathrm{d} \alpha}{\sqrt{1-k'^2 \sin^2 \alpha }}= \frac{\mathcal{K}(k')-\mathcal{E}(k')}{k'^2}
$$
Now, we need to find out what $\lambda$ is. To do this we use an equation for $\mathcal{K}$ and $\mathcal{E}$ (GR \S 8.122)

\begin{equation} \label{elipticidentity}
    \mathcal{E}(k) \mathcal{K}(k') + \mathcal{E}(k') \mathcal{K}(k) -\mathcal{K}(k) \mathcal{K}(k') = \pi/2
\end{equation}
If we insert the functions we found into \eqref{eq1}

$$
\lambda \mathcal{E}(k) \mathcal{K}(k') - \lambda k'^2 \mathcal{K}(k) \left(\frac{\mathcal{K}(k')-\mathcal{E}(k')}{k'^2} \right)=1
$$
It is easy to see that $ \lambda = 2/\pi$ with the use of \eqref{elipticidentity}.

\subsubsection{High-Temperature Case : \texorpdfstring{$k>1$}{}}

In the elliptic integrals that we are using we must have $k^2<1$, so we will use variables $l=1/k$ and $l'^2=1-l^2$. Then the solution of $c$ becomes (GR \S 8.128.3)

\begin{equation}
    c(k)=l (\lambda' \mathcal{K}(l) + \mu' \mathcal{K}(l'))
\end{equation}
As $k\xrightarrow{ } \infty$, $\mathcal{K}(l) \xrightarrow{ } \pi/2$ and $\mathcal{K}(l') \xrightarrow{ } \infty$. So, due to the same reasons in the low temperature case, $\mu'$ must be zero.
\medskip
$a(k)$ and $b(k)$ in \eqref{aveb} becomes (GR \S 8.127)
\begin{equation}
    a(k)= \frac{\lambda ' (\mathcal{E}(l)-l'^2 \mathcal{K}(l))}{l} \,\,\,\,\,\,\,\,\,\,\,\,\,\,\,\,\,\,\,\,\,\,\,\,\,\,\,\,\,\,\,\,\, b(k)= \frac{-\lambda l'^2 \mathcal{K} (l)}{l}
\end{equation}
For the integrals in \eqref{integrals}, this time we change our variable as $\tan\alpha = k \sinh{x}$.

$$
A(K,k) = \int \frac{1}{k} \frac{\mathrm{d} \alpha}{\sqrt{\cos^2{\alpha}+k^{-2}\sin^2{\alpha}}}
$$
Instead of $k$, if we use $l$ and $l'$, and choose an upper boundary proper to the complete elliptic integral, just like in the low-temperature case.

\begin{equation}
    A(\infty,k)= \int _{0}^{\pi/2} \frac{l \mathrm{d}\alpha}{\sqrt{1-l'^2\sin^2{\alpha}}} = l \mathcal{K}(l')
\end{equation}
Doing the same variable change, we get for $B(K,k)$
\begin{equation}
    B(\infty,k) = l(\mathcal{E}(l')-l^2\mathcal{K}(l'))/l'^2
\end{equation}
Again, we need to find out what $\lambda'$ is. First, let's write the \eqref{elipticidentity} in terms of $l$ and $l'$ (GR \S 8.122.4, \S 8.127, \S 8.128.3)
$$
(\mathcal{E}(l)-l'^2 \mathcal{K}(l))\mathcal{K}(l') +\mathcal{E}(l') \mathcal{K}(l) -\mathcal{K}(l') \mathcal{K}(l) = \pi/2
$$
Now if we put the variables we found into \eqref{eq1} we get
$$
\frac{\lambda ' (\mathcal{E}(l)-l'^2 \mathcal{K}(l))}{l} l \mathcal{K}(l') + \frac{\lambda l'^2 \mathcal{K} (l)}{l} \frac{l(\mathcal{E}(l')-l^2\mathcal{K}(l'))}{l'^2}=1
$$
It is clear that $\lambda '$ is also $2/\pi$. Thus, the value of this coefficient is the same in the two different cases.

\subsection{Conclusion}

To be able to write the functions $a$ and $b$ in a single form for both low and high-temperature cases, we will use the Landen transformation, which relates the parameters of the elliptic integral $(k,k')$.

\begin{equation}
    k_1= \frac{2\sqrt{k}}{1+k}
\end{equation}

Using (GR \S 8.126), the complete elliptic integrals can be written as

\begin{subequations}\label{k1form}
\begin{equation}
\mathcal{K}(k) = \frac{\mathcal{K}(k_1)}{1+k} 
\end{equation}
\begin{equation}
\mathcal{E}(k) = \frac{\mathcal{E}(k_1)(1+k)}{2} + \frac{(1-k)\mathcal{K}(k_1)}{2} 
\end{equation}
\end{subequations}
Then $a$ and $b$ can be written as below
$$
a(k)= \frac{(1+k)\mathcal{E}(k_1)+ (1-k)\mathcal{K}(k_1)}{\pi} \,\,\,\,\,\,\,\,\,\,\,\,\,\,\,\,\,\,\,\,\,\,\,\, b(k) = \frac{2(1-k)\mathcal{K}(k_1)}{\pi}
$$
If we eliminate $a(k)$ using the \eqref{eq1} and insert it into \eqref{ffuncab} we get an equation

\begin{equation}
    f(K,k) = \frac{A(K,k)}{A(\infty , k)}-b(k) \left (B(K,k)- \frac{B(\infty,k) A(K,k)}{A(\infty ,k)} \right)
\end{equation}
In this function above, the only non-analytic term $b(k)$ and it has a singularity at $k=1$. Then, this situation can be interpreted as the singularity in internal energy does not depend on whether our lattice is honeycomb, triangular, or square because the singularity is independent of the interaction coefficients. The equation for the internal energy can be derived as follows

\begin{equation}
    u=-T^2 \frac{\partial}{\partial T} \left (\frac{f_r}{T} \right)
\end{equation}

Instead of $f_r$, we can use $\psi$ with the \eqref{psifr}.

\begin{equation}
    u=T^2 \frac{\partial (k_{B} \psi)}{\partial T} = T^2 k_{B} \left (\frac{\partial \psi}{\partial K} \frac{\partial K}{\partial T} + \frac{\partial \psi}{ \partial L} \frac{ \partial L}{ \partial T} \right) 
\end{equation}

Using \eqref{l}, \eqref{k}, \eqref{ffunc} and \eqref{corrwithpsi} we derive the relation between $f(K,k)$ and the internal energy.
$$
u= \coth{2K}f(K,k)J+\coth{2L}f(L,k)J'
$$
For the $b(k)$ near the $k=1$ we will use another series representation of $\mathcal{K}$ (GR \S 8.113.3)

\begin{equation}
    \mathcal{K}= \ln{\frac{4}{k'}} + \left(\frac{1}{2} \right)^2 \left (\ln{\frac{4}{k'}} - \frac{2}{1 \cdot 2} \right) k'^2 + \left ( \frac{1 \cdot 3}{2 \cdot 4}\right) ^2 \left ( \ln{\frac{4}{k'}} - \frac{2}{1 \cdot 2} - \frac{2}{3 \cdot 4}\right)k'^4 + \cdots
\end{equation}

Then $b(k)$ near $k=1$ can be written using with the \eqref{k1form}
 \begin{equation}
     b(k) \approx \frac{(1-k^2)}{\pi} \ln{(16/|1-k^2|)}
 \end{equation}
It is known that the 2-dimensional Ising model has a symmetric logarithmic divergence in its specific heat. This can be seen in the solution of our case. It is also shown that the phase transition point does not depend on the shape of the lattice whether it is square, triangular or honeycomb lattice. This solution shows a proper method to understand all these lattices with only star-triangle relation.

\section{Fermionic Formulation \textit{(Ali Mert Yetkin and Reyhan Yumuşak)}}\label{chapter:5}

In this work, the solution of the Ising model via the fermionic formulation is studied. The Ising model is an essential model for statistical physics, and its partition function initially includes spin variables. A remarkable feature is that the Ising model can be reformulated in terms of fermionic variables \cite{Itzykson1989sx,Berezin:1966nc}. To transform them one needs to know some basic information about Grassmann variables and their properties. The study begins with this mathematical framework. After the partition function has a new form with systematic replacement, the representations of Bloch walls, corners, and monomers can be reached. The quadratic Grassmann action is diagonalized after some Fourier transformations. The determinant of the resulting matrix provides the exact solution of the Ising model.

\subsection{Integrals of Grassmann Variables}
In this section we provide a basic treatment of Grassmann algebra to provide sufficient mathematical background, for a more comprehensive review of the subject one may consult \cite{Peskin:1995ev}. 

A set of variables $\boldsymbol{\eta}=\{\eta_i\}=\{\eta_1,\eta_2,\dots,\eta_N\}$, which satisfies the relation $\eta_\alpha\eta_\beta = -\eta_\beta\eta_\alpha$, is defined as Grassmann variables. These variables define an algebra over the complex field $\mathbb{C}$, such that complex numbers $z\in \mathbb{C}$ and the variables $\eta_i$ satisfy the following relation: $z\eta_i = \eta_i z$. One immediate consequence of the anti-commutation relation is that squared Grassmann variables vanish.
\begin{equation}
     \eta_\alpha^2 = \eta_{\alpha}\eta_{\alpha} = -\eta_\alpha\eta_\alpha = 0
\end{equation} 
This property provides us with an opportunity to expand any polynomial $f(\eta_{\alpha})$ in the following form:
\begin{equation}
    \label{2} f=\theta_0 +\sum_\alpha \theta_\alpha\eta_\alpha+\sum_{\alpha \leq \beta} \theta_{\alpha\beta} \eta_\alpha \eta_\beta + \dots + \theta_{123\dots N}\eta_1 \eta_2 \dots \eta_N 
\end{equation}

The differentiation rules for Grassmann variables are approximately the same as in the usual case. To be consistent with the anti-commutation relation, we must define two new differential operators $\overleftarrow{\frac{d}{d\eta}}$ and $\overrightarrow{\frac{d}{d\eta}}$, where arrows indicate the direction of differential operators act. A simple illustration will help to clarify this point, 
\begin{align}
    f({\eta}) = C_1\eta_1\eta_2 + C_2 \\
    \nonumber \overrightarrow{\frac{d}{d\eta_1}}f = C_1\eta_2 \\
    \nonumber \overrightarrow{\frac{d}{d\eta_2}}f = -C_1\eta_1 
\end{align} 
in the latter term, we introduced a minus sign since the right operator is only able to act after changing the order of terms. 

When attempting to define integration for Grassmann variables, the usual Riemann integral does not apply since the square of a Grassmann variable is zero. However, the properties of the definite Riemann integral, namely translation invariance, and linearity, can be used. Let $f(x)=a+bx$ be a function of Grassmann variables, where $a,b \in \mathbb{C}$. Then, the properties
\begin{align}
     \nonumber \int f(x) dx= \int f(x+y) dx
     \\ \nonumber \int (a+bx) dx = \int (a+b(x+y))dx = \int (a+bx)dx + \int by dx 
\end{align}
It can be assumed that $by\int dx$ is zero. For the remaining part, $\int(a+bx)dx=b\int xdx$. Berezin chose the convention that $\int xdx =1$. In general, integration for Grassmann variables can be thought of as usual differentiation, namely
 \begin{align}
     \label{4} \int \: d\boldsymbol{x}f(\boldsymbol{x}) = \frac{d}{d\boldsymbol{x}}f(\boldsymbol{x})
\end{align} where $d\boldsymbol{x}$ denotes whole set of Grassmann variables. Simply we have,
\begin{align}
     \int d\boldsymbol{x}\boldsymbol{x} = 1 \text{ and} \int d\boldsymbol{x} 1 = 0 
\end{align}
The equations presented in Equations \eqref{2} and \eqref{4} provide us with a very powerful tool for calculating Gaussian integrals.
\begin{subequations}
    \begin{equation} 
     \int d\boldsymbol{x} e^{f(\boldsymbol{x} )} = \theta_{1234\dots n}
   \end{equation}\label{6a}
   \begin{equation}
\int d\boldsymbol{x} d\boldsymbol{x}^{\dag} \:\boldsymbol{x}^{\dag}\boldsymbol{x}  = 1
\end{equation}\label{6b}
\begin{equation}
\int d\boldsymbol{x} d\boldsymbol{x}^{\dag} e^{\boldsymbol{x} \boldsymbol{x}^{\dag}} = \int d\boldsymbol{x}  d\boldsymbol{x} ^{\dag}(1+\boldsymbol{x} \boldsymbol{x}^{\dag}) = -1
\end{equation}\label{6c}
\begin{equation}
    \int d\boldsymbol{x}  d\boldsymbol{x}^{\dag} e^{\lambda \boldsymbol{x}^{\dag}\boldsymbol{x} } = \lambda,\: \: \lambda \in \mathbb{C}
\end{equation}\label{6d}
\end{subequations}

Furthermore, the change of the variable needs to be reconsidered. In the context of Riemann integration, changing a variable affects the integration by a quotient 
\begin{equation}
    \int f(ax)dx=\frac{1}{a}\int f(x)dx 
\end{equation}
where $dy=d(ax)=adx$. Since Berezin integration behaves as differentiation, the effect of changing variables is observed as a multiplicative factor.
\begin{align}
    \label{8} \int y dy =\int x dx =1 = \int (ax) dy = \int x dx
\end{align}
so $a dy=dx$ and changing variables in Berezin integration results in 
\begin{align}
   \int f(ax)dx=a\int f(x)dx 
\end{align}. 
These cases can be generalised for $N$ generators $y_i$ where  $y_i=\sum_j a_{ij}x_j$ and $y_1y_2...y_N= \det (a) x_1x_2...x_N $
\begin{align}
    \text{Riemann integral} \:\:\:\: dy_1 \: dy_2 ... \: dy_N= \frac{1}{\det(a)}dx_1 \: dx_2 ... \: dx_N \rightarrow \int f(\boldsymbol{y})d\boldsymbol{y}=\det(a)\int f(\boldsymbol{x})d\boldsymbol{x} \\ \nonumber 
    \text{Berezin integral} \:\:\:\: dy_1 \: dy_2 ... \: dy_N= \det (a) dx_1 \: dx_2 ... \: dx_N \rightarrow \int f(\boldsymbol{y})d\boldsymbol{y}=\frac{1}{\det(a)}\int f(\boldsymbol{x})d\boldsymbol{x} 
\end{align}

For quadratic and quadratic-like actions, Gaussian integrals should also be considered for Berezin integration. For Riemann integration, one can write
\begin{subequations}
    \begin{equation}
        \label{11a}\int_{\mathbb{R}}e^{-ax^2}dx =\sqrt{\frac{\pi}{a}}
    \end{equation}
    \begin{equation}
        \int_{\mathbb{R}^2}e^{-a(x^2+y^2)}dxdy=\frac{\pi}{a}
    \end{equation}
\end{subequations}.
The analogous \eqref{11a} gives zero for Berezin integration since 
\begin{align}
    \int e^{-ax^2}dx=\int(1-ax^2)dx=\int (1-a . 0) \:dx=0
\end{align} and the analogous for the latter is the following
\begin{align}
    \int \int e^{-a(x^2+y^2)}dxdy=\int \int e^{-2axy}dxdy=\int \int (1-\Tilde{a}xy)dxdy \\ \nonumber =\int -\Tilde{a}xy \: dxdy  = \int \Tilde{a}yx \: dxdy =\Tilde{a}
\end{align}
with using $(x-y)^2=0=x^2-2xy+y^2$ and $\Tilde{a}=2a$.

If x and y are generators, there are $2N$ generators.

\begin{align}
    d\boldsymbol{x}d\boldsymbol{y}=dx_1dy_1...dx_Ndy_N
\end{align} and if one considers the Berezin Gaussian integral, firstly variables must be changed to diagonalize matrix $A$
\begin{align}
    \int e^{-\sum_{i,j}x_iA_{ij}y_j}d\boldsymbol{x}d\boldsymbol{y}=\int e^{-\sum_{i}x_i'A_{ii}'y_i'}d\boldsymbol{x}d\boldsymbol{y}=\prod_iA_{ii}'=\det A'= \det A
\end{align}
where $A'$ is unitary transformation of $A$. For Riemann integration 
\begin{align}
    \int_{\mathbb{R}^N}e^{-x^TAx}d^Nx=\sqrt{\frac{\pi^N}{\det A}}
\end{align}

Gaussian integration also can be expressed in terms of Pfaffian. 
\begin{subequations}
    \begin{equation}
        x_i=\frac{1}{\sqrt{2}}\left( z^{(1)}_j+iz^{(2)}_j\right)
    \end{equation}
    \begin{equation}
        y_i=\frac{1}{\sqrt{2}}\left( z^{(1)}_j-iz^{(2)}_j\right)
    \end{equation}
\end{subequations} where $x$ and $y$ are Grassmann variables. The Jacobian of the generators affects like \eqref{8}.
\begin{align}
   \label{18} dx_idy_i=\det  
    \begin{bmatrix}
        \frac{1}{\sqrt{2}} & \frac{i}{\sqrt{2}} \\
        \frac{1}{\sqrt{2}} & -\frac{i}{\sqrt{2}} \\
    \end{bmatrix}^{-1} dz^{(1)}_i dz^{(2)}_i=idz^{(1)}_i dz^{(2)}_i
\end{align}
Let A be an $N \times N$ dimensional antisymmetric matrix where $N$ is even. As a result, the cross terms cancel each other
\begin{align}
    \label{19}x_iA_{ij}y_j=\frac{A_{ij}}{2}\left( z^{(1)}_iz^{(1)}_j-iz^{(1)}_iz^{(2)}_j+iz^{(2)}_iz^{(1)}_j+z^{(2)}_iz^{(2)}_j\right)\\
    \nonumber = \frac{A_{ij}}{2}\left(z^{(1)}_iz^{(1)}_j+z^{(2)}_iz^{(2)}_j \right)
\end{align}
since $iz^{(1)}_iz^{(2)}_j=-iz^{(1)}_jz^{(2)}_i=iz^{(2)}_iz^{(1)}_j$ due to antisymmetry 
property of Grassmann variables. By using \eqref{18} and \eqref{19}, integral becomes
\begin{align}
    \label{sor1}\int e^{-\sum_{i,j}x_iA_{ij}y_j}d\boldsymbol{x}d\boldsymbol{y}=
    i^N\int e^{\sum_{i,j}\frac{-1}{2}\left(z^{(1)}_iA_{ij}z^{(1)}_j+z^{(2)}_iA_{ij}z^{(2)}_j \right)}dz^{(1)}_1dz^{(2)}_1...dz^{(1)}_Ndz^{(2)}_N \\
    \nonumber = i^N(-1)^{N(N-1)/2}\int e^{\sum_{i,j}\frac{-1}{2}\left(z^{(1)}_iA_{ij}z^{(1)}_j+z^{(2)}_iA_{ij}z^{(2)}_j \right)}dz^{(1)}_1...dz^{(1)}_Ndz^{(2)}_1...dz^{(2)}_N \\
    \nonumber =  (-1)^{N/2}(-1)^{N(N-1)/2}\int e^{\sum_{i,j}\frac{-1}{2}z^{(1)}_iA_{ij}z^{(1)}_j}dz^{(1)}_1...dz^{(1)}_N\int e^{\sum_{i,j}\frac{-1}{2}z^{(2)}_iA_{ij}z^{(2)}_j }dz^{(2)}_1...dz^{(2)}_N \\
    \nonumber = \left[\int e^{\sum_{i,j}\frac{-1}{2}z_iA_{ij}z_j}d\boldsymbol{z}\right]^2=\left(Pf(A)\right)^2
\end{align}
since $Pf(A)=\sqrt{\det A}$.

Finally, integration by parts can be considered as a useful tool. At the beginning, the differentiation rule for multiplication has the following form 
\begin{align}
    \frac{d}{dx} \left(f.g\right)=\left( \frac{d}{dx}f\right)g +f\left( \frac{d}{dx}g\right)= \left(\overrightarrow{\frac{d}{dx}}f\right)g -g \overleftarrow{\frac{d}{dx}}f
\end{align}
Then, integration by parts becomes
\begin{align}
    \int dx f \overrightarrow{\frac{d}{dx}}g= \int dx f \overleftarrow{\frac{d}{dx}}g
\end{align}

\subsection{Partition Function for Ising Model}

The Ising model is a lattice spin model in which only nearest-neighbor interactions are considered. Spins are located at the vertices of the lattice and are allowed to only take $\sigma_{i,j} = \pm$. The partition function of the model is as follows 
\begin{align}
    \mathcal{Z} = \sum_{\{\sigma\}}e^{-\beta\mathcal{H}}, \ \ \beta = \frac{1}{k_B T}
\end{align}
where Hamiltonian $ \mathcal{H} = \sum_{i,j}  J_h\sigma_{i,j}\sigma_{i+1,j} + J_v\sigma_{i,j}\sigma_{i,j+1}$. $k_B$ is the Boltzmann constant, $T$ is the temperature, and $\{\sigma\}$ corresponds to possible spin configurations. Using formula : $e^{\sigma_i \sigma_j\theta} = \cosh{\theta}+\sigma_i\sigma_j\sinh{\theta}$, the partition function becomes \\
\begin{equation}
    Z = \sum_{\{\sigma\}}\prod_{i,j}cosh(\beta J_h)cosh(\beta J_v)(1+\sigma_{i,j}\sigma_{i+1,j}\tanh(\beta J_h))\times(1+\sigma_{i,j}\sigma_{i,j+1}tanh(\beta J_v)) 
\end{equation} and,
\begin{equation}
     \mathcal{Z}:=\sum_{\{\sigma\}}\prod_{i,j}(1+\sigma_{i,j}\sigma_{i+1,j}\tanh(\beta J_h))\times(1+\sigma_{i,j}\sigma_{i,j+1}tanh(\beta J_v))
\end{equation} is the reduced partition function. When $\mathcal{Z}$ is expanded as powers of $tanh(\beta J_v)$ and $tanh(\beta J_h)$, we obtain
\begin{equation}
    \mathcal{Z} = \sum_{\{\sigma\}} \ \ \ \sum_{u,z} \prod_{\substack{(\sigma_i, \sigma_j) \in \mathcal{N}_v \\ (\sigma_k, \sigma_l) \in \mathcal{N}_h}}  \sigma_i\sigma_jtanh(\beta J_h)^u\times \sigma_{k} \sigma_{l} tanh(\beta J_v)^z
\end{equation}
where $\mathcal{N}_v$ and $\mathcal{N}_h$ denote vertical and horizontal neighboring spins respectively. \\ 

The visual interpretation of the above equation is straightforward. Each term in the sum corresponds to a polygonal configuration on the grid with $u$ horizontal and $v$ vertical lines. These vertical and horizontal lines are counted in the product. If there is a horizontal or vertical spin interaction between two neighboring sites, it contributes to the product with a factor $tanh(\beta J_v)$ or $tanh(\beta J_h)$ respectively. From this point of view, the partition function calculation is just a problem of counting all possible polygon configurations on the lattice.
\\
\subsection{Fermionic Representations of Partition Function} 

To obtain a purely fermionic Gaussian integral, one needs to eliminate the spin variables present at each site and replace them with fermionic variables \cite{Plechko:2000py}. The elimination of spin variables can be achieved after the outer summation has been performed; however, this operation is not directly feasible because spin variables connect neighboring sites. Therefore, at the moment of summation, all spin variables must be grouped belonging to the same site. \\

At each site, a set of Grassmann variables $\{\eta_(\alpha,\beta)^{\{h^{X,O},v^{X,O}}\}$, must be defined where $(\alpha,\beta)$ represents the position of the spin on the lattice, $h,v$ represents the direction of the interaction and $X,O$ distinguishes two different types of variables. To calculate the Gaussian integral, the Boltzmann weights must be factorized corresponding to the horizontal and vertical interactions.

\begin{equation}
    1+t_h\sigma_{i,j}\sigma_{i+1,j}=\int d\eta^{h^O} d\eta^{h^X} e^{\eta^{h^O} \eta^{h^X}}(1+\sigma_{i,j}\eta^{h^O})(1+t_h\sigma_{i+1,j}\eta^{h^X})
\end{equation}
\begin{equation}
    1+t_v\sigma_{i,j}\sigma_{i,j+1}=\int d\eta^{v^O} d\eta^{v^X} e^{\eta^{v^O} \eta^{v^X}}(1+\sigma_{i,j}\eta^{v^O})(1+t_v\sigma_{i,j+1}\eta^{v^X})
\end{equation}
For shorthand, we write $\tanh{\beta J_h} := t_h$ and $\tanh{\beta J_v} := t_v$. For the sake of notation, each Boltzmann factor in the integrand can be renamed,
    
\begin{equation}
        1+\sigma_{i,j}\eta^{h^O}_{i,j} = H_{i,j}^O
\end{equation}
\begin{equation}
          1+t_h\sigma_{i+1,j}\eta^{h^X}_{i,j} = {H}_{i+1,j}^X
\end{equation}
\begin{equation}
        1+\sigma_{i,j}\eta^{v^O}_{i,j} = V_{i,j}^O
\end{equation}
\begin{equation}
          1+t_v\sigma_{i,j+1}\eta^{h^X}_{i,j} = {V}_{i,j+1}^X
\end{equation}
and partition function takes the following form,
\begin{equation}
    \mathcal{Q} = \sum_{\{\sigma\}}\int d\eta^{h^O}d\eta^{h^X} d\eta^{v^O}\eta^{v^X}\prod_{i=1}^{N} \prod_{j=1}^{N} H_{i,j}^X
    {H}_{i+1,j}^OV_{i,j}^X{V}_{i,j+1}^O
\end{equation} 

\subsection{Spin Averaging}

Spin variables $\sigma_{i,j}$ with an odd number of occurrences do not contribute to the partition function, i.e. only even powers $\sigma_{i,j}^{2}$ and $\sigma_{i,j}^{4}$ are subject to calculation. So one can replace all even power terms in the product with 2 and odd power terms with 0. However, as mentioned, this isn't possible in the actual form of the product. First, the terms must be rearranged:

\begin{align}
     H_{i,j}^X
    {H}_{i+1,j}^OV_{i,j}^X{V}_{i,j+1}^O \rightarrow V_{i,j}^O H_{i,j}^O V_{i,j}^X H_{i,j}^X
\end{align}

The structure of the horizontal factors $H_{i,j}^X
{H}_{i+1,j}^O$ is such that the product over $i$ should be taken for fixed $j$, the structure of the vertical factors $V_{i,j}^X{V}_{i,j+1}^O$ is such that the product over $j$ should be taken for fixed $i$. So the rearrangement needs attention. One can start by expanding the inner product,

\begin{align}
    \prod_{i=1}^{N} \prod_{j=1}^{N} H_{i,j}^X
    {H}_{i+1,j}^OV_{i,j}^X{V}_{i,j+1}^O = \prod_{i=1}^N H_{i,1}^X
    {H}_{i+1,1}^OV_{i,1}^X{V}_{i,2+1}^O H_{i,2}^X
    {H}_{i+1,2}^OV_{i,2}^X{V}_{i,3+1}^O...
\end{align}
Since vertical and horizontal factors appear as pairs and even products of Grassmann variables commute with all elements of the algebra, one can group them.
\begin{align}
    \prod_{i=1}^{N} (H_{i,1}^X
    {H}_{i+1,1}^O)(H_{i,2}^X
    {H}_{i+1,2}^O)...(H_{i,L}^X
    {H}_{i+1,L}^O )V_{i,1}^X({V}_{i,2}^OV_{i,2}^X)({V}_{i,3}^OV_{i,3}^X)...(V_{i,L}^OV_{i,L})^X{V}_{i,L+1}^O
\end{align} 
Since the term $V_{i,1}^{X}$ is not paired as well as the last one, one needs to impose boundary conditions on both spin variables and fermionic variables before proceeding. Our lattice is placed on a torus, so the spin variables have periodic $\sigma_{i,L+1} = \sigma_{i,1}$ and $\sigma_{L+1,j} = \sigma_{1,j}$ boundary conditions. However, it's not enough to match the first and last terms.
\begin{align}
    V_{i,L+1}^O = 1+\sigma_{i,L+1}\eta_{i,L+1}^{h^O} \xrightarrow{pbc} 1+\sigma_{i,1}\eta_{i,L+1}^{h^O} \overset{\mathrm{!}}{=} 1-\sigma_{i,1}\eta_{i,1}^{h^O} 
\end{align}
Here the minus sign in front of the latter term is due to odd times of displacement. As a result spin periodic boundary conditions imply aperiodic fermionic boundary conditions, and our product becomes:

\begin{align}
    \prod_{i=1}^{N} (H_{i,1}^X
    {H}_{i+1,1}^O)(H_{i,2}^X
    {H}_{i+1,2}^O)...(H_{i,L}^X
    {H}_{i+1,L}^O )({V}_{i,1}^OV_{i,1}^X)({V}_{i,2}^OV_{i,2}^X)({V}_{i,3}^OV_{i,3}^X)...(V_{i,L}^OV_{i,L}^X)
\end{align}

Now, there is a one-unit shift in the first index between the terms $H^X$ and $H^O$, which is problematic. So they should be separated. The last two terms become:
\begin{align}
     H_{i,L-1}^X
    {H}_{i+1,L-1}^O H_{i,L}^X
    {H}_{i+1,L}^O  = H_{i,L-1}^X H_{i,L}^X
    {H}_{i+1,L}^O 
    {H}_{i+1,L-1}^O 
\end{align} 
which is legal, since ${H}_{i+1,L-1}^O$ changes its sign even times. Repeating the same procedure for both horizontal and vertical factors, one ends up with the following fully ordered product:
\begin{align}
    \prod_{i=1}^{N} (H_{i,1}^X
    H_{i,2}^X...H_{i,L-1}^X H_{i,L}^X)
    ({H}_{i+1,L}^O{H}_{i+1,L-1}^O...{H}_{i+1,2}^O{H}_{i+1,1}^O) (V_{i,1}^X
    V_{i,2}^X...V_{i,L-1}^X V_{i,L}^X)
    ({V}_{i,L}^O{V}_{i,L-1}^O...{V}_{i,2}^O{V}_{i+1,1}^O) 
\end{align}
and will denote mirror-ordered pairs as

\begin{align}
    H_{i,1}^X
    H_{i,2}^X...H_{i,L-1}^X H_{i,L}^X = \{H\}_{i}^{X} \\
    {H}_{i+1,L}^O{H}_{i+1,L-1}^O...{H}_{i+1,2}^O{H}_{i+1,1}^O = \{H\}_{i+1}^{O} \\
    V_{i,1}^X
    V_{i,2}^X...V_{i,L-1}^X V_{i,L}^X = \{V\}_{i}^{X} \\
    {V}_{i,L}^O{V}_{i,L-1}^O...{V}_{i,2}^O{V}_{i+1,1}^O =\{V\}_{i}^{O}
\end{align}.

After all these steps, one should expand the second product over $i$,
\begin{align}
    \prod_{i}^{L} \{H\}_{i}^{X} \{H\}_{i+1}^{O} \{V\}_{i}^{X} \{V\}_{i}^{O}
\end{align}
and obtain nested mirror-ordered products using the same procedure (imposing constraints and making arrangements).
\begin{align}
    \{H\}_{1}^{O} \{H\}_{2}^{O}..\{H\}_{L-1}^{O} \{H\}_{L}^{O} \{H\}_{L}^{X} \{H\}_{L-1}^{X} ...\{H\}_{2}^{X}\{H\}_{1}^{X}  \{V\}_{1}^{O} \{V\}_{2}^{O}..\{V\}_{L-1}^{O} \{V\}_{L}^{O} \{V\}_{L}^{X} \{V\}_{L-1}^{X} ...\{V\}_{2}^{X}\{V\}_{1}^{X}
\end{align} 

It hasn't reached the end yet. To finish, one needs to bring the terms at either end to the middle, but this is done without any sign problem, thanks to symmetry: minus signs on either side cancel each other out. 
\begin{align}
     \{H\}_{1}^{O} \{H\}_{1}^{X}  \{V\}_{1}^{O}  \{V\}_{1}^{X} = H_{1,1}^OH_{1,2}^OH_{1,3}^O.... H_{1,L}^OH_{1,L}^XH_{1,L-1}^X...H_{1,2}^XH_{1,1}^X    
     V_{1,1}^OV_{1,2}^OV_{1,3}^O.... V_{1,L}^OV_{1,L}^XV_{1,L-1}^X...V_{1,2}^XV_{1,1}^X
\end{align}
Then, one has to repeat the same procedure for each mirrored pair, but for the second index. The final expression in compact form follows:

\begin{align}
    V_{i,L+1}^O = V_{i,L+1}^O \implies 1+\sigma_{i,j}\eta_{i,j}^{v^O}\: \vline_{\eta_{i,j}}
\end{align}
\begin{align}
    \sum_{\{\sigma = \pm 1\}} \sigma_{i,j} = 0 \: \: \: \sum_{\{\sigma = \pm 1\}} \sigma_{i,j}^2 = 2
\end{align}
\begin{equation}
     H_{i,j} \overline{H}_{i,j}V_{i,j}\overline{V}_{i,j}.
\end{equation}

\begin{equation}
    (1+\sigma_{i,j}\eta^{h^O}_{i,j}) \times (1+t_h\sigma_{i,j}\eta^{h^X}_{i-1,j}) \times (1+\sigma_{i,j}\eta^{v^O}_{i,j}) \times
    ( 1+t_v\sigma_{i,j}\eta^{h^X}_{i,j-1})
    \nonumber =  1 
\end{equation}
\begin{equation}+ \sigma_{i,j}^{2}(\eta^{h^O}_{i,j}t_h\eta^{h^X}_{i-1,j}+\eta^{h^O}_{i,j}\eta^{v^O}_{i,j}+\eta^{h^O}_{i,j}t_v\eta^{h^x}_{i,j-1}+\eta^{h^X}_{i-1,j}\eta^{v^O}_{i,j}+\eta^{v^O}_{i,j}t_v\eta^{h^X}_{i,j-1} \nonumber
    +t_ht_v\eta^{h^X}_{i-1,j}\eta^{h^X}_{i,j-1})+\sigma_{i,j}^{4}(\eta^{h^O}_{i,j}\eta^{h^X}_{i-1,j}\eta^{v^O}_{i,j}\eta^{h^O}_{i,j-1})
\end{equation}
Even products of Grassmann variables commute with any element of algebra, 
\begin{equation}
    \eta\eta^{\dag}\overline{\eta} - \overline{\eta}\eta\eta^{\dag} = 0
\end{equation}.
Thus, products of $H_{i,j} \overline{H}_{i+1,j}$ can $V_{i,j}\overline{V}_{i,j+1}$ commute freely, as spin averaging doesn't seem odd Grassmann terms.

\subsection{Linear Ordering}

The linear ordering preserves the algebraic structure and allows simplifications by arranging the terms in the partition function. For a fixed $j$,
 $H_{i,j} \overline{H}_{i+1,j}$ terms can be written as following form:
\begin{equation}
        \prod_{i=1}^{N,j} H_{i,j}\overline{H}_{i+1,j} = H_{1,j}(\overline{H}_{2,j}H_{2,j})(\overline{H}_{3,j}H_{3,j})(\overline{H}_{4,j}H_{4,j})\dots
        H_{N,j}\overline{H}_{N+1,j}
\end{equation}
Periodic boundary conditions for spins imply $\sigma_{N+1,j} = \sigma_{1,j}$, and $\overline{H}_{N+1,j} = H_{1,j}$. One should be careful bringing $\overline{H}_{N+1,j}$ next to $H_{1,j}$ due to minus sign.
\begin{equation}
        \overline{H}_{N+1,j} = (1+t_h\sigma_{N+1,j}\eta_{N+1,j}^{h^X}) = (1+t_h\sigma_{1,j}\eta_{0,j}^{h^X}) = \overline{H}_{1,j}
\end{equation}
After moving $\overline{H}_{N+1,j}$ to next to $H_{1,j}$, 
\begin{equation}
            \overline{H}_{N+1,j} = 1-t_t\sigma_{1}\eta_{0,j}^{h^X}
\end{equation}.
One ends up with an aperiodic boundary condition $\eta_{0,j} = -\eta_{N+1,j}$ for fermionic variables. After reordering, the product takes the following form:
\begin{equation}
        \prod_{i=1}^{N} = (\overline{H}_{i,j}H_{i,j})
\end{equation}.
One can be sure that all terms are properly ordered for integration and repeat the process for every row of the lattice.

\subsection{Mirror Ordering}
 
 The mirror ordering reorders the Grassmann variables vertically and horizontally while consistency and symmetry are protected. Consider the following arrangement: 
\begin{equation}
        \prod_{i=1}^{N}  = H_{i,j}\overline{H}_{i,j}
         = (H_{i,j}\overline{H}_{i,j})
\end{equation} 
where,
\begin{equation}
        (H_{i,j}\overline{H}_{i,j})\dots H_{N-1,j}\overline{H}_{N-1,j}H_{N,j}\overline{H}_{N,j}.
\end{equation}
One can get the following expression using the ordering property of even products of Grassmann Variables.
\begin{equation}
        \dots H_{N-2,j}\overline{H}_{N-2,j}H_{N-1,j}H_{N,j}\overline{H}_{N,j}\overline{H}_{N-1,j}
\end{equation} 
Eventually, one obtains the mirror-ordered product
\begin{equation}
        H_{1,j}H_{2,j}H_{3,j}\dots H_{N,j}\overline{H}_{N,j}\overline{H}_{N-1,j}\overline{H}_{N-2,j}\dots
\end{equation}.

Definitions of $V_{i,j}\overline{V}_{i,j+1}$ also allow one to mirror order them but this time concerning the first index,
\begin{equation}
        \prod_{i=1}^{N} V_{i,j}\overline{V}_{i,j+1} = \prod_{i=1}^{\overleftarrow{N}} V_{i,j} \prod_{i=1}^{\overrightarrow{N}} \overline{V}_{i,j+1} .
\end{equation} 
Then using the linear ordering rule, it becomes
\begin{equation}
        \prod_{j=1}^{\overrightarrow{N}} \{ { \prod_{i=1}^{\overrightarrow{N}} \overline{{V}}_{i,j} \prod_{i=1}^{\overleftarrow{N}} V_{i,j}} \} .
\end{equation}.
One puts $H_{i,j} \overline{H}_{i+1,j}$ into,
\begin{equation}
        \prod_{j=1}^{\overrightarrow{N}} \{ { \prod_{i=1}^{\overrightarrow{N}} \overline{{V}}_{i,j} H_{i,j} \overline{H}_{i+1,j} \prod_{i=1}^{\overleftarrow{N}} V_{i,j}} \}
\end{equation} 
and linear ordering for an outer product is,
\begin{equation}
        \prod_{j=1}^{\overrightarrow{N}} \{ { \prod_{i=1}^{\overrightarrow{N}} \overline{H}_{i,j}\overline{{V}}_{i,j} H_{i,j}  \prod_{i=1}^{\overleftarrow{N}} V_{i,j}} \}.
\end{equation} 
The final expression doesn't mix neighbor spin terms, so one can expand the following expression 
\begin{equation}
        \overline{H}_{i,j}\overline{{V}}_{i,j} H_{i,j}V_{i,j}
\end{equation}
and after averaging the final expression, it becomes 
\begin{equation}
       \nonumber \prod_{i,j}^{N} \{1 + 2(\eta^{h^O}_{i,j}t_h\eta^{h^X}_{i-1,j}+\eta^{h^O}_{i,j}\eta^{v^O}_{i,j}+\eta^{h^O}_{i,j}t_v\eta^{h^X}_{i,j-1}+\eta^{h^X}_{i-1,j}\eta^{v^O}_{i,j}+\eta^{v^O}_{i,j}t_v\eta^{h^X}_{i,j-1}  \nonumber
\end{equation}
\begin{equation}
        +t_ht_v\eta^{h^X}_{i-1,j}\eta^{h^X}_{i,j-1})+2(\eta^{h^O}_{i,j}\eta^{h^X}_{i-1,j}\eta^{v^O}_{i,j}\eta^{h^O}_{i,j-1})\} .
\end{equation}

\subsection{Low Temperature Expansion of Ising Model}

In low-temperature expansion, the Ising model can be analyzed by its partition function which is
\begin{equation}
   \mathcal{Z}(K) = e^{2N(\beta J_v + \beta J_h)}(1+Ne^{-4\times2(K_1+K_2)}+2Ne^{-6\times2K(K_1+K_2)}...)
\end{equation}
\begin{equation}
    = 2e^{2N(\beta J_v+ \beta J_h)}\sum_{\text{islands}} e^{2l_v K_1 + 2l_h K_2}
\end{equation} 
where $K_1:= -\beta J_h \ K_2:= -\beta J_v$, $l_{v,h}$ stands for the vertical and horizontal length of the islands. Each excitation from the ground state costs $e^{-2K_{1,2}\text{broken bonds}}$ and they draw non-intersecting polygons which are called domain walls.

\begin{figure}[H]
    \centering
    \includegraphics{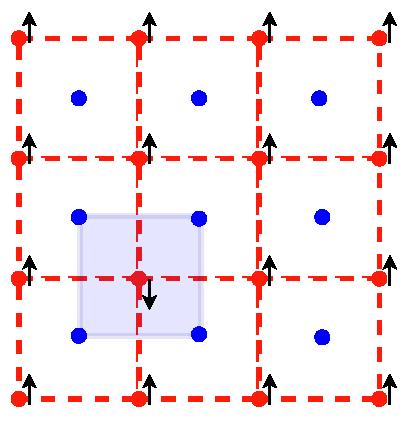}
    \caption{The dual lattice and domain walls}
    \label{fig:diaagfer6}
\end{figure}

The spins in the lattice are aligned in the ground state such that all of them are either up or down at low temperatures. If any thermal fluctuations appear, spins flip in the corresponding region. The boundaries of these regions are named as $domain$ $walls$ which form non-interacting polygons. 
In Figure \ref{fig:diaagfer6}, lattice points are represented as red dots, and red dashed lines are used for domain walls. The blue dots represent dual lattice which will be used for the fermionic formalism. 

\subsection{Quadratic Grassmanian Action}

Anti-commuting variables are assigned to include spins,
atoms, bonds, etc.
Following the notation in the original paper by Stuart Samuel \cite{Samuel:1978zx}, one can define four types of Grassmann variables for each site $\boldsymbol{x}(\alpha,\beta)$ of the dual lattice (blue dots in \ref{fig:diaagfer6}).
\begin{equation}
    \{\eta_{\alpha,\beta}^{r}\} = \{\eta_{\alpha,\beta}^{h^O},\eta_{\alpha,\beta}^{h^X},\eta_{\alpha,\beta}^{v^O},\eta_{\alpha,\beta}^{v^X}\}
\end{equation} where $O$ and $X$ stand for conjugate pairs and $h,v$ for horizontal and vertical variables, respectively. It is important to remember that a contribution to an integral only occurs if each location is covered by one $O$ and one $X$ of each type. The partition function can be expressed as a fermionic Gaussian integral as follows:
\begin{equation}
    \mathcal{Z} = \int \prod_{\alpha,\beta} d\eta d\eta^{\dag} e^{\eta^{\dag}_{\alpha}A_{\alpha,\beta}\eta_{\beta}}
\end{equation} 
where $A$ is a 4-tensor representing site interactions. $A(\boldsymbol{x}(\alpha,\beta))_{i,j}$ is in explicit form and composed of the following parts which will be examined individually. 
\begin{equation}
    A = A_{\text{Bloch Wall}}+A_{\text{Corner}}+A_{\text{Monomer}}
\end{equation}

\subsection{Bloch Wall}

A Bloch wall is the element that produces unit action in one direction; horizontal or vertical. 
\begin{figure}[H]
    \centering
    \includegraphics{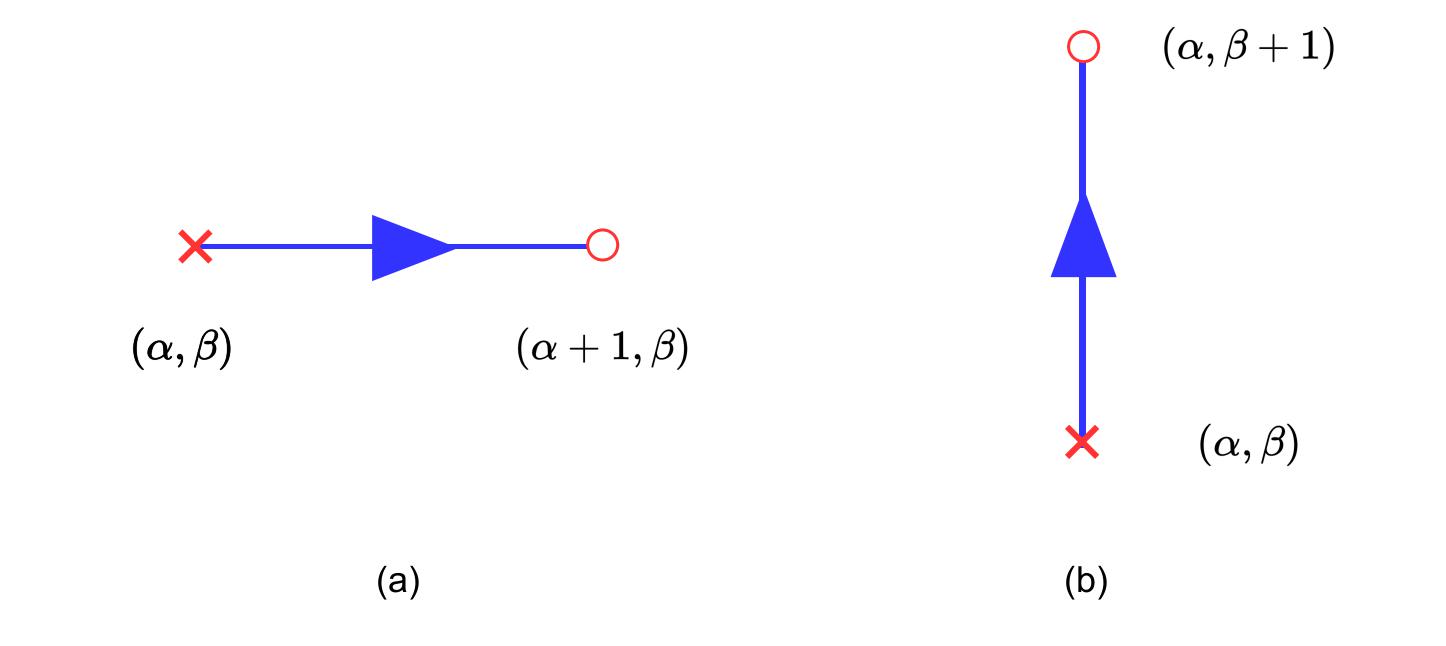}
    \caption{Bloch wall representation}
    \label{fig:blochwall}
\end{figure}
The first and second images correspond to the following product of Grassmann variables contributing as, respectively,

\begin{equation}
    z_h\eta_{\alpha,\beta}^{h^X}\eta_{\alpha+1,\beta}^{h^O} \   \: \:\ z_v\eta_{\alpha,\beta}^{v^X}\eta_{\alpha,\beta+1}^{v^O}
\end{equation} where, $z_{h,v} = e^{-2\beta J_H}$ , $e^{-2\beta J_v}$
\begin{equation}
    A_{\text{Bloch Wall}} = \sum_{\alpha\beta}\left(z_h\eta_{\alpha\beta}^{h^X}\eta_{\alpha+1,\beta}^{h^O}+z_v\eta_{\alpha\beta}^{v^X}\eta_{\alpha,\beta+1}^{v^O}\right)
\end{equation}

\subsection{Corner}

To remain consistent with $X,O$ notation, corner terms are drawn,
\begin{figure}[H]
    \centering
    \includegraphics{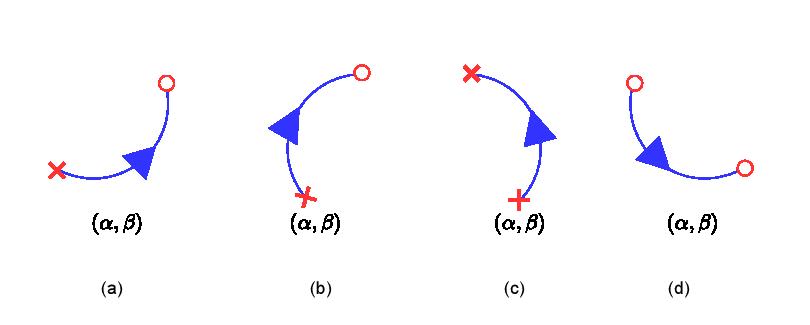}
    \caption{4 possible corner contribution}
    \label{fig:diagcornercont}
\end{figure}
It can be expressed as respectively,
\begin{equation}
        a_1\eta_{\alpha\beta}^{h^X}\eta_{\alpha\beta}^{v^O} \ \ a_3\eta_{\alpha\beta}^{v^X}\eta_{\alpha\beta}^{h^O} \ \ a_2\eta_{\alpha\beta}^{v^X}\eta_{\alpha\beta}^{h^X} \ \ a_4\eta_{\alpha\beta}^{v^O}\eta_{\alpha\beta}^{h^O} 
\end{equation}
\begin{equation}
A_{corner}=\sum_{\alpha\beta}\left(a_1\eta_{\alpha\beta}^{h^X}\eta_{\alpha\beta}^{v^O}+a_3\eta_{\alpha\beta}^{v^X}\eta_{\alpha\beta}^{h^O}+a_2\eta_{\alpha\beta}^{v^X}\eta_{\alpha\beta}^{h^X}+a_4\eta_{\alpha\beta}^{v^O}\eta_{\alpha\beta}^{h^O}\right)
\end{equation}

\begin{figure}[H]
       \centering
       \includegraphics{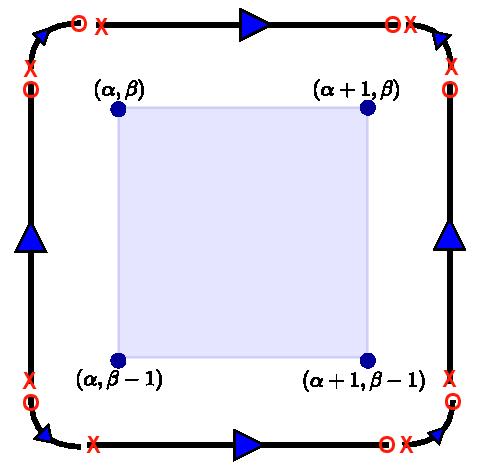}
       \caption{A consistent example}
       \label{fig:consistentexa}
\end{figure}

After these two contribution representations, one can understand that consistency means with Figure \ref{fig:consistentexa}.  \\
\begin{equation}
        \rightarrow \eta_{\alpha,\beta}^{v^X} \eta_{\alpha,\beta}^{h^O}\eta_{\alpha,\beta}^{h^X} \: \: \eta_{\alpha+1,\beta}^{h^O}\eta_{\alpha+1,\beta}^{h^X}\eta_{\alpha+1,\beta}^{v^X} \eta_{\alpha+1,\beta}^{v^O}  \:\: \nonumber
\end{equation}
\begin{equation}
    \eta_{\alpha+1,\beta-1}^{v^X}\eta_{\alpha+1,\beta-1}^{v^O}\eta_{\alpha+1,\beta-1}^{h^X}\eta_{\alpha+1,\beta-1}^{h^O} 
\end{equation}
\begin{equation}
        \eta_{\alpha,\beta-1}^{h^X}\eta_{\alpha,\beta-1}^{h^O}\eta_{\alpha,\beta-1}^{v^O}\eta_{\alpha,\beta-1}^{v^X} \:\: \eta_{\alpha,\beta}^{v^O}
        \times(a_1a_2a_3a_4z_v^2z_h^2)\footnote{for Ising model set $a_i=b_h=b_v=-1$}\nonumber
\end{equation}

\subsection{Monomer}
A monomer term represents the absence of a bond. If a monomer represents the absence of a vertical bond it's called vertical monomer, and vice versa.
\begin{equation} b_h\eta_{\alpha\beta}^{h^o}\eta_{\alpha\beta}^{h^x} \ \ b_v\eta_{\alpha\beta}^{v^o}\eta_{\alpha\beta}^{v^x}
\end{equation}
A vertical monomer can connect a horizontal line to another horizontal line and a horizontal monomer connects a vertical line to another vertical line. The contribution coming from the monomer can be expressed as 
\begin{equation}
A_{monomer}=\sum_{\alpha\beta}\left(b_h\eta_{\alpha\beta}^{h^o}\eta_{\alpha\beta}^{h^x}+b_v\eta_{\alpha\beta}^{v^o}\eta_{\alpha\beta}^{v^x}\right).
\end{equation}

The quadratic Grassmann action combines these three contributions. The use of anticommuting variables simplifies the computation.

\subsection{Spin Configurations of Monomers and Corners}

Here, line elements are not presented since their spin configuration is determined by two unit squares.
\begin{figure}[H]
    \centering
    \includegraphics{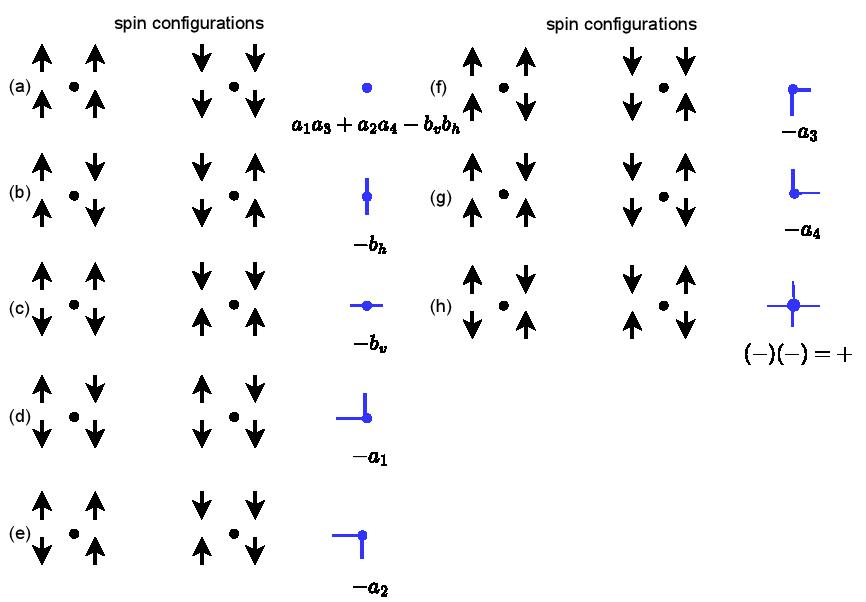}
    \caption{}
    \label{fig:enter-label}
\end{figure}
To saturate the Grassmann integral, we need to assign 4 types of Grassmann variables for each site. The following are possible configurations of structural elements to satisfy this condition.
\begin{itemize}
        \item 2 lines 1 corner
        \item 2 lines 1 monomer
        \item 2 monomers
        \item 4 lines
        \item 2 corner
    \end{itemize}
The third and final case requires special attention.
The main goal is to count the sites that are on the non-overlapping polygons. However, in this setup, locations outside the polygon are also summed. Consider the expression for two vertices,
    \begin{equation}
        a_1\eta_{\alpha,\beta}^{h^X}\eta_{\alpha,\beta}^{v^O}a_3\eta_{\alpha\beta}^{v^X}\eta_{\alpha\beta}^{h^O} \ \ a_2\eta_{\alpha,\beta}^{h^X}\eta_{\alpha,\beta}^{v^X}a_4\eta_{\alpha\beta}^{v^O}\eta_{\alpha\beta}^{h^O}
\end{equation} rearranging,
\begin{equation}
        -a_1a_3\eta_{\alpha,\beta}^{h^X}\eta_{\alpha\beta}^{h^O}\eta_{\alpha\beta}^{v^X}
        \eta_{\alpha,\beta}^{v^O}  \ \
        -a_2a_4\eta_{\alpha,\beta}^{h^X}\eta_{\alpha\beta}^{h^O}\eta_{\alpha,\beta}^{v^X}
        \eta_{\alpha\beta}^{v^O}
\end{equation}
\begin{equation}
        b_vb_h\eta_{\alpha\beta}^{h^o}\eta_{\alpha\beta}^{h^x} \eta_{\alpha\beta}^{v^o}\eta_{\alpha\beta}^{v^x}
\end{equation}

The total contribution is $a_1a_3 + a_2a_4-b_vb_h$, for the Ising case, this contribution equals to $-1$, which adds a total factor $(-1)^N$ to the integral.

\subsection{Visual Representation of Partition Function}

The final form of the partition function in anticommuting variables is the following,
    \begin{equation}
        \int d\eta_{\alpha,\beta}d\eta_{\alpha,\beta}^{\dag}exp\big({\sum_{\alpha\beta}\left(a_1\eta_{\alpha\beta}^{h^X}\eta_{\alpha\beta}^{v^O}+a_3\eta_{\alpha\beta}^{v^X}\eta_{\alpha\beta}^{h^O}+a_2\eta_{\alpha\beta}^{v^X}\eta_{\alpha\beta}^{h^X}+a_4\eta_{\alpha\beta}^{v^O}\eta_{\alpha\beta}^{h^O}\right)} \nonumber
    \end{equation} 
    \begin{equation}     +\left(z_h\eta_{\alpha\beta}^{h^X}\eta_{\alpha+1,\beta}^{h^O}+z_v\eta_{\alpha\beta}^{v^X}\eta_{\alpha,\beta+1}^{v^O}\right)
      +\left(b_h\eta_{\alpha\beta}^{h^O}\eta_{\alpha\beta}^{h^X}+b_v\eta_{\alpha\beta}^{v^O}
      \eta_{\alpha\beta}^{v^X}\right)  \big).
\end{equation} 
All possible terms can be generated by expanding the exponential.
\begin{align}
        \nonumber \int d\eta_{\alpha,\beta}d\eta_{\alpha,\beta}^{\dag}\bigg( 1+ \prod_{\alpha,\beta} ( \left(a_1\eta_{\alpha\beta}^{h^X}\eta_{\alpha\beta}^{v^O}+a_3\eta_{\alpha\beta}^{v^X}\eta_{\alpha\beta}^{h^O}+a_2\eta_{\alpha\beta}^{v^X}\eta_{\alpha\beta}^{h^X}+a_4\eta_{\alpha\beta}^{v^O}\eta_{\alpha\beta}^{h^O}\right) \\
        + \left(z_h\eta_{\alpha\beta}^{h^X}\eta_{\alpha+1,\beta}^{h^O}+z_v\eta_{\alpha\beta}^{v^X}\eta_{\alpha,\beta+1}^{v^O}\right)+\left(b_h\eta_{\alpha\beta}^{h^o}\eta_{\alpha\beta}^{h^x}+b_v\eta_{\alpha\beta}^{v^o}
      \eta_{\alpha\beta}^{v^x} \right) \bigg)
\end{align} 
The product over all the grids gives the above configurations. Since the Grassmann integral over a constant is zero, a second term can be obtained which contains the product.

Since the lattice is placed on a torus, there is a periodic boundary condition.
\begin{equation}
        \eta_{\alpha,0}^r = \eta_{\alpha,2N+1}^r \ \ \ \eta_{0,\beta}^r = \eta_{2M+1,\beta}^r
\end{equation}
Transitionally invariant quadratic action is diagonalized in site variables by going the momentum space with discrete Fourier transformation. Transformation$\left( \frac{2\pi s}{2M+1},\frac{2\pi t}{2N+1}\right) \longleftrightarrow (p_x,p_y)$ between site variables  $(\alpha,\beta) \longleftrightarrow (x,y)$ in generic form is following,
\begin{align}
        \eta_{\alpha\beta}^r=\sum_{s,t}\frac{1}{\sqrt{(2M+1)(2N+1)}}e^{\left(\frac{2\pi i \alpha s}{2M+1}+\frac{2 \pi i \beta t}{2N+1}\right)}a_{st}^r
        \\ \nonumber \eta_{\alpha\beta}^{r^\dagger} =\sum_{s,t}\frac{1}{\sqrt{(2M+1)(2N+1)}}e^{\left(-\frac{2\pi i \alpha s}{2M+1}-\frac{2 \pi i \beta t}{2N+1}\right)}a_{st}^{r^\dagger}
\end{align}

\subsection{Fourier Transformations In Explicit Form}
Fourier transformation simplifies the partition function by diagonalizing the quadratic Grassmann action. Using orthogonality relation $\delta_s^{ \: s'} = \frac{1}{\sqrt{L}} \sum_{\alpha} e^{\frac{2\pi i (s-s')}{{L}}}$ between Fourier eigenfunctions, terms connecting the neighboring sites are transformed as follows:

\textbf{1.Horizontal Interaction}
\begin{equation}
            \eta_{\alpha\beta}^{h^X}\eta_{\alpha+1\beta}^{h^O}=\sum_{s,t}\frac{e^{\left(\frac{2\pi i \alpha s}{2M+1}+\frac{2 \pi i \beta t}{2N+1}\right)}}{\sqrt{(2M+1)(2N+1)}}a_{st}^{h^X} 
             \sum_{s',t'}\frac{e^{\left(-\frac{2\pi i (\alpha+1) s'}{2M+1}-\frac{2 \pi i \beta t'}{2N+1}\right)}}{\sqrt{(2M+1)(2N+1)}}a_{st}^{h^O}
            = e^{\frac{2\pi i s}{2M+1}}a_{st}^{h^X}a_{st}^{h^O} 
\end{equation}

\textbf{2.Vertical Interaction}
\begin{equation}
            \eta_{\alpha\beta}^{v^X}\eta_{\alpha\beta+1}^{v^O}=\sum_{s,t}\frac{e^{\left(\frac{2\pi i \alpha s}{2M+1}+\frac{2 \pi i \beta t}{2N+1}\right)}}{\sqrt{(2M+1)(2N+1)}}a_{st}^{v^X} 
             \sum_{s',t'}\frac{e^{\left(-\frac{2\pi i \alpha s'}{2M+1}-\frac{2 \pi i (\beta+1) t'}{2N+1}\right)}}{\sqrt{(2M+1)(2N+1)}}a_{st}^{v^O} 
            = e^{\frac{2\pi i t}{2N+1}}a_{st}^{v^X}a_{st}^{v^O}
\end{equation}

\textbf{Mixed Terms}
\begin{equation}
            \eta_{\alpha\beta}^{v^X}\eta_{\alpha\beta}^{h^X}=\sum_{s,t}\frac{e^{\left(\frac{2\pi i \alpha s}{2M+1}+\frac{2 \pi i \beta t}{2N+1}\right)}}{\sqrt{(2M+1)(2N+1)}}a_{st}^{h^X}  
             \sum_{s',t'}\frac{e^{\left(\frac{2\pi i \alpha s'}{2M+1}+\frac{2 \pi i \beta t'}{2N+1}\right)}}{\sqrt{(2M+1)(2N+1)}}a_{st}^{h^O}  \nonumber 
\end{equation}
\begin{equation}
             = \frac{1}{(2M+1)(2N+1)}   \sum_{s,t}\sum_{s',t'} e^{\frac{2\pi i \alpha (s+s')}{2M+1}+\frac{2 \pi i \beta (t+t')}{2N+1}} a_{st}^{v^X}  a_{st}^{h^X}   
            = a_{st}^{v^X}a_{-s-t}^{h^X}
\end{equation}
Similarly,
\begin{equation}
            \eta_{\alpha\beta}^{v^O}\eta_{\alpha\beta}^{h^O}=\sum_{s,t}\frac{e^{\left(-\frac{2\pi i \alpha s}{2M+1}-\frac{2 \pi i \beta t}{2N+1}\right)}}{\sqrt{(2M+1)(2N+1)}}a_{st}^{h^O}  
             \sum_{s',t'}\frac{e^{\left(-\frac{2\pi i \alpha s'}{2M+1}-\frac{2 \pi i \beta t'}{2N+1}\right)}}{\sqrt{(2M+1)(2N+1)}}a_{st}^{h^O} \nonumber
\end{equation}
\begin{equation}
             = \frac{1}{(2M+1)(2N+1)}   \sum_{s,t}\sum_{s',t'} e^{-\frac{2\pi i \alpha (s+s')}{2M+1}-\frac{2 \pi i \beta (t+t')}{2N+1}} a_{st}^{v^O}  a_{st}^{h^O}  
            = a_{st}^{v^O}  a_{-s-t}^{h^O}
\end{equation}
Using the transformation mentioned above, the action in momentum space takes the following form,
\begin{equation}
         A_{free \: fermion} = \sum_{st} z_h \: e^{\frac{2\pi i s}{2M+1}} a_{st}^{h^X} a_{st}^{h^O} + z_v \: e^{\frac{2 \pi i t}{2N+1}} a_{st}^{v^X} a_{st}^{v^O} + a_1 a_{st}^{h^X} a_{st}^{v^O} +a_3 a_{st}^{v^X} a_{st}^{h^O} \nonumber 
\end{equation}
\begin{equation}
        + a_2 a_{st}^{v^X} a_{-s-t}^{h^X} + a_4 a_{st}^{v^O} a_{-s-t}^{h^O} + b_h a_{st}^{h^O} a_{st}^{h^X} + b_v a_{st}^{v^O} a_{st}^{v^X}
\end{equation}
where terms are coupled. The notation can be manipulated and $k<0$ can be used to separate coupled variables, since if one is positive, the other must be negative. 

\begin{align}
            a A_{corner} a^\dag = a A_{st} a^\dag + a A_{-s-t} a^\dag
\end{align}
Explicit forms of $A_{st}$ and $A_{-s-t}$ are following,
\begin{align}
         A_{st} = \sum_{st} z_h \: e^{\frac{2\pi i s}{2M+1}} a_{st}^{h^X} a_{st}^{h^O} + z_v \: e^{\frac{2 \pi i t}{2N+1}} a_{st}^{v^X} a_{st}^{v^O} + a_1 a_{st}^{h^X} a_{st}^{v^O} +a_3 a_{st}^{v^X} a_{st}^{h^O} \nonumber \\ 
        + a_2 a_{st}^{v^X} a_{-s-t}^{h^X} + a_4 a_{st}^{v^O} a_{-s-t}^{h^O} + b_h a_{st}^{h^O} a_{st}^{h^X} + b_v a_{st}^{v^O} a_{st}^{v^X}
\end{align} 
\begin{align}
        \nonumber A_{-s-t} = \sum_{ -s-t} z_h \: e^{\frac{-2\pi i s}{2M+1}} a_{ -s-t}^{h^X} a_{ -s-t}^{h^O} + z_v \: e^{\frac{-2 \pi i t}{2N+1}} a_{ -s-t}^{v^X} a_{ -s-t}^{v^O}+ a_1 a_{ -s-t}^{h^X} a_{ -s-t}^{v^O} \nonumber \\ +a_3 a_{ -s-t}^{v^X} a_{ -s-t}^{h^O} 
        + a_2 a_{ -s-t}^{v^X} a_{st}^{h^X} + a_4 a_{ -s-t}^{v^O} a_{st}^{h^O} + b_h a_{ -s-t}^{h^O} a_{ -s-t}^{h^X} + b_v a_{ -s-t}^{v^O} a_{ -s-t}^{v^X}.
\end{align}

\subsection{Matrix Form}

First, the action can be expressed as matrix form.
\begin{align}
       \nonumber \begin{bmatrix}
                a_{st}^{h^O} \:& &
                a_{st}^{h^X} \:& &
                a_{st}^{v^O} \:& &
                a_{st}^{v^X} \:& &
                a^{h^O}_{-s-t}\: & &
                a^{h^X}_{-s-t}\:& &
                a^{v^O}_{-s-t}\:& &
                a^{v^X}_{-s-t}\:& 
\end{bmatrix} 
        \: \:\:\ \times \textbf{      }\textbf{      }\textbf{      }\\ \nonumber 
\begin{bmatrix}
                0 & b_h & 0&0 &0& 0& 0& 0\\
                z_he^{\frac{2\pi i s}{2M+1}} & 0 & a_1 &0 &0& 0& 0& 0 \\
                0 & 0 & 0&b_v &a_4& 0& 0& 0 \\
                a_3 & 0 & z_ve^{\frac{2\pi i t}{2N+1}}&0 &0& a_2 & 0& 0 \\
                0 & 0 & 0&0 &0& b_h& 0& 0 \\
                0 & 0 & 0&0 &z_he^{-\frac{2\pi i s}{2M+1}}& 0& a_1& 0\\
                a_4 & 0 & 0&0 &0& 0& 0& b_v\\
                0 & a_2 & 0&0 &a_3& 0& z_v e^{\frac{-2\pi i t}{2N+1}}& 0\\
\end{bmatrix}
\begin{bmatrix}
                a_{st}^{h^O} \\
                a_{st}^{h^X} \\
                a_{st}^{v^O} \\
                a_{st}^{v^X} \\
                a^{h^O}_{-s-t}\\
                a^{h^X}_{-s-t}\\
                a^{v^O}_{-s-t}\\
                a^{v^X}_{-s-t}\\
\end{bmatrix} \nonumber 
\end{align} 
To get rid of the 8 by 8 matrix one can group O's and X's to get the bilinear form $\eta\eta^\dag$. The bilinear action can be written in terms of smaller matrices. To get the bilinear form, each term must contain only one $X$ and one $O$. Note that only (s,t) and (-s,-t) terms are coupled. To get one O and one X for every term change terms 
\begin{align}
             a_2 a_{st}^{v^X} a_{-s-t}^{h^X} = a_2 a_{st}^{v^O} a_{-s-t}^{h^X} 
\end{align}
\begin{align}
 a_4 a_{st}^{v^O} a_{-s-t}^{h^O} = a_4 a_{st}^{v^X}  a_{-s-t}^{h^O}= -a_4 a_{-s-t}^{h^O} a_{st}^{v^X}
\end{align}
After changing the order and these two terms, two actions become
\begin{align}
        \nonumber A_{st} = \sum_{st} \left(b_h-z_he^{\frac{2\pi i s}{2M+1}}\right) a_{st}^{h^O} a_{st}^{h^X} + \left( b_v - z_v e^{\frac{2\pi i t}{2N+1}}\right) a_{st}^{v^O} a_{st}^{v^X} \\- a_1 a_{st}^{v^O} a_{st}^{h^X} -a_3 a_{st}^{h^O} a_{st}^{v^X}+a_2 a_{st}^{v^O} a_{-s-t}^{h^X}-a_4 a_{-s-t}^{h^O}a_{st}^{v^X}   \\
        \nonumber A_{-s-t} = \sum_{ -s-t} (b_h-z_h \: e^{\frac{-2\pi i s}{2M+1}}) a_{ -s-t}^{h^O} a_{ -s-t}^{h^X} +( b_v- z_v \: e^{\frac{-2 \pi i t}{2N+1}} )a_{ -s-t}^{v^O} a_{ -s-t}^{v^X} \\ \nonumber - a_1 a_{ -s-t}^{v^O} a_{ -s-t}^{h^X} -a_3 a_{ -s-t}^{h^O} a_{ -s-t}^{v^X} \nonumber \\
        +a_2 a_{ -s-t}^{v^O} a_{st}^{h^X} - a_4 a_{ st}^{h^O} a_{-s-t}^{v^X} + b_h a_{ -s-t}^{h^O} a_{ -s-t}^{h^X} + b_v a_{ -s-t}^{v^O} a_{ -s-t}^{v^X}
        \nonumber 
\end{align}
 The reason one can define these is that they have the same $\eta$'s in different order $h_s=b_h-z_he^{\frac{2\pi i s}{2M+1}} \:\:\: v_t= b_v - z_v e^{\frac{2\pi i t}{2N+1}}$
\begin{equation}
    a A_{monomer+blochwall}a^\dag= 
\begin{bmatrix}
     a_{st}^{h^O} \\
     a_{st}^{v^O} \\ 
     a_{-s-t}^{h^O} \\
     a_{-s-t}^{v^O} 
    \end{bmatrix}^T
    \begin{bmatrix}
    h_s & 0 & 0 & 0 \\
    0 & v_t & 0 & 0 \\
    0 & 0 & h_{-s} &  0 \\
    0 & 0 & 0 & v_{-t}
\end{bmatrix}
\begin{bmatrix}
           a_{st}^{h^X} \\            
           a_{st}^{v^X} \\            
           a_{-s-t}^{h^X}\\            
           a_{-s-t}^{v^X}
\end{bmatrix} 
\end{equation}
\begin{equation} 
    a A_{corner}a^\dag = 
\begin{bmatrix}
     a_{st}^{h^O} \\
     a_{st}^{v^O} \\ 
     a_{-s-t}^{h^O} \\
     a_{-s-t}^{v^O} 
\end{bmatrix}^T
\begin{bmatrix}
    0 & -a_3 & 0 & -a_4 \\
    -a_1 & 0 & a_2 & 0 \\
    0 & -a_4 & 0 & -a_3 \\
    a_2 & 0 & -a_1 & 0
\end{bmatrix}
\begin{bmatrix}
          a_{st}^{h^X} \\            
          a_{st}^{v^X} \\            
          a_{-s-t}^{h^X}\\           
          a_{-s-t}^{v^X}
\end{bmatrix} 
\end{equation} 

\begin{frame}{the Sum of the matrices}
\begin{align}
        a(A_{monomer+blochwall}+A_{corner})a^\dag \\ 
       =\begin{bmatrix}
        h_s & -a_3 & 0 & -a_4 \\
        -a_1 & v_t & a_2 & 0 \\
        0 & -a_4 & h_{-s} & -a_3 \\
        a_2 & 0 & -a_1 & v_{-t}
        \end{bmatrix}
\end{align} 
which is a squared reduced partition function.
\begin{align}
        \nonumber detA=h_sv_th_{-s}v_{-t}-a_1a_3(h_sv_t+h_{-s}v_{-t})-a_2a_4(h_sv_{-t}+h_{-s}v_t)
        +(a_1a_3+a_2a_4)^2\nonumber 
\end{align}
\end{frame}
plug the values for $b_h=b_v=a_i=-1$ for Ising model and $h_s=1-z_he^{ip_x}$ and $v_t=1-z_ve^{ip_y}$ which simplifies to
\begin{align}
          detA=(1+z_h^2)(1+z_v^2)+2(1-z_v^2)z_h cos(\frac{2\pi s}{2M+1}) +2(1-z_h^2)z_v cos(\frac{2\pi t}{2N+1}).
\end{align}

\subsection{Exact Calculation of Partition Function}
The partition function in Grassmann variables is expressed as
\begin{align}
    \label{39}\int \prod_{m=1}^nd\eta_m d\eta^\dagger_m\: exp\left(\sum_{i=1}^n\eta_i \lambda_i\eta^\dagger_i\right) = \prod_{m=1}^n \lambda_m
\end{align}
The proof can be done by starting from the expanding the exponential. Since $\eta_i^2=0$, there remain only the terms where each Grassmann variable appears.
\begin{align}
     =\int d\eta_1d\eta^\dagger_1\dots d\eta_nd\eta^\dagger_n \frac{1}{n!}\left(\sum_{i=1}^n \eta_i \lambda_i \eta^\dagger_i\right)^n         = \int \eta_1\eta^\dagger_1\dots \eta_n\eta^\dagger_n 
\end{align}
Here, $\lambda$'s has the relation $\Lambda_{kl} = \sum_{i=1}^n M_{ki} M_{li} \lambda_i$. To understand better, the multinomial theorem can be used.
\begin{align}
    \left( x_1+x_2+\dots +x_m \right)^n= \sum_{k_1+k_2+\dots+k_m=n} \begin{pmatrix}
         n \\
         k_1,k_2,\dots,k_m 
\end{pmatrix}x_1^{k_1}x_2^{k_2}\dots x_m^{k_m} \\
    \nonumber = \sum \frac{n!}{k_1!k_2!...k_m!}x_1^{k_1}x_2^{k_2}\dots x_m^{k_m}
\end{align}
Since the square of any Grassmann variable gives zero all $k_i$ must equal 1 
\begin{align}
        \int d\eta_1 d\eta^\dagger_1 ... d\eta_n d\eta^\dagger_n \frac{1}{n!}n!\eta_n\lambda_n\eta^\dagger_n...\eta_1\lambda_1\eta^\dagger_1 =\prod_{m=1}^n \lambda_m
\end{align}
For a Gaussian Grassmann integral, this result is related to the determinant.
\begin{align}
     I = \int d\boldsymbol{\eta}d\boldsymbol{\eta}^\dagger e^{\sum_{\alpha,\beta}\eta_{\alpha}A_{\alpha,\beta}\eta_{\beta}^{\dag}} = \int d\boldsymbol{\eta} d\boldsymbol{\eta}^\dagger e^{\sum_{\alpha}\eta_{\alpha}\sum_{\beta}A_{\alpha,\beta}\eta_{\beta}^{\dag}}\\
     = \nonumber \sum_{\sigma}\textbf{sgn}({\sigma})\prod_{\alpha}A_{\alpha,\sigma(\alpha)} = \textbf{Det(A)}
\end{align} 
where the expansion of the exponential is
\begin{align}
     e^{\sum_{\alpha}\eta_{\alpha}\sum_{\beta}A_{\alpha,\beta}
      \eta_{\beta}^{\dag}} = \prod_{\alpha}\eta_{\alpha}\sum_{\beta}A_{\alpha,\beta}\eta_{\beta}^{\dag}.
\end{align}
When integrated with respect to Grassmann variables, the result includes all permutations of the matrix elements weighted by a sign. This is the definition of the determinant.
\begin{equation}  \sum_{\sigma}\textbf{sgn}({\sigma})\prod_{\alpha}A_{\alpha,\sigma(\alpha)} = \textbf{Det(A)}
\end{equation}
So, our reduced partition function for $A\prime = -A^T_{-s,-t}+A(s,t)$, gives
\begin{equation}
     Z_{reduced} = \prod_{s,t}(det(A^\prime))^{\frac{1}{2}}
\end{equation}
If one takes the logarithm and divides it by the number of sites for finding free energy per site, for $N\rightarrow\infty$ the result is 
\begin{equation}
     \lim_{N \to \infty}\frac{1}{N}log(\mathcal{Z}_{reduced}) = \frac{1}{2N}log(det(A^\prime)).
\end{equation} 
In the continuum limit, the expression becomes an integral
\begin{equation}
     \frac{1}{2}\int_{-\pi}^{\pi}\int_{-\pi}^{\pi} \frac{dp_x}{2\pi} \frac{dp_y}{2\pi} log(det(A^{\prime}))
\end{equation} 
which is the solution to the 2-dimensional Ising model. 

The exact calculation of the partition function with Grassmann variables provides an elegant way to solve. The antisymmetry property of Grassmann variables simplifies the expression and provides a direct pathway for calculations. These steps can be extended not only to the Isıing model but also to other two-dimensional fermionic statistical mechanics models.

\section{Partition Function Evaluated by Spinor Analysis \textit{(Cansu Özdemir and Sinan Ulaş Öztürk)}}
\label{chapter:6.5}

In this section, we will review the paper \cite{PhysRev.76.1232}. We aim to calculate the partition function of the two-dimensional Ising model by rewriting it in terms of the transfer matrix and then expressing it as a representation of the rotation group. We convert the calculation to an eigenvalue problem and use the rotation and spinor representations to find the eigenvalues of the transfer matrix.

\subsection{The Model and its Transfer Matrix}

Let us consider a two-dimensional Ising model on a lattice with $m$ rows and $n$ columns where the interaction energy between rows is $J_b$ and interaction energy between columns is given as $J_a$. Then, the total energy of the system is given as follows,
\begin{equation}
    E = J_b\sum_{\mu=1}^{m}s_{\mu\nu}s_{\mu+1,\nu}+J_a\sum_{n=1}^{n}s_{\mu\nu}s_{\mu,\nu+1}
\end{equation}
If we impose cylindrical boundary conditions, then the partition function will take the form,
\begin{equation*}
Z(a,b) = \sum_{\{s\}}\exp (-\beta E) = \sum_{\{s\}}\prod_{\mu=1}^m\left(\prod_{\nu=1}^n \exp(b s_{\mu\nu}s_{\mu+1,\nu})\exp(a s_{\mu\nu}s_{\mu,\nu+1})\right) 
\end{equation*}
\begin{equation}\label{transfer}
    \sum_{\{s\}}\prod_{\mu=1}^{m}T_{\mu,\mu+1} = \tr T^m
\end{equation}
Here, $T$ is called the transfer matrix, and $a,b$ are defined as $a=-\beta J_a$ and $b = -\beta J_b$. To understand the behavior of the transfer matrix, we need to investigate its components. Each component of the transfer matrix will correspond to a microstate of rows $\mu$ and $\mu+1$. Since there are $n$ spin sites at any given row, and each spin takes two values, namely $\pm1$, there are $2^n$ microstates for each row. One possible basis to represent these microstates can be constructed by introducing a 2 component vector for each spin value. Let $i$ denote one of these microstates and let $|s^{(i)}_\nu\rangle$ denote the value of the $\nu$th spin value in the $i$th microstate. We assign,
\begin{flalign}
    s^{(i)}_\nu = +1\Rightarrow |s^{(i)}_\nu\rangle =\ket{+} = \begin{pmatrix}
        1 \\ 0
    \end{pmatrix}\\
    s^{(i)}_\nu = -1\Rightarrow |s^{(i)}_\nu\rangle = \ket{-} = \begin{pmatrix}
        0 \\ 1
        \end{pmatrix}
\end{flalign}
Then, each microstate $i$ can be expressed as follows,
\begin{equation}
    \ket{i} = \bigotimes_{\nu=1}^n|s^{(i)}_\nu\rangle
\end{equation}
Where $i=1\dots 2^n$. Here, each basis vector is the standard basis vector in $\mathbb{R}^{2^n}$. For example,
$$
\ket{1}= \begin{pmatrix}
    1\\ 0 \\ \vdots \\ 0
\end{pmatrix} = \bigotimes_{\nu=1}^n\begin{pmatrix}
    1 \\ 0
\end{pmatrix}
$$
Corresponds to a state where each spin value in the row is $+1$. This way of expressing the microstate is equivalent to representing the state as a number in base 2, where +1 corresponds to 0 and -1 corresponds to 1. Then, the sum over all possible microstates in the partition function simply becomes a sum over $i$ for each row $\mu$. We can rewrite the partition function as follows,
\begin{equation}
    Z = \sum_{i_1\dots i_m}\prod_{\mu=1}^m\left[\left(\prod_{\nu=1}^n\exp(bs^{(i_\mu)}_\nu s_\nu^{(i_{\mu+1})})\right)\left(\prod_{\nu=1}^n\delta_{i_\mu,i_{\mu+1}}\exp(as_{\nu}^{(i_\mu)}s_{\nu+1}^{(i_{\mu+1})})\right)\right]
\end{equation}
This is completely identical with eq.\ref{transfer}. At this point, let us introduce two intermediary matrices $V^a$ and $V^b$ by their components in the following way,
\begin{flalign}\label{comp}
    \begin{cases}
        V^a_{ij} = \prod_{\nu}\delta_{ij}\exp(as_{\nu}^{(i)}s_{\nu+1}^{(j)}) = \bra{i}V^a\ket{j} \\\\
        V^b_{ij} = \prod_{\nu}\exp(bs^{(i)}_\nu s_\nu^{(j)}) = \bra{i}V^b\ket{j}
    \end{cases}
\end{flalign}
Then the partition function is,
$$
Z = \sum_{i_1\dots i_m}\prod_{\mu=1}^m V^a_{i_1i_2}V^a_{i_2i_3}\dots V^a_{i_mi_1}V^b_{i_1i_2}V^b_{i_2i_3}\dots V^b_{i_mi_1}
$$
Here again, we used the fact that we have cylindrical boundary conditions. Thus $i_{m+1} = i_1$. They are in the same microstate. If we define $T=V^aV^b$ and calculate its components, we get,
$$
T_{ij}=\sum_kV^a_{ik}V^b_{kj}=\sum_k\prod_\nu\exp(as^{(i)}_\nu s^{(k)}_\nu)\exp(bs^{(k)}_\nu s^{(j)}_{\nu+1})\delta_{kj}=\prod_\nu\exp(as^{(i)}_\nu s^{(j)}_\nu+bs^{(j)}_\nu s^{(j)}_{\nu+1})
$$
This is the exact expression given in \ref{transfer}. Therefore, $T$ is the transfer matrix. This also means that $T_{ij}=V^a_{ij}V^b_{ij}$. Hence, the partition function becomes,
\begin{equation}
  Z = \sum_{i_1\dots i_m}\prod_{\mu=1}^m T_{i_1 i_2}T_{i_2 i_3}\dots T_{i_m i_1} = \sum_{i_1} T^m_{i_1 i_1} = \tr T^m
\end{equation}

It turns out that the matrices $V^a$ and $V^b$ can be expressed using the matrices $X_\nu$ and $Z_\nu$ which are defined in terms of the Pauli matrices and are given as follows,

\begin{equation}
    X_{\nu}= \underbrace{\id\otimes\cdots\id\otimes
    \sigma_x}_{\nu\textrm{th place}}\otimes\id\dots\otimes\id \,,
\end{equation}
And $Z_\nu$ and $Y_\nu$ are defined analogously. Here are the Pauli matrices,
$$
\sigma_x=\begin{pmatrix}
0 & 1\\
1 & 0
\end{pmatrix},\quad 
\sigma_y=\begin{pmatrix}
    0 & -i\\  
    i & 0
\end{pmatrix},\quad
\sigma_z= \begin{pmatrix}
    1 & 0\\
    0 & -1
\end{pmatrix}
$$
Along with the 2-dimensional identity matrix $\id$, these matrices already form a representation for the 3-dimensional rotation group $SO(3)$. We will later discuss this in great detail but it turns out that the matrices $X_\nu,Y_\nu,Z_\nu$ can be used to create the generators of $n$-dimensional rotation group. Defining $\Bar{b}$ by the equation $\sinh{2b} \sinh 2 \Bar{b}=1$ with $\Bar{b}>0$, we can also express $V^a$ and $V^b$ as, 

\begin{flalign}
\begin{cases}
    V^b = (2 \sinh{2b})^{n/2}\exp({\Bar{b} \sum_{\nu=1}^n X_\nu}) \\\\
    V^a = \exp (a \sum_{\nu=1}^n Z_\nu Z_{\nu+1}) 
\end{cases}
\end{flalign}
We define,
\begin{align}
    &B= \sum_{\nu=1}^n X_\nu & A= \sum_{\nu=1}^n Z_\nu Z_{\nu+1} \label{ab}
\end{align}

It is not immediately obvious why these matrices are the same as those defined in eq.\ref{comp}. To do that we have to investigate the components of $V^a$ and $V^b$ in the basis that we have constructed earlier.
\subsubsection{Components of \texorpdfstring{$V^a$}{}}

The components of $V^a$ are expressed as,
$$
V^a_{ij} = \bra{i}V^a\ket{j} = \bra{i}\prod_{\nu=1}^n\exp (a Z_\nu Z_{\nu+1})\ket{j}
$$
This must be equal to the expression in eq.\ref{comp}. To understand how this matrix acts on the basis vectors, we first have to see how $\sigma_z$ acts on vectors $\ket{+}$ and $\ket{-}$, representing the state of one spin site. Notice,

\begin{flalign}
    \begin{cases}
        \sigma_z\ket{+}=\begin{pmatrix}
            1 & 0 \\
            0 & -1
        \end{pmatrix}\begin{pmatrix}
            1 \\
            0
        \end{pmatrix}=\begin{pmatrix}
            1 \\
            0
        \end{pmatrix}=\ket{+}
    \\\\
    \sigma_z\ket{-}=\begin{pmatrix}
            1 & 0 \\
            0 & -1
        \end{pmatrix}\begin{pmatrix}
            0 \\
            1
        \end{pmatrix}=\begin{pmatrix}
            0 \\
            -1
        \end{pmatrix}=-\ket{-}
        \end{cases}
\end{flalign}
Or, more concisely,
\begin{equation}
    \sigma_z|s^{(i)}_\nu\rangle = s^{(i)}_\nu|s^{(i)}_\nu\rangle
\end{equation}
Then it follows that,
$$
Z_\nu\ket{i}=\left[\bigotimes_{k=1}^{\nu-1}\id\otimes\sigma_z\otimes\left(\bigotimes_{k=\nu+1}^{n}\id\right)\right]\bigotimes_{\lambda=1}^n|s^{(i)}_\lambda\rangle = \bigotimes_{\lambda=1}^{\nu-1}|s^{(i)}_\lambda\rangle\otimes\sigma_z|s^{(i)}_\nu\rangle\otimes\left(\bigotimes_{\lambda=\nu+1}^{n}|s^{(i)}_\lambda\rangle\right)
$$
$$
=\bigotimes_{\lambda=1}^{\nu-1}|s^{(i)}_\lambda\rangle\otimes s^{(i)}_\nu|s^{(i)}_\nu\rangle\otimes\left(\bigotimes_{\lambda=\nu+1}^{n}|s^{(i)}_\lambda\rangle\right)=s^{(i)}_\nu\ket{i}
$$
Hence, it turns out that $|i\rangle$ is an eigenvector of the matrix $Z_\nu$ with the eigenvalue $s^{(i)}_\nu$. $Z_\nu$ is diagonalized in this basis. Then, 
$$
Z_{\nu}Z_{\nu+1}\ket{i} =  s^{(i)}_\nu s^{(i)}_{\nu+1}\ket{i}
$$
The eigenvalue $s^{(i)}_\nu s^{(i)}_{\nu+1}$ equals $+1$ if $s^{(i)}_\nu$ is equal to $s^{(i)}_{\nu+1}$ and equals $-1$ if they have the opposite sign. It "checks" whether two adjacent spin sites have the same spin value. Since $\ket{i}$ is an eigenvector of $Z_\nu Z_{\nu+1}$ then it is also an eigenvector of a $\exp(bZ_\nu Z_{\nu+1})$ with the eigenvalue $\exp(as^{(i)}_\nu s^{(i)}_{\nu+1})$. Which indicates,

\begin{equation}
    \bra{i}V^a\ket{j} = \bra{i}\prod_{\nu=1}^n\exp(aZ_\nu Z_{\nu+1})\ket{j} = \prod_{\nu=1}^n\exp(as^{(i)}_\nu s^{(i)}_{\nu+1})\bra{i}\ket{j} = \prod_{\nu=1}^n\exp(as^{(i)}_\nu s^{(i)}_{\nu+1})\delta_{ij}
\end{equation}
Which is exactly what we needed to show. 

\subsubsection{ Components of \texorpdfstring{$V^b$}{}}  
Components of $V^b$ can be found as follows,
$$
V^b_{ij} =(2\sinh2b)^{n/2}\bra{i}\prod_\nu^n \exp(\Bar{b}X_\nu)\ket{j} 
$$
And again, this must be equal to the expression given in eq.\ref{comp}. Similar to the calculation in the previous part, we begin by denoting what $\sigma_x|s^{(i)}_\nu\rangle$ is,

\begin{flalign*}
    \begin{cases}
        \sigma_x\ket{+}=\begin{pmatrix}
            0 & 1 \\
            1 & 0
        \end{pmatrix}\begin{pmatrix}
            1 \\
            0
        \end{pmatrix}=\begin{pmatrix}
            0 \\
            1
        \end{pmatrix}=\ket{-}
    \\\\
    \sigma_x\ket{-}=\begin{pmatrix}
            0 & 1 \\
            1 & 0
        \end{pmatrix}\begin{pmatrix}
            0 \\
            1
        \end{pmatrix}=\begin{pmatrix}
            1 \\
            0
        \end{pmatrix}=\ket{+}
        \end{cases}
\end{flalign*}
Or, in short,
$$
\sigma_x|s^{(i)}_\nu\rangle = |-s^{(i)}_\nu\rangle
$$
Then we have,
$$
X_\nu|i\rangle=\left[\bigotimes_{k=1}^{\nu-1}\id\otimes\sigma_x\otimes\left(\bigotimes_{k=\nu+1}^{n}\id\right)\right]\bigotimes_{\lambda=1}^n|s^{(i)}_\lambda\rangle = \bigotimes_{\lambda=1}^{\nu-1}|s^{(i)}_\lambda\rangle\otimes\sigma_x|s^{(i)}_\nu\rangle\otimes\left(\bigotimes_{\lambda=\nu+1}^{n}|s^{(i)}_\lambda\rangle\right)
$$
$$
=\bigotimes_{\lambda=1}^{\nu-1}|s^{(i)}_\lambda\rangle\otimes|-s^{(i)}_\nu\rangle\otimes\left(\bigotimes_{\lambda=\nu+1}^{n}|s^{(i)}_\lambda\rangle\right)
$$
The result is another microstate of the row where the $\nu$th spin has the opposite sign. This vector also corresponds to another basis vector in the $\ket{i}$ basis. Effectively, the matrix $X_\nu$ "flips" the $\nu$th spinsite of $\ket{i}$ and produces another basis vector within all possible microstates of the row. We can imagine this as a type of permutation $P_\nu:\{1,2,\dots 2^n\}\to\{1,2,\dots 2^n\}$  with,
$$
X_\nu\ket{i} = |P_\nu(i)\rangle
$$
Now we can investigate $\exp(\Bar{b}X_\nu)$. Let us expand this matrix,
$$
\exp(\Bar{b}X_\nu) = \sum_{k=0}^\infty\dfrac{(\Bar{b}X_\nu)^k}{k!}
$$
Where we have $X_\nu^2=\id$. Evidently, we get,
$$
\exp(\Bar{b}X_\nu) = \cosh\Bar{b}\id + \sinh\Bar{b} X_\nu = \cosh\Bar{b}(\id+\tanh\Bar{b}X_\nu)
$$
Notice that we have the following identities from the definition of $\bar{b}$, which will help us simplify these results.

$$
\tanh b = e^{-2\Bar{b}} \quad ; \quad \tanh \Bar{b} = e^{-2b}
$$
$$
\prod_{\nu=1}^n\exp(\Bar{b}X_\nu) = \cosh^n\Bar{b}\prod_{\nu=1}^n(\id+e^{-2b}X_\nu)
$$
Then,
$$
V^b = (2\sinh{2b})^{n/2}\cosh^n\Bar{b}\prod_{\nu=1}^n(\id+e^{-2b}X_{\nu})
$$
If we consider the coefficient of this expression, we see that,
$$
\sqrt{2\sinh2b}\cosh^n\Bar{b} = \sqrt{\dfrac{2}{\sinh2\Bar{b}}}\cosh \Bar{b} = \sqrt{\dfrac{\cosh^2\Bar{b}}{\sinh \bar{b}\cosh \Bar{b}}}=\sqrt{\dfrac{1}{\tanh\Bar{b}}} = e^{b} 
$$
Finally, we arrive at the final expression for $V^b$, which is,

\begin{equation}
    V^b = e^{nb}\prod_{\nu=1}^n(\id +e^{-2b}X_\nu) = e^{nb}\left(\id+e^{-2b}\sum_{\nu=1}^nX_\nu+e^{-4b}\sum_{\nu_1<\nu_2}X_{\nu_1}X_{\nu_2}+\dots e^{-2nb}\prod_{\nu=1}^n X_\nu\right)
\end{equation}
Its components are,
$$
\bra{i}V^b\ket{j} = e^{nb}\left(\delta_{ij}+e^{-2b}\sum_\nu\delta_{i,P_\nu(j)}+\dots e^{-2nb}\delta_{i,P_1\dots P_n(j)}\right)
$$
At this point, we can see how this matrix compares the energy of interaction between two adjacent rows. If the microstates $i$ and $j$ of these rows are the same, then the energy is $e^{nb}$, and from the Kronecker deltas in this expression, only the first term survives. However, if only one spin site has the opposite sign between $i$ and $j$, then the difference in energy between the first case will be $e^{-2b}$. The coefficient of the second term reflects this. Accordingly, only one of the Kronecker deltas will survive from the second term. The same is true for any number of different spin sites between $i$ and $j$. Kronecker deltas will pick the necessary combinations of $P_\nu$ to get from $j$ to $i$, and the coefficient of that term will adjust the energy in the exponent accordingly. This way, $V^b$ accounts for all possible energies between different rows.

Up to this point, we introduced the physical background of the problem of calculating the partition function for the two-dimensional Ising model. We transformed the problem into calculating the trace of the transfer matrix and how to express the transfer matrix in terms of the generators of rotation groups. Now we investigate the problem from a representation theory perspective to find the eigenvalues of the transfer matrix. 

\subsection{Spin Representations}
Before moving on with doing spinor analysis of the transfer matrix, we should examine the relations between $2n$ and $2^n$ dimensional spaces. This section will construct the spin representations of $2n$-dimensional rotation groups using a set of $2n$ anti-commuting matrices. \par
Spin representations are double covers of $O(n)$ or $SO(n)$ groups with arbitrary dimension. In the physical sense, spin representations take the "number of full turns before a rotation" into consideration. As clear from their names, we will see that a $2 \pi$ rotation would be equivalent to the initial state. Thus, contrary to "normal" representations, every element can be represented with 2 distinct elements of their spin representations. \par
We start by writing a set of $2n$ matrices $\Gamma_k$ that anti-commute, which we will use to construct the $2^n$ dimensional matrix algebra. 
\begin{align}
    \{ \Gamma_k , \Gamma_l \} = 2 \delta_{kl}
\end{align}
To construct such quantities, we will use the faithful representations of the quaternion group as building blocks, with the smallest one being $2$-dimensional.  The reader may refer to \cite{dirac1928quantum} for the physical interpretation of this construction. Which representation we will use is redundant. We may refer to the representations as $\{\id, \sigma_x, \sigma_y, \sigma_z\}$. Let us realize this set as the Pauli spin matrices for convenience since they already form a set of anti-commuting matrices. given,
\begin{equation}
    X_\nu= \underbrace{\id\otimes\cdots\id\otimes
    \sigma_x}_{\textrm{$\nu$th place}}\otimes\id\dots\otimes\id \, ,
\end{equation}
and similarly, $Y_\nu$ and $Z_\nu$. Now, we may construct a set of $2n$ anti-commuting matrices $\Gamma_k$ as
\begin{align}
    \Gamma_{2r-1} = X_1 \cdots X_{r - 1} \cdot Z_r \cdot \id_{r+1} \cdot \id_{n} \nonumber \\
    \Gamma_{2r} = X_1 \cdots X_{r - 1} \cdot Y_r \cdot \id_{r+1} \cdot \id_{n}  \, .
\end{align}
Note that $\Gamma_k$ are of the form of tensor product of $n$ $2 \times 2$ matrices, thus are of dimension $2^n$. Now, let us take all the possible products of these matrices. Since these matrices are anti-commuting, we know that $\Gamma_k^2 = 1$ and changing the order of $\Gamma_k$ matrices with different indices only adds a factor of $(-1)$, that is $\Gamma_k \Gamma_l = - \Gamma_l \Gamma_k$ for $k \neq l$. Thus we can see that $\Gamma$ matrices can only appear in first power and changing the order does not result in independent matrices, leading to only $2^{2n}$ independent matrices, which are of the form 
\begin{align*}
    1, \, \Gamma_1, \, \Gamma_2, \, \cdots, \, \Gamma_1 \Gamma_2, \, \cdots, \, \Gamma_1 \Gamma_2 \Gamma_3, \, \cdots, \, \Gamma_1 \Gamma_2 \Gamma_3 \Gamma_4, \, \cdots, \, \Gamma_1 \Gamma_2 \cdots \Gamma_{2n} \, .
\end{align*}
Any $2^n$-dimensional matrix can be written as a linear combination of these $2^{2n}$ independent matrices. Thus base matrices span the complete algebra of $2^n$ dimensional matrices. For the proof, the reader may refer to \cite{brauer1935spinors}. \par
Now let us suppose we apply a similarity transformation to our $\Gamma_k$ matrices to obtain a new converted set of $\Gamma_k^*$ matrices,
\begin{equation}
    \Gamma_k^* = S \Gamma_k S^{-1} \, .
\end{equation}
Notice that the anti-commuting property is conserved by this transformation, 
\begin{align}
    \{\Gamma_k^*, \Gamma_l^*\} &= \Gamma_k^* \Gamma_l^* + \Gamma_l^*\Gamma_k^* \nonumber \\ 
    &= S \Gamma_k S^{-1} S \Gamma_l S^{-1} + S \Gamma_l S^{-1} S \Gamma_k S^{-1} \nonumber \\
    &= S (\Gamma_k \Gamma_l + \Gamma_l \Gamma_k) S^{-1} = 2\delta_{kl} \, .
\end{align}
Hence, we can obtain other sets of anti-commuting matrices by applying similarity transformations to a set of anti-commuting matrices. These matrices are equivalently as good as other ones to obtain bases for $2^n$-dimensional matrices. \par
We have said that $\Gamma_i$ matrices constitute a basis for the $2^n$-dimensional matrix algebra, and every automorphism of a matrix algebra is inner. Thus we know that \textit{every} automorphism can be written as a similarity transformation. For a more detailed discussion on this topic, the reader may refer to \cite{Jordan:1928wi} and \cite{brauer1935spinors}. \par
We now look at what linear combinations of $\Gamma_k$ matrices yield another set of $2n$ anti-commuting matrices. By explicitly writing the anti-commuting relations, we will obtain constraints on the linear combination coefficients of $\Gamma_k$'s. \par
Let us write $\Gamma_k^*$ in terms of $\Gamma_k$,
\begin{equation}
    \Gamma_k^* = \sum_{j=1}^{2n} o_{kj} \Gamma_j
\end{equation}
We want this new set to be anti-commuting, that is, satisfy $(\Gamma_k^*)^2 = \id$ and $\Gamma_k^* \Gamma_l^* = \Gamma_l^* \Gamma_k^*$. We can explicitly write $\Gamma_k^*$ matrices to see the constraints $o_{kj}$ need to satisfy. 
\begin{align*}
    (\Gamma_k^*)^2 =& \left(\sum_{j=1}^{2n} o_{kj} \Gamma_j \right) \cdot \left(\sum_{l=1}^{2n} o_{kl} \Gamma_l \right) \\
    =& \sum_{j=1}^{2n} \sum_{l=1}^{2n} o_{kj} o_{kl} \Gamma_j  \Gamma_l
\end{align*}
We know that $\Gamma_k$ anti-commute, thus for $j \neq l$ we have $o_{kj} o_{kl} \Gamma_j \Gamma_l + o_{kl} o_{kj} \Gamma_l \Gamma_j = 0$. We are now left with
\begin{align*}
    \sum_{j=1}^{2n} o_{kj}^2 \id \, .
\end{align*}
Hence we see that $\sum_{j=1}^{2n} o_{kj}^2 = 1$ should be satisfied. \par
We now look at the constraint on $o_{kj}$ to satisfy the anti-commuting property. Suppose $k \neq l$.
\begin{align*}
    \{\Gamma_k^* , \Gamma_l^*\} =& \Gamma_k^* \Gamma_l^* + \Gamma_l^* \Gamma_k^* \\
    =& \sum_{j=1}^{2n} \sum_{i=1}^{2n} o_{kj} o_{li} \Gamma_j \Gamma_i + \sum_{i=1}^{2n} \sum_{j=1}^{2n} o_{li} o_{kj} \Gamma_i \Gamma_j \\
    =& \sum_{j=1}^{2n} \sum_{i=1}^{2n} o_{kj} o_{li} ( \Gamma_j \Gamma_i + \Gamma_i \Gamma_j ) \\
    =& 2 \cdot \sum_{j=1}^{2n} o_{kj} o_{lj} \id = 0
\end{align*}
Combining the two constraints we found on $o_{kj}$'s, we can say that in order to obtain a $2n$ anti-commuting set from linear combinations of $\Gamma_k$, the coefficients $o_{kj}$ must satisfy
\begin{equation}
    \sum_{j=1}^{2n} o_{kj} o_{lj} = \delta_{kl} \, .
\end{equation}
We can interpret $o_{kj}$ as members of a $2n$-dimensional matrix. We can understand that the matrix is orthogonal from the properties of rotation matrices. We thus say that the rotation characterized by $o_{kj}$ is $\boldsymbol{o}$ which relates $\Gamma_k$ and $\Gamma_k^*$,
\begin{equation}
    \boldsymbol{o}: \Gamma_k \rightarrow \Gamma_k^* = \sum_{j=1}^{2n} o_{kj} \Gamma_j \, .
\end{equation}
\subsubsection{Spin representation of the orthogonal group}
Since we know that there exists a transformation matrix $S$ linking two sets, we write the similarity transformation matrix linking the sets of anti-commuting matrices obtained by applying $\boldsymbol{o}$ as $S(\boldsymbol{o})$,
\begin{equation}
    \Gamma_k^* = S(\boldsymbol{o}) \Gamma_k S(\boldsymbol{o})^{-1} \, .
\end{equation}
We now consider the full group of orthogonal matrices $\boldsymbol{o}$ in $2n$-dim space. We will show that the collection of $S(\boldsymbol{o})$ matrices forms a $2^n$-dimensional representation of this group. The first thing we need to show is that it is a group isomorphism, that is, for $\boldsymbol{o} \cdot \boldsymbol{o'} = \boldsymbol{o''}$,
\begin{equation}
     S(\boldsymbol{o}) \cdot S(\boldsymbol{o'}) = S(\boldsymbol{o} \cdot \boldsymbol{o'}) = S(\boldsymbol{o''}) \, . \label{homS}
\end{equation}
We can show this by explicitly writing the transformations. We first write the transformations $\boldsymbol{o}$ and $\boldsymbol{o'}$.
\begin{align*}
    &\boldsymbol{o}: \Gamma_k \rightarrow \sum_{j=1}^{2n} o_{kj} \Gamma_j = S(\boldsymbol{o}) \Gamma_k S(\boldsymbol{o})^{-1} \\
    &\boldsymbol{o'}: \Gamma_k \rightarrow \sum_{i=1}^{2n} {o'}_{ki} \Gamma_i = S(\boldsymbol{o'}) \Gamma_k S(\boldsymbol{o'})^{-1} 
\end{align*}
Let us combine the transformations.
\begin{align*}
    \boldsymbol{o''} = \boldsymbol{o} \cdot \boldsymbol{o'} : \Gamma_k \rightarrow  & \sum_{j=1}^{2n} o_{kj} \left( \sum_{i=1}^{2n} {o'}_{ji} \Gamma_i \right) = \sum_{j=1}^{2n} \sum_{i=1}^{2n} o_{kj} {o'}_{ji} \Gamma_i \\
    &= S(\boldsymbol{o}) S(\boldsymbol{o'}) \Gamma_k S(\boldsymbol{o'})^{-1} S(\boldsymbol{o})^{-1}
\end{align*}
By defining 
\begin{equation}
    {o''}_{ki} = \sum_{j=1}^{2n} o_{kj} {o'}_{ji}
\end{equation}
we can write
\begin{align*}
    \boldsymbol{o''}: \Gamma_k \rightarrow  & \sum_{i=1}^{2n} {o''}_{ki} \Gamma_i \\
    &= S(\boldsymbol{o''}) \Gamma_k S(\boldsymbol{o''})^{-1} \, .
\end{align*}
Thus, we conclude that eq. \ref{homS} is satisfied. \par
We want to explicitly write what $S(\boldsymbol{o})$ is. Notably, this is not a simple task on all occasions, but since we can realize the rotations of $2n$ $\Gamma$ matrices as plane rotations, let's see what we can do. \par
Consider the plane rotation K. It rotates $\Gamma_k$ and  $\Gamma_l$ while fixing others. 
\begin{align*}
    \boldsymbol{K}: & \Gamma_k \rightarrow \cos{\theta} \cdot \Gamma_k - \sin{\theta} \cdot \Gamma_l \equiv \Gamma_k^* \\
    & \Gamma_l \rightarrow \sin{\theta} \cdot \Gamma_k + \cos{\theta} \cdot \Gamma_l \equiv \Gamma_l^* \\
    & \Gamma_i \rightarrow \Gamma_i \, \, \, \, \, \textrm{for $i\neq k,l$}
\end{align*}
We will now show that 
\begin{equation}
    S(\boldsymbol{K}) = \exp{\frac{\theta}{2} \cdot \Gamma_k \Gamma_l} \, . \label{s(k)matrix}
\end{equation}
We can write the series expansion of $S(\boldsymbol{K})$.
\begin{align*}
    \exp{\frac{\theta}{2} \cdot \Gamma_k \Gamma_l} =& \id + \left( \frac{\theta}{2} \right) \Gamma_k \Gamma_l + \frac{1}{2!} \left( \frac{\theta}{2} \right)^2 (\Gamma_k \Gamma_l)^2 + \frac{1}{3!} \left( \frac{\theta}{2} \right)^3 (\Gamma_k \Gamma_l)^3 + \frac{1}{4!} \left( \frac{\theta}{2} \right)^4 (\Gamma_k \Gamma_l)^4 + \cdots \\
    =&\left( \id + \frac{1}{2!} \left( \frac{\theta}{2} \right)^2 (\Gamma_k \Gamma_l)^2 + \frac{1}{4!} \left( \frac{\theta}{2} \right)^4 (\Gamma_k \Gamma_l)^4 + \cdots \right) \\ & + \left( \left( \frac{\theta}{2} \right) \Gamma_k \Gamma_l + \frac{1}{3!} \left( \frac{\theta}{2} \right)^3 (\Gamma_k \Gamma_l)^3 + \frac{1}{5!} \left( \frac{\theta}{2} \right)^5 (\Gamma_k \Gamma_l)^5 + \cdots \right) \, .
\end{align*}
Let us look at the second half and explicitly write the terms; the logic is the same for the first half.
\begin{align*}
    \left( \left( \frac{\theta}{2} \right) \Gamma_k \Gamma_l + \frac{1}{3!} \left( \frac{\theta}{2} \right)^3 \Gamma_k \Gamma_l \Gamma_k \Gamma_l \Gamma_k \Gamma_l + \frac{1}{5!} \left( \frac{\theta}{2} \right)^5 \Gamma_k \Gamma_l \Gamma_k \Gamma_l \Gamma_k \Gamma_l \Gamma_k \Gamma_l \Gamma_k \Gamma_l  + \cdots \right) 
\end{align*}
We are working with anti-commuting matrices. Thus we know that $\Gamma_k \Gamma_l = - \Gamma_l \Gamma_k$, so we can write
\begin{align*}
    &\left( \left( \frac{\theta}{2} \right) \Gamma_k \Gamma_l + (-1) \frac{1}{3!} \left( \frac{\theta}{2} \right)^3 \Gamma_k \Gamma_k \Gamma_l \Gamma_l \Gamma_k \Gamma_l + (-1)^2 \frac{1}{5!} \left( \frac{\theta}{2} \right)^5 \Gamma_k \Gamma_k \Gamma_l \Gamma_l \Gamma_k \Gamma_k \Gamma_l \Gamma_l \Gamma_k \Gamma_l  + \cdots \right) \\
    &=\left( \left( \frac{\theta}{2} \right) \Gamma_k \Gamma_l - \frac{1}{3!} \left( \frac{\theta}{2} \right)^3 \Gamma_k \Gamma_l + \frac{1}{5!} \left( \frac{\theta}{2} \right)^5 \Gamma_k \Gamma_l  + \cdots \right) \\
    &=\left( \left( \frac{\theta}{2} \right) - \frac{1}{3!} \left( \frac{\theta}{2} \right)^3 + \frac{1}{5!} \left( \frac{\theta}{2} \right)^5 + \cdots \right) \Gamma_k \Gamma_l = \sin{\frac{\theta}{2}} \cdot \Gamma_k \Gamma_l 
\end{align*}
The calculations in the first half are similar. 
\begin{align*}
    &\left( \id + \frac{1}{2!} \left(\frac{\theta}{2} \right)^2 \Gamma_k \Gamma_l \Gamma_k \Gamma_l  + \frac{1}{4!} \left( \frac{\theta}{2} \right)^4 \Gamma_k \Gamma_l \Gamma_k \Gamma_l \Gamma_k \Gamma_l \Gamma_k \Gamma_l  + \cdots \right) \\
    &=\left( \id + (-1) \frac{1}{2!} \left(\frac{\theta}{2} \right)^2 \Gamma_k \Gamma_k \Gamma_l \Gamma_l + (-1)^2 \frac{1}{4!} \left( \frac{\theta}{2} \right)^4 \Gamma_k \Gamma_k \Gamma_l \Gamma_l \Gamma_k \Gamma_k \Gamma_l \Gamma_l + \cdots \right)  \\
    &=\left( \id + (-1) \frac{1}{2!} \left(\frac{\theta}{2} \right)^2 \id + (-1)^2 \frac{1}{4!} \left( \frac{\theta}{2} \right)^4 \id  + \cdots \right) = \cos{\frac{\theta}{2}} \cdot \id  
\end{align*}
Thus,
\begin{equation}
    \exp{\frac{\theta}{2} \cdot \Gamma_k \Gamma_l} = \cos{\frac{\theta}{2}} \cdot \id + \sin{\frac{\theta}{2}} \cdot \Gamma_k \Gamma_l \, .
\end{equation}
We can easily see that $S(\bm{K})^{-1} = \cos{\frac{\theta}{2}} \id - \sin{\frac{\theta}{2}}\Gamma_k \Gamma_l $,
\begin{align*}
    S(\boldsymbol{K}) \cdot S(\boldsymbol{K})^{-1} &= (\cos{\frac{\theta}{2}} \id + \sin{\frac{\theta}{2}}\Gamma_k \Gamma_l) \cdot ( \cos{\frac{\theta}{2}} \id - \sin{\frac{\theta}{2}}\Gamma_k \Gamma_l) \\
    &= \cos^2{\frac{\theta}{2}} \id - \sin{\frac{\theta}{2}}\cos{\frac{\theta}{2}}\Gamma_k \Gamma_l + \cos{\frac{\theta}{2}} \sin{\frac{\theta}{2}}\Gamma_k \Gamma_l) - \sin^2{\frac{\theta}{2}}(\Gamma_k \Gamma_l)^2 \\
    &= \cos^2{\frac{\theta}{2}} \id - (-1) \sin^2{\frac{\theta}{2}}\Gamma_k \Gamma_k \Gamma_l \Gamma_l = \id \, .
\end{align*}
We can now explicitly write $\Gamma_k^*$,
\begin{align*}
    S(\boldsymbol{K}) \cdot \Gamma_k \cdot S(\boldsymbol{K})^{-1} &= (\cos{\frac{\theta}{2}} \id + \sin{\frac{\theta}{2}}\Gamma_k \Gamma_l) \cdot \Gamma_k \cdot ( \cos{\frac{\theta}{2}} \id - \sin{\frac{\theta}{2}}\Gamma_k \Gamma_l) \\
    &= \left( \cos{\frac{\theta}{2}} \Gamma_k - \sin{\frac{\theta}{2} \Gamma_l} \right) \cdot \left( \cos{\frac{\theta}{2}} \id - \sin{\frac{\theta}{2}} \Gamma_k \Gamma_l \right) \\
    &= \cos^2{\frac{\theta}{2}} \Gamma_k -\cos{\frac{\theta}{2}} \sin{\frac{\theta}{2}} \Gamma_l - \sin{\frac{\theta}{2}} \cos{\frac{\theta}{2}} \Gamma_l - \sin^2{\frac{\theta}{2}} \Gamma_k \\
    &= \left(  \cos^2{\frac{\theta}{2}} - \sin^2{\frac{\theta}{2}}\right) \Gamma_k - \left( 2 \cos{\frac{\theta}{2}} \sin{\frac{\theta}{2}} \right) \Gamma_l \\
    &= \cos{\theta} \cdot \Gamma_k - \sin{\theta} \cdot \Gamma_l = \Gamma_k^* \, .
\end{align*}
Similarly, $S(\boldsymbol{K}) \cdot \Gamma_l \cdot S(\boldsymbol{K})^{-1} = \sin{\theta} \cdot \Gamma_k + \cos {\theta} \cdot \Gamma_l = \Gamma_l^*$
The half angle seen in eq. \ref{s(k)matrix} provides a double-valued representation, which is a characteristic feature of spin representations. To understand it more deeply, let us do a little experiment. Let $\boldsymbol{K'}$ be the rotation characterized by $\theta + 2 \pi$. Since adding $2 \pi$ does not affect the rotation, this might seen like a redundant thing to do, however, if we look at $S(\boldsymbol{K'})$ we see that it does get affected. 
\begin{align*}
    S(\boldsymbol{K'}) &= \cos{\frac{\theta + 2 \pi}{2}} \cdot \id + \sin{\frac{\theta + 2 \pi}{2}} \cdot \Gamma_k \Gamma_l \\
    &= - \cos{\frac{\theta}{2}} \cdot \id - \sin{\frac{\theta}{2}} \cdot \Gamma_k \Gamma_l = - S(\boldsymbol{K})
\end{align*}
Thus, it can be seen that the number of full turns done before the rotation changes the value. \par
\subsubsection{Eigenvalues of the spin representatives}
We want to explicitly write what $S(\boldsymbol{o})$ is. It is worth noting that this is not a simple task on all occasions, but we can use it since we are working with plane rotations. \par
Consider the plane rotation K, as defined in the previous sections. We can write $S(\boldsymbol{K})$ easily if it acts on $\Gamma^*_{2r-1}$ and $\Gamma^*_{2r}$. 
\begin{align*}
    S(\boldsymbol{K}) &= \exp{\frac{\theta}{2} \cdot \Gamma^*_{2r-1} \Gamma^*_{2r}} = \exp{\frac{\theta}{2} \cdot i Z_r} \\
    &= \cos{\frac{\theta}{2}} \cdot \id + i \sin{\frac{\theta}{2}} \cdot Z_r \\
    &= (\id_1 \cdot \id_2 \cdots \begin{pmatrix} \cos{\frac{\theta}{2}} &  \\  & \cos{\frac{\theta}{2}} \end{pmatrix}) \cdots \id_{2n}) + (\id_1 \cdot \id_2 \cdots \begin{pmatrix} i\sin{\frac{\theta}{2}} &  \\  & - i \sin{\frac{\theta}{2}} \end{pmatrix}) \cdots \id_{n}) \\
    &= \id_1 \cdot \id_2 \cdots \begin{pmatrix} e^{i \theta / 2} &  \\  &  e^{- i \theta / 2}\end{pmatrix} \cdots \id_{n}
\end{align*}
Thus we see that the eigenvalues of $S(\bm{K})$ are of the form $e^{i \theta / 2}$ and $e^{- i \theta / 2}$ with each having $2^{n-1}$ degeneracy. This is a consequence of the $n-1$ $\id$ matrices in the direct product. Remark that $\bm{K}$ has eigenvalues $e^{i \theta / 2}$, $e^{- i \theta / 2}$ and $1$, where $1$ has $2(n-1)$ degeneracy because of the $2n-2$ $\Gamma$ matrices it fixes. \par
We now consider a product of $n$ commuting matrices. To obtain $n$ commuting matrices from rotations acting on $2n$ $\Gamma$ matrices, we pair $\Gamma$ matrices and consider the rotations acting on them. Because no $\Gamma_i$ appears on two rotations, we can clearly see that they commute. We can write this product of rotations as
\begin{equation}
    \boldsymbol{K} = \prod_{r=1}^n \boldsymbol{K}_r \, .
\end{equation}
The eigenvalues of $\boldsymbol{K}$ are 
\begin{equation}
    e^{\pm i\theta_1/2}, \, e^{\pm i\theta_2/2}, \cdots, e^{\pm i\theta_n/2} \,.
\end{equation}
We can explicitly write $S(\boldsymbol{K})$ since we now the representations of $S(\boldsymbol{K_i})$,
\begin{equation}
    S(\boldsymbol{K}) = \prod_{r=1}^n S(\boldsymbol{K_r}) =  \prod_{r=1}^n \exp{\frac{\theta_r}{2} \cdot \Gamma_{r_1} \Gamma_{r_2}} \, . \label{formofsk}
\end{equation}
If we consider the case $\Gamma_{r_1} = \Gamma_{2r-1}$ and $\Gamma_{r_2} = \Gamma_{2r}$, then we can write 
\begin{align}
    S(\boldsymbol{K}) &=  \prod_{r=1}^n \exp{\frac{\theta_r}{2} \cdot \Gamma_{2r-1} \Gamma_{2r}} \nonumber \\
    &= \begin{pmatrix} e^{i \theta_1 / 2} &  \\  &  e^{- i \theta_1 / 2}\end{pmatrix} \times \begin{pmatrix} e^{i \theta_2 / 2} &  \\  &  e^{- i \theta_2 / 2}\end{pmatrix} 
    \times \cdots \begin{pmatrix} e^{i \theta_n / 2} &  \\  &  e^{- i \theta_n / 2}\end{pmatrix}
\end{align}
With $2^n$ eigenvalues
\begin{align*}
    \lambda = \exp{\frac{i}{2}\cdot \left(\pm \theta_1 \pm \theta_2 \pm \cdots \pm \theta_n \right)}
\end{align*}
where every combination of $\pm$ is taken. \par
Remember that we had said there exists a basis transformation between any given $2n$ anti-commuting $\Gamma$ matrices. Since we know that eigenvalues are invariant under basis transformations, we can conclude that for any rotation $\boldsymbol{o}$ given with eigenvalues
\begin{align}
    e^{- i \theta_1}, \, e^{- i \theta_2}, \cdots , e^{- i \theta_n}
\end{align}
the spin representative $S(\boldsymbol{o})$ has the eigenvalues
\begin{align*}
    \exp{\frac{i}{2} \cdot \left(\pm \theta_1 \pm \theta_2 \pm \cdots \pm \theta_n \right)} \, .
\end{align*}
\subsection{Spinor analysis of the transfer matrix}
We now have the necessary tools to find the eigenvalues of the $V$ matrix. \par
We now write another set of $\Gamma_r^*$ matrices by performing a similarity transformation $g$ which changes the places of $X$ and $Z$ in $\Gamma_r$. We can write the $g$ matrix explicitly as
\begin{equation}
    g= 2^{\frac{-n}{2}} \left( (X_1+Z_1) \cdot (X_2 + Z_2) \cdots (X_n+Z_n) \right) \, . \label{basischangeg}
\end{equation}
We can see that $g=g^{-1}$ easily, since
\begin{align*}
    \frac{1}{2} (X+Z) (X+Z) = \frac{1}{2} X^2 + XZ + ZX + Z^2 = \frac{1}{2} 2 \id +iY -iY = \id \, .
\end{align*}
Now, let's look at how $g$ would act on spin matrices. 
\begin{align*}
    \frac{1}{2} \left( (X+Z) X (X+Z) \right) = \frac{1}{2} \left( X + ZX^2 + Z + ZXZ \right) = Z \\
    \frac{1}{2} \left( (X+Z) Y (X+Z) \right) = \frac{1}{2} \left( iZX - iX^2 + iZ^2 -iXZ \right) = -Y \\
    \frac{1}{2} \left( (X+Z) Z (X+Z) \right) = \frac{1}{2} \left( XZX + X + XZ^2 +Z \right) = X 
\end{align*}
By applying $g$ on $\Gamma_r$ matrices, we can write 
\begin{align}
    g \Gamma_{2r-1} g^{-1} = \Gamma_{2r-1}^* = Z_1 \cdots Z_{r - 1} \cdot X_r \cdot \id_{r+1} \cdot \id_{n}\equiv P_r^* \nonumber \\
    g \Gamma_{2r} g^{-1} = \Gamma_{2r}^* = Z_1 \cdots Z_{r - 1} \cdot Y_r \cdot \id_{r+1} \cdot \id_{n} \equiv Q_r^* 
\end{align}
which is a perfectly good realization of a set of $2n$ anti-commuting matrices. \par
We can express our $V_a$ and $V_b$ matrices in terms of $\Gamma_r^*$ matrices. \par
Let us look at the structure of these matrices. We can explicitly write the matrix $i\Gamma_{2r-1}^* \Gamma_{2r}^*$,
\begin{align*}
    i\Gamma_{2r-1}^* \Gamma_{2r}^* &= i (Z_1 Z_2 \cdots X_r \cdot \id_{r+1} \cdots \id_{n}) \cdot (Z_1 Z_2 \cdots Y_r \cdot \id_{r+1} \cdots \id_{n}) \\
    &=(\id_{1} \cdots \id_{r-1} \cdot (-Z_{r}) \cdot \id_{r+1} \cdots \id_{n}) \, ,
\end{align*}
which is seen to have a diagonal form. Whereas, the matrix $-i\Gamma_{2r+1}^* \Gamma_{2r}^*$ does not have a diagonal form.
\begin{align*}
    -i\Gamma_{2r+1}^* \Gamma_{2r}^* &= -i (Z_1 Z_2 \cdots \cdots Z_{r} X_{r+1} \cdot \id_{r+1} \cdots \id_{n}) \cdot (Z_1 Z_2 \cdots Y \cdot \id_{r+1} \cdots \id_{n}) \\
    &=(\id_{1} \cdots \id_{r-1} \cdot (-X_{r}) \cdot (x_{r+1}) \cdot \id_{r+1} \cdots \id_{n}) \, .
\end{align*}
Before moving further, it is worth pointing out that we can write a transformation between $V_a$ and $V_b$ through the rotation $D$ that acts as
\begin{align}
    D: &\Gamma_{2r-1} \rightarrow \Gamma_{2r+1} \\
    &\Gamma_{2r} \rightarrow \Gamma_{2r} \, .
\end{align}
If we write the matrices $V_a$ and $V_b$ depending on the interactions between rows and columns respectively,
that is $V_a(a)$ and $V_b(b)$, then 
\begin{align}
    D: &V_a(a) \rightarrow V_b(a') \\
    &V_b(b) \rightarrow V_a(b') \, .
\end{align}
Thus for a lattice $V(a,b)$, there corresponds a dual lattice $V(b', a')$ which has the same partition function except for the factor of $(\sinh{2a} \cdot \sinh{2b})^{-mn/2}$ caused by the boundary conditions.
\subsection{The eigenvalues and eigenvectors of \texorpdfstring{$V$}{}} \label{sec: eigens of v}
We will now use the fact that $V_a$ and $V_b$ are spin representations of certain rotations, say, $R_a$ and $R_b$. Then clearly, $V$ is the spin representative of $R_a \cdot R_b$. Thus if we can find the eigenvalues of $R_a \cdot R_b$, then we easily obtain the eigenvalues of $V$. The similarity between $V_a$ and $V_b$ mentioned in the previous section necessitates a relation between $R_a$ and $R_b$, which we will see that one can be obtained from the other by a permutation of rows and columns. \par
We now express $V_a$ and $V_b$ in the form of eq. \ref{formofsk}.
\begin{align}
    & V_b = \prod_{r=1}^n \exp{i b' \Gamma_{2r-1} \Gamma_{2r}} \nonumber \\
    & V_a = \prod_{r=1}^{n-1} \exp{-i a \Gamma_{2r+1} \Gamma_{2r}} \cdot \exp{ia\Gamma_1 \Gamma_{2n}}
\end{align}
In this notation of $V_b$ and $V_a$), $\frac{1}{2} \theta_{b_r} = i b'$ and  $\frac{1}{2} \theta_{a_r} = i a$) for $r=1, \cdots, n$. Hence we can say that all of the factors of $V_b$ and $V_a$ except $\exp{ia\Gamma_1 \Gamma_{2n}}$ are representatives of plane rotations. The last factor of $V_a$ indicates a boundary condition on the system. Since we are working on a 2-dimensional lattice model, putting boundary conditions on one of the sides gives a cylinder. \par
We know define an involution $U$, to factor in the boundary factor $\exp{ia\Gamma_1 \Gamma_{2n}}$. We define projection operators $P^+= \frac{1}{2}(1+U)$ and $P^- = \frac{1}{2} (1-U)$. We then define
\begin{equation}
    V^\pm = P^\pm V \, .
\end{equation}
Now, we can change our original coordinate by the transformation matrix $g$ that interchanges $X$ and $Z$ matrices, which is given in eq. \ref{basischangeg}. Let us now look at what happens to $B$ and $A$ matrices given in eq. \ref{ab} under this coordinate change. 
\begin{align}
    &g\left( \sum_{r=1}^n X_r \right) g = \sum_{r=1}^n Z_r \equiv B' \\
    &g\left( \sum_{r=1}^n Z_r Z_{r+1} \right) g = \sum_{r=1}^n X_r X_{r+1} \equiv A' 
\end{align}
Our involution $U$ has the form now has the form
\begin{equation}
    U' = Z_1 \cdot Z_2 \cdots Z_n
\end{equation}
in this basis. \par
Since $U'$ is a diagonal matrix with half of its entries $1$ and the other half $-1$, we can reorganize the rows and columns to obtain a matrix of the form $\begin{pmatrix} \id & \\ & - \id \end{pmatrix}$. \par
We remark that $U'$ commutes with $A'$ and $B'$.
\begin{align*}
    &U' \cdot A' = \sum_{r=1}^n (Z_1 \cdot Z_2 \cdots Z_n) \cdot  X_r X_{r+1} \\
    &= \sum_{r=1}^n  X_r X_{r+1} \cdot (Z_1 \cdot Z_2 \cdots Z_n) = A' \cdot U'
\end{align*}
since
\begin{align*}
    &(Z_1 \cdot Z_2 \cdots Z_n) \cdot X_r X_{r+1} = (-i)^2 Z_1 \cdots Y_r Y_{r+1} \cdots Z_n \\
    &= (i)^2 Z_1 \cdots Y_r Y_{r+1} \cdots Z_n = X_r X_{r+1} \cdot (Z_1 \cdot Z_2 \cdots Z_n) 
\end{align*}
Similarly, 
\begin{align*}
    U' \cdot B' = \sum_{r=1}^n (Z_1 \cdot Z_2 \cdots Z_n) \cdot  Z_r = \sum_{r=1}^n Z_r \cdot (Z_1 \cdot Z_2 \cdots Z_n) = B' = U'
\end{align*}
Thus, $[U', B'] = 0 $ and $[U' , A'] = 0$. In general, if a matrix $M$ commutes with $U'$, then it has the form
\begin{align*}
    M = \begin{pmatrix} M_1 & 0 \\ 0 & M_2 \end{pmatrix} \, ,
\end{align*}
and if a matrix $N$ anti-commutes with $U'$, then it has the form
\begin{align*}
    N = \begin{pmatrix} 0 & N_1 \\ N_2 & 0 \end{pmatrix} \, .
\end{align*}
From this, we can conclude that the form $V'$ is 
\begin{align*}
    V' = \begin{pmatrix} V'_1 & 0 \\ 0 & V'_2 \end{pmatrix} = \begin{pmatrix} \id & 0 \\ 0 & 0 \end{pmatrix} \cdot \begin{pmatrix} V^{'+}_1 & 0 \\ 0 & V^{'+}_2 \end{pmatrix} + \begin{pmatrix} 0 & 0 \\ 0 & \id \end{pmatrix} \cdot \begin{pmatrix} V^{'-}_1 & 0 \\ 0 & V^{'-}_2 \end{pmatrix} \, .
\end{align*}
We now have separated the upper and lower halves of the matrix $V'$. Thus, we can diagonalize them separately and then choose the relevant half of the eigenvalues of $V^{'+}$ and $V^{'-}$ matrices. \par
Separating $V'$ into two parts gives us the freedom to diagonalize each part separately. It will be later seen that the transformations involved in diagonalizing will be representations of orthogonal rotations, say $T^+$ and $T^-$, which will act on $U'$ as
\begin{align}
    &T^+: U' \rightarrow U' &T^-: U' \rightarrow -U' \, . 
\end{align}
Thus, we can see that
\begin{align}
    &S(T^+) \cdot \left(\frac{1}{2}(1+U')\cdot V^{'+}\right) S(T^+)^{-1} \\
    &= \frac{1}{2}(1+U')\cdot S(T^+) \cdot V^{'+} \cdot S(T^+)^{-1} \\
    &= \frac{1}{2}(1+U')\cdot S(T^+ \cdot R^{+} \cdot (T^+)^{-1})
\end{align}
and 
\begin{align}
    &S(T^-) \cdot \left(\frac{1}{2}(1-U')\cdot V^{'-}\right) S(T^-)^{-1} \\
    &= \frac{1}{2}(1+U')\cdot S(T^-) \cdot V^{'-} \cdot S(T^-)^{-1} \\
    &= \frac{1}{2}(1+U')\cdot S(T^- \cdot R^{-} \cdot (T^-)^{-1})
\end{align}
Now, suppose that we have already found the $2n$-eigenvalues of $R^-$, $\exp{\pm \gamma_{2r}}$. Then, we know that the diagonal matrix 
\begin{equation}
    S(T^-)\cdot V^- \cdot S(T^-)^{-1} = \prod_{r=1}^n \exp{\frac{i}{2} \gamma_{2r} \Gamma_{2r-1}^* \Gamma_{2r}^*} \equiv \Lambda^-
\end{equation}
which has the eigenvalues 
\begin{equation}
    \exp{\frac{1}{2}(\pm \gamma_2 \pm \gamma_4 \pm \cdots \pm \gamma_{2n})} \, .
\end{equation}
The sign of each component depends on the value of the components of $\Gamma_{2r-1}^* \Gamma_{2r}^*$. We know that the factor $\frac{1}{2}(1+U')$ keeps only half of the eigenvalues that fall in the upper half. Thus
\begin{equation*}
    U'_{ii} = 1 = \prod_{r=1}^n (\Gamma_{2r-1}^* \Gamma_{2r}^*)_{ii} \, .
\end{equation*}
Yet, $(\Gamma_{2r-1}^* \Gamma_{2r}^*)_{ii}= \pm 1$, and for $U'_{ii}=1$, only an even number of  $(\Gamma_{2r-1}^* \Gamma_{2r}^*)_{ii}$ can be negative. Hence we can conclude that the eigenvalues that we should keep are the ones with an even number of negative $\gamma_{2r}$'s. \par
The case is similar for $V^{'+}$. Let the eigenvalues of $R^+$ be $\exp{\pm \gamma_{2r-1}}$, then the eigenvalues of $V^+$ are of the form  \begin{equation}
    \exp{\frac{1}{2}(\pm \gamma_1 \pm \gamma_3 \pm \cdots \pm \gamma_{2n-1})} \, .
\end{equation}
Again, by taking the factor $\frac{1}{2}(1+U')$ we can say that the only the eigenvalues with an even number of $\gamma_{2r-1}$ will survive. \par
We remark that the angles $\gamma_{2r}$ and $\gamma_{2r-1}$ are the angles of rotations of $R^-$ and $R^+$ respectively, which are found by diagonalizing the respective $2n$-dimensional matrices. 
\subsection{The Complete Partition Function}
We can now write the complete partition function 
\begin{equation}
    Z=\frac{1}{2}(2 \sinh{2b})^{nm/2} \cdot \sum_{i=1}^{2^n} \lambda_i^m
\end{equation}
as
\begin{equation}
    Z=\frac{1}{2}(2 \sinh{2b})^{nm/2} \cdot \left( \sum  \exp{\frac{m}{2}(\pm \gamma_2 \pm \cdots \pm \gamma_{2n})} + \sum  \exp{\frac{m}{2}(\pm \gamma_1 \pm \cdots \pm \gamma_{2n-1})} \right) \, .
\end{equation}
To write in a more compact form, we can write
\begin{align}
    Z = \frac{1}{2}(2\sinh{2b})^{\frac{nm}{2}}\cdot \left(  \prod_{r=1}^n (2 \cosh{\frac{m}{2} \gamma_{2r}}) + \prod_{r=1}^n (2 \sinh{\frac{m}{2} \gamma_{2r}}) +  \prod_{r=1}^n (2 \cosh{\frac{m}{2} \gamma_{2r-1}}) + \prod_{r=1}^n (2 \sinh{\frac{m}{2} \gamma_{2r-1}})\right)
\end{align}
\subsubsection{Explicitly calculating the partition function}
We can now move to explicitly calculating the eigenvalues of $V^+$ and $V^-$. Since working with a matrix in a symmetric form 
\begin{equation*}
    V_b^{1/2} V_a V_b^{1/2} \equiv V_0
\end{equation*}
is easier, we will first find the eigenvectors of $V_0$ and then obtain the eigenvectors of $V$ by applying a $V_b^{1/2} = \exp{\frac{1}{2}bB}$.
We have
\begin{equation*}
    V_0^- = \prod_{r=1}^n \exp{\frac{i}{2} b \Gamma_{2r-1} \Gamma_{2r}} \cdot \prod_{r=1}^n \exp{-ia \Gamma_{2r-1} \Gamma_{2r}} \cdot \prod_{r=1}^n \exp{\frac{i}{2} b \Gamma_{2r-1} \Gamma_{2r}}
\end{equation*}
where
\begin{equation*}
    \prod_{r=1}^n \exp{\frac{i}{2}b \Gamma_{2r-1} \Gamma_{2r}} = \begin{pmatrix}
    \begin{bmatrix} R_{b} \end{bmatrix} & 0 & \cdots & 0 \\
    0 & \begin{bmatrix} R_{b} \end{bmatrix} & \cdots & 0  \\
    \vdots & \vdots & \ddots & \vdots  \\
    0 & 0 & \cdots & \begin{bmatrix} R_{b} \end{bmatrix} 
\end{pmatrix} \, 
\end{equation*}
and
\begin{equation}
    \prod_{r=1}^n \exp{-ia \Gamma_{2r-1} \Gamma_{2r}} = \begin{pmatrix}
    \cosh{2a} & 0 & \cdots & 0 & \pm i \sinh{2a} \\
    0 & \begin{bmatrix} R_{2a} \end{bmatrix} & \cdots & 0 & 0 \\
    \vdots & \vdots & \ddots & \vdots & \vdots \\
    0 & 0 & \cdots & \begin{bmatrix} R_{2a} \end{bmatrix} & 0 \\
    \mp i \sinh{2a} & 0 & \cdots & 0 & \cosh{2a}
\end{pmatrix} \, ,
\end{equation}
where 
\begin{equation}
     \begin{bmatrix} R_{i} \end{bmatrix} =  \begin{bmatrix} \cosh{i} & i \sinh{i} \\
     -i \sinh{i} & \cosh{i} \end{bmatrix} \, .
\end{equation}
Now we can write $V_0^{-}$,
\begin{equation}
    V_0^{\pm} = \begin{pmatrix}
    A & B & 0 & \cdots & 0 & \mp B^\dagger \\
    B^\dagger & A & B & \cdots & 0 & 0 \\
    0 & B^\dagger & A & \cdots & 0 & 0 \\
    \vdots & \vdots & \vdots & \ddots & \vdots & \vdots \\
    0 & 0 & 0 & \cdots & A & B \\
    \mp B & 0 & 0 & \cdots & B^\dagger & A 
\end{pmatrix} \, ,
\end{equation}
where 
\begin{align}
   & A = \cosh{2a} \begin{pmatrix}
    \cosh{2 \Bar{b}} & i \sinh{2 \Bar{b}} \\
     -i \sinh{2 \Bar{b}} & \cosh{2 \Bar{b}} \end{pmatrix}
     & B = \sinh{2a} \begin{pmatrix}
    -\frac{1}{2}\sinh{2 \Bar{b}} & -i \sinh^2{\Bar{b}} \\
     i \cosh^2{ \Bar{b} } & -\frac{1}{2}\sinh{2\Bar{b}} \end{pmatrix} \, .
\end{align}
By using the properties of cyclic matrices, we can deduce the eigenvalues and eigenvectors of this matrix. The form of the eigenvectors and eigenvalues of circulant matrices are well known. Using this ansatz, we can say that 
\begin{equation}
\phi = \frac{1}{\sqrt{n}} 
\begin{pmatrix} z u \\ z^2 u \\ \vdots \\ z^n u \end{pmatrix} \, , \label{eigenvectorphi}
\end{equation}
where $z \in \mathbb{C}$ and $u$ is a two-component vector such that $u^\dagger u = 1$. 
From the eigenvector condition $\mathcal{V}^{\pm} \phi = \lambda \phi $ we get $n$ equations 
\begin{align}
    (zA + z^2 B \mp z^n B^\dagger)u &= z \lambda u \\
    (z^2 A + z^3 B + z B^\dagger)u &= z^2 \lambda u \\
    (z^3 A + z^4 B + z^2 B^\dagger)u &= z^3 \lambda u \\
    & \vdots \\
    (z^{n-1} A + z^n B + z^{n-2} B^\dagger)u &= z^{n-1} \lambda u \\
    (z^n A \mp z B + z^{n-1} B^\dagger)u &= z^n \lambda u \, .
\end{align}
The equations apart from the first and last ones are identical. Therefore these $n$ equations can be reduced to three independent equations:
\begin{align}
    (A + z B \mp z^{n-1} B^\dagger)u &= \lambda u \\
    (A + z B + z^{-1} B^\dagger)u &= \lambda u \\
    (A \mp z^{1-n} B + z^{-1} B^\dagger)u &= \lambda u \, .
\end{align}
If we set $z^n=\mp 1$, then these three equations become identical,
\begin{equation}
    (A+ z B + z^{-1} B^\dagger )u = \lambda u \, .
\end{equation}
We can see that if $z_k = e^{i \pi k / n}$, for $k=\Bar{0,1, \cdots , 2n-1}$, then there are a total of $2n$ values that solve $z=\mp 1$, where odd $k$ lead to $-1$ and even $k$ lead to $1$. For each $k$, we now need to find the associated eigenvalues $\lambda^{\uparrow \downarrow}_k$ and eigenvectors $u^{\uparrow \downarrow}_k$. Now, let us define a matrix 
\begin{equation}
    M_k = A + e^{i \pi k/n} B + e^{-i \pi k/n} B^\dagger = \begin{pmatrix} d_k & o_k \\ o^*_k & d_k \end{pmatrix} \, .
\end{equation}
Here, 
\begin{equation}
   d_k = \cosh{2a} \cosh{2\Bar{b}} - \cos{w_k} \sinh{2\Bar{b}} \sinh{2a}
\end{equation}
and
\begin{equation}
   o_k = - \sin{w_k} \sinh{2\Bar{b}} + i (\sinh{2a}\cosh{2\Bar{b}} - \cos{\frac{\pi k}{n}} \cosh{2 a} \sinh{2\Bar{b}}) \,,
\end{equation}
where $w_k = \frac{\pi k}{n} $.

We can explicitly calculate the determinant of $M_k$ matrix
\begin{align*}
    \det M_k =& (\cosh{2a} \cosh{2\Bar{b}} - \cos{\frac{\pi k}{n}} \sinh{2a} \sinh{2\Bar{b}})^2 \\ 
    &- (- \sin{\frac{\pi k}{n}} \sinh{2\Bar{b}} + i (\sinh{2a}\cosh{2\Bar{b}} - \cos{\frac{\pi k}{n}} \cosh{2a} \sinh{2\Bar{b}})) \\ & \cdot (- \sin{\frac{\pi k}{n}} \sinh{2\Bar{b}} - i (\sinh{2a}\cosh{2\Bar{b}} - \cos{\frac{\pi k}{n}} \cosh{2 a} \sinh{2\Bar{b}})) \\
    =& (\cosh{2a} \cosh{2\Bar{b}})^2 - (\sinh{2 a} \cosh{2\Bar{b}})^2 + (\cos{\frac{\pi k}{n}})^2 (\sinh{2 a} \sinh{2\Bar{b}})^2 \\ & - (\sin{\frac{\pi k}{n}})^2 (\sinh{2\Bar{b}})^2 + 2(\sinh{a}\cosh{2\Bar{b}}\cos{\frac{\pi k}{n}}\cosh{2a}\sinh{2\Bar{b}} \\ &- \cosh{2a}\cosh{2\Bar{b}}\cos{\frac{\pi k}{n}}\sinh{a}\sinh{2\Bar{b}}) - (\cos{\frac{\pi k}{n}})^2(\cosh{2a} \sinh{2\Bar{b}})^2 \\
    =& \cosh^2{2\Bar{b}} (\cosh^2{2a}-\sinh^2{2a}) + (\cos{\frac{\pi k}{n}})^2 (\sinh^2{2\Bar{b}} (\sinh^2{a}-\cosh^2{a})) - (\sin{\frac{\pi k}{n}})^2 (\sinh^2{2\Bar{b}}) \\
    =& \cosh^2{2\Bar{b}} - \sinh^2{2\Bar{b}} ((\cos{\frac{\pi k}{n}})^2 + (\sin{\frac{\pi k}{n}})^2) = 1
\end{align*}
The unit determinant implies eigenvalues of the form 
\begin{align}
    & \lambda^{\uparrow}_k = e^{+ \gamma} & \lambda^{\downarrow}_k = e^{- \gamma}
\end{align}
with $\gamma \in \mathbb{R}$. We can use the trace to calculate the value of $\gamma$.
\begin{equation}
    tr (M_k) = 2(\cosh{2a} \cosh{2\Bar{b}} - \cos{\frac{\pi k}{n}} \sinh{2a} \sinh{2\Bar{b}}) = e^{+ \gamma} +  e^{- \gamma} = 2 \cosh{ \gamma} \label{coshgamma_k}
\end{equation}

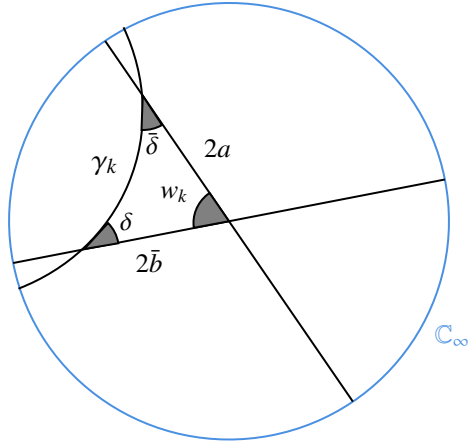
\begin{figure}
    \centering
    \tikzset{every picture/.style={line width=0.75pt}} 
    \begin{tikzpicture}[x=0.75pt,y=0.75pt,yscale=-1,xscale=1]

\draw  [color={rgb, 255:red, 74; green, 144; blue, 226 }  ,draw opacity=1 ] (312.11,16.39) .. controls (371.61,4.79) and (429.25,43.62) .. (440.85,103.12) .. controls (452.44,162.63) and (413.61,220.26) .. (354.11,231.86) .. controls (294.61,243.46) and (236.97,204.63) .. (225.37,145.13) .. controls (213.78,85.62) and (252.61,27.99) .. (312.11,16.39) -- cycle ;
\draw    (271.29,33.43) -- (394.93,214.82) ;
\draw    (440.85,103.12) -- (225.37,145.13) ;
\draw  [draw opacity=0] (280.46,26.59) .. controls (291.97,51.3) and (293.65,79.59) .. (282.78,105.53) .. controls (272.35,130.4) and (252.29,148.55) .. (228.01,158.03) -- (181.25,62.96) -- cycle ; \draw   (280.46,26.59) .. controls (291.97,51.3) and (293.65,79.59) .. (282.78,105.53) .. controls (272.35,130.4) and (252.29,148.55) .. (228.01,158.03) ;  
\draw  [color={rgb, 255:red, 0; green, 0; blue, 0 }  ,draw opacity=1 ][fill={rgb, 255:red, 128; green, 128; blue, 128 }  ,fill opacity=1 ] (315.55,127.42) .. controls (314.67,121.38) and (317.06,114.9) .. (322.36,110.67) .. controls (322.7,110.4) and (323.05,110.14) .. (323.4,109.9) -- (333.11,124.12) -- cycle ;
\draw  [color={rgb, 255:red, 0; green, 0; blue, 0 }  ,draw opacity=1 ][fill={rgb, 255:red, 128; green, 128; blue, 128 }  ,fill opacity=1 ] (299.66,75.71) .. controls (296.68,77.47) and (293.13,78.38) .. (289.43,78.19) -- (290.05,60.96) -- cycle ;
\draw  [color={rgb, 255:red, 0; green, 0; blue, 0 }  ,draw opacity=1 ][fill={rgb, 255:red, 128; green, 128; blue, 128 }  ,fill opacity=1 ] (272.71,125.02) .. controls (275.37,127.68) and (277.23,131.22) .. (277.85,135.21) -- (260.85,138.09) -- cycle ;

\draw (297.34,105.16) node [anchor=north west][inner sep=0.75pt]  [font=\small]  {$w_{k}$};
\draw (285.03,137.34) node [anchor=north west][inner sep=0.75pt]  [font=\small]  {$2\Bar{b}$};
\draw (319.26,81.3) node [anchor=north west][inner sep=0.75pt]  [font=\small]  {$2a$};
\draw (263.09,89.64) node [anchor=north west][inner sep=0.75pt]    {$\gamma _{k}$};
\draw (434.42,174.7) node [anchor=north west][inner sep=0.75pt]  [font=\small,color={rgb, 255:red, 74; green, 144; blue, 226 }  ,opacity=1 ]  {$\mathbb{C}_{\infty }$};
\draw (290.3,78.17) node [anchor=north west][inner sep=0.75pt]  [font=\small]  {$\Bar{\delta}$};
\draw (277.22,118.39) node [anchor=north west][inner sep=0.75pt]  [font=\small]  {$\delta$};

\end{tikzpicture}

    \caption{The hyperbolic triangle.}
    \label{hyperbolictriangle}
\end{figure}

It can be clearly seen that this equation is the law of cosine for the hyperbolic triangles. From a geometrical perspective, $\gamma_{2r}$ is the third side of the hyperbolic triangle whose other sides are $2a$ and $2\Bar{b}$, which includes the angle $w_{k}= \frac{k \pi}{n}$, this hyperbolic triangle is illustrated in Fig. \ref{hyperbolictriangle}. 

We can use the properties of hyperbolic triangles to simplify the $M_k$ matrix. These properties are explained in \cite{klein1928geometrie}. The equations we will use are
\begin{align}
    &\cosh{\gamma_k} = \cosh{2a} \cdot \cosh{2\Bar{b}}- \sinh{2a} \cdot \sinh{2\Bar{b}} \cdot \cos{w_k} \label{hyptrigid1} \\
    &\sinh{\gamma_k} \cdot \cos{\delta'} = \sinh{2\Bar{b}} \cdot \cosh{2} - \cosh{2\Bar{b}} \cdot \sinh{2a} \cdot \cos{w_k} \label{hyptrigid2} \\
   &\frac{\sin{w_k}}{\sinh{\gamma_k}} = \frac{\sin{\delta'}}{\sinh{2a}} = \frac{\sin{\delta^*}}{\sinh{2\Bar{b}}} \label{hyptrigid3}
\end{align}
Using these equations, we can write 
\begin{align*}
    d_k &= \cosh{2a} \cdot \cosh{2\Bar{b}} - \cos{w_k} \cdot \sinh{2a} \cdot \sinh{2\Bar{b}} = \cosh{\gamma_k} \\
    o_k &= - \sin{w_k} \sinh{2\Bar{b}} + i (\sinh{2a}\cosh{2\Bar{b}} - \cos{\frac{\pi k}{n}} \cosh{2a} \sinh{2\Bar{b}}) \\
    &= \sin{\delta'} \cdot \sinh{\gamma_k} - i \cosh{2a} \cdot \sinh{2\Bar{b}} + i \cos{w_k} \cdot \sinh{2a} \cdot \cosh{2\Bar{b}} \\
    &= \sinh{\gamma_k} \cdot (\sin{\delta} - i \cos{\delta})  . 
\end{align*}
Thus, we can write
\begin{align*}
    M_k = \begin{pmatrix}
    \cosh{\gamma_k} & \sinh{\gamma_k}(\sin{\delta_k}-i\cos{\delta_k})\\
     \sinh{\gamma_k}(\sin{\delta_k}+i\cos{\delta_k}) & \cosh{\gamma_k} \end{pmatrix} = \cosh{\gamma_k} \id + i \sinh{\gamma_k} \begin{pmatrix}
     0 & -e^{i\delta_k}\\
     e^{-i\delta_k} & 0 \end{pmatrix} \, .
\end{align*}
Now $M_k$'s eigenvectors $u_k$'s can be easily found. We see that 
\begin{align*}
    &\frac{1}{\sqrt{2}} \begin{pmatrix} e^{i\frac{\delta_k}{2}}\\ i e^{-i\frac{\delta_k}{2}} \end{pmatrix}
    &\frac{1}{\sqrt{2}} \begin{pmatrix} i e^{i\frac{\delta_k}{2}}\\ e^{-i\frac{\delta_k}{2}} \end{pmatrix} \, .
\end{align*}
Thus, we can now write the eigenvectors \eqref{eigenvectorphi} of $V_0^\pm$ explicitly,
\begin{align}
    &\phi_k^{\uparrow} = \frac{1}{\sqrt{n}} 
\begin{pmatrix} e^{i\left( w_k + \frac{\delta_k}{2} \right)} \\ i e^{i\left( w_k - \frac{\delta_k}{2} \right)}  \\ e^{i\left( 2w_k + \frac{\delta_k}{2} \right)}  \\ i e^{i\left( 2w_k - \frac{\delta_k}{2} \right)} \\ \vdots \\ e^{i\left( n w_k + \frac{\delta_k}{2} \right)} \\ i e^{i\left( n w_k - \frac{\delta_k}{2} \right)} \end{pmatrix} 
&\phi_k^{\downarrow} = \frac{1}{\sqrt{n}} 
\begin{pmatrix} ie^{i\left( w_k + \frac{\delta_k}{2} \right)} \\ e^{i\left( w_k - \frac{\delta_k}{2} \right)}  \\ i e^{i\left( 2w_k + \frac{\delta_k}{2} \right)}  \\ e^{i\left( 2w_k - \frac{\delta_k}{2} \right)} \\ \vdots \\ i e^{i\left( n w_k + \frac{\delta_k}{2} \right)} \\ e^{i\left( n w_k - \frac{\delta_k}{2} \right)} \end{pmatrix} \, .
\end{align}
To diagonalize $R^\pm$, we write $E^\pm$ matrices such that 
\begin{align}
    E_{+}^{-1}=(\phi^{\uparrow }_1 , \phi^{ \downarrow}_{2n-1}, \phi^{\uparrow}_{3}, \phi^{ \downarrow}_{2n-3}, \cdots, \phi^{\uparrow}_{2n-3}\phi^{ \downarrow}_{3}, \phi^{\uparrow }_{2n-1}, \phi^{\downarrow}_{1}) \\
    E_{-}^{-1}=(\phi^{\uparrow }_0 , \phi^{ \downarrow}_{0}, \phi^{\uparrow}_{2}, \phi^{ \downarrow}_{2n-2}, \cdots, \phi^{\uparrow}_{2n-4}\phi^{ \downarrow}_{4}, \phi^{\uparrow }_{2n-2}, \phi^{\downarrow}_{2}) \, .
\end{align}
Our aim is to represent these matrices in the spin space. Yet neither $E^\pm$ nor $\Lambda^\pm$ is in orthogonal form. Thus we need a further transformation $T_0$ so that 
\begin{equation*}
    T^\pm_0 E^\pm R^\pm (E^\pm)^{-1} {T^{\pm}_0}^{-1} = K^\pm
\end{equation*}
where $K^\pm$ is a rotation. Let us define $T^\pm = T_0 E^\pm$. We choose $T^\pm_0$ such that it brings $\Lambda$ to its canonical form and simultaneously makes $T^\pm$ orthogonal. \par
Since we have shown that $V^\pm$ can be written as spin representatives, we are certain that such $T$ can be written, we only to need to verify that $T$ is also in the spin space, thus we check if $T^\pm$ is orthogonal. 
The transformation $T^-$ acts as
\begin{align}
    T^-: & \Gamma_{2a-1} \rightarrow \frac{1}{\sqrt{n}} \left( \sum_{r=1}^{n} \cos{\left( w_a \cdot 2r + \frac{\delta_{2r}'}{2}\right)} \Gamma_{2r-1} - \sum_{r=1}^{n} \sin{\left( w_a \cdot 2r + \frac{\delta_{2r}'}{2}\right)} \Gamma_{2r} \right) \\
    &\Gamma_{2a} \rightarrow \frac{1}{\sqrt{n}} \left( \sum_{r=1}^{n} \sin{\left( w_a \cdot 2r - \frac{\delta_{2r}'}{2}\right)} \Gamma_{2r-1} - \sum_{r=1}^{n} \cos{\left( w_a \cdot 2r - \frac{\delta_{2r}'}{2}\right)} \Gamma_{2r} \right) \, .
\end{align}
We can see that the determinant of $T^-$ is $-1$, an improper rotation; thus, T is orthogonal. Therefore,
\begin{equation*}
    T^{\pm} \cdot R^\pm \cdot (T^\pm)^{-1} = K^\pm \, ,
\end{equation*}
and in the spin space,
\begin{equation}
    S(T^-) \cdot V_0^- \cdot S((T^-))^{-1} = S(K^-) = \prod_{r=1}^n \exp{i(\frac{\gamma_r}{2})\Gamma_{2r-1}\Gamma_{2r}}
\end{equation}
Remark that because of our basis choice, $S(K^\pm)$ is not diagonal since $i \Gamma_{2r-1} \Gamma_{2r} = X_r$. Thus, we need to apply one more transformation to $S(K^\pm)$ in order to obtain a diagonal spin representative, the basis change transformation \eqref{basischangeg}. Then, we obtain
\begin{equation}
    g\cdot S(T^\pm) \cdot V_0^\pm \cdot S(T^\pm)^{-1} \cdot g = \Lambda^\pm
\end{equation}

\subsection{Discussion of the Partition Function}

\begin{figure}[h!]
    \begin{center}
    \resizebox{10cm}{!}{\includegraphics{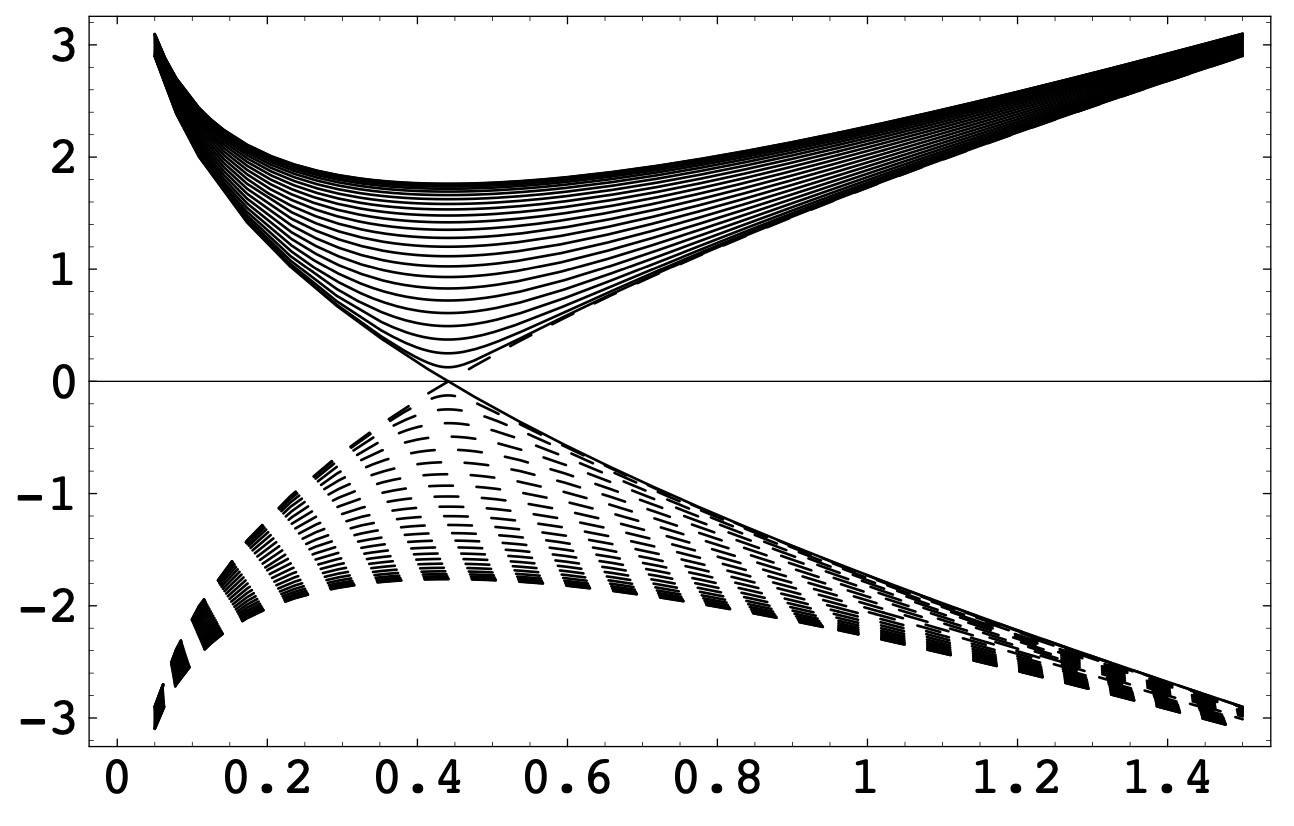}}
    \caption{$\gamma_k$ graph.}
    \label{graph}
    \end{center}
\end{figure}

We consider the partition function 
\begin{equation}
    Z= \sum_{i=1}^{2n} \lambda_i^m \, .
\end{equation}
It is clear from the formulation of the partition function that as $m$ increases, the contribution coming from the biggest eigenvalue $\lambda_{max}$ will increase exponentially. In fact, for sufficiently large $m$,
\begin{equation}
    Z \sim f \cdot \lambda_{max}^m \, ,
\end{equation}
where $f$ is the degeneracy of $\lambda_{max}$. Thus, it is important to examine the ordering of the eigenvalues $\lambda_{k}$.
We have two sets of eigenvalues $\lambda^{\pm}$, such that
\begin{align}
    \ln \lambda^+ &= \frac{1}{2} (\pm \gamma_1 \pm \gamma_3 \pm \gamma_5 \pm \cdots \pm \gamma_{2n-1} ) \\
    \ln \lambda^- &= \frac{1}{2} (\pm \gamma_0 \pm \gamma_2 \pm \gamma_4 \pm \cdots \pm \gamma_{2n-2} ) 
\end{align}
with an even number of minus signs at each combination, as explained in section \ref{sec: eigens of v}. Obviously, the biggest eigenvalues in each set will be the ones with no minus signs since we defined $\gamma_k$ to be non-negative.
\begin{align}
    \lambda^+_{max} &= \exp{\frac{1}{2} (\gamma_1 + \gamma_3 + \gamma_5 + \cdots + \gamma_{2n-1})} \\
    \lambda^-_{max} &= \exp{\frac{1}{2} (\gamma_0 + \gamma_2 + \gamma_4 + \cdots + \gamma_{2n-2})}
\end{align}
For large $n$, it follows from their definition that $\gamma_{2r}=\gamma_{2r-1}$, hence $\lambda^+_{max}$ and $\lambda^-_{max}$ are almost equal for low temperatures up to the critical point. However, $\gamma_0$ falls of very rapidly relative to $\gamma_1$, thus $\lambda^-_{max}$ becomes negligible compared to $\lambda^+_{max}$ for temperatures higher than the critical point. This can be seen clearly in Fig. \ref{graph}.
Let us look at the case where the interaction energy is the same for the row and column interactions, namely the case of a quadratic crystal that is, the case $J=J'$ and therefore $H=H'$. Thus, from eq. \ref{coshgamma_k} we have
\begin{equation}
    \cosh{\gamma_k} = \cosh{2\Bar{b}} \cdot \cosh{2b} - \sinh{2b} \cdot \sinh{2\Bar{b}} \cos{\frac{\pi k}{n}} \, .
\end{equation}
From the definition of $b$ and $\Bar{b}$, it follows that
\begin{equation}
    \cosh{\gamma_k} = \coth{2\Bar{b}} \cdot \cosh{2b} - \cos{\frac{\pi k}{n}} \, .
\end{equation}
We can deduce the value of the critical point $H_c$ using this formulation. Every $\gamma_k$ should have a minimum at the critical point, hence
\begin{equation}
    \coth{2H_c} \cdot \sinh{2H_c} - \frac{\cosh{2H_c}}{\sinh^2{2H_c}} = 0\, ,
\end{equation}
thus $H_c$ satisfies
\begin{equation}
    \sinh^2{2H_c} = 1 
\end{equation}
With a simple calculation,
\begin{equation}
    H_c \approx 0.4406867935\cdots
\end{equation}

We have said that $\gamma_0$ acts differently from the rest of $\gamma_k$'s. For $\gamma_0$ we have
\begin{equation}
    \cosh{\gamma_0} =  \cosh{2\Bar{b}} \cdot \cosh{2b} - \sinh{2\Bar{b}} \cdot \sinh{2b} \cos{\frac{0 \cdot \pi}{n}} = \cosh{2(\Bar{b}-b)} \, .
\end{equation}
As it can also been seen from Fig. \ref{graph}, $\gamma_0$ changes sign at the critical point. The other $\gamma_k$'s are almost symmetric about the $H_c$ point. This causes $\lambda_{max}^-$ to increase rapidly compared to $\lambda_{max}^+$. Thus
\begin{align*}
\frac{\lambda_{max}^-}{\lambda_{max}^+} = 
\begin{cases}
    1 & \text{for } b > H_c \text{ and } T<T_c \, ,\\ 
    e^{2(b - \Bar{b}} & \text{for } b < H_c \text{ and } T>T_c \, .
\end{cases}
\end{align*}

When $b > H_c$,
\begin{equation}
    2 \sinh{2b}^{-mn/2} \cdot Z \sim 2 \lambda_{max} = 2 \exp{\frac{m}{2}(\gamma_1 + \gamma_3 + \cdots + \gamma_{2n-1})} \, ,
\end{equation}
and when $b < H_c$,
\begin{equation}
    \sinh{2b}^{-mn/2} \cdot Z \sim \lambda_{max} = 2 \exp{\frac{m}{2}(\gamma_1 + \gamma_3 + \cdots + \gamma_{2n-1})} \, .
\end{equation}

The exact partition function does not differ much from the approximate result. For the exact solution, we have
\begin{equation}
    Z = (\sinh{2b})^{mn/2}\cdot \left[ \prod_{k=1}^{n} \left( 2 \cosh{\frac{m}{2} \gamma_{2k-1}} \right) + \left( 2 \sinh{\frac{m}{2} \gamma_{2k-1}} \right) + \prod_{k=1}^{n} \left( 2 \cosh{\frac{m}{2} \gamma_{2k}} \right) + \left( 2 \sinh{\frac{m}{2} \gamma_{2k}} \right) \right] \, . \label{exactpartitionfunc}
\end{equation}
We can take $\gamma_{2r}= \gamma_{2r-1}$ where $r=1, \cdots , n-1$, for large $n$. We must point out that the case is different for $\gamma_0$,
\begin{equation*}
    \gamma_0 = \begin{cases}
        \gamma_1 & \text{for }T<T_c \, , \\
        -\gamma_1 & \text{for }T>T_c \, .
    \end{cases}
\end{equation*}
Therefore, the two terms with products of $\cosh$ in eq. \ref{exactpartitionfunc}
 are equal regardless of the critical point whereas the terms with products are equal for $T<T_c$ and cancel each other for $T \geq T_c$. Hence
 \begin{align}
     (2\sinh{2b})^{-mn/2} \cdot Z  \cong \begin{cases}
         & \prod_{k=1}^{n} \left( 2 \cosh{\frac{m}{2} \gamma_{2k-1}} \right) \text{for } T \geq T_c \, , \\
         & \prod_{k=1}^{n} \left( 2 \cosh{\frac{m}{2} \gamma_{2k-1}} \right) + \left( 2 \sinh{\frac{m}{2} \gamma_{2k-1}} \right) \\
         & =  \prod_{k=1}^{n} \left( 2 \cosh{\frac{m}{2} \gamma_{2k-1}} \right) \cdot \left[ 1+ \prod_{k=1}^{n} \left(  \tanh{\frac{m}{2} \gamma_{2k-1}} \right)\right] \text{for } T < T_c \, .
     \end{cases}
 \end{align}
Since $m$ is very large,
\begin{equation*}
    2 \cosh{\frac{m}{2} \gamma_k} \sim \exp{\frac{m}{2} \gamma_k}
\end{equation*}
with an exception only if $T$ is in a close neighbourhood of $T_c$. So we can write 
\begin{equation}
    (2\sinh{2b})^{-mn/2} \cdot Z \sim \exp{\frac{m}{2}(\gamma_1 + \gamma_3 + \cdots + \gamma_{2n-1})} \equiv \lambda_{max}^m \text{ for } T> T_c
\end{equation}
and 
\begin{equation}
    (2\sinh{2b})^{-mn/2} \cdot Z \sim \eta \cdot \lambda_{max}^m \text{ for } T < T_c\, ,
\end{equation}
where
\begin{equation*}
    \eta = \left[ 1+ \prod_{k=1}^{n} \left(  \tanh{\frac{m}{2} \gamma_{2k-1}} \right)\right] \, .
\end{equation*}
We remark that the product of factors $\tanh{\frac{m}{2} \gamma_k}$ is never larger than $1$, thus $1< \eta < 2$. For large $m$, $\eta \sim 2$ rather close to $T_c$. The appearance of $\eta$ here is equivalent to the two-fold degeneracy of $\lambda_{max}$ for $T<T_c$ in the approximation solution of $Z$.

\section{Simplifying Kaufman’s Solution \textit{(Cansu Ozdemir and Sinan Ulaş Öztürk)}}
\label{chapter:6}

We review \cite{kastening2001simplifying}. Our aim is to find the partition function of the two-dimensional Ising model on an $m\times n$ square lattice with zero magnetic field and toroidal boundary conditions by expressing the transfer matrix in terms of two commuting representations for the generators of $SO(2n,\mathbb{C})$ algebra. After that, we diagonalize the transfer matrix by finding its eigenvectors and eigenvalues.
 
\subsection{The Model and its Transfer Matrix}

The Energy for the two-dimensional Ising model with $m$ rows and $n$ columns given a particular microstate $\{s_{\mu\nu}\}$ and $J_a$ as interaction  energy parameter between rows and $J_b$ between columns is given by,
$$
E=J_a\sum^{m}_{\mu=1}\sum_{\nu=1}^ns_{\mu\nu}s_{\mu+1,\nu}+J_b\sum^{m}_{\mu=1}\sum_{\nu=1}^ns_{\mu\nu}s_{\mu,\nu+1}
$$
Then the partition function is as follows,

$$
Z(a,b)=\sum_{\{s\}}\exp\left(a\sum^{m}_{\mu=1}\sum_{\nu=1}^ns_{\mu\nu}s_{\mu+1,\nu}+b\sum^{m}_{\mu=1}\sum_{\nu=1}^ns_{\mu\nu}s_{\mu,\nu+1}\right)=
$$
$$
\sum_{\{s\}}\prod_{\mu=1}^m\left(\prod_{\nu=1}^n\exp(as_{\mu\nu}s_{\mu+1,\nu})\exp(bs_{\mu\nu}s_{\mu,\nu+1})\right)=\sum_{\{s\}}\prod_{\mu=1}^m\langle\pi_\mu|T|\pi_{\mu+1}\rangle
$$
$$
=\textrm{Tr}T^m
$$
Where $T$ is called \textit{the transfer matrix} and $a=-\beta J_a$, $b=-\beta J_b$. Elements of the transfer matrix are given by

$$
\langle\pi_\mu|T|\pi_{\mu+1}\rangle=\prod_{\nu=1}^n\exp(as_{\mu\nu}s_{\mu+1,\nu})\exp(bs_{\mu\nu}s_{\mu,\nu+1})
$$
\subsubsection{Matrix Elements of \texorpdfstring{$T$}{}}
To be able to see how $|\pi_\mu\rangle$ behaves and elements of $T$ explicitly, let us explore the notion of "all possible microstates" deeper. On a lattice site of size $(m,n)$ there are $m$ rows and $n$ columns. Then in a given row, there are $n$ spin sites. If we find a way to list all of the possible values that the spin sites for a given row can take, then we can index each of these microstates, say with the letter $i$, and indicate the value of any spin at $\nu$th place of a row in $i$th microstate as $s_\nu^{(i)}=\pm1$. 

One way to do that is to switch the sign of the $\nu$th spin in every $i=2^{n-\nu}$ value. For example for $n=4$, The table below is the way to list all $2^4=16$ microstates,

\begin{center}
\begin{tabular}{cc}
    $++++\quad i=1$ & $-+++\quad i=9$\\
    $+++-\quad i=2$ & $-++-\quad i=10$\\
    $++-+\quad i=3$ & $-+-+\quad i=11$\\
    $++--\quad i=4$ & $-+--\quad i=12$\\
    $+-++\quad i=5$ & $--++\quad i=13$\\
    $+-+-\quad i=6$ & $--+-\quad i=14$\\
    $+--+\quad i=7$ & $---+\quad i=15$\\
    $+---\quad i=8$ & $----\quad i=16$\\   
\end{tabular}
\end{center}

Where $+$ and $-$ are short for $+1$ and $-1$. What we are doing is representing these spin sites as mod 2 numbers where 0 represents $+$ and 1 represents $-$. Let $\chi(m)$ be a function that is defined to be $-1$ when $m$ is even and $+1$ when $m$ is odd. That is,
\begin{flalign*}
\chi(m)=
    \begin{cases}
        -1,\quad\textrm{$m$ is even}\\
        +1,\quad\textrm{$m$ is odd}
    \end{cases}
\end{flalign*}
So, in essence, the function $\chi$ plays the role of taking modulo 2. Then the value of $\nu$th spin in the $i$th configuration is given by 
$$
s_\nu^{(i)}=\chi\left(\left\lceil\dfrac{i}{2^{n-\nu}}\right\rceil\right)
$$
These $i$ values which represent spin configurations can be viewed as an ordering for the elements of the following set

$$
S(n)=\{+,-\}^n=\underbrace{\{+,-\}\times\{+,-\}\dots\times\{+,-\}}_{n\textrm{ times}}
$$
Let us name these elements $r_i$ such that $r_i\in S(n)$ and $r_i$ is the $i$th element of $S(n)$. For example for $n=4$, $r_{12}$ is the following ,

$$
r_{12}=(-,+,-,-)=-+--
$$
Finally, let us define a vector $|r\rangle_n=|r\rangle$ for any given $n$. We omit $n$ for brevity. 
$$
|r\rangle=\sum_ir_i|i\rangle
$$
In this representation, we can consider vectors $|i\rangle$ as standard basis vectors. Hence,

$$
|r\rangle=\underbrace{(++\dots+)}_{r_1}\begin{bmatrix}
    1\\
    0\\
    0\\
    \vdots\\
    0
\end{bmatrix}+
\underbrace{(++\dots+-)}_{r_2}\begin{bmatrix}
0\\1\\0\\\vdots\\0
\end{bmatrix}+\dots+\underbrace{(--\dots-)}_{r_{2^n}}\begin{bmatrix}
0\\0\\\vdots\\0\\1
\end{bmatrix}
$$
If we represent spin states by vectors, as follows,

\begin{flalign*}
   |s^{(i)}_\nu\rangle=  
\begin{cases}
    |+\rangle=\begin{bmatrix}
        1\\0
    \end{bmatrix}\quad\textrm{ if $s^{(i)}_\nu$}=+1\\\\
    |-\rangle=\begin{bmatrix}
        0\\1
    \end{bmatrix}\quad\textrm{ if $s^{(i)}_\nu$}=-1
\end{cases}
\end{flalign*}
Using these vectors we can construct $2^n$ number of vectors in the standard basis by taking the Kronecker product of spin vectors.
\begin{equation}\label{ii}
|i\rangle=\bigotimes_{\nu=1}^n|s^{(i)}_\nu\rangle   
\end{equation}
To give the same example as above, for $n=4$ we can construct $|12\rangle$ as below,

$$
|12\rangle=|-\rangle\otimes|+\rangle\otimes|-\rangle\otimes|-\rangle=\begin{bmatrix}
        0\\1
    \end{bmatrix}\otimes\begin{bmatrix}
        1\\0
    \end{bmatrix}\otimes\begin{bmatrix}
        0\\1
    \end{bmatrix}\otimes\begin{bmatrix}
        0\\1
    \end{bmatrix}
$$
Let us get back to calculating the partition function. Imagine there are two rows, i.e. $\mu=1,2$. Then partition function would take the form,

$$
Z=\sum_{i,j}\exp\left(\sum_\nu as^{(i)}_\nu s^{(j)}_\nu\right)\exp(\sum_\nu bs^{(j)}_\nu s^{(j)}_{\nu+1})\exp(\sum_\nu bs^{(i)}_\nu s^{(i)}_{\nu+1})=\sum_{i,j}V^a_{ij}V^b_iV^b_j
$$
Where $i$ represents a sum over all the possible states of the first row ($\mu=1$) and $j$ represents the same thing for the second row ($\mu=2$). The contribution due to Interaction between adjacent columns $V^b$, depends only on the state of the row that makes the contribution so it depends only on either $i$ or $j$. If we generalize this for $m$ rows, that is $\mu=1,\dots m$, we get,

$$
Z=\sum_{i_1,i_2,\dots i_m}V^a_{i_1i_2}V^a_{i_2i_3}\dots V^a_{i_mi_1}V^b_{i_1}\dots V^b_{i_m}=\sum_{i_1\dots i_m}\prod_\mu V^a_{i_\mu i_{\mu+1}} V^b_{i_\mu}
$$
Where we have toroidal boundary conditions $i_{m+1}=i_1$. Here,

\begin{align}
\begin{cases}
V^a_{ij}=\exp(a\sum_\nu s_\nu^{(i)}s_\nu^{(j)})=\prod_\nu\exp(as_\nu^{(i)}s_\nu^{(j)})\\\\
V_i^b=\prod_\nu\exp(bs^{(i)}_\nu s^{(i)}_{\nu+1})
\end{cases}
\end{align}
Let us slightly alter the representation of $V_i^b$ to look like matrix elements. 

$$
V_{ij}^b=\delta_{ij}\prod_\nu\exp(bs_\nu^{(i)}s^{(j)}_{\nu+1})
$$
Now we can regard $V^a_{ij}$ and $V^b_{ij}$ as matrices. Then we can think of their matrix elements in terms of basis vectors,

$$
V^{a,b}_{ij}=\langle i|V^{a,b}|j\rangle
$$
If we define $T=V^aV^b$, elements of $T$ are
$$
T_{ij}=\sum_kV^a_{ik}V^b_{kj}=\sum_k\prod_\nu\exp(as^{(i)}_\nu s^{(k)}_\nu)\exp(bs^{(k)}_\nu s^{(j)}_{\nu+1})\delta_{kj}=\prod_\nu\exp(as^{(i)}_\nu s^{(j)}_\nu+bs^{(j)}_\nu s^{(j)}_{\nu+1})
$$
Which is exactly the summand in the partition function. $T$ is called \textit{transfer matrix.} From here, we can rewrite the partition function,

$$
Z=\sum_{i_1}\sum_{i_2\dots i_m}\prod_\mu\langle i_\mu|T|i_{\mu+1}\rangle=\sum_{i_1}\langle i_1|T^m|i_1\rangle=\Tr T^m \, .
$$
It is given in the article that the matrices $V^a$ and $V^b$ can be expressed using newly defined matrices $X_\nu$, $Y_\nu$ and $Z_\nu$ in terms of Pauli matrices. Let us state Pauli matrices,
$$
\sigma_x=\begin{bmatrix}
0 & 1\\
1 & 0
\end{bmatrix},\quad 
\sigma_y=\begin{bmatrix}
    0 & -i\\  
    i & 0
\end{bmatrix},\quad
\sigma_z= \begin{bmatrix}
    1 & 0\\
    0 & -1
\end{bmatrix}
$$
Then define
\begin{equation}
    X_\nu=\underbrace{\id\otimes\cdots\id\otimes
    \sigma_x}_{\textrm{$\nu$th place}}\otimes \id \dots\otimes\id
    =\bigotimes_{k=1}^{\nu-1} \id \otimes\sigma_x\otimes\left(\bigotimes_{k=\nu+1}^{n} \id \right) \, .
\end{equation}
$Y_\nu$, $Z_\nu$ are defined in the same way by replacing $\sigma_x$ by $\sigma_y$ and $\sigma_z$ respectively. Then, further define $\bar{a}>0$ as by $\sinh2\Bar{a}\sinh2a=1$ and say,
$$
V^a=(2\sinh2a)^{n/2}\prod_{\nu=1}^n\exp(\bar{a}X_\nu)
$$
Notice,
$$
\tanh a=e^{-2\bar{a}}\quad;\quad\tanh\bar{a} = e^{-2a}
$$
and $V^b$ is given by
$$
V^b=\prod_\nu\exp(bZ_\nu Z_{\nu+1}) 
$$
Let us investigate these matrices separately and why these products of matrices are equal to counting interaction energies for rows and columns separately. 

\subsubsection{Elements of \texorpdfstring{$V^a$}{}}

Our aim now is to investigate and understand the relationship between the given product 

\begin{equation}\label{va}
 \langle i|V^a|j\rangle=(2\sinh2a)^{n/2}\langle i|\prod_{\nu=1}^n\exp(\bar{a}X_\nu)|j\rangle
\end{equation}
and the contribution to the partition function given in terms of different configurations and spin values,

\begin{equation}\label{va2}
 \langle i|V^a|j\rangle=\prod_\nu\exp(as_\nu^{(i)}s_\nu^{(j)})
\end{equation}
To see this connection, let us first look at how $X_\nu$ acts on a state $|i\rangle$. Any spin is either $+1$ or $-1$ thus, $|s_\nu^{(i)}\rangle=|\pm\rangle$. If we act on $|s_\nu^{(i)}\rangle$ we get,

$$
\sigma_x|+\rangle=\begin{bmatrix}
    0 & 1 \\
    1 & 0
\end{bmatrix}\begin{bmatrix}
    1 \\ 0
\end{bmatrix}=\begin{bmatrix}
    0 \\ 1
\end{bmatrix} = |-\rangle
$$
Similarly,
$$
\sigma_x|-\rangle=|+\rangle
$$
We can write these two equations in one expression,

\begin{equation}
    \sigma_x|s^{(i)}_\nu\rangle = |-s^{(i)}_\nu\rangle
\end{equation}
Since $|i\rangle$ can be written in terms of $|s_\nu^{(i)}\rangle$ by the \eqref{ii}, we have,

$$
X_\nu|i\rangle=\left[\bigotimes_{k=1}^{\nu-1}\id\otimes\sigma_x\otimes\left(\bigotimes_{k=\nu+1}^{n}\id\right)\right]\bigotimes_{\lambda=1}^n|s^{(i)}_\lambda\rangle = \bigotimes_{\lambda=1}^{\nu-1}|s^{(i)}_\lambda\rangle\otimes\sigma_x|s^{(i)}_\nu\rangle\otimes\left(\bigotimes_{\lambda=\nu+1}^{n}|s^{(i)}_\lambda\rangle\right)
$$
$$
=\bigotimes_{\lambda=1}^{\nu-1}|s^{(i)}_\lambda\rangle\otimes|-s^{(i)}_\nu\rangle\otimes\left(\bigotimes_{\lambda=\nu+1}^{n}|s^{(i)}_\lambda\rangle\right)
$$
Which corresponds to another state. In fact, this is the state where the sign of the $\nu$th spin site is flipped! This is very nice since what $\langle i |V^a|j\rangle$ actually does is to compare each spin site on the $\nu$th position for the $|i\rangle$ and $|j\rangle$ states. If the spin sites have different signs, multiply by $e^{-a}$ and if they are the same, multiply by $e^{a}$ for all $\nu\in\{1,2\dots n\}$. $X_\nu$ allows us to flip the $\nu$th spin site for a given configuration. Let us have a look at this new configuration which is also an element of $S(n)$. It is not obvious to see what will be the order of this new state in relation to $i$. $X_\nu$ will take $|i\rangle$ where $i\in\{1,2,\dots,2^n\}$ and give another element in the same set as an output.  This is a kind of permutation. Let us denote this permutation by $P_\nu$ thus,
\begin{equation}\label{perm}
X_\nu|i\rangle = |P_\nu(i)\rangle    
\end{equation}
Where,
$$
P_\nu:\{1,\dots2^n\}\to\{1,\dots2^n\}\quad;\quad P_\nu\in\textrm{Sym}(2^n)
$$
It is not necessary to explicitly show what $P_\nu$ is. Nevertheless, in this equation, we can give an  expression for $P_\nu$ which is,

$$
P_\nu(i)=i+\chi\left(\left\lceil\dfrac{i}{2^{n-\nu}}\right\rceil\right)2^{n-\nu}
$$
\vspace{0.5cm}
Now Let us expand the \eqref{va},
$$
\exp(\bar{a}X_\nu)=\sum_k\dfrac{(\Bar{a}X_\nu)^k}{k!}
$$
Notice that $X_\nu^2=\id_{2^n}$ since $\sigma_x^2=\id_2$. Thus,

$$
\exp(\Bar{a}X_\nu)=\cosh\Bar{a}\id+\sinh\Bar{a}X_\nu=\cosh\Bar{a}(\id+\tanh\bar{a}X_\nu)=\cosh\Bar{a}(\id+e^{-2a}X_\nu)
$$
From here,
$$
\prod_\nu\exp(\Bar{a}X_\nu)=\prod_\nu\cosh\Bar{a}(\id+e^{-2a}X_\nu)=\cosh^n\Bar{a}\prod_\nu(\id+e^{-2a}X_\nu)
$$
Then, $V_a$ is,
$$
V^a=(2\sinh2a)^{n/2}\cosh^n\Bar{a}\prod_\nu(\id+e^{-2a}X_\nu)
$$
The coefficient of this expression is, 

$$
\sqrt{2\sinh2a}\cosh\Bar{a}=\dfrac{\sqrt{2}}{\sqrt{\sinh2\Bar{a}}}\cosh\Bar{a}=\sqrt{\dfrac{2\cosh^2\Bar{a}}{2\sinh\Bar{a}\cosh\Bar{a}}}=\sqrt{\dfrac{\cosh\Bar{a}}{\sinh\Bar{a}}}=\dfrac{1}{\sqrt{\tanh\Bar{a}}}=e^a
$$
Thus we arrive at the final expression,

\begin{equation}\label{va3}
    V^a=e^{na}\prod_\nu(\id+e^{-2a}X_\nu)
\end{equation}
This is just another expression for \eqref{va}. We can already see how this expression will count all spin sites. If we expand this product we see the explicit terms that lead to a transition between $\ket{i}$ and $\ket{j}$,
$$
e^{na}\left(\id+e^{-2a}\sum_\nu X_\nu+e^{-4a}\sum_{\nu_1,\nu_2;\nu_1<\nu_2}X_{\nu_1}X_{\nu_2}+\dots+e^{-2na}\prod_\nu X_\nu\right)
$$
The first term corresponds to $\ket{i}$ and $\ket{j}$ configurations being the same. Then all the energy terms will add up to $e^{na}$ since each $\nu$ will contribute $e^a$ to the final product. If however, $\ket{i}$ and $\ket{j}$ differ only by one spin site, say the spin at $\nu^*$, then $V^a_{ij}$ will be $e^{(n-2)a}$. Also notice, $P_{\nu^*}\ket{i}=\ket{j}$. Hence every flip in the configuration corresponds to a $e^{-2a}$ difference from $e^{na}$.  With this information in mind, we can interpret the matrix elements,
$$
\bra{i}V^a\ket{j}=e^{na}\left(\langle i|j\rangle+e^{-2a}\sum_\nu\bra{i}X_\nu\ket{j}+\dots+e^{-2na}\bra{i}\prod_\nu X_\nu\ket{j}\right)
$$
$$
=e^{na}\left(\delta_{ij}+e^{-2a}\sum_\nu\delta_{i,P_\nu(j)}+\dots+e^{-2na}\delta_{i,P_1\dots P_n(j)}\right)
$$

\subsubsection{Elements of \texorpdfstring{$V^b$}{}}
As for $V^b$, we have a very similar expression with $V^a$. However, its interpretation will be different. Expression for $V^b$ is given by,
\begin{equation}\label{vb}
\bra{i}V^b\ket{j}=\delta_{ij}\prod_\nu\exp(bs^{(i)}_\nu s^{(j)}_{\nu+1})
\end{equation}
Which is said to be equal to,
\begin{equation}\label{vb2}
    \bra{i}V^b\ket{j}=\bra{i}\prod_\nu\exp(bZ_\nu Z_{\nu+1})\ket{j}
\end{equation}
let us first observe how $Z_\nu$ affects a basis vector $\ket{i}$. Just as above, 

\begin{flalign*}
    \begin{cases}
        \sigma_z\ket{+}=\begin{bmatrix}
            1 & 0 \\
            0 & -1
        \end{bmatrix}\begin{bmatrix}
            1 \\
            0
        \end{bmatrix}=\begin{bmatrix}
            1 \\
            0
        \end{bmatrix}=\ket{+}
    \\\\
    \sigma_z\ket{-}=\begin{bmatrix}
            1 & 0 \\
            0 & -1
        \end{bmatrix}\begin{bmatrix}
            0 \\
            1
        \end{bmatrix}=\begin{bmatrix}
            0 \\
            -1
        \end{bmatrix}=-\ket{-}
        \end{cases}
\end{flalign*}
Which is in short,
\begin{equation}
\sigma_z\ket{s^{(i)}_\nu}=s^{(i)}_\nu\ket{s^{(i)}_\nu}
\end{equation}
Then, using \eqref{ii},

$$
Z_\nu \ket{i}=\left[\bigotimes_{k=1}^{\nu-1}\id\otimes\sigma_z\otimes\left(\bigotimes_{k=\nu+1}^{n}\id\right)\right]\bigotimes_{\lambda=1}^n|s^{(i)}_\lambda\rangle = \bigotimes_{\lambda=1}^{\nu-1}|s^{(i)}_\lambda \rangle \otimes\sigma_z|s^{(i)}_\nu\rangle\otimes\left(\bigotimes_{\lambda=\nu+1}^{n}|s^{(i)}_\lambda\rangle\right)
$$
$$
=\bigotimes_{\lambda=1}^{\nu-1}|s^{(i)}_\lambda\rangle\otimes s^{(i)}_\nu|s^{(i)}_\nu\rangle\otimes\left(\bigotimes_{\lambda=\nu+1}^{n}|s^{(i)}_\lambda\rangle\right)=s^{(i)}_\nu\ket{i}
$$
Thus, the effect of $Z_\nu$ on a basis vector is multiplying the same vector by $\nu$th spin value. i.e.,
\begin{equation}
    Z_\nu\ket{i}=s^{(i)}_\nu\ket{i}
\end{equation}
From here, we have
$$
Z_\nu Z_{\nu+1}\ket{i}=s^{(i)}_\nu s^{(i)}_{\nu+1}\ket{i}
$$
At this point, let us introduce $W_\nu=Z_\nu Z_{\nu+1}$ and $w_\nu=s^{(i)}_\nu s^{(i)}_{\nu+1}$ for brevity. The equation above becomes

$$
W_\nu\ket{i}=w_\nu\ket{i}
$$
Where $w_\nu$ takes the values $\pm1$. $w_n$ is essentially an indicator to tell if $\nu$th and $(\nu+1)$th spin sites have the same value. $w_\nu$ is $+1$ when the $\nu$th spin site and $(\nu+1)$th spin site has the same sign. It equals $-1$ when they have different signs.  $\ket{i}$ is an eigenvector of $W_\nu$ with eigenvalue $w_n$. Then we have the following relation,

$$
\exp(bW_\nu)\ket{i}=\exp(bw_\nu)\ket{i}\Rightarrow\prod_\nu\exp(bW_\nu)\ket{i}=\prod_\nu\exp(bw_\nu)\ket{i}
$$
Finally, we have,
$$
\bra{i}V^b\ket{j}=\bra{i}\prod_\nu\exp(bW_\nu)\ket{j}=\prod_\nu\exp(bw_\nu)\bra{i}\ket{j}=\delta_{ij}\prod_\nu\exp(bw_\nu)
$$
Plugging in the expressions for $W_\nu$ and $w_\nu$ we get,

\begin{equation}
    \bra{i}\prod_\nu\exp(bZ_\nu Z_{\nu+1})\ket{j}=\delta_{ij}\prod_\nu\exp(bs^{(i)}_\nu s^{(j)}_{\nu+1})
\end{equation}

\subsection{
\texorpdfstring{$\Gamma_\nu$}{} Matrices and Algebras of Their Projected Bilinears}
Spin representations of $2^n$-dimensional rotation groups can be obtained by defining $2n$ matrices with a basis for the $2 \times 2$ matrix algebra. For convenience, let us choose the basis as the Pauli matrices with the identity. It is noteworthy that the basis can be chosen arbitrarily, and there always exist transformations between two arbitrary bases. For a detailed discussion, see \cite{brauer1935spinors} and chapter 10 of \cite{murnaghan1938theory}. Let us define \(2n\) matrices such that
\begin{align}
    \Gamma_{2\nu - 1} & = X_1 \cdots X_{\nu - 1} \cdot Z_\nu = \bigotimes_{k=1}^{\nu-1}\sigma_X\otimes\sigma_Z\otimes\left(\bigotimes_{k=\nu+1}^n\id\right)\\
    \Gamma_{2\nu } & = X_1 \cdots X_{\nu - 1} \cdot Y_\nu = \bigotimes_{k=1}^{\nu-1}\sigma_X\otimes\sigma_Y\otimes\left(\bigotimes_{k=\nu+1}^n\id\right)
\end{align}
The defined matrices obey \(\{\Gamma_\mu, \Gamma_\nu \} = 2 \delta_{\mu \nu}\). Let us define a matrix that is the product of all \(X_i\) matrices. This matrix can be expressed using can be expressed using \(\Gamma_\nu\) matrices. 
\begin{align}
    U_{X} = X_1 \cdots X_{n} = i^n \Gamma_{1} \cdots  \Gamma_{2n} 
\end{align}
Using the properties of the spin matrices, this equation can be explicitly written as
\begin{align}
   i^n \Gamma_{1} \cdots  \Gamma_{2n} &= i^n Z_1 Y_1 X_1 Z_2 X_1 Y_2 \cdots  X_1 \cdots X_{n - 1} Z_n X_1 \cdots X_{n - 1} Y_n \\
   &= i^n Z_1 Y_1 Z_2 Y_2 \cdots Z_n Y_n \\
   &= i^n (-i) X_1 (-i) X_2 \cdots (-i) X_n =  X_1 \cdots X_{n}.
\end{align}
By the properties of the spin matrices, it is clear that \(U_{X}^2 = X_1^2 \cdots X_{n}^2 = 1\). The \(U_{X}\) matrix has the property of anti-commuting with every \(\Gamma_\mu\), 
\begin{align}
    \{ \Gamma_\mu , U_{X} \} & = \Gamma_\mu U_{X} +  U_{X} \Gamma_\mu \\
    &= (X_1 \cdots X_{\mu- 1} T_{\mu} I_{\mu+1} \cdots I_n)(X_1 \cdots X_n) + (X_1 \cdots X_{\mu- 1} T_{\mu} I_{\mu+1} \cdots I_n)(X_1 \cdots X_n) \\
    &= \{T_{\mu}, X_{\mu} \} = 0
\end{align}
Where $T=Z$ or $T=Y$. Using only \(U_{X}\) and bilinears \(\Gamma_\alpha \Gamma_\beta\) we can express \(X_\nu\) and \(Z_\nu Z_{\nu+1}\)
which were previously used in the formulation of \(V\).
\begin{align}
   X_\nu & = - \frac{i}{2} [\Gamma_{2\nu}, \Gamma_{2\nu-1}] \\
   Z_\nu Z_{\nu+1} & = - \frac{i}{2} [\Gamma_{2\nu +1 }, \Gamma_{2\nu}] \\
   Z_n Z_1 & = \frac{i}{2} U_X [\Gamma_{1}, \Gamma_{2n}]
\end{align}
These relations will lead to two commuting algebras of projected bilinears of the \(\Gamma_\mu\). \par
Let us define the projectors
\begin{equation}
    P^{\pm} \equiv \frac{1}{2} (1 \pm U_X)
\end{equation}
which has the properties
\begin{align}
   &P^{+} + P^{-} =1 = \frac{1}{2} (2 + U_X - U_X) \\
   &P^{\pm}P^{\pm}=P^{\pm} = \frac{1}{4} (1 \pm 2 U_X + U_X^2) = \frac{1}{2} (1 \pm U_X) \\
   &P^{+}P^{-}=0=P^{-}P^{+} = \frac{1}{4} (1 \pm U_X \mp U_X - U_X^2) \\
   &P^{\pm}U_X = U_X P^{\pm} =\pm P^{\pm} = \frac{1}{2} (1 \pm U_X) U_X = \frac{1}{2} (U_X \pm 1) = \pm \frac{1}{2} (1 \pm U_X) 
\end{align}
And satisfy 
\begin{align}
   &[P^{\pm}, \Gamma_\alpha \Gamma_\beta] = P^\pm \Gamma_\alpha \Gamma_\beta - \Gamma_\alpha \Gamma_\beta P^\pm \\
   &=(\frac{1}{2} (1 \pm U_X) \Gamma_\alpha \Gamma_\beta) - ( \frac{1}{2} \Gamma_\alpha \Gamma_\beta (1 \pm U_X)) \\
   &= \frac{1}{2}(\Gamma_\alpha \Gamma_\beta \pm U_X \Gamma_\alpha \Gamma_\beta - \Gamma_\alpha \Gamma_\beta \mp \Gamma_\alpha \Gamma_\beta U_X) \\
   &= \frac{1}{2}(\pm U_X \Gamma_\alpha \Gamma_\beta \mp \Gamma_\alpha \Gamma_\beta U_X)
\end{align}
We know that $\{\Gamma_\mu , U_X\} = 0$, thus $\Gamma_\mu U_X = -U_X \Gamma_\mu $. Hence,
\begin{align}
   &[P^{\pm}, \Gamma_\alpha \Gamma_\beta] =\frac{1}{2} (\pm \Gamma_\alpha \Gamma_\beta U_X \mp \Gamma_\alpha \Gamma_\beta U_X) = 0 
\end{align}
With these matrices, define
\begin{align}
   J_{\alpha \beta} = - \frac{i}{4} [\Gamma_\alpha , \Gamma_\beta] \\
   J_{\alpha \beta}^{\pm} = P^{\pm} J_{\alpha \beta} \, .
\end{align}
Notice that 
\begin{align*}
   &J_{\alpha \beta} = P^{+} J_{\alpha \beta} + P^{-} J_{\alpha \beta} = \frac{1}{2}(1 + U_X)(- \frac{i}{4} [\Gamma_\alpha , \Gamma_\beta]) + \frac{1}{2}(1 - U_X)(- \frac{i}{4} [\Gamma_\alpha , \Gamma_\beta]) = J_{\alpha \beta}^{+} + J_{\alpha \beta}^{-} \\
   &U_X J_{\alpha \beta}^{\pm} = (U_X P^{\pm}) J_{\alpha \beta} = \pm P^{\pm} J_{\alpha \beta}^{\pm} = \pm P^{\pm} J_{\alpha \beta}^{\pm}
\end{align*}
Clearly, $J_{\alpha \beta}^{\pm} = - \frac{i}{4} [\Gamma_\alpha , \Gamma_\beta] = \frac{i}{4} [\Gamma_\beta , \Gamma_\alpha] = - J_{\beta \alpha}^{\pm}$.
Now let us discuss the number of independent $J_{\alpha \beta}^{\pm}$ matrices there are. Remark that, since $\{\Gamma_\mu, \Gamma_\nu \} = 2 \delta_{\mu \nu}$, for $\mu \neq \nu $, $\Gamma_\mu \Gamma_\nu = - \Gamma_\nu \Gamma_\mu$, hence for each $\nu$, there are $(2n-1)$ $\Gamma_\mu$ matrices such that $J_{\nu \mu} \neq 0 $. Thus, there are a total of $n(2n-1)$ independent matrices of each type $J_{\alpha \beta}^{\pm}$. Their algebra decomposes into two commuting parts 
\begin{align}
    &[ J_{\alpha \beta}^{+}, J_{\gamma \delta}^{-}] \\
    &= \frac{1}{2}(1 + U_X)(- \frac{i}{4} [\Gamma_\alpha , \Gamma_\beta]) \cdot \frac{1}{2}(1 - U_X)(- \frac{i}{4} [\Gamma_\gamma , \Gamma_\delta]) \\
    &= \left( \frac{1}{2} \cdot \frac{- i}{4} \right)^2 ( ([\Gamma_\alpha , \Gamma_\beta]+ U_X [\Gamma_\alpha , \Gamma_\beta] ) \cdot ([\Gamma_\gamma , \Gamma_\delta] - U_X [\Gamma_\gamma , \Gamma_\delta] ) \\ &- ([\Gamma_\gamma , \Gamma_\delta]- U_X [\Gamma_\gamma , \Gamma_\delta] ) \cdot ([\Gamma_\alpha , \Gamma_\beta]+ U_X [\Gamma_\alpha , \Gamma_\beta] ) ) \\
    &= \left( \frac{1}{2} \cdot \frac{- i}{4} \right)^2 (([\Gamma_\alpha , \Gamma_\beta] [\Gamma_\gamma , \Gamma_\delta] + U_X [\Gamma_\alpha , \Gamma_\beta] [\Gamma_\gamma , \Gamma_\delta] - U_X [\Gamma_\alpha , \Gamma_\beta][\Gamma_\gamma , \Gamma_\delta] - [\Gamma_\alpha , \Gamma_\beta] [\Gamma_\gamma , \Gamma_\delta] ) \\ &- ([\Gamma_\gamma , \Gamma_\delta][\Gamma_\alpha , \Gamma_\beta]- U_X [\Gamma_\gamma , \Gamma_\delta][\Gamma_\alpha , \Gamma_\beta] + U_X [\Gamma_\gamma , \Gamma_\delta] [\Gamma_\alpha , \Gamma_\beta] - [\Gamma_\gamma , \Gamma_\delta] [\Gamma_\alpha , \Gamma_\beta] )) \\
    &=0 
\end{align}
which obey the identical algebras 
\begin{equation}
    [ J_{\alpha \beta}^{\pm}, J_{\gamma \delta}^{\pm}]=i(\delta_{\alpha \gamma}J_{\beta \delta}^{\pm}+\delta_{\beta \delta}J_{\alpha \gamma}^{\pm} - \delta_{\alpha \delta}J_{\beta \gamma}^{\pm} - \delta_{\beta \gamma}J_{\alpha \delta}^{\pm})
\end{equation}
This relation can be explicitly shown,
\begin{align*}
    &[ J_{\alpha \beta}^{\pm}, J_{\gamma \delta}^{\pm}] = (J_{\alpha \beta}^{\pm} J_{\gamma \delta}^{\pm}-J_{\gamma \delta}^{\pm} J_{\alpha \beta}^{\pm}) \\
    &= (P^{\pm}J_{\alpha \beta} P^{\pm}J_{\gamma \delta} P^{\pm}J_{\alpha \beta}) \\
    &= \left( \frac{- i}{4} \right)^2 ((P^{\pm}\Gamma_\alpha \Gamma_\beta - P^{\pm} \Gamma_\beta \Gamma_\alpha )\cdot (P^{\pm}\Gamma_\gamma \Gamma_\delta - P^{\pm} \Gamma_\delta \Gamma_\gamma ))
\end{align*}
Since $[P^{\pm} ,\Gamma_\mu \Gamma_\nu] = 0$ and $P^{\pm} P^{\pm} = P^{\pm}$,
\begin{align}
    &= \left( \frac{- i}{4} \right)^2 P^{\pm} (( \Gamma_\alpha \Gamma_\beta \Gamma_\gamma \Gamma_\delta - \Gamma_\alpha \Gamma_\beta \Gamma_\delta \Gamma_\gamma - \Gamma_\beta \Gamma_\alpha \Gamma_\gamma  \Gamma_\delta + \Gamma_\beta \Gamma_\alpha \Gamma_\delta  \Gamma_\gamma) \\ & - (\Gamma_\gamma \Gamma_\delta\Gamma_\alpha \Gamma_\beta - \Gamma_\gamma \Gamma_\delta \Gamma_\beta \Gamma_\alpha -  \Gamma_\delta \Gamma_\gamma \Gamma_\alpha \Gamma_\beta + \Gamma_\delta \Gamma_\gamma \Gamma_\beta \Gamma_\alpha))\\
    &= \left( \frac{-i}{4} \right)^2 P^{\pm} ((\Gamma_\alpha \Gamma_\beta \Gamma_\gamma \Gamma_\delta - \Gamma_\delta \Gamma_\gamma \Gamma_\beta \Gamma_\alpha) + (\Gamma_\delta \Gamma_\gamma \Gamma_\alpha \Gamma_\beta - \Gamma_\beta \Gamma_\alpha \Gamma_\gamma  \Gamma_\delta) \\ & + (\Gamma_\gamma \Gamma_\delta \Gamma_\beta \Gamma_\alpha - \Gamma_\alpha \Gamma_\beta \Gamma_\delta \Gamma_\gamma ) + (\Gamma_\beta \Gamma_\alpha \Gamma_\delta \Gamma_\gamma - \Gamma_\gamma \Gamma_\delta \Gamma_\alpha \Gamma_\beta ) ) \label{jabalgebra}
\end{align}
Now, let us take a closer look at each term in the sum. Without loss of generality, look at the term
\begin{equation*}
    (\Gamma_\alpha \Gamma_\beta \Gamma_\gamma \Gamma_\delta - \Gamma_\delta \Gamma_\gamma \Gamma_\beta \Gamma_\alpha) \, .
\end{equation*}
If $\beta = \gamma $, then it is equal to 
\begin{equation*}
    =(\Gamma_\alpha \Gamma_\beta^2 \Gamma_\delta -\Gamma_\delta \Gamma_\beta^2 \Gamma_\alpha ) \, .
\end{equation*}
Now, using the anti-commutativity property of $\Gamma_mu$ matrices, we can write
\begin{equation*}
    = (\Gamma_\alpha \Gamma_\beta^2 \Gamma_\delta - \Gamma_\delta \Gamma_\alpha \Gamma_\beta^2) \, ,
\end{equation*}
and since $\Gamma_\mu^2 = 1$
\begin{equation*}
    = \Gamma_\alpha \Gamma_\delta -  \Gamma_\delta \Gamma_\alpha = [\Gamma_\alpha, \Gamma_\delta]\, .
\end{equation*}
For the case $\beta \neq \gamma $, only by using the anti-commutativity, we can easily show that $ (\Gamma_\alpha \Gamma_\beta \Gamma_\gamma \Gamma_\delta -\Gamma_\gamma \Gamma_\delta \Gamma_\alpha \Gamma_\beta ) = 0$:
\begin{align*}
     (\Gamma_\alpha \Gamma_\beta \Gamma_\gamma \Gamma_\delta - \Gamma_\delta \Gamma_\gamma \Gamma_\beta \Gamma_\alpha) \\
     =  (\Gamma_\alpha \Gamma_\beta \Gamma_\gamma \Gamma_\delta - (-1)^6 \Gamma_\alpha \Gamma_\beta \Gamma_\gamma \Gamma_\delta) = 0 
\end{align*}
Thus, we have shown that 
\begin{equation*}
    (\Gamma_\alpha \Gamma_\beta \Gamma_\gamma \Gamma_\delta - \Gamma_\delta \Gamma_\gamma \Gamma_\beta \Gamma_\alpha) = \delta_{\beta \gamma} [\Gamma_\alpha , \Gamma_\delta ] \, .
\end{equation*}
Going back to \eqref{jabalgebra}, 
\begin{align}
    &\left( \frac{- i}{4} \right)^2 P^{\pm} ((\Gamma_\alpha \Gamma_\beta \Gamma_\gamma \Gamma_\delta - \Gamma_\delta \Gamma_\gamma \Gamma_\beta \Gamma_\alpha) + (\Gamma_\delta \Gamma_\gamma \Gamma_\alpha \Gamma_\beta - \Gamma_\beta \Gamma_\alpha \Gamma_\gamma  \Gamma_\delta) \\ & + (\Gamma_\gamma \Gamma_\delta \Gamma_\beta \Gamma_\alpha - \Gamma_\alpha \Gamma_\beta \Gamma_\delta \Gamma_\gamma ) + (\Gamma_\beta \Gamma_\alpha \Gamma_\delta  \Gamma_\gamma - \Gamma_\gamma \Gamma_\delta \Gamma_\alpha \Gamma_\beta ) ) \\ 
    &= \left( \frac{- i}{4} \right)^2 P^{\pm} (\delta_{\beta \gamma} [\Gamma_\alpha , \Gamma_\delta ] + \delta_{\beta \delta} [\Gamma_\gamma , \Gamma_\alpha] + \delta_{\alpha \gamma} [\Gamma_\delta , \Gamma_\beta] + \delta_{\alpha \delta} [\Gamma_\beta , \Gamma_\gamma]) \\
    &= \frac{i}{4} (\delta_{\alpha \gamma} J_{\beta \delta}^{\pm}+\delta_{\beta \delta} J_{\alpha \gamma}^{\pm} - \delta_{\alpha \delta}J_{\beta \gamma}^{\pm} - \delta_{\beta \gamma} J_{\alpha \delta}^{\pm})
\end{align}
We can write \(X_\nu\) and \(Z_\nu Z_{\nu+1}\) matrices in terms of projected $J_{\alpha \beta}^{\pm}$ matrices.
\begin{align*}
    X_\nu &= 2(J_{2\nu, 2\nu-1}^{+} + J_{2\nu, 2\nu-1}^{-}) \\
   &= 2( \frac{1}{2}(1 + U_X + 1 - U_X ) J_{2\nu, 2\nu-1}) = 2 J_{2\nu, 2\nu-1} = \frac{-i}{2}[\Gamma_{2\nu}, \Gamma_{2\nu-1}]
\end{align*}
And similarly, 
\begin{align}
    Z_\nu Z_{\nu +1} &= 2(J_{2\nu +1, 2\nu}^{+} + J_{2\nu+1, 2\nu}^{-}) \\
    Z_n Z_1 &= -2U_X(J_{1,2n}^{+} + J_{1, 2n}^{-}) = -2(J_{1,2n}^{+} - J_{1, 2n}^{-})
\end{align}
Using these equations we can express \(V_{a/2}\) and \(V_{b}\) in terms of \(J_{\alpha \beta}^{\pm}\):
\begin{align}
    V_{a/2} &= \prod_{\nu=1}^{n} \exp[\bar{a} X_\nu /2] \\
    &= \prod_{\nu=1}^{n} \exp[\bar{a} (J_{2\nu, 2\nu-1}^{+} + J_{2\nu, 2\nu-1}^{-})] \\
    &= \left(  \prod_{\nu=1}^{n} \exp[\bar{a} J_{2\nu, 2\nu-1}^{+}] \right) \cdot \left(  \prod_{\nu=1}^{n} \exp[\bar{a} J_{2\nu, 2\nu-1}^{-}] \right) \\
    & = V_{a/2}^{+}V_{a/2}^{-}
\end{align}
where 
\begin{equation}
    V_{a/2}^{\pm} = \prod_{\nu=1}^{n} \exp[\bar{a} J_{2\nu, 2\nu-1}^{\pm}] \, .
\end{equation}
For the \(V_b\) matrix,
\begin{align}
    V_{b} &= \prod_{\nu=1}^{n} \exp[b Z_\nu Z_{\nu+1}] \\
    &= \exp(b Z_n Z_1) \prod_{\nu=1}^{n-1} \exp[b Z_\nu Z_{\nu+1}] \\
    &= \exp[-2b(J_{1,2n}^{+} - J_{1, 2n}^{-})] \prod_{\nu=1}^{n-1} \exp[2b(J_{2\nu +1, 2\nu}^{+} + J_{2\nu+1, 2\nu}^{-})]=V_{b}^{+}V_{b}^{-} \\
    &= \left( \exp(-2b J_{1,2n}^{+}) \prod_{\nu=1}^{n-1} \exp[2b J_{2\nu +1, 2\nu}^{+}]\right) \cdot \left( \exp(+2b J_{1, 2n}^{-}) \prod_{\nu=1}^{n-1} \exp[2b J_{2\nu+1, 2\nu}^{-}] \right) \\
    &=V_{b}^{+}V_{b}^{-} \, ,
\end{align}
where
\begin{equation}
    V_{b}^{\pm} = \exp(\mp 2b J_{1,2n}^{\pm}) \prod_{\nu=1}^{n-1} \exp(J_{2\nu+1, 2\nu}^{\pm}) \, .
\end{equation}
Then the matrix \(V\) can be written as
\begin{equation}
    V=V^+ V^-
\end{equation}
where
\begin{equation}
    V^{\pm} = V_{a/2}^{\pm} V_b^{\pm} V_{a/2}^{\pm}
\end{equation}
and \([V^+, V^-] = 0\). 

\subsection{Diagonalization of \texorpdfstring{$\mathcal{V}^{\pm}$}{} Matrices}
In this section, we will construct the algebra of $\mathcal{V}^{\pm}$, which has similarities with $V^{\pm}$ matrices, along with periodicity properties that allow for explicit diagonalization. \par
Define,
\begin{equation}
    (\mathcal{J}_{\alpha \beta})_{ij}=-i (\delta_{\alpha i}\delta_{\beta j}-\delta_{\beta i}\delta_{\alpha j}) \, .
\end{equation}
The $\mathcal{J}$ matrices have the form 
\begin{equation}
    \mathcal{J}_{\alpha \beta} = \begin{pmatrix}
    &&&& \\
    &&-i&& \\
    &i&&& \\
    &&&& \\
    &&&&
\end{pmatrix}
\end{equation}
for $\alpha \neq \beta$. If $\alpha = \beta$, then it is clear that $\mathcal{J}_{\alpha \alpha} = 0$. Since $\mathcal{J}_{\alpha \beta} = - \mathcal{J}_{\beta \alpha }$, it can be easily seen that there are $n(2n-1)$ different $\mathcal{J}_{\alpha \beta}$ matrices. These matrices, too, obey the algebra $[ \mathcal{J}_{\alpha \beta}^{\pm}, \mathcal{J}_{\gamma \delta}^{\pm}]=i(\delta_{\alpha \gamma}\mathcal{J}_{\beta \delta}^{\pm}+\delta_{\beta \delta}\mathcal{J}_{\alpha \gamma}^{\pm} - \delta_{\alpha \delta}\mathcal{J}_{\beta \gamma}^{\pm} - \delta_{\beta \gamma}\mathcal{J}_{\alpha \delta}^{\pm})$. To show this algebra is obeyed, let us look at the elements of the $\mathcal{J}_{\alpha \beta}^{\pm} \mathcal{J}_{\gamma \delta}^{\pm} - \mathcal{J}_{\gamma \delta}^{\pm} \mathcal{J}_{\alpha \beta}^{\pm}$ matrix.
\begin{equation}
    (\mathcal{J}_{\alpha \beta}^{\pm} \mathcal{J}_{\gamma \delta}^{\pm} - \mathcal{J}_{\gamma \delta}^{\pm} \mathcal{J}_{\alpha \beta}^{\pm})_{ij}= (\mathcal{J}_{\alpha \beta}^{\pm} \mathcal{J}_{\gamma \delta}^{\pm} )_{ij} - (\mathcal{J}_{\gamma \delta}^{\pm} \mathcal{J}_{\alpha \beta}^{\pm})_{ij} \label{fancyjalgebra}
\end{equation}
Let us first look at the components of $(\mathcal{J}_{\alpha \beta}^{\pm} \mathcal{J}_{\gamma \delta}^{\pm} )_{ij}$, 
\begin{align}
    &(\mathcal{J}_{\alpha \beta}^{\pm} \mathcal{J}_{\gamma \delta}^{\pm} )_{ij} \\
    &=\sum_k (-i (\delta_{\alpha i}\delta_{\beta k}-\delta_{\beta i}\delta_{\alpha k})) (-i (\delta_{\gamma k}\delta_{\delta j}-\delta_{\delta j}\delta_{\gamma k})) \\
    &= - \sum_k \delta_{\alpha i}\delta_{\beta k}\delta_{\gamma k}\delta_{\delta j} - \delta_{\alpha i}\delta_{\beta k}\delta_{\delta j}\delta_{\gamma k} - \delta_{\beta i}\delta_{\alpha k}\delta_{\gamma k}\delta_{\delta j} + \delta_{\beta i}\delta_{\alpha k} \delta_{\delta j}\delta_{\gamma k} \\
    &= - (\delta_{\alpha i}\delta_{\beta \gamma }\delta_{\delta j} - \delta_{\alpha i}\delta_{\beta \gamma}\delta_{\delta j} - \delta_{\beta i}\delta_{\alpha \gamma }\delta_{\delta j} + \delta_{\beta i}\delta_{\alpha \gamma} \delta_{\delta j} ) \, .
\end{align}
The components $( \mathcal{J}_{\alpha \beta}^{\pm} \mathcal{J}_{\gamma \delta}^{\pm} -\mathcal{J}_{\gamma \delta}^{\pm} \mathcal{J}_{\alpha \beta}^{\pm})_{ij}$ can be obtained in a similar way. We can now write the components of the $\mathcal{J}_{\alpha \beta}^{\pm} \mathcal{J}_{\gamma \delta}^{\pm} - \mathcal{J}_{\gamma \delta}^{\pm} \mathcal{J}_{\alpha \beta}^{\pm}$ matrix explicitly:
\begin{align}
    &(\mathcal{J}_{\alpha \beta}^{\pm} \mathcal{J}_{\gamma \delta}^{\pm} - \mathcal{J}_{\gamma \delta}^{\pm} \mathcal{J}_{\alpha \beta}^{\pm})_{ij} \\ 
    &= (-\delta_{\alpha i}\delta_{\beta \gamma }\delta_{\delta j} + \delta_{\alpha i}\delta_{\beta \delta}\delta_{\gamma j} + \delta_{\beta i}\delta_{\alpha \gamma }\delta_{\delta j} - \delta_{\beta i}\delta_{\alpha \delta} \delta_{\gamma j}) + (\delta_{\gamma i} \delta_{\delta \alpha} \delta_{\beta j} - \delta_{\gamma i} \delta_{\delta \beta} \delta_{\alpha j} - \delta_{\delta i} \delta_{\gamma \alpha} \delta_{\beta j} + \delta_{\delta i} \delta_{\gamma \beta} \delta_{\alpha j} ) \\
    &= \delta_{\beta \gamma}(\delta_{\alpha i}\delta_{\delta j}-\delta_{\alpha j}\delta_{\delta i} ) +
    \delta_{\alpha \gamma }(\delta_{\beta i}\delta_{\delta j} - \delta_{\delta i} \delta_{\beta j}) + 
    \delta_{\alpha \delta}(\delta_{\gamma i} \delta_{\beta j} - \delta_{\beta i} \delta_{\gamma j}) + 
    \delta_{\beta \delta}(\delta_{\alpha i} \delta_{\gamma j} - \delta_{\gamma i} \delta_{\alpha j}) \\
    &= \delta_{\beta \gamma}(\mathcal{J}_{\alpha \delta})_{ij} +
    \delta_{\alpha \gamma}(\mathcal{J}_{\beta \delta})_{i j} + 
    \delta_{\alpha \delta}(\mathcal{J}_{\gamma \beta})_{ij} + 
    \delta_{\beta \delta}(\mathcal{J}_{\alpha \gamma})_{ij} \\
\end{align}
Thus, 
\begin{equation}
    [ \mathcal{J}_{\alpha \beta}^{\pm}, \mathcal{J}_{\gamma \delta}^{\pm}]=i(\delta_{\alpha \gamma}\mathcal{J}_{\beta \delta}^{\pm}+\delta_{\beta \delta}\mathcal{J}_{\alpha \gamma}^{\pm} - \delta_{\alpha \delta}\mathcal{J}_{\beta \gamma}^{\pm} - \delta_{\beta \gamma}\mathcal{J}_{\alpha \delta}^{\pm}) \, .
\end{equation}
Consider the matrices,
\begin{equation}
    S = \exp(ic_{\alpha \beta} \mathcal{J}_{\alpha \beta} )
\end{equation}
It is easily seen that
\begin{equation}
    S^T = \exp(ic_{\alpha \beta} \mathcal{J}_{\alpha \beta} )^T= \exp(ic_{\alpha \beta} \mathcal{J}_{\alpha \beta} ^T ) = \exp(-ic_{\alpha \beta} \mathcal{J}_{\alpha \beta} )= S^{-1} \, .
\end{equation}
Thus, 
\begin{equation}
    \det(S)^2 =\det(S) \det(S^T) = \det(S) \det(S^{-1}) = 1 
\end{equation}
S is smoothly connected to the unit matrix, for real parameters $c_{\alpha \beta}$ the $S$ matrices are also real and form the group $SO(N)$
of orthogonal matrices with unit determinant. Therefore, the algebra of equation \eqref{fancyjalgebra} is called the Lie algebra of $SO(N)$. Here, we let $c_{\alpha \beta}$ be arbitrary complex numbers; thus, the $S$ matrices form the group $SO(N, \mathbb{C})$. \par
Let us define the $SO(N ,\mathbb{C})$ matrices
\begin{equation}
    \mathcal{V}^{\pm} = \mathcal{V}_{a/2}  \mathcal{V}_b^{\pm} \mathcal{V}_{a/2}
\end{equation}
where 
\begin{align}
    &\mathcal{V}_{a/2} = \prod_{\nu=1}^{n} \exp(\bar{a} \mathcal{J}_{2\nu, 2\nu-1}^{\pm}) 
    & \mathcal{V}_{b}^{\pm} = \exp(\mp 2b \mathcal{J}_{1,2n}^{\pm}) \prod_{\nu=1}^{n-1} \exp(\mathcal{J}_{2\nu+1, 2\nu}^{\pm}) \, .
\end{align}
It can be seen that these matrices are defined in analogy with $V$ matrices. Since $\bar{a}$ and $b$ are real, the matrices $\mathcal{V}$, $\mathcal{V}_{a/2}$, and $\mathcal{V}_{b}^{\pm}$ are hermitian. Thus they only have real eigenvalues and a complete set of orthonormal eigenvectors. Now let us show the structures of $\mathcal{V}_{a/2}$ and $\mathcal{V}_{b}^{\pm}$. \par
We can write 
\begin{equation}
    \mathcal{V}_{a/2} = \begin{pmatrix}
    \begin{bmatrix} R_{\Bar{a}} \end{bmatrix} & 0 & \cdots & 0 \\
    0 & \begin{bmatrix} R_{\Bar{a}} \end{bmatrix} & \cdots & 0  \\
    \vdots & \vdots & \ddots & \vdots  \\
    0 & 0 & \cdots & \begin{bmatrix} R_{\Bar{a}} \end{bmatrix} 
\end{pmatrix} \, ,
\end{equation}
where 
\begin{equation}
     \begin{bmatrix} R_{\Bar{a}} \end{bmatrix} =  \begin{bmatrix} \cosh{\Bar{a}} & i \sinh{\Bar{a}} \\
     -i \sinh{\Bar{a}} & \cosh{\Bar{a}} \end{bmatrix}
\end{equation}
and 
\begin{equation}
    \mathcal{V}_{b}^{\pm} = \begin{pmatrix}
    \cosh{2b} & 0 & \cdots & 0 & \pm i \sinh{2b} \\
    0 & \begin{bmatrix} R_{2b} \end{bmatrix} & \cdots & 0 & 0 \\
    \vdots & \vdots & \ddots & \vdots & \vdots \\
    0 & 0 & \cdots & \begin{bmatrix} R_{2b} \end{bmatrix} & 0 \\
    \mp i \sinh{2b} & 0 & \cdots & 0 & \cosh{2b}
\end{pmatrix} \, ,
\end{equation}
where
\begin{equation}
     \begin{bmatrix} R_{2b} \end{bmatrix} =  \begin{bmatrix} \cosh{2b} & i \sinh{2b} \\
     -i \sinh{2b} & \cosh{2b} \end{bmatrix} \, .
\end{equation}
Now let us write $\mathcal{V}^{\pm}$,
\begin{equation}
    \mathcal{V}^{\pm} = \begin{pmatrix}
    A & B & 0 & \cdots & 0 & \mp B^\dagger \\
    B^\dagger & A & B & \cdots & 0 & 0 \\
    0 & B^\dagger & A & \cdots & 0 & 0 \\
    \vdots & \vdots & \vdots & \ddots & \vdots & \vdots \\
    0 & 0 & 0 & \cdots & A & B \\
    \mp B & 0 & 0 & \cdots & B^\dagger & A 
\end{pmatrix} \, ,
\end{equation}
where 
\begin{align}
   & A = \cosh{2b} \begin{pmatrix}
    \cosh{2 \Bar{a}} & i \sinh{2 \Bar{a}} \\
     -i \sinh{2 \Bar{a}} & \cosh{2 \Bar{a}} \end{pmatrix}
     & B = \sinh{2b} \begin{pmatrix}
    -\frac{1}{2}\sinh{2 \Bar{a}} & -i \sinh^2{\Bar{a}} \\
     i \cosh^2{\Bar{a}} & -\frac{1}{2}\sinh{2 \Bar{a}} \end{pmatrix}
\end{align}
Using this information, we want to find the eigenvectors of $\mathcal{V}^{\pm}$, thus using the analogy between $\mathcal{V}^{\pm}$ and $V^{\pm}$ we can diagonalize $V^{\pm}$. To find the eigenvectors of $\mathcal{V}^{\pm}$, we will make an ansatz. Let us assume the eigenvectors of $\mathcal{V}^{\pm}$ are of the form
\begin{equation}
\phi = \frac{1}{\sqrt{n}} 
\begin{pmatrix} z u \\ z^2 u \\ \vdots \\ z^n u \end{pmatrix} \, ,
\end{equation}
where $z \in \mathbb{C}$ and $u$ is a two-component vector such that $u^\dagger u = 1$. From the eigenvector condition $\mathcal{V}^{\pm} \phi = \lambda \phi $ we get $n$ equations 
\begin{align}
    (zA + z^2 B \mp z^n B^\dagger)u &= z \lambda u \\
    (z^2 A + z^3 B + z B^\dagger)u &= z^2 \lambda u \\
    (z^3 A + z^4 B + z^2 B^\dagger)u &= z^3 \lambda u \\
    & \vdots \\
    (z^{n-1} A + z^n B + z^{n-2} B^\dagger)u &= z^{n-1} \lambda u \\
    (z^n A \mp z B + z^{n-1} B^\dagger)u &= z^n \lambda u \, .
\end{align}
The equations apart from the first and last ones are identical. Therefore, these $n$ equations can be reduced to three independent equations:
\begin{align}
    (A + z B \mp z^{n-1} B^\dagger)u &= \lambda u \\
    (A + z B + z^{-1} B^\dagger)u &= \lambda u \\
    (A \mp z^{1-n} B + z^{-1} B^\dagger)u &= \lambda u \, .
\end{align}
If we set $z^n=\mp 1$, then these three equations become identical,
\begin{equation}
    (A+ z B + z^{-1} B^\dagger )u = \lambda u \, .
\end{equation}
We can see that if $z_k = e^{i \pi k / n}$, for $k=\Bar{0,1, \cdots , 2n-1}$, then there are a total of $2n$ values that solve $z=\mp 1$, where odd $k$ lead to $-1$ and even $k$ lead to $1$. For each $k$, we now need to find the associated eigenvalues $\lambda^{\uparrow \downarrow}_k$ and eigenvectors $u^{\uparrow \downarrow}_k$. Now, let us define a matrix 
\begin{equation}
    M_k = A + e^{i \pi k/n} B + e^{-i \pi k/n} B^\dagger = \begin{pmatrix} d_k & o_k \\ o^*_k & d_k \end{pmatrix} \, .
\end{equation}
Here, 
\begin{equation}
   d_k = \cosh{2\Bar{a}} \cosh{2b} - \cos{\frac{\pi k}{n}} \sinh{2\Bar{a}} \sinh{2b}
\end{equation}
and
\begin{equation}
   o_k = - \sin{\frac{\pi k}{n}} \sinh{2b} + i (\sinh{2\Bar{a}}\cosh{2b} - \cos{\frac{\pi k}{n}} \cosh{2 \Bar{a}} \sinh{2b}) \, \label{ok1}.
\end{equation}
We can explicitly calculate the determinant of $M_k$ matrix. 
\begin{align*}
    \det M_k =& (\cosh{2\Bar{a}} \cosh{2b} - \cos{\frac{\pi k}{n}} \sinh{2\Bar{a}} \sinh{2b})^2 \\ 
    &- (- \sin{\frac{\pi k}{n}} \sinh{2b} + i (\sinh{2\Bar{a}}\cosh{2b} - \cos{\frac{\pi k}{n}} \cosh{2 \Bar{a}} \sinh{2b})) \\ & \cdot (- \sin{\frac{\pi k}{n}} \sinh{2b} - i (\sinh{2\Bar{a}}\cosh{2b} - \cos{\frac{\pi k}{n}} \cosh{2 \Bar{a}} \sinh{2b})) \\
    =& \cosh^2{2b} (\cosh^2{2\Bar{a}}-\sinh^2{2\Bar{a}}) + (\cos{\frac{\pi k}{n}})^2 (\sinh^2{2b} (\sinh^2{\Bar{a}}-\cosh^2{\Bar{a}})) - (\sin{\frac{\pi k}{n}})^2 (\sinh^2{2b}) \\
    =& \cosh^2{2b} - \sinh^2{2b} ((\cos{\frac{\pi k}{n}})^2 + (\sin{\frac{\pi k}{n}})^2) = 1
\end{align*}
The unit determinant implies eigenvalues of the form 
\begin{align}
    & \lambda^{\uparrow}_k = e^{+ \gamma} & \lambda^{\downarrow}_k = e^{- \gamma}
\end{align}
with $\gamma \in \mathbb{R}$. We can use the trace to calculate the value of $\gamma$.
\begin{equation}
    tr (M_k) = 2(\cosh{2\Bar{a}} \cosh{2b} - \cos{\frac{\pi k}{n}} \sinh{2\Bar{a}} \sinh{2b}) = e^{+ \gamma} +  e^{- \gamma} = 2 \cosh{ \gamma}
\end{equation}
We can write $o_k$ in a simpler form, depending on $\gamma_k$. Since $d_k=\cosh{\gamma_k}$ and $\det M_k = 1$,
\begin{equation}
    \cosh^2{\gamma_k-o_k^*o_k = 1} \, .
\end{equation}
From here we can see that $o_k^*o_k=\sinh^2{\gamma_k}$, thus $o_k$ has the form 
\begin{equation}
    o_k = i e^{i \delta_k} \sinh{\gamma_k} \, \label{ok2}.
\end{equation}
The phase in $e^{i \delta_k}$ will be discussed later. \par
From the forms of the eigenvalues, it can be seen that both $\gamma_k$ and $-\gamma_k$'s are solutions. For convenience, let us fix the sign of $\gamma_k$ by defining $\gamma_k=2\Bar{a}$ for $b=0$. Which means, $\gamma_0 = 2(\Bar{a} - b)$. \par
It can be clearly seen that for $k=1 \cdots , n-1$, $\gamma_k = \gamma_{2n-k} $, which means the eigenvalues $\lambda^{\uparrow \downarrow}$
Now let us look at $\frac{\partial \gamma_k }{\partial k}$ by extending $\cosh{ \gamma_k} = \cosh{2\Bar{a}} \cosh{2b} - \cos{\frac{\pi k}{n}} \sinh{2\Bar{a}} \sinh{2b}$ to non-integer values of $k$ and then looking at $\frac{\partial \cosh{ \gamma_k} }{\partial k}$:
\begin{align*}
\frac{\partial \cosh{\gamma_k} }{\partial k} &= \frac{\partial \cosh{\gamma_k} }{\partial \gamma_k} \frac{\partial \gamma_k }{\partial k} = \sinh{\gamma_k} \frac{\partial \gamma_k }{\partial k} \\
&=\frac{ \partial (\cosh{2\Bar{a}} \cosh{2b} - \cos{\frac{\pi k}{n}} \sinh{2\Bar{a}} \sinh{2b})}{\partial k} \\
&= \sinh{2 \Bar{a}} \sinh{2b} \sin{\frac{\pi k}{n}} \frac{\pi}{n} \, .
\end{align*}
Thus, we can say,
\begin{equation}
    \frac{\partial \gamma_k }{\partial k} = \left(\frac{\pi}{n} \right) \frac{\sinh{2\Bar{a}} \sinh{2b}}{\sinh{\gamma_k}} \sin{\frac{\pi k}{n}} \, > 0.
\end{equation}
Hence, we have
\begin{equation}
    0 < |\gamma_0| < \gamma_1 < \cdots < \gamma_n
\end{equation}
more precisely,
\begin{align}
    &0 < + \gamma_0 < \gamma_1 < \cdots < \gamma_n \, \, \, , \Bar{a}> b \\
    &0 < - \gamma_0 < \gamma_1 < \cdots < \gamma_n \, \, \, , \Bar{a} < b \, .
\end{align}
Now let us turn back to the phase in $o_k$. Comparing the two equations \eqref{ok1} and \eqref{ok2}, we can obtain information about $\delta_k$
\begin{align}
    o_k =& i e^{i \delta_k} \sinh{\gamma_k} \\
    =&i \cos{\delta_k} \sinh{\gamma_k} - \sin{\delta_k} \sinh{\gamma_k}\\
    =& - \sin{\frac{\pi k}{n}} \sinh{2b} + i (\sinh{2\Bar{a}}\cosh{2b} - \cos{\frac{\pi k}{n}} \cosh{2 \Bar{a}} \sinh{2b}) \, .
\end{align}
It can be clearly seen that,
\begin{align}
    & \cos{\delta_k} \sinh{\gamma_k} = \sinh{2\Bar{a}}\cosh{2b} - \cos{\frac{\pi k}{n}} \cosh{2 \Bar{a}} \sinh{2b} \\
    & \sin{\delta_k} \sinh{\gamma_k} = \sin{\frac{\pi k}{n}} \sinh{2b} \, .
\end{align}
For $k=0$, 
\begin{align}
    \cos{\delta_0} \sinh{\gamma_0} &= \sinh{2\Bar{a}}\cosh{2b} - \cosh{2 \Bar{a}} \sinh{2b} \\
    &= \sinh{2(\Bar{a}-b)}
\end{align} 
and since $\gamma_0 = 2(\Bar{a}-b)$
\begin{align}
    \sin{\delta_0} \sinh{\gamma_0} &= \sin{\frac{0 \cdot \pi}{n}} \sinh{2b} = 0 \, ,
\end{align} 
we can conclude that $\delta_0=0$. \par
We can now write 
\begin{align}
    & u_k^{\uparrow} = \frac{1}{\sqrt{2}} \begin{pmatrix} e^{\frac{i}{2}\delta_k} \\ -ie^{\frac{-i}{2}\delta_k} \end{pmatrix} 
    & u_k^{\downarrow} = \frac{1}{\sqrt{2}} \begin{pmatrix} -ie^{\frac{i}{2}\delta_k} \\ e^{\frac{-i}{2}\delta_k} \end{pmatrix}
\end{align}

The $\gamma^{\uparrow \downarrow}_k$ matrices are evidently orthonormal, which can be shown by explicit calculation. For distinct eigenvalues for $\mathcal{V}^\pm$, we have the condition that $k-l = 2m$ for some whole number $m$. This is true since $\mathcal{V}^+$ and $\mathcal{V}^-$ commute and can be diagonalized simultaneously. If we examine $\mathcal{V}^+$ for example, eigenvalues $k,l=1,3,5\dots2n-1$. Then,
$$
\phi_k^{\uparrow\downarrow\dagger}\phi_l^{\uparrow\downarrow}=\dfrac{1}{n}u_k^{\uparrow\downarrow\dagger}u_l^{\uparrow\downarrow}\sum_\alpha\left[e^{i\pi(l-k)/n}\right]^\alpha=\dfrac{1}{n}u_k^{\uparrow\downarrow\dagger}u_l^{\uparrow\downarrow}\sum_{\alpha=1}^n\left[e^{i\pi2m/n}\right]^\alpha
$$
Here, 
$$
\sum_{\alpha=1}^n[e^{i\pi2m/n}]^\alpha=\sum_\alpha q^\alpha
$$
with $q=e^{i\pi2m/n}$. We know that the result of this sum is given by,

$$
S = \dfrac{q(q^n-1)}{q-1}=\dfrac{e^{i\pi2m/n}(e^{i\pi2m}-1)}{e^{i\pi2m/n}-1}
$$
Notice that $e^{i\pi2m}=1$ for all $m$. Thus, we have $S=0$ i.e.,
$$
\phi^{\uparrow\downarrow\dagger}_k\phi^{\uparrow\downarrow}_l=0
$$
The same argument is valid for $\mathcal{V}^-$. We can also show that the eigenvectors are orthonormal for degenerate eigenvalues, $\gamma_k$ and $\gamma_{2n-k}$, $k=1,\dots n-1$ by explicit calculation,
$$
\phi^{\uparrow\downarrow\dagger}_k\phi_{2n-k}^{\uparrow\downarrow}=\dfrac{1}{n}u^{\uparrow\downarrow\dagger}_{k}u^{\uparrow\downarrow}_{2n-k}\sum_\alpha[e^{-ik\pi/n}e^{i(2n-k)\pi/n}]^\alpha=\dfrac{1}{n}u^{\uparrow\downarrow\dagger}_{k}u^{\uparrow\downarrow}_{2n-k}\left(\dfrac{1-e^{-2\pi i(n+1)k/n}}{1-e^{-2\pi ik/n}}-1\right)=0
$$

Now let us write the $R_{\pm}$ matrices do diagonalize $\mathcal{V}^{\pm}$, such that
\begin{align}
   R_{+}\mathcal{V}^{+} R_{+}^{-1}= \begin{pmatrix}
    e^{+ \gamma_1} &  & & & & & \\
     & e^{- \gamma_1} & & & & & \\
     &  & e^{+ \gamma_3} & & & & \\
     &  &  & e^{- \gamma_3} & & & \\
     &  &  & & \ddots & & \\ 
     &  &  & & & e^{+ \gamma_{2n-1}} & \\ 
     &  &  & & & &e^{- \gamma_{2n-1}} 
     \end{pmatrix}
\end{align}
and
\begin{align}
   R_{-}\mathcal{V}^{-} R_{-}^{-1}= \begin{pmatrix}
    e^{+ \gamma_0} &  & & & & & \\
     & e^{- \gamma_0} & & & & & \\
     &  & e^{+ \gamma_2} & & & & \\
     &  &  & e^{- \gamma_2} & & & \\
     &  &  & & \ddots & & \\ 
     &  &  & & & e^{+ \gamma_{2n-2}} & \\ 
     &  &  & & & &e^{- \gamma_{2n-2}} 
     \end{pmatrix} \, .
\end{align}
Now, let us explicitly write the $R_{\pm}$ matrices,
\begin{align}
    R_{+}^{-1}=(\phi^{\uparrow }_1 , \phi^{ \downarrow}_{2n-1}, \phi^{\uparrow}_{3}, \phi^{ \downarrow}_{2n-3}, \cdots, \phi^{\uparrow}_{2n-3}\phi^{ \downarrow}_{3}, \phi^{\uparrow }_{2n-1}, \phi^{\downarrow}_{1}) \\
    R_{-}^{-1}=(\phi^{\uparrow }_0 , \phi^{ \downarrow}_{0}, \phi^{\uparrow}_{2}, \phi^{ \downarrow}_{2n-2}, \cdots, \phi^{\uparrow}_{2n-4}\phi^{ \downarrow}_{4}, \phi^{\uparrow }_{2n-2}, \phi^{\downarrow}_{2}) \, .
\end{align}
They can even be written more explicitly; let us write $R_+^{-1}$ since the $R_-^{-1}$ has the same form.
\begin{equation}
    R_{+}^{-1}= \begin{pmatrix}
    e^{\frac{i \pi }{n}} (e^{\frac{i}{2} \delta_{1}}) & e^{\frac{i \pi (2n-1) }{n}} (- i e^{\frac{i}{2} \delta_{2n-1}}) & \cdots & e^{\frac{i \pi (2n-1) }{n}} (e^{\frac{i}{2} \delta_{2n-1}}) & e^{\frac{i \pi }{n}} (- i e^{\frac{i}{2} \delta_{1}})\\
    
    e^{\frac{i \pi }{n}} (- i e^{\frac{-i}{2} \delta_{1}})& e^{\frac{i \pi (2n-1) }{n}} (e^{\frac{-i}{2} \delta_{2n-1}}) & \cdots & e^{\frac{i \pi (2n-1) }{n}} (- i e^{\frac{-i}{2} \delta_{2n-1}})& e^{\frac{i \pi }{n}} (e^{\frac{-i}{2} \delta_{1}}) \\

    e^{\frac{2i \pi }{n}} (e^{\frac{i}{2} \delta_{1}}) & e^{\frac{2 i \pi (2n-1) }{n}} (- i e^{\frac{i}{2} \delta_{2n-1}}) & \cdots & e^{\frac{2i \pi (2n-1) }{n}} (e^{\frac{i}{2} \delta_{2n-1}}) & e^{\frac{2 i \pi }{n}} (- i e^{\frac{i}{2} \delta_{1}})\\
    
    e^{\frac{2i \pi }{n}} (- i e^{\frac{-i}{2} \delta_{1}})& e^{\frac{2i \pi (2n-1) }{n}} (e^{\frac{-i}{2} \delta_{2n-1}}) & \cdots & e^{\frac{2i \pi (2n-1) }{n}} (- i e^{\frac{-i}{2} \delta_{2n-1}})& e^{\frac{2i \pi }{n}} (e^{\frac{-i}{2} \delta_{1}}) \\

    e^{\frac{3i \pi }{n}} (e^{\frac{i}{2} \delta_{1}}) & e^{\frac{3i \pi (2n-1) }{n}} (- i e^{\frac{i}{2} \delta_{2n-1}}) & \cdots & e^{\frac{3i \pi (2n-1) }{n}} (e^{\frac{i}{2} \delta_{2n-1}}) & e^{\frac{3 i \pi }{n}} (- i e^{\frac{i}{2} \delta_{1}})\\
    
    e^{\frac{3 i \pi }{n}} (- i e^{\frac{-i}{2} \delta_{1}})& e^{\frac{3i \pi (2n-1) }{n}} (e^{\frac{-i}{2} \delta_{2n-1}}) & \cdots & e^{\frac{3i \pi (2n-1) }{n}} (- i e^{\frac{-i}{2} \delta_{2n-1}})& e^{\frac{3i \pi }{n}} (e^{\frac{-i}{2} \delta_{1}}) \\
    
    \vdots & \vdots & \ddots & \vdots & \vdots \\
    
    e^{\frac{n i \pi }{n}} (e^{\frac{i}{2} \delta_{1}}) & e^{\frac{ni \pi (2n-1) }{n}} (- i e^{\frac{i}{2} \delta_{2n-1}}) & \cdots & e^{\frac{n i \pi (2n-1) }{n}} (e^{\frac{i}{2} \delta_{2n-1}}) & e^{\frac{ni \pi }{n}} (- i e^{\frac{i}{2} \delta_{1}})\\

    e^{\frac{n i \pi }{n}} (- i e^{\frac{-i}{2} \delta_{1}})& e^{\frac{ni \pi (2n-1) }{n}} (e^{\frac{-i}{2} \delta_{2n-1}}) & \cdots & e^{\frac{n i \pi (2n-1) }{n}} (- i e^{\frac{-i}{2} \delta_{2n-1}})& e^{\frac{n i \pi }{n}} (e^{\frac{-i}{2} \delta_{1}})
     \end{pmatrix}
\end{equation}
The $R_{\pm}$ matrices can be written in a more compact form by separating them to $2 \times 2$ blocks $D_{kl}$, 
\begin{equation}
     D_{kl} =  \frac{1}{\sqrt{2}} \begin{pmatrix} \exp{ \frac{ k i \pi l }{n} + \frac{-i}{2} \delta_{l}} & - i \exp{ \frac{ k i \pi (2n-l)}{n} + \frac{-i}{2} \delta_{2n-l}} \\
     -i \exp{\frac{ k i \pi l }{n} - \frac{-i}{2} \delta_{l}} & \exp{ \frac{ k i \pi (2n-l)}{n} - \frac{-i}{2} \delta_{2n-l}} \end{pmatrix} \, .
\end{equation}
Since, $\exp{ \frac{ k i \pi (2n-l)}{n}} =  \exp{ - \frac{ k i \pi l)}{n}} $ and $\delta_{2n-k}= -\delta_k$ as shown before, we can write $D_{kl}$ as
\begin{equation}
     D_{kl} =  \frac{1}{\sqrt{2}} \begin{pmatrix} \exp{i (\frac{kl\pi}{n} + \frac{\delta_l}{2})} & - i \exp{i ( -\frac{kl\pi}{n} - \frac{\delta_l}{2})} \\
     -i \exp{i (\frac{kl\pi}{n} - \frac{\delta_l}{2})} & \exp{i ( - \frac{kl\pi}{n} + \frac{\delta_l}{2})} \end{pmatrix} \, .
\end{equation}
Which can be written in terms of $z$ and $u^{\uparrow \downarrow}$:
\begin{equation}
    D_{kl} = (z_l^k u_l^{\uparrow}, z_{2n-l}^k u_{2n-l}^{\downarrow} ) \, .
\end{equation}
With this formulation, we can return to $R_{\pm}$ matrices and write them as
\begin{align}
    &R_{+}^{-1}= \frac{1}{\sqrt{n}} \begin{pmatrix} D_{11} & D_{13} & \cdots & D_{1,(2n-1)} \\
    D_{21} & D_{23} & \cdots & D_{2,(2n-1)} \\
    \vdots & \vdots & \ddots & \vdots \\
    D_{n1} & D_{n3} & \cdots & D_{n(2n-1)} \end{pmatrix}
    & R_{-}^{-1}= \frac{1}{\sqrt{n}} \begin{pmatrix} D_{10} & D_{12} & \cdots & D_{1,(2n-2)} \\
    D_{20} & D_{22} & \cdots & D_{2,(2n-2)} \\
    \vdots & \vdots & \ddots & \vdots \\
    D_{n0} & D_{n2} & \cdots & D_{n(2n-2)} \end{pmatrix} \, .
\end{align}
We want to reach the $SO(n, \mathbb{C})$ structure mentioned earlier, thus we define the $R^{\pm}_X$ matrices as
\begin{equation}
    R^{\pm}_X= \begin{pmatrix}
    \exp{\mp i \pi \sigma_x / 4} & & & \\
    & \exp{\mp i \pi \sigma_x / 4} & & \\
    &  & \ddots &  \\
    & & & \exp{\mp i \pi \sigma_x / 4}
     \end{pmatrix}
\end{equation}
where \begin{equation}
    \exp{\mp i \pi \sigma_x / 4} = \frac{1}{\sqrt{2}} \begin{pmatrix}
    1 & \mp i \\
    \mp i & 1
     \end{pmatrix} \, .
\end{equation}
Now we can write the transformation
\begin{equation}
    \mathcal{V}^\pm_\mathcal{S} = R_X R_\pm \mathcal{V}^\pm R_\pm^{-1}R_X^{-1} = \mathcal{S}_\pm \mathcal{V}^\pm \mathcal{S}_\pm^{-1} \, .
\end{equation}
With explicit calculation, it can be seen that 
\begin{align}
    \mathcal{V}^+_\mathcal{S} &= \begin{pmatrix}
    \exp{-\gamma_1 \sigma_y} & & & \\
    & \exp{-\gamma_3 \sigma_y} & & \\
    &  & \ddots &  \\
    & & & \exp{-\gamma_{2n-1} \sigma_y}
     \end{pmatrix} \\
     &= \exp \begin{pmatrix}
    -\gamma_1 \sigma_y & & & \\
    & -\gamma_3 \sigma_y & & \\
    &  & \ddots &  \\
    & & & -\gamma_{2n-1} \sigma_y
     \end{pmatrix} \\
     &=\exp{\sum_{\nu=1}^n \gamma_{2\nu - 1} \mathcal{J}_{2\nu, 2\nu-1}}
\end{align}
and similarly,
\begin{equation}
    \mathcal{V}^-_\mathcal{S} = \exp{\sum_{\nu=1}^n \gamma_{2\nu - 2} \mathcal{J}_{2\nu, 2\nu-1}} \, .
\end{equation}
Let us write $S_\pm^{-1}$ matrices explicitly by carrying out the multiplication
\begin{align}
    R_+^{-1}R_X^{-1} &= \frac{1}{\sqrt{n}} \begin{pmatrix} D_{11} & D_{13} & \cdots & D_{1,(2n-1)} \\
    D_{21} & D_{23} & \cdots & D_{2,(2n-1)} \\
    \vdots & \vdots & \ddots & \vdots \\
    D_{n1} & D_{n3} & \cdots & D_{n(2n-1)} \end{pmatrix} \cdot \begin{pmatrix}
    \exp{\mp i \pi \sigma_x / 4} & & \\
    &  \ddots &  \\
    & & \exp{\mp i \pi \sigma_x / 4}
     \end{pmatrix} \\
    &= \frac{1}{\sqrt{n}} \begin{pmatrix} D_{11} \exp{\mp i \pi \sigma_x / 4} & D_{13} \exp{\mp i \pi \sigma_x / 4} & \cdots & D_{1,(2n-1)} \exp{\mp i \pi \sigma_x / 4} \\
    D_{21} \exp{\mp i \pi \sigma_x / 4} & D_{23} \exp{\mp i \pi \sigma_x / 4} & \cdots & D_{2,(2n-1)} \exp{\mp i \pi \sigma_x / 4}\\
    \vdots & \vdots & \ddots & \vdots \\
    D_{n1} \exp{\mp i \pi \sigma_x / 4} & D_{n3} \exp{\mp i \pi \sigma_x / 4}& \cdots & D_{n(2n-1)} \exp{\mp i \pi \sigma_x / 4}\end{pmatrix} \, .
\end{align}
Let us look at each $2 \times 2$ block. For convenience, let us call each block $\Tilde{D_{kl}}$.
\begin{align}
    \Tilde{D_{kl}}  &= \begin{pmatrix} \exp{i (\frac{kl\pi}{n} + \frac{\delta_l}{2})} & - i \exp{i ( -\frac{kl\pi}{n} - \frac{\delta_l}{2})} \\
     -i \exp{i (\frac{kl\pi}{n} - \frac{\delta_l}{2})} & \exp{i ( - \frac{kl\pi}{n} + \frac{\delta_l}{2})} \end{pmatrix} \cdot \begin{pmatrix}
    1 & \mp i \\
    \mp i & 1
     \end{pmatrix} \\
     &= \begin{pmatrix} \exp{i (\frac{kl\pi}{n} + \frac{\delta_l}{2})} + \exp{i ( -\frac{kl\pi}{n} - \frac{\delta_l}{2})} & i \exp{i (\frac{kl\pi}{n} + \frac{\delta_l}{2})} - i \exp{i ( -\frac{kl\pi}{n} - \frac{\delta_l}{2})} \\ 
     -i \exp{i (\frac{kl\pi}{n} - \frac{\delta_l}{2})} + i \exp{i (\frac{- kl\pi}{n} + \frac{\delta_l}{2})} &  \exp{i (\frac{kl\pi}{n} - \frac{\delta_l}{2})} +  \exp{i (\frac{- kl\pi}{n} + \frac{\delta_l}{2})}
     \end{pmatrix} \\
     &= \begin{pmatrix} \cosh{i (\frac{kl\pi}{n} + \frac{\delta_l}{2})} & i \sinh{i (\frac{kl\pi}{n} + \frac{\delta_l}{2})} \\  
     -i \sinh{i (\frac{kl\pi}{n} - \frac{\delta_l}{2})} &  \cosh{i (\frac{kl\pi}{n} - \frac{\delta_l}{2})} \end{pmatrix} \\
     &= \frac{1}{2} \begin{pmatrix} e^{i (\frac{kl\pi}{n} + \frac{\delta_l}{2})} + e^{-i (\frac{kl\pi}{n} + \frac{\delta_l}{2})}& i( e^{i (\frac{kl\pi}{n} + \frac{\delta_l}{2})} - e^{-i (\frac{kl\pi}{n} + \frac{\delta_l}{2})})\\
     -i (e^{i (\frac{kl\pi}{n} - \frac{\delta_l}{2})} - e^{- i (\frac{kl\pi}{n} - \frac{\delta_l}{2})})&  e^{i (\frac{kl\pi}{n} - \frac{\delta_l}{2})} + e^{-i (\frac{kl\pi}{n} - \frac{\delta_l}{2})} \end{pmatrix} \\
     &= \begin{pmatrix} \cos{(\frac{kl\pi}{n} + \frac{\delta_l}{2})} & - \sin{(\frac{kl\pi}{n} + \frac{\delta_l}{2})} \\
     \sin{ (\frac{kl\pi}{n} - \frac{\delta_l}{2})} &  \cos{(\frac{kl\pi}{n} - \frac{\delta_l}{2})} \end{pmatrix}
\end{align}
Now, we can write $\mathcal{S}^{-1}_\pm$ matrices in a more compact way,
\begin{align}
    &\mathcal{S}^{-1}_+ = R_+^{-1}R_X^{-1} = \frac{1}{\sqrt{n}} \begin{pmatrix} \Tilde{D}_{11} & \Tilde{D}_{13} & \cdots & \Tilde{D}_{1,(2n-1)} \\
    \Tilde{D}_{21} & \Tilde{D}_{23} & \cdots & \Tilde{D}_{2,(2n-1)} \\
    \vdots & \vdots & \ddots & \vdots \\
    \Tilde{D}_{n1} & \Tilde{D}_{n3} & \cdots & \Tilde{D}_{n,(2n-1)} \end{pmatrix} \\
    &S^{-1}_- = R_-^{-1}R_X^{-1} = \frac{1}{\sqrt{n}} \begin{pmatrix} \Tilde{D}_{10} & \Tilde{D}_{12} & \cdots & \Tilde{D}_{1,(2n-2)} \\
    \Tilde{D}_{20} & \Tilde{D}_{22} & \cdots & \Tilde{D}_{2(2n-2)} \\
    \vdots & \vdots & \ddots & \vdots \\
    \Tilde{D}_{n0} & \Tilde{D}_{n2} & \cdots & \Tilde{D}_{n(2n-2)} \end{pmatrix} \, .
\end{align}
We can now discuss whether these $\mathcal{S}_\pm$ matrices are elements of $SO(n, \mathbb{C})$. Since $R_\pm$ and $R_X$ are unitary, $\mathcal{S}_\pm$ are also unitary. Since the $\mathcal{S}_\pm$ are also real, it follows that they are also orthogonal. Their orthogonality leads to 
\begin{equation}
    (\det \mathcal{S}_\pm)^2 = \det \mathcal{S}_\pm \det \mathcal{S}_\pm^T = \det \mathcal{S}_\pm \det \mathcal{S}_\pm^{-1}=1
\end{equation}
thus $\det \mathcal{S}_\pm=\pm1$. What we need to show is that $\det \mathcal{S}_\pm=1$. We know that $\mathcal{S}_\pm$ are analytic in $b$. We can start with $b=0$. Thus the phase is $\delta_k = 0$, then analytically continues to non-zero $b$, upon which $\det \mathcal{S}_\pm$ cannot change discontinuously and therefore not change at all. 
We can write $\det \mathcal{S}_\pm^{-1}$ as 
\begin{equation}
    \det \mathcal{S}_\pm^{-1} = \det R_X^{-1} \mathcal{S}_\pm^{-1} R_X =  \det R_X^{-1} (R_\pm^{-1} R_X^{-1}) R_X = \det R_X^{-1} R_\pm^{-1} \, .
\end{equation}
Now let us calculate $R_X^{-1} R_\pm^{-1}$ first in order to calculate its determinant.
\begin{align}
    R_X^{-1} R_\pm^{-1} &=  \frac{1}{\sqrt{n}}  \begin{pmatrix}
    \exp{\mp i \pi \sigma_x / 4} & & \\
    &  \ddots &  \\
    & & \exp{\mp i \pi \sigma_x / 4}
     \end{pmatrix}\cdot  \begin{pmatrix} D_{11} & D_{13} & \cdots & D_{1,(2n-1)} \\
    D_{21} & D_{23} & \cdots & D_{2,(2n-1)} \\
    \vdots & \vdots & \ddots & \vdots \\
    D_{n1} & D_{n3} & \cdots & D_{n(2n-1)} \end{pmatrix}
\end{align}
Looking at each $2 \times 2$ block,
\begin{align}
    \Bar{D_{kl}}&= \frac{1}{2} \begin{pmatrix}
    1 & \mp i \\
    \mp i & 1
     \end{pmatrix}  \cdot \begin{pmatrix} \exp{i (\frac{kl\pi}{n} + \frac{\delta_l}{2})} & - i \exp{i ( -\frac{kl\pi}{n} - \frac{\delta_l}{2})} \\
     -i \exp{i (\frac{kl\pi}{n} - \frac{\delta_l}{2})} & \exp{i ( - \frac{kl\pi}{n} + \frac{\delta_l}{2})} \end{pmatrix}\\
     &= \frac{1}{2} \begin{pmatrix} \exp{i (\frac{kl\pi}{n} + \frac{\delta_l}{2})} + \exp{i (\frac{kl\pi}{n} - \frac{\delta_l}{2})} & - i \exp{i ( -\frac{kl\pi}{n} - \frac{\delta_l}{2})} + i \exp{i ( -\frac{kl\pi}{n} + \frac{\delta_l}{2})}\\
     i \exp{i (\frac{kl\pi}{n} - \frac{\delta_l}{2})} - i \exp{i (\frac{kl\pi}{n} + \frac{\delta_l}{2})} & \exp{i ( - \frac{kl\pi}{n} + \frac{\delta_l}{2})} + \exp{i ( - \frac{kl\pi}{n} - \frac{\delta_l}{2})} \end{pmatrix}
\end{align}
We had said $\delta_k=0$, thus 
\begin{align}
     &= \frac{1}{2} \begin{pmatrix} 2 \exp{i (\frac{kl\pi}{n} )} & - i \exp{i ( -\frac{kl\pi}{n})} + i \exp{i ( -\frac{kl\pi}{n} )}\\
     i \exp{i (\frac{kl\pi}{n})} - i \exp{i (\frac{kl\pi}{n})} & \exp{i ( - \frac{kl\pi}{n})} + \exp{i ( - \frac{kl\pi}{n})} \end{pmatrix} \\
     &= \begin{pmatrix} \exp{ \frac{ikl\pi}{n}} & 0\\
     0 & \exp{\frac{-ikl\pi}{n}} \end{pmatrix}
\end{align}
We can now write $\mathcal{S}^{-1}_\pm$ in terms of $\Bar{D}_{kl}$'s,
\begin{align}
    &\mathcal{S}^{-1}_+ = \frac{1}{\sqrt{n}} \begin{pmatrix} \Bar{D}_{11} & \Bar{D}_{13} & \cdots & \Bar{D}_{1,(2n-1)} \\
    \Bar{D}_{21} & \Bar{D}_{23} & \cdots & \Bar{D}_{2,(2n-1)} \\
    \vdots & \vdots & \ddots & \vdots \\
    \Bar{D}_{n1} & \Bar{D}_{n3} & \cdots & \Bar{D}_{n,(2n-1)} \end{pmatrix}
    & \mathcal{S}^{-1}_- = \frac{1}{\sqrt{n}} \begin{pmatrix} \Bar{D}_{10} & \Bar{D}_{12} & \cdots & \Bar{D}_{1,(2n-2)} \\
    \Bar{D}_{20} & \Bar{D}_{22} & \cdots & \Bar{D}_{2(2n-2)} \\
    \vdots & \vdots & \ddots & \vdots \\
    \Bar{D}_{n0} & \Bar{D}_{n2} & \cdots & \Bar{D}_{n(2n-2)} \end{pmatrix} \, ,
\end{align}
We can reorganise the block matrices in $S^{-1}_\pm$ as
\begin{equation}
    \mathcal{S}^{-1}_\pm = \begin{pmatrix}
    R_\pm^+ &  \\
     & R_\pm^-
     \end{pmatrix}
\end{equation}
by grouping all $\exp{ \frac{ikl\pi}{n}}$ to form $R_\pm^+$ and $\exp{ \frac{-ikl\pi}{n}}$ to form $R_\pm^-$. The equal number of changes in rows and columns allows us to do this reorganization without varying the determinant. Now, remark that we have $R_\pm^{-}=R_\pm^{+*}$, thus
\begin{equation}
    \det \mathcal{S}_\pm^{-1} = \det R_X^{-1} R_\pm^{-1} = \det R_\pm^+ R_\pm^- = |\det \det R_\pm^+|^2 > 0 \, ,
\end{equation}
which clearly shows that 
\begin{equation}
    \det \mathcal{S}_\pm = 1 \, ,
\end{equation}
This means that $\mathcal{S}_\pm$ matrices are elements of $SO(n, \mathbb{C})$, thus can be represented as 
\begin{equation}
    \mathcal{S}_\pm = \exp{i c_{\alpha \beta}^\pm \mathcal{J}_{\alpha \beta}}
\end{equation}
with certain unknown parameters $c_{\alpha \beta}^\pm \in \mathbb{C}$.

\subsection{Diagonalization of \texorpdfstring{$V$}{}}

Now, we use the parameter $c_{\alpha \beta}^{\pm}$ from the previous section to define the $2^n \times 2^n$ dimensional transformation matrix
\begin{align}
    &S=S^+ S^-  &S^\pm = \exp{i c_{\alpha \beta}^\pm J_{\alpha \beta}^\pm}
\end{align}
and write 
\begin{equation}
    V_S = SVS^- = S_+ V^+ S_+^{-1}S_V^-S_-^{-1} \equiv V_\mathcal{S}^+ V_\mathcal{S}^-
\end{equation}
where
\begin{equation}
    V_\mathcal{S}^\pm = \exp{ic_{\alpha \beta}^\pm J_{\alpha \beta}^\pm} V_{a/2}^\pm V_b^\pm V_{a/2}^\pm \exp{- ic_{\alpha \beta}^\pm J_{\alpha \beta}^\pm} \, . 
\end{equation}
It is worth pointing out that this expression has the same form as 
\begin{equation}
    \mathcal{V}_\mathcal{S}^\pm = \exp{ic_{\alpha \beta}^\pm \mathcal{J}_{\alpha \beta}^\pm} \mathcal{V}_{a/2}^\pm \mathcal{V}_b^\pm \mathcal{V}_{a/2}^\pm \exp{- ic_{\alpha \beta}^\pm \mathcal{J}_{\alpha \beta}^\pm}
\end{equation}
from the previous section. 
\begin{equation}
    V_S = \exp{ic_{\alpha \beta}^+ J_{\alpha \beta}^+} V_{a/2}^+ V_b^+ V_{a/2}^+ \exp{- ic_{\alpha \beta}^+ J_{\alpha \beta}^+} \cdot \exp{ic_{\alpha \beta}^- J_{\alpha \beta}^-} V_{a/2}^- V_b^- V_{a/2}^- \exp{- ic_{\alpha \beta}^- J_{\alpha \beta}^-}
\end{equation}
We can use the Baker-Campbell-Hausdorff formula \cite{campbell1896law, baker1905alternants, hausdorff1906symbolische} 
\begin{equation}
    \exp{A}\exp{B}= \exp{A+B-\frac{1}{2}[B,A] + \frac{1}{12}([A, [A,B]]+[B, [B,A]]) + \cdots}
\end{equation}
and the commutation property $[J_{\alpha \beta}^+,J_{\gamma \delta}^-]=0$ to obtain
\begin{align}
    = \exp{ic_{\alpha \beta}^+ J_{\alpha \beta}^+} V_{a/2}^+ V_b^+ V_{a/2}^+ V_{a/2}^- V_b^- V_{a/2}^- \exp{- ic_{\alpha \beta}^- J_{\alpha \beta}^-} \\
    = \exp{ic_{\alpha \beta}^+ J_{\alpha \beta}^+} V_{a/2}V_b V_{a/2} \exp{- ic_{\alpha \beta}^- \mathcal{J}_{\alpha \beta}^-} \, .
\end{align}
Thus, we can write
\begin{equation}
    V_S=V_S^+V_S^- \, ,
\end{equation}
with
\begin{align}
    &V_S^+ = \exp{\sum_{\nu-1}^n \gamma_{2\nu -1} J_{2\nu, 2 \nu-1}^+ } &V_S^- = \exp{\sum_{\nu-1}^n \gamma_{2\nu -2} J^{-}_{2\nu, 2\nu-1}}
\end{align}
so that
\begin{align}
    V_S &= \exp{\sum^n_{\nu=1}(\gamma_{2\nu-1}J^+_{2\nu, 2\nu-1}+ \gamma_{2\nu-2}J^-_{2\nu,2\nu-1})} \\
    &= \exp{\frac{1}{4}\sum^n_{\nu=1}(\gamma_{2\nu-1}(1+U_X)X_\nu)+ \frac{1}{4}\sum^n_{\nu=1}(\gamma_{2\nu-2}(1-U_X)X_\nu))} 
\end{align}
In order to diagonalize $V_S$, we define another similarity transform,
$$
V_Y=R_YV_SR_Y^{-1}
$$
Where,
\begin{equation}
    R_Y^{\pm1}=2^{-n/2}\prod_{\nu=1}^{n}(\id \pm iY_\nu)
\end{equation}
and,
$$
R_YX_\nu R_Y^{-1}=Z_\nu
$$
This transformation takes the following form,
\begin{equation}
    V_Y=R_YV_SR_Y^{-1}= \exp \left[ \dfrac{1}{4}\sum_{\nu=1}^n\gamma_{2\nu -1}(\id+U_Z)Z_\nu + \dfrac{1}{4}\sum_{\nu=1}^n\gamma_{2\nu-2}(\id-U_Z)Z_\nu \right]
\end{equation}
Where we defined,
\begin{equation}
    U_Z=Z_1\dots Z_n
\end{equation}
Thus we obtain the partition function as obtained in \cite{kaufman1949crystal},
\begin{align*}
    Z(a,b) =& [2 \sinh{2a}]^{mn/2} \left[ \sum_e \exp{\frac{m}{2} \sum^n_{\nu=1}\pm \gamma_{2\nu-1}} + \sum_e \exp{\frac{m}{2} \sum^n_{\nu=1}\pm \gamma_{2\nu-2}} \right] \\
    =& \frac{1}{2} [2 \sinh{2a}]^{mn/2} \times \{ \prod^n_{k=1} \left[ 2 \cosh{\frac{m}{2}} \gamma_{2\nu-1} \right] +  \prod^n_{k=1} \left[ 2 \sinh{\frac{m}{2}} \gamma_{2\nu-1} \right] \\& + \prod^n_{k=1} \left[ 2 \cosh{\frac{m}{2}} \gamma_{2\nu-2} \right] - \prod^n_{k=1} \left[ 2 \sinh{\frac{m}{2}} \gamma_{2\nu-2} \right] \} \, .
\end{align*}

\section{Solution of the Ising Model via Dirac Algebra \textit{(Oğuz Alp Ağırbaş, Anıl Ata and Yunus Emre Yıldırım)}}
\label{chapter:7}

In his paper \cite{vergeles2009}, Vergeles demonstrated that the partition function of the Ising model on a two-dimensional regular lattice can be calculated using the matrix representation of the Dirac algebra. This approach involves a set of generators corresponding to the total number of lattice sites. Notably, the partition function can be represented as the trace of a polynomial constructed from Dirac matrices. This polynomial is a member of the rotation group in its spinor representation, implying that the partition function depends on a character in the spinor representation of an orthogonal group.

\subsection{Formulation of the Problem}
\begin{figure}[htpb]
    \centering
    \includegraphics[width=0.40\textwidth]{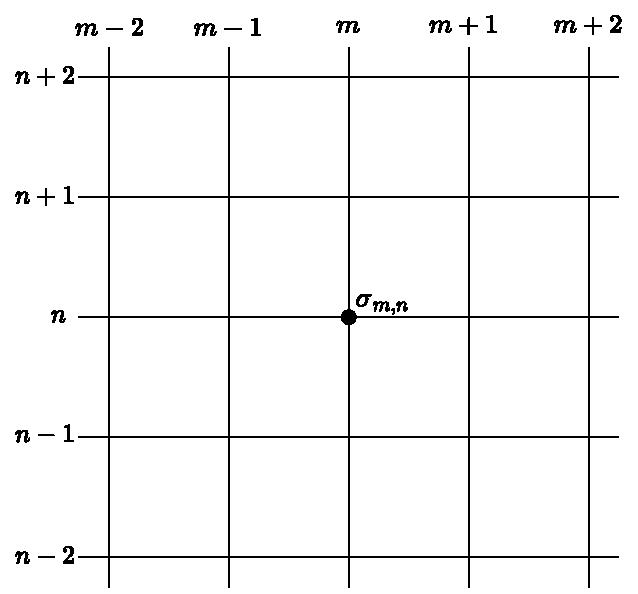}\\
    \caption{$(m,n)$ denotes $m^{th}$ column and $n^{th}$ line of the lattice.}
\end{figure}

To outline the different setups, we follow this approach \cite{landau}. For every lattice point characterized by integral coordinates (m, n), we assign a variable $\sigma_{m,n}$ that can assume one of two values, +1 or -1, representing the two potential orientations of the dipole. When considering solely the interaction between neighboring dipoles, the configuration's energy can be expressed as follows:
\hspace{0.3cm}

\begin{align}
        H&(\sigma)=-J\sum_{m=1}^{M-1}\sum_{n=1}^{N-1}[\sigma_{m,n}(\sigma_{m+1,n}+\sigma_{m,n+1})]\\
        &m=1,\ldots,M \hspace{1cm}
        n=1,\ldots,N \nonumber 
\end{align}

Then the partition function is

\begin{equation} \label{eq:2}
    Z=\sum_{\{\sigma\}}e^{-H(\sigma)/T}=\sum_{\{\sigma\}}\exp\{
    {\theta\sum_{m=1}^{M-1}\sum_{n=1}^{N-1}\sigma_{m,n}(\sigma_{m+1,n}+\sigma_{m,n+1})\}}
\end{equation}
where $\theta=J/T$.
\newline
\break
Expanding $\exp(\theta \sigma_{m,n}\sigma_{m',n'})$, we obtain
\begin{equation}
    \exp(\theta \sigma_{m,n}\sigma_{m',n'})=1+ \theta \sigma_{m,n}\sigma_{m',n'} + \frac{(\theta \sigma_{m,n}\sigma_{m',n'})^2}{2!} + \frac{(\theta \sigma_{m,n}\sigma_{m',n'})^3}{3!}+\ldots \nonumber
\end{equation}

Since $\sigma_{m,n}^0=\sigma_{m,n}^2=\sigma_{m,n}^4=1$, then
\begin{equation}
    \exp(\theta \sigma_{m,n}\sigma_{m',n'})=\left(1+\frac{\theta^{2}}{2!}+\frac{\theta^{4}}{4!}+ \ldots\right) + \sigma_{m,n}\sigma_{m',n'} \left(\theta +\frac{\theta^{3}}{3!}+\frac{\theta^{5}}{5!}+ \ldots\right) \nonumber
\end{equation}
Thus we can easily see that
\begin{equation}
    \exp(\theta \sigma_{m,n}\sigma_{m',n'})=\cosh\theta+ \sigma_{m,n}\sigma_{m',n'}\sinh\theta \nonumber
\end{equation}

Then we can express the partition function (\ref{eq:2}) in the form
\begin{equation}\label{eq:3}
    Z=(\cosh^{2}\theta)^{(M-1)(N-1)/2}\prod_{m=1}^{M-1}\prod_{n=1}^{N-1}[(1 + \sigma_{m,n} \sigma_{m+1,n}\tanh\theta)(1 + \sigma_{m,n} \sigma_{m,n+1}\tanh\theta)]
\end{equation}

This expression is a polynomial in the variables $\{ \sigma_{m,n} \}$, where the degree of each variable $\sigma_{m,n}$ cannot exceed 4 since it has 4 neighbors. So, each line in the diagonal is assigned a factor of $\tanh\theta$ and each end of the line a factor of $\sigma_{m,n}$.

\vspace{6mm}

\begin{figure}[htpb]
    \centering
    \begin{subfigure}[t]{0.46\textwidth}
        \centering
        \includegraphics[width=0.7\textwidth]{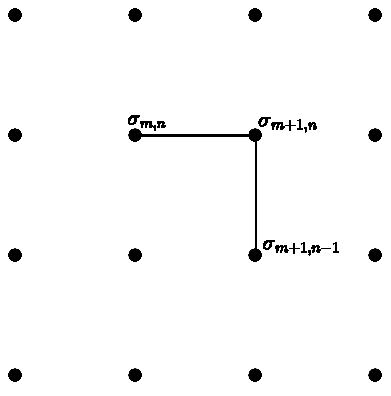}
        \caption{Diagram of the polynomial
        $\tanh^{2}\theta\sigma_{m,n}\sigma^{2}_{m+1,n}\sigma_{m+1,n-1}$.}
        \label{fig:diag2}
    \end{subfigure}
    \hfill
    \begin{subfigure}[t]{0.46\textwidth}
        \centering
        \includegraphics[width=0.7\textwidth]{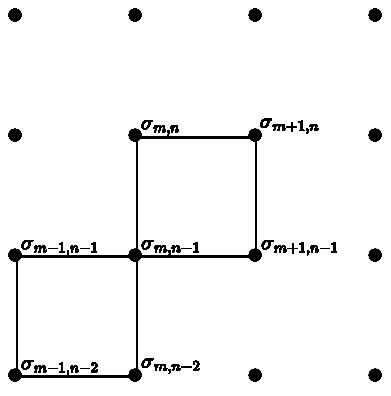}
        \caption{Diagram of the polynomial   \\
    $\tanh^{8}\theta\sigma^{2}_{m,n}\sigma^{2}_{m+1,n}\sigma^{2}_{m+1,n-1}\sigma^{4}_{m,n-1}\sigma^{2}_{m-1,n-1}\sigma^{2}_{m+1,n-2}\sigma^{2}_{m,n-2}$.}
        \label{fig:diag1}
    \end{subfigure}
    \caption{}
    \label{fig:sidebyside}
\end{figure}
\newpage
Non-zero contributions to the partition function come only from the terms in the polynomial which contain all the $\sigma_{m,n}$ in even powers, which signifies geometrically that either $2$ or $4$ bonds must end at each point in the diagram. \textit{Hence, the summation is taken only over closed diagrams and the partition function (\ref{eq:3}) can be written as}

\begin{equation}
    Z=(\cosh^{2}\theta)^{(M-1)(N-1)}2^{MN}\sum_{loops}\tanh^{\nu}\theta
\end{equation}
where $\nu$ is the total number of bonds.

\subsection{Theory of Gamma Matrices in Large Dimensions}

\subsubsection{Some calculations}

Consider the Clifford-Dirac algebra with \textit{MN} generators in the matrix representation.

\begin{equation}\label{eq:4}
    \begin{aligned}
        \gamma_x\gamma_y + \gamma_y\gamma_x = 2\delta_{xy}, \hspace{0.5cm} x, y = 1,..., MN
    \end{aligned}
\end{equation}
\\
where $\{ \gamma_x \}$ are Hermitian matrices of dimension $2^{MN/2}$.
\\
Euclidean anti-commutation relationship is given as
\begin{equation}\label{eq:5}
    \begin{aligned}
        \{\gamma_x, \gamma_y \} = 2\delta_{xy}I
    \end{aligned}
\end{equation}
and we can write
\begin{equation}\label{eq:6}
        \gamma_z\gamma_v = \gamma_z\gamma_v + \gamma_v\gamma_z - \gamma_v\gamma_z  = \{\gamma_z, \gamma_v \} - \gamma_v\gamma_z
\end{equation}\\

It follows directly from (\ref{eq:4}) that the trace of any product of an odd number of $\gamma$-matrices vanishes, and we have the following trace identities.\\

\begin{equation}\label{eq:7}
    \begin{aligned}
        Tr\gamma_x\gamma_y =& \frac{1}{2}(Tr\gamma_x\gamma_y + Tr\gamma_x\gamma_y) 
        = \frac{1}{2}Tr(\gamma_x\gamma_y + \gamma_x\gamma_y)\\ 
        =& \frac{1}{2}Tr \{\gamma_x, \gamma_y \} 
        = \frac{1}{2}2\delta_{xy}Tr(I) \\
        =& 2^{MN/2}\delta_{xy}
    \end{aligned}    
\end{equation}

Using (\ref{eq:5}) and (\ref{eq:6})
\begin{align*}
\centering
Tr\gamma_x\gamma_y\gamma_z\gamma_v =& Tr(\gamma_x\gamma_y(2\delta_{zv}I - \gamma_v\gamma_z)) \\
=& 2\delta_{zv}Tr(\gamma_x\gamma_y) - Tr(\gamma_x\gamma_y\gamma_v\gamma_z)
\\[5pt]
Tr(\gamma_x\gamma_y\gamma_v\gamma_z) =& Tr(\gamma_x(2\delta_{yv}I - \gamma_v\gamma_y)\gamma_z)\\
=& 2\delta_{yv}Tr\gamma_x\gamma_z - Tr(\gamma_x\gamma_v\gamma_y\gamma_z) \\[5pt]
Tr(\gamma_x\gamma_v\gamma_y\gamma_z) =& Tr((2\delta_{xv}-\gamma_v\gamma_x)\gamma_y\gamma_z) \\
=& 2\delta_{xv}Tr\gamma_y\gamma_z - Tr\gamma_v\gamma_x\gamma_y\gamma_z \\[5pt]
Tr\gamma_x\gamma_y\gamma_z\gamma_v =& 2\delta_{zv}Tr\gamma_x\gamma_y - 2\delta_{yv}Tr\gamma_x\gamma_z + 2\delta_{xv}Tr\gamma_y\gamma_z - Tr\gamma_v\gamma_x\gamma_y\gamma_z
\end{align*}
Using \eqref{eq:7} and the cyclic property of the trace, we get:
\begin{align}
    2Tr\gamma_x\gamma_y\gamma_z\gamma_v =& 2 (\delta_{zv}2^{MN/2}\delta_{xy} - \delta_{yv}2^{MN/2}\delta_{xz} + \delta_{xv}2^{MN/2}\delta_{yz}) \nonumber 
    \\ 
    Tr\gamma_x\gamma_y\gamma_z\gamma_v =& 2^{MN/2}(\delta_{zv}\delta_{xy} - \delta_{yv}\delta_{xz} + \delta_{xv}\delta_{yz})
\end{align} 

\subsubsection{Switching to Gamma Matrix Representation}

From now on, the indices x and y will be treated as integer vectors in the plane: x=(m, n); It is assumed that the index m increases from left to right, while
n increases vertically. Consequently, the
matrix $\gamma_x$ = $\gamma_{m,n}$ corresponds to each node (m, n). The lattice basis
vectors are $e_{1}$ = (1, 0) and $e_2$ = (0, 1), accordingly x=m$e_1$ + n$e_2$.

\hspace{0.3cm}

Now, under the transformations
\begin{gather}
    \sigma_{m,n} \xrightarrow{}\gamma_{m,n}, \hspace{0.3cm} \sum_{\sigma = \pm 1} \xrightarrow{}2^{MN/2} Tr
\end{gather}

the partition function (\ref{eq:3}) can be written as 

\begin{equation}
    \begin{gathered}
    \label{eq:8}
    Z = 2 ^ {MN/2}(cosh2\theta)^{(M-1)(N-1)}Tr\biggl\{\prod_{n}^{\xrightarrow{}}\prod_{m}^{\xrightarrow{}} U_{m,n} \biggl\}
    \\[5pt] 
    U_{m,n} = (\lambda + \mu\gamma_{m,n}\gamma_{m,n+1})(\lambda + \mu\gamma_{m,n}\gamma_{m+1,n})
    \end{gathered}
\end{equation}
\vspace{5mm}
where $\lambda$ = cos$\dfrac{\psi}{2}$ = $\dfrac{cosh\theta}{\sqrt{cosh2\theta}}$ and
$\mu$ = sin$\dfrac{\psi}{2}$ = $\dfrac{sinh\theta}{\sqrt{cosh2\theta}}$.

The symbol $\prod\limits_{m}^{\xrightarrow{}}$ denotes the product of all matrices $U_{m,n}$ while n is held constant
\begin{gather}
    \mathscr{U}^{(n)} \equiv \prod_{m}^{\xrightarrow{}}U_{m,n} \equiv \ldots U_{m-1,n}U_{m,n}U_{m+1,n}\ldots
\end{gather}
Similarly,
\begin{equation}
    \begin{aligned}
        \label{eq:9}
        \mathscr{U} \equiv \prod_{n}^{\xrightarrow{}}\mathscr{U}^{(n)} \equiv \ldots \mathscr{U}^{(n -1)}\mathscr{U}^{(n)}\mathscr{U}^{(n+1)}\ldots
    \end{aligned}
\end{equation}

\hspace{0.4cm}

To verify (\ref{eq:8}) we simply put in the values 
inside the product which becomes $cosh2\theta$
\begin{align}
     Z =& 2 ^ {MN/2}Tr\biggl\{\prod_{n}^{\xrightarrow{}}\prod_{m}^{\xrightarrow{}}(\cosh2\theta)\left(\frac{\cosh\theta}{\sqrt{\cosh2\theta}} + \frac{\sinh\theta}{\sqrt{\cosh2\theta}}\gamma_x\gamma_{x+e_2}\right)\left(\frac{\cosh\theta}{\sqrt{\cosh2\theta}} + \frac{\sinh\theta}{\sqrt{\cosh2\theta}}\gamma_x\gamma_{x+e_1}\right)\biggl\} \nonumber \\
     =& 2 ^ {MN/2}(cosh2\theta)^{(M-1)(N-1)} \nonumber\\
     \times&
Tr\biggl\{\prod_{n}^{\xrightarrow{}} \prod_{m}^{\xrightarrow{}} \left(\frac{\cosh\theta}{\sqrt{\cosh2\theta}} + \frac{\sinh\theta}{\sqrt{\cosh2\theta}}\gamma_{m,n}\gamma_{m,n+1}\right) \left(\frac{\cosh\theta}{\sqrt{\cosh2\theta}} + \frac{\sinh\theta}{\sqrt{\cosh2\theta}}\gamma_{m,n}\gamma_{m+1,n} \right) \biggl\}
\end{align}

Observe that the matrix factor corresponding to the
unit loop of the unit cell whose vertices
are
\begin{gather}
    \pmb{x}, \; \pmb{x+e_1},  \; \pmb{x+e_2}, \; \pmb{x+e_1+e_2} \nonumber
\end{gather}
is unity. If we write gamma matrices as 
\begin{equation}
    \gamma_x = \begin{pmatrix}
    0 & \sigma_x \\
    -\sigma_x & 0
    \end{pmatrix} \nonumber
\end{equation}
using $\sigma_x^2 = 1$, the unit loop is calculated as 

\begin{gather}
    \left(\gamma_x \gamma_{x+e_2}\right)\left(\gamma_x\gamma_{x+e_1} \right)  
\left(\gamma_{x+e_1}\gamma_{x+e_1+e_2}\right)\left(\gamma_{x+e_2}\gamma_{x+e_1+e_2}\right) =   
     \begin{pmatrix}
         \sigma^2_x\sigma^2_{x+e_1}\sigma^2_{x+e_2}\sigma^2_{x+e_1+e_2} & 0
         \\ 0 & \sigma^2_x\sigma^2_{x+e_1}\sigma^2_{x+e_2}\sigma^2_{x+e_1+e_2}
         \end{pmatrix} = 1
\end{gather}

\begin{figure}[htpb]
    \centering
    \includegraphics[width=0.9\textwidth]{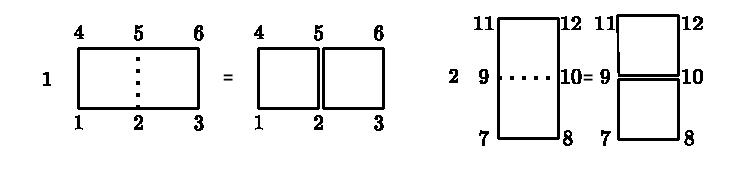}
    \caption{Loop examples 1 and 2}
    \label{fig:diag3}
\end{figure}

Figure \ref{fig:diag3} effectively demonstrates the structure by using solid lines to depict loops, while broken lines serve to separate the interiors of these loops into discrete unit cells. The matrix factors are systematically calculated based on the product ordering convention defined in (\ref{eq:9}).

\vspace{7mm}

For example, equalities 1 and 2 in Figure \ref{fig:diag3} are
\begin{align}
        &\left[(\gamma_1\gamma_4)(\gamma_1\gamma_2)(\gamma_2\gamma_3)(\gamma_3\gamma_6)(\gamma_4\gamma_5)(\gamma_5\gamma_6)\right] \nonumber \\
        &= \left[(\gamma_1\gamma_4)(\gamma_1\gamma_2)(\gamma_2\gamma_5)(\gamma_4\gamma_5)\right] \nonumber \\ &\times\left[(\gamma_2\gamma_5)(\gamma_2\gamma_3)(\gamma_3\gamma_6)(\gamma_5\gamma_6)\right] \nonumber 
\end{align}

\begin{align}
        &\left[(\gamma_7\gamma_9)(\gamma_7\gamma_8)(\gamma_8\gamma_{10})(\gamma_9\gamma_{11})(\gamma_{10}\gamma_{12})(\gamma_{11}\gamma_{12})\right] \nonumber \\
        &= \left[(\gamma_7\gamma_9)(\gamma_7\gamma_8)
        (\gamma_8\gamma_{10})(\gamma_9\gamma_{10})\right] \nonumber \\ &\times\left[(\gamma_9\gamma_{11})(\gamma_9\gamma_{10})(\gamma_{10}\gamma_{12})(\gamma_{11}\gamma_{12})\right] \nonumber 
\end{align}
Note that each factor $\lambda + \mu\gamma_x\gamma_{x+e_1}$ in (\ref{eq:8}) is the orthogonal matrix of a rotation in the MN-dimensional Euclidean space by an angle of $\psi$ in the plane represented by a pair of $\gamma$-matrices in the spinor representation. Therefore the product in (\ref{eq:9}) is the orthogonal matrix of a rotation in the MN-dimensional Euclidean space in the spinor representation.
\subsubsection{Rotation Operators}
A finite rotation $\mathscr{O}$ can be constructed starting from an infinitesimal rotation  $\bm\omega$  \cite{Dirac}:
\begin{gather}
    e^{\bm\omega} = \mathscr{O}
\end{gather}

If $\bm\omega$ is antisymmetric, $\mathscr{O}$ must be orthogonal, since:
\begin{gather*}
    \mathscr{O}^T = (e^{\bm{\omega}})^T = e^{{\bm{\omega}}^T} = e^{-\bm{\omega}} = \mathscr{O}^{-1}
\end{gather*}

Let $u_\lambda$, $v_\mu$ be eigenvectors of $\mathscr{O}$ with the eigenvalues $\lambda$, $\mu$ respectively:
\begin{gather*}
    \mathscr{O}u_\lambda = \lambda u_\lambda \\
    \mathscr{O}v_\mu = \mu v_\mu
\end{gather*}
Now, doing some calculations we can see that
\begin{gather*}
    v_\mu ^T \mathscr{O}^T = \mu v_\mu ^T \\
    v_\mu ^T u_\lambda = v_\mu ^T \mathscr{O}^T \mathscr{O} u_\lambda = \mu v_\mu ^T \lambda u_\lambda \\
    v_\mu ^T u_\lambda = \mu \lambda v_\mu ^T u_\lambda \xrightarrow{} (1-\lambda\mu)v_\mu ^T u_\lambda = 0
\end{gather*}
if $\lambda$ is an eigenvalue, then $\mu$ = $\lambda^{-1}$ must be another.

We begin by examining the infinitesimal case, where the rotation matrices can be expressed as  $R=1+\epsilon A$, with  $\mathcal{A}$  being an anti-symmetric matrix. Then the operator  $\mathcal{A}$  is defined as follows:
\begin{equation}
    \mathcal{A}=\dfrac{1}{4}\gamma_{x}A_{xy}\gamma _{y}
\end{equation}
Now with the gamma matrices, we have the following identity 
\begin{equation}\label{eqn:9}
    \begin{aligned}\mathcal{A}\gamma _{z}-\gamma _{z}\mathcal{A}=&\dfrac{1}{4}A_{xy}\left( \gamma _{x}\gamma _{y}\gamma _{z}-\gamma _{z}\gamma _{y}\gamma _{x}\right) \\
=&\dfrac{1}{4}A_{xy}\left\{ \gamma _{x}\left( \gamma _{y}\gamma _{z}+\gamma _{z}\gamma _{y}\right) -\left( \gamma _{z}\gamma _{x}+\gamma _{x}\gamma _{z}\right) \gamma _{y}\right\} \\
=&\dfrac{1}{4}A_{xy}\left\{ \gamma _{x}2\delta _{yz}-2\delta _{zx}\gamma _{y}\right\} \\
=&\dfrac{1}{2}\left( A_{xz}\gamma _{x}-A_{zy}\gamma _{y}\right) \\
=&\gamma _{x}A_{xz}\end{aligned}
\end{equation}
Therefore, to the first order in $\varepsilon$, the following is satisfied
\begin{equation}
    e^{\varepsilon \mathcal{A}}\gamma _{z}e^{-\varepsilon \mathcal{A}}=\gamma _{x}\left( 1+\varepsilon A\right) _{xz}
    \nonumber
\end{equation}
To build the finite case, assume that
\begin{equation}\label{eqn:10}
    e^{\theta \mathcal{A}}\gamma _{z}e^{-\theta \mathcal{A}}=\gamma _{x}\left( e^{\theta A}\right) _{xz}=\left( \gamma e^{\theta A}\right) _{z}
\end{equation}
holds for some particular value of $\theta$. Then one can show that it holds for a neighboring $\theta$ by differentiating both sides of the equality with respect to $\theta$. The derivatives of both the LHS and the RHS  
\begin{equation}
    e^{\theta \mathcal{A}}\left( \mathcal{A}\gamma _{z}-\gamma _{z}\mathcal{A}\right) e^{-\theta \mathcal{A}},\hspace{10mm} \left( \gamma e^{\theta A}A\right) _{z}=\left( e^{\theta A}\gamma \right) _{x}A_{xz} 
\end{equation}
Now applying equalities \ref{eqn:9} and \ref{eqn:10} in turn to the LHS with the original value for $\theta$, we see that the equality for the derivatives also holds. Taking $\theta=1$ yields
\begin{equation}
    e^{\mathcal{A}}\gamma _{z}e^{-\mathcal{A}}=\left( \gamma e^{A}\right) _{z}
\end{equation}
In our case, $\exp\mathcal{A}=\mathscr{U}$ and the skew matrix $A=\boldsymbol\omega$. 

\begin{gather}\label{eqn:11}
    \mathscr{U} = \exp{\left(\frac{1}{4}\boldsymbol\omega_{x,y}\gamma_x\gamma_y\right)}, \hspace{10mm} \boldsymbol\omega_{x,y} = -\boldsymbol\omega_{y,x}
\end{gather}

and,
\begin{gather}\label{eqn:18}
    \mathscr{U}^\dag \gamma_x \mathscr{U} = \mathscr{O}_{x,y}\gamma_y, 
\end{gather}
\begin{gather} \label{eqn:12}
    \mathscr{O}_{x,y} = (e^{\boldsymbol\omega})_{x,y} = \delta_{x,y} + \boldsymbol\omega_{x,y} + \frac{1}{2!}\boldsymbol\omega_{x,z}\boldsymbol\omega_{z,y} + ...
\end{gather}

\subsubsection{Trace of \texorpdfstring{$\mathscr{U}$}{} in terms of the eigenvalues of \texorpdfstring{$\mathscr{O}_{x,y}$}{}}

Let $\left\{\nu_{x}^{(k)}, \overline{\nu_{x}^{(k)}}\right\}$ be the complete orthonormal set of eigenvectors of the rotation matrix $\mathscr{O}_{x,y}$ and let $\rho_k$ and $\overline{\rho_k}$ be their respective eigenvalues, where $k = 1,..., MN/2$. Based on these definitions, the following holds

\begin{gather}
    \nu^{(k)T} \nu^{(k')} = 0, \hspace{0.5cm} \nu^{(k)\dag} \nu^{(k')} = \delta_{k k'}
\end{gather}
Redefining the set of eigenvectors as
\begin{gather}
    \left\{\nu_{x}^{(k)}, \overline{\nu_{x}^{(k)}}\right\} \equiv \left\{\nu_x^a\right\}, \hspace{0.5cm} a = 1, ..., MN
\end{gather}
Then we can diagonalize $\mathscr{O}$
\begin{gather} \label{eqn:14}
      U_{xa} \equiv \nu^a_x, \hspace{0.5cm} U \equiv \left( \nu^{(1)}, \overline{\nu^{(1)}}, ...\right) 
\end{gather}

\begin{gather}
    \left(U ^{\dag} \mathscr{O} U\right)_{ab} = diag\left(\rho_1, \overline{\rho_1},...\right) \equiv D_{ab}
\end{gather}

In \cite{Dirac} it was shown that
\begin{gather} \label{eqn:13}
    \left(U ^{\dag} \boldsymbol\omega U\right)_{ab} =
    diag\left(\ln\rho_1, - \ln\rho_1, ...\right) \equiv \Delta_{ab}
\end{gather}
Using \ref{eqn:14} and \ref{eqn:13}, we obtain
\begin{gather}\label{eqn:16}
    \frac{1}{4}\gamma_x\boldsymbol\omega_{xy}\gamma_y = \frac{1}{4}\gamma_xU_{xa}\Delta_{ab}\left(U^{\dag}_{by}\gamma_y\right)
\end{gather}
Now if we define $2^{MN/2} -by- 2^{MN/2}$ matrices
\begin{gather} \label{eqn:15}
    c_k^\dag = \gamma_x v_x^{(k)}, c_k = \gamma_x \Bar{v_x^k}
\end{gather}
with the properties of $\nu^{(k)}$s and the gamma matrices, it is easy to see that they carry all the properties of fermionic creation/annihilation operators

\begin{gather}
    [c_k, c_{k'}^\dag] = \delta_{kk'}, \hspace{0.3cm} [c_k, c_{k'}]_+ = [c_k^\dag, c_{k'}^\dag]_+ = 0
\end{gather}
It follows from \ref{eqn:14} and \ref{eqn:15} that:
\begin{align}
    \gamma_x U_{xa} = (c_1^\dag, c_1, ..., c_{MN/2}^\dag,c_{MN/2}) 
\end{align}
Now we can rewrite \ref{eqn:16}
\begin{gather}
     \frac{1}{4}\gamma_x\boldsymbol\omega_{xy}\gamma_y = \frac{1}{2} \sum_{k=1}^{MN/2} [\ln \rho_{k}(c_k^\dag c_k - c_k c_k^\dag)] \\
     = 
    \frac{1}{2} \sum_{k=1}^{MN/2} [\ln \rho_{k}(c_k^\dag c_k + c_k^\dag c_k - c_k c_k^\dag - c_k^\dag c_k)] \nonumber \\
    = 
    \frac{1}{2} \sum_{k=1}^{MN/2} [\ln \rho_{k}(2c_k^\dag c_k -[c_k, c_{k}^\dag]] \nonumber \\
    = \sum_{k=1}^{MN/2} [\ln \rho_{k}c_k^\dag c_k - \frac{1}{2}\ln \rho_{k}]
\end{gather}

And,
\begin{equation}
    \begin{aligned}
    Tr\mathscr{U}=&Tr\exp \left( \sum ^{MN/2}_{k=1}\left[ \ln \rho_{k}c_{k}^{\dag}c_{k}-\dfrac{1}{2}\ln \rho_{k}\right] \right) \end{aligned}
\end{equation}
Then we first put the sum in the power of the exponential as a product in front then expand the matrices in the power and do some straightforward calculations
\begin{equation}
    \begin{aligned}
        Tr\mathscr{U}=&Tr\left(\prod^{MN/2}_{k=1}\exp \left[\ln \rho_{k}c_{k}^{\dag}c_{k}-\dfrac{1}{2}\ln \rho_{k}\right] \right) \\
        =& Tr\left(\prod^{MN/2}_{k=1}\sum_n \frac{1}{n!}\left[(\ln \rho_{k}c_{k}^{\dag}c_{k}-\dfrac{1}{2}\ln \rho_{k}\right]^n\right)
        \end{aligned}
\end{equation}
First, we notice that $\left(c_{k}^{\dag}c_{k}\right)^2 = \left(I - c_{k}c_{k}^{\dag}\right)\left(c_{k}^{\dag}c_{k}\right)= c_{k}^{\dag}c_{k}$ thus for any integer n, $\left(c_{k}^{\dag}c_{k}\right)^n = c_{k}^{\dag}c_{k}$. Then:
\begin{equation}
    \begin{aligned}
       Tr\mathscr{U}&=Tr\left(\prod^{MN/2}_{k=1}\sum_n\frac{1}{n!}\ln \rho_{k}^n\left(c_{k}^{\dag}c_{k}-\dfrac{1}{2}\right)^n \right) \\
       &=Tr\left(\prod^{MN/2}_{k=1}\sum_n\frac{1}{n!}\ln \rho_{k}^n \left(c_{k}^{\dag}c_{k}\sum_m^{n-1}\binom{n}{m}\left(\frac{-1}{2}\right)^m\right) + \frac{1}{2^n}\right)
        \end{aligned} 
\end{equation}

\vspace{3mm}
Using Mathematica, we confirmed our calculations after expanding to the 7th power. The results indicate that for even $n$, the sum involving $c_{k}^{\dag}c_{k}$ evaluates to zero. For odd $n$, the entire bracket simplifies to $\left(c_{k}^{\dag}c_{k} - \frac{1}{2}\right)\left(\frac{1}{2}\right)^{n-1}$. Using the properties of the creation and annihilation operators, along with their commutative behavior under the trace operation, we deduce that this expression also evaluates to zero. Consequently, only the even terms remain
\begin{equation}\label{2.25}
    \begin{aligned}
        &=Tr\left(\prod^{MN/2}_{k=1}\sum_n\left(\frac{\ln \rho_{k}}{2}\right)^{2n}\frac{1}{2n!}\right) \\
        &=Tr\left(\prod^{MN/2}_{k=1}\cosh{\frac{\ln\rho_k}{2}}\right) = \prod^{MN/2}_{k=1}2\cosh{\frac{\ln\rho_k}{2}} \\
        &= \prod^{MN/2}_{k=1}\left(\sqrt{\rho_k} + \sqrt{\Bar{\rho_k}}\right)
    \end{aligned}
\end{equation}
We will use the last equality for calculating the partition function.

\subsection{Calculating The Eigenvalues}

\begin{figure}[htpb]
    \centering
    \includegraphics[width=0.7\textwidth]{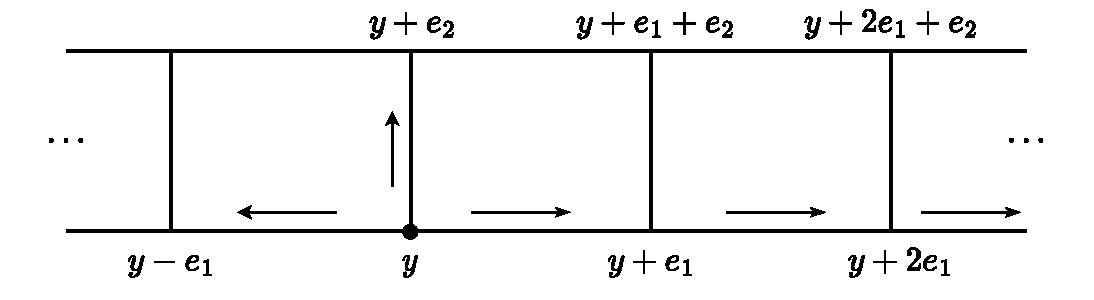}\\
    \caption{The dot represents the initial location of a $\gamma$-matrix. Arrows indicate the directions of displacements of the $\gamma$-matrix under linear transformations}
\end{figure}

Since $(\gamma_x \gamma_{x+e})^{\dag}$ = $\gamma_{x+e} \gamma_x$:
\begin{gather} 
    (\lambda + \mu \gamma_x \gamma_{x+e})^{\dag} \gamma_x (\lambda + \mu \gamma_x \gamma_{x+e}) = \lambda^2\gamma_x + 2\lambda\mu \gamma_{x+e} + \mu^2\gamma_{x+e}\gamma_x \gamma_{x+e} \nonumber 
    \\
    = \lambda^2\gamma_x + 2\lambda\mu \gamma_{x+e} - \mu^2 \gamma_x = (cos\psi) \gamma_x + (sin\psi)\gamma_{x+e}
\end{gather}

\begin{align}\label{8.42}
    (\lambda + \mu \gamma_x \gamma_{x+e})^{\dag} \gamma_{x+e} (\lambda + \mu \gamma_x \gamma_{x+e}) &= \lambda^2\gamma_x + 2\lambda\mu \gamma_{x+e} + \mu^2\gamma_{x+e}\gamma_x \gamma_{x+e} \nonumber 
    \\
    = \lambda^2\gamma_{x+e} - 2\lambda\mu \gamma_x - \mu^2 \gamma_{x+e} &= (cos\psi) \gamma_{x+e} - (sin\psi)\gamma_{x}
\end{align}

\begin{gather} \label{eqn:17}
    (\lambda + \mu \gamma_x \gamma_{x+e})^{\dag} \gamma_{z} (\lambda + \mu \gamma_x \gamma_{x+e}) =\lambda^2 \gamma_{z} + \lambda \mu \gamma_z\gamma_x\gamma_{x+e} + \lambda\mu\gamma_{x+e}\gamma_x\gamma_z + \mu^2 \gamma_{x+e}\gamma_x \gamma_z \gamma_{x+e} \nonumber \\
    = (\lambda^2 + \mu^2) \gamma_z \nonumber 
    \\
    z \neq x,x+e
\end{gather}

Here $e$ is either $e_1$ or $e_2$.\\
First, let y = (m, n). In this case, all spin rotation matrices in $\mathscr{U}^{(n)\dag}$ placed on the right of $$(\lambda + \mu \gamma_x\gamma_{x+e})^{\dag}$$ and  all spinor rotation matrices $\mathscr{U}^{(n)}$ on the left of $(\lambda + \mu \gamma_x \gamma_{x+e})$ cancel out because they commute with the matrix $\gamma_{y}$. The matrix factors transform $\gamma_{y}$ according to \ref{8.42} with $e = e_1$ and $x +
e = y$. This yields the matrix element
\begin{gather}
\mathscr{O}_{y,y-e_1}^{(n)}=-\sin{\psi}, \\ y=(m,n)\nonumber.
\end{gather}

First, let e = $e_2$, followed by e = $e_1$, for each $m'$. This procedure is illustrated above, where the dot denotes the initial position of the $\gamma$-matrix on the left-hand side of \ref{eqn:18}, and the arrows represent the directions of displacement resulting from the described linear transformations. Each transformation introduces a factor of $\sin{\psi}$ when the matrix is shifted upward or to the right, or $\cos{\psi}$ when the matrix remains stationary. Once the matrix is shifted upward, it becomes invariant to any subsequent transformations. The final result is the following

\begin{gather}
\mathscr{O}_{y,y+m'e_1}^{(n)}=[\sin{\psi}\cos{\psi}]^{m'}(\cos{\psi})^3  \\
\mathscr{O}_{y,y+m'e_1+e_2}^{(n)}=[\sin{\psi}\cos{\psi}]^{m'+1} \nonumber
\end{gather}

For instance, when transitioning from $y$ to $y+m'e_1$, the first step contributes a factor of $\cos^2\psi$ for not moving left or upward, and a factor of $\sin\psi$ for moving to the right. Subsequently, during each step from $y + e_1$ to $y + m'e_1$, a factor of $(\sin\psi \cos\psi)^{m'-1}$ arises due to not moving upward while shifting right $m' - 1$ times. Finally, when you reach the point $y + m'e_1$, a factor of $\cos^2\psi$  is introduced for not moving further upward or to the right. As a result, the overall expression simplifies to $[\sin{\psi}\cos{\psi}]^{m'}(\cos{\psi})^3$.

\begin{figure}[htpb]
    \centering
    \includegraphics[width=0.5\textwidth]{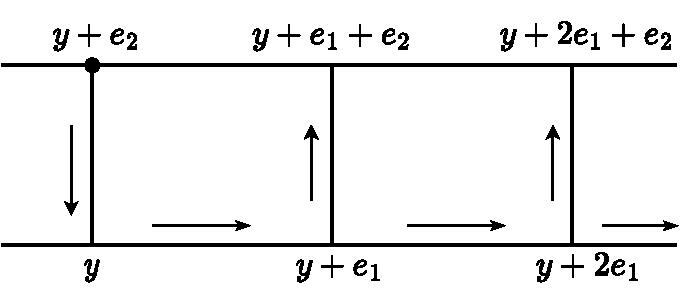}\\
    \caption{The dot represents the initial location of a $\gamma$-matrix. Arrows indicate the directions of displacements of the $\gamma$-matrix under linear transformations.}
    \label{fig62}
\end{figure}
 
The matrix elements of the form $\mathscr{O}_{y+e_2, z}^{(n)}$ are calculated as shown in Figure \ref{fig62}:
\begin{gather}
    \mathscr{O}_{y+e_2, y+e_2}^{(n)} = cos\psi, \\
    \mathscr{O}_{y+e_2, y+(m'+1)e_1 + e_2}^{(n)} = -sin^3\psi[sin\psi cos\psi]^{m'}, \\
    \mathscr{O}_{y+e_2, y+m'e_1}^{(n)} = -[sin\psi cos\psi]^{m'+1}
\end{gather}

For example, as shown in the figure, when moving from $y +e_2$ to $y+ (m' +1)e_1 + e_2$ , the first step introduces a factor of $-\sin\psi$ for moving downward, followed by a factor of $\sin\psi$ for shifting to the right. Subsequently, during each step from $y + e_1$ to $y + (m{\prime} + 1)e_1$, a factor of $(\sin\psi \cos\psi)^{m'}$ is contributed as a result of not moving upward while shifting to the right $m'$ times. Finally, an additional factor of $\sin\psi$ arises from moving up.

\vspace{0.5cm}

\subsubsection{Diagonalization of \texorpdfstring{$\mathscr{O}_{x,y}$}{}}
In the statistical limit, as both M and N approach infinity, the matrices $\mathscr{O}_{x, x+z}$ and $w_{x,x+z}$ only depend on the value of z when we are relatively close to the lattice's boundary. This characteristic is referred to as translational invariance, and it allows us to diagonalize these matrices by changing to a quasi-momentum representation.

For this purpose, following \cite{Shankar}, we will use the orthonormal set of functions on the lattice given below:

\begin{gather}
    |p\rangle \equiv \psi_p(m) = \frac{1}{\sqrt{M}}e^{ipm}, \nonumber \\
    \hspace{0.2cm}
    |q\rangle \equiv \psi_q(n) = \frac{1}{\sqrt{N}}e^{iqn}, \nonumber \\
    \hspace{0.2cm}
    p = -\frac{\pi(M-2)}{M}, -\frac{\pi(M-4)}{M}, ..., 0, ..., \frac{2\pi}{M}, \pi, \nonumber \\
    \hspace{0.2cm}
    q = -\frac{\pi(N-2)}{N}, -\frac{\pi(N-4)}{N}, ..., 0, ..., \frac{2\pi}{N}, \pi, \nonumber \\
    \hspace{0.2cm}
    |k\rangle \equiv \psi_k(x) = \psi_p(m) \psi_q(n), k = (p,q), \nonumber \\
    \hspace{0.2cm}
    \sum_x \Bar{\psi}_k(x)\psi_{k'}(x) = \delta_{kk'} \longleftrightarrow
     \sum_k \psi_k(x)\Bar{\psi}_{k}(x') = \delta_{xx'}\label{3.9} 
\end{gather}

Now, the orthogonal matrix undergoes partial diagonalization by applying Fourier transforms to the elements of each row:
\begin{gather}
    \gamma_n(p) = \sum_m\Bar{\psi}_m(p)\gamma_{m,n} = \psi^{\dag}_n(-p) \\
    [\gamma_n(p), \gamma^{\dag}_{n'}(p')] = 2\delta_{n,n'}\delta_{p,p'}
\end{gather}

Due to translational invariance, \ref{eqn:18} can be expressed using the Fourier-transformed components of the rows as follows:

\begin{gather}
    \mathscr{U}^{(n)\dag} \gamma_{n'}(p) \mathscr{U}^{(n)} = \Biggr[\sum_{m'}  \mathscr{O}^{(n)}_{m,n';m+m',n''}e^{ipm'} \Biggr]\gamma_{n''}(p)
\label{Opn}
\end{gather}
 where:
\begin{gather}
    \mathscr{O}^{(n)}_{p;n',n''} = \begin{pmatrix}
        . & . & . & . & . &  . & .    \\
        . & 1 & 0 & 0 & 0 & 0 & .       \\
        . & 0 & 1 & 0 & 0 & 0 & .   \\
        . & 0 & 0 & a_p & b_p & 0 & .   \\
        . & 0 & 0 & -b_p & c_p & 0 & .  \\
        . & 0 & 0 & 0 & 0 & 1 & .       \\
        . & . & . & . & . & . & . & 
    \end{pmatrix}
\end{gather}
where: 
\begin{equation}
    \begin{aligned}a_{p}=\dfrac{\cos \psi -e^{-ip}\sin \psi }{1-e^{ip}\sin \psi {\cos }\psi }\\
b_{p}=\dfrac{\sin \psi \cos \psi }{1-e^{ip}\sin \psi {\cos }\psi }\\
c_{p}=\dfrac{\cos \psi -e^{+ip}\sin \psi }{1-e^{ip}\sin \psi {\cos }\psi }\end{aligned}
\label{apcp}
\end{equation}

\vspace{0.5cm}
Here $a_p$ and $c_p$ are the n-th and (n + 1)-th diagonal elements, respectively. The common term in the denominator is the normalization factor that is multiplied to ensure that $\mathscr{O}$ satisfies the condition of being a unitary matrix. 

\begin{equation*}
    \begin{aligned}
        a_{p}\overline{a}_{p}+b_{p}\overline{b_{p}}
        =&\dfrac{\left( \cos \psi -e^{-ip}{\sin }\psi \right) \left( \cos \psi -e^{ip}\sin \psi \right) +\left( \sin \psi {\cos }\psi \right) ^{2}}{\left( 1-e^{ip}{\sin }\psi{\cos }\psi \right) \left( 1-e^{-ip}{\sin }\psi {\cos }\psi \right)} \\    =&\dfrac{\cos ^{2}\psi +\sin ^{2}\psi -\left( e^{ip}+e^{-ip}\right) \sin \psi {\cos }\psi +\sin ^{2}\psi \cos ^{2}\psi }{1-\left( e^{ip}+e^{-ip}\right) \sin \psi {\cos }\psi +\sin ^{2}\psi \cos ^{2}\psi }\\
        =&1 \\
        c_{p}\overline{c}_{p}+b_{p}\overline{b_{p}}
        =&\dfrac{\left( \cos \psi -e^{-ip}{\sin }\psi \right) \left( \cos \psi -e^{ip}\sin \psi \right) +\left( \sin \psi {\cos }\psi \right) ^{2}}{\left( 1-e^{ip}{\sin }\psi{\cos }\psi \right) \left( 1-e^{-ip}{\sin }\psi {\cos }\psi \right)} \\
         =&\dfrac{\cos ^{2}\psi +\sin ^{2}\psi -\left( e^{ip}+e^{-ip}\right) \sin \psi {\cos }\psi +\sin ^{2}\psi \cos ^{2}\psi }{1-\left( e^{ip}+e^{-ip}\right) \sin \psi {\cos }\psi +\sin ^{2}\psi \cos ^{2}\psi }\\
          =&1 \\
          -\overline{a}_{p}b_{p}+c_{p}\overline{b}_{p}=&-\overline{c}_{p}b_{p}+a_{p}\overline{b}_{p}=0
    \end{aligned}
\end{equation*}
Thus, 
\begin{equation}
    \mathscr{O}_{p;n',n"}\equiv \left[ \sum _{m'}\mathscr{O}_{m,n';m+m',n"}\Psi _{p}\left( m'\right) \right] =\left\{ \ldots \mathscr{O}_{p}^{\left( n-1\right) }\mathscr{O}_{p}^{\left( n\right) }\mathscr{O}_{p}^{\left( n+1\right) }\ldots \right\} _{n',n''}
\end{equation}
And it is calculated easily as
\begin{gather}
     \mathscr{O}_{p;n,n + n'} =
    \begin{cases}
        0 &: n' < -1 \\
        -b_p &: n' = -1 \\
        c_pb_p^{n'}a_p &: n' \ge 0
    \end{cases}
\end{gather}

\begin{align}\label{3.17}
    \rho_{p,q}&= \mathscr{O}_{p,q}=\sum_{n'}\mathscr{O}_{p;n,n+n'}e^{ipm'} \nonumber
    \\
    &=\dfrac{a_{p}c_{p}+b_{p}^{2}-e^{-iq}b_{p}}{1-e^{iq}b_{p}}=\dfrac{\eta _{p,q}}{\overline{\eta}_{p,q}}
    \\
    \eta _{p,q}&=1-\left( e^{-iq}+e^{-ip}\right) \left( \sin \psi {\cos}\psi\right) \nonumber
\end{align}

\subsection{Partition Function}
It follows from \ref{2.25},\ref{3.9} and \ref{3.17} that the Helmholtz free energy is proportional to
\begin{equation}\label{4.1}
    \begin{aligned}
        \ln{\prod_{p,q}\left(\sqrt{\rho_k} + \sqrt{\Bar{\rho_k}}\right)} = \sum_{p,q}\ln{\left(\sqrt{\rho_k} + \sqrt{\Bar{\rho_k}}\right)} \\
        = \frac{MN}{4\pi^2}\int\limits^{\pi}_{-\pi}dp\int\limits^{\pi}_{-\pi}dq\ln{\left(\sqrt{\rho_k} + \sqrt{\Bar{\rho_k}}\right)} \\
        \frac{MN}{4\pi^2}\int\limits^{\pi}_{-\pi}dp\int\limits^{\pi}_{-\pi}dq \ln{\left(\frac{\eta_{p,q} + \Bar{\eta}_{p,q}}{\sqrt{\eta_{p,q} \Bar{\eta}_{p,q}}}\right)} \\
        = \frac{MN}{4\pi^2}\int\limits^{\pi}_{-\pi}dp\int\limits^{\pi}_{-\pi}dq \ln{\left(\eta_{p,q} + \Bar{\eta}_{p,q}\right)}
    \end{aligned}       
\end{equation}
Where the last equality comes from
\begin{equation}
    \begin{aligned}
        \eta_{p,q} &= 1 - \left(e^{-ip} + e^{-iq}\right)\sin{\psi}\cos{\psi} \\
        &=\left(1 - \frac{1}{2}\sin{2\psi}e^{-ip}\right)\left(1 - \frac{sin{2\psi}e^{-iq}}{2-sin{2\psi}e^{-ip}}\right) \\
        &= \left(1-\alpha e^{-ip}\right)\left(1 - \beta_p e^{-iq}\right) \\
        &\alpha=\frac{1}{2}\sin{2\psi} \hspace{0.3cm} , \hspace{0.3cm} \beta_p = \frac{sin{2\psi}e^{-iq}}{2-sin{2\psi}e^{-ip}}
    \end{aligned}
\end{equation}
Then we calculate the integral of $\ln{\eta_{p,q}}$ in terms of $\alpha$ and $\beta_p$ and expand them respectively
\begin{equation}
    \begin{aligned}
       \int\limits^{\pi}_{-\pi}dp\int\limits^{\pi}_{-\pi}dq\ln{\eta_{p,q}} = \int\limits^{\pi}_{-\pi}dp\ln{\left(1-\alpha e^{ip}\right)} + \int\limits^{\pi}_{-\pi}dp\int\limits^{\pi}_{-\pi}dq\ln{\left(1 - \beta_p e^{-iq}\right)} \\
       = -\sum^{\infty}_{n=1}\frac{1}{n}\left\{\alpha^n\int\limits^{\pi}_{-\pi}dp\int\limits^{\pi}_{-\pi}dqe^{-inp}+\int\limits^{\pi}_{-\pi}dp\beta^{n}_{p}\int\limits^{\pi}_{-\pi}dqe^{-inq}\right\} = 0
    \end{aligned}
\end{equation}
The integral of the conjugate is verified analogously. \\
Finally, an expression for the free energy is obtained by combining \ref{4.1}
and putting $2^{MN/2} \cosh^{(M-1)(N-1)}$ in the product for large MN
\begin{equation}
    \begin{aligned}
        F &= -T\ln Z = -T\ln{2^{MN/2}cosh^{(M-1)(N-1)}2\theta \prod^{MN/2}_{k=1}\left(\sqrt{\rho_k} + \sqrt{\Bar{\rho_k}}\right)} \\
        &= -T\ln\prod^{MN/2}_{k=1}2cosh^22\theta\left(\sqrt{\rho_k} + \sqrt{\Bar{\rho_k}}\right) = - \frac{T}{2} \ln\prod_{p,q}2cosh^22\theta\left(\sqrt{\rho_k} + \sqrt{\Bar{\rho_k}}\right)
    \end{aligned}
\end{equation}
Then we just evaluate with respect to \ref{4.1} and \ref{3.17}
\begin{equation}
    \begin{aligned}
        F &= -MNT \int\limits^{\pi}_{-\pi}dp\int\limits^{\pi}_{-\pi}dq\frac{1}{8\pi^2}\ln\left[2cosh^22\theta\left(2-4\sin\psi \cos\psi(\cos p + \cos q)\right)\right] \\
        &= -MNT \left\{ \ln2 + \frac{1}{8\pi^2}\int\limits^{\pi}_{-\pi}dp\int\limits^{\pi}_{-\pi}dq \ln\left[\cosh^22\theta - \sinh2\theta(\cos p + cos q)\right]\right\}
    \end{aligned}
\end{equation}
Where we used $\sin2\psi = {\sinh2\theta}/{\cosh^22\theta}$. Then we write all the terms with $\theta$ with $\tanh\theta$ and get the well-known solution

\begin{equation*}
        F = -MNT \left\{ \ln 2 -\ln (1-x^2) + \frac{1}{8\pi^2} \int\limits^{\pi}_{-\pi}dp\int\limits^{\pi}_{-\pi}dq \ln \left[ 
        (1 + x^2)^2 - 2x(1-x^2) (\cos p + \cos q) \right] \right\}, \nonumber \\
\end{equation*}
\begin{equation}
    x = \tanh\theta
\end{equation}

To find the temperature of phase transition we must first find the point where the Helmholtz free energy is singular. This can be seen in \ref{4.1}.  

\begin{gather}
    \eta_{p,q} + \Bar{\eta}_{p.q} \longrightarrow 0, \hspace{0.3cm} 2 - 2\sin{\psi}_c\cos{\psi}_c(\cos p + \cos q) = 0 \nonumber \\
    p,q = 0, \hspace{0.3cm} \sin{\psi}_c = \cos{\psi}_c = \frac{1}{\sqrt{2}} 
\end{gather}

From the definition of the partition function, $\theta = J/T$; thus the critical temperature can be found as:

\begin{gather}
    \tanh{\frac{J}{T_c}} = \frac{1- \cos{\psi_c}}{\sin{\psi_c}} = \sqrt{2}-1
\end{gather}

\appendix  
\counterwithin*{equation}{section} 
\renewcommand\theequation{\thesection\arabic{equation}} 
\section{Appendix 1: Recursive Method in One-Dimensional Ising Model \textit{(Zehra Özcan)}}
\label{chapter:8}

This section will present a recursive method for solving nearest-neighbor interactions exactly. We will follow the paper \cite{Marchi1980RecursiveMI}. The one-dimensional Ising model with only nearest neighbor interactions is one of the simplest models to solve exactly in statistical mechanics. It can be solved exactly by various methods, including the transfer matrix, generating function, and induction. However, when second or third-nearest-neighbor interactions are considered, the problem becomes significantly more complicated. This method is more useful in higher dimensions than other methods for finding exact analytical solutions. 

\subsection{First Neighbour-Interactions}
Firstly, for a configuration ($\sigma$), the interaction energy and partition function are given by:
\begin{align}
        E(\sigma)=-J\sum_{i=1}^{n-1} \sigma_{i}\sigma_{i+1} -h\sum_{i=1}^{n}\sigma_{i}
\end{align} 

\begin{align}
       Z=\sum_{\{\sigma\}} e^{-\beta E(\sigma)}=\sum_{\{\sigma\}} e^{\beta J\sum_{i=1}^{n-1} \sigma_{i}\sigma_{i+1} + \beta h\sum_{i=1}^{n}\sigma_{i}}
\end{align}
\
Where $\beta=1/kT$ and $\sigma = \pm 1$. The partition function can be expressed using the product operator.
\begin{align}
    Z=\sum_{\{\sigma\}} \prod_{i=1}^{n-1} e^{\beta J \sigma_{i}\sigma_{i+1} } \prod_{i=1}^{n} e^{\beta h\sigma_{i}}
\end{align}
\
The exponential function can be expressed in terms of hyperbolic functions. Following \cite{thompson2015mathematical} we simply do a case analysis for $\sigma_{i}\sigma_{i+1}$:
\
\begin{equation}
\begin{aligned}
    \sigma_{i}\sigma_{i+1} &=1,        \hspace{0.7cm}   &       &e^{\beta J \sigma_{i}\sigma_{i+1}}= e^{{\beta J }}=\cosh{(+\beta J)}+\sinh{(+\beta J)} \\ 
    \sigma_{i}\sigma_{i+1} & =-1,     \hspace{0.7cm}      &     &e^{\beta J \sigma_{i}\sigma_{i+1}}= e^{-{\beta J }}=\cosh{(+\beta J) }-\sinh{(+\beta J)} 
\end{aligned}
\end{equation}
\
Therefore, it is apparent that the following equalities hold, where \ $\bar{J}=\beta J, \  \omega_{\bar{J}}=\tanh{\bar{J}}$:
\
\begin{equation}
\begin{aligned}
         e^{{\beta J \sigma_{i}\sigma_{i+1}}} =\cosh{\bar{J}}\biggl(1+ \tanh{(\bar{J} )} \sigma_{i}\sigma_{i+1}\biggr)  =\cosh{\bar{J}}\biggl(1+ \omega_{\bar{J} } \sigma_{i}\sigma_{i+1}\biggr)
\end{aligned}
\end{equation}
\
Likewise, where \ $ \bar{h}=\beta h, \ \omega_{\bar{h}}=\tanh{\bar{h}}$:
\begin{equation}
    \begin{gathered}
       e^{{\beta h \sigma_{i}}}=\cosh{\bar{h}}\biggl(1+ \tanh{(\bar{h}) } \sigma_{i}\biggr) 
    =\cosh{\bar{h}}\biggl(1+ \omega_{\bar{h} } \sigma_{i}\biggr)  
    \end{gathered}
\end{equation}
\
In terms of these, the partition function can be defined in the following way:
\
\begin{equation}
    \begin{aligned}
        Z&=\sum_{\{\sigma\}}\prod_{i=1}^{n-1}\cosh{\bar{J}}(1+ \omega_{\bar{J} } \sigma_{i}\sigma_{i+1})\prod_{i=1}^{n} \cosh{\bar{h}}(1+ \omega_{\bar{h} } \sigma_{i}) \\
       & =\cosh{(\bar{J})}^{n-1}\cosh{(\bar{h})}^{n}\sum_{\{\sigma\}}\prod_{i=1}^{n-1}(1+ \omega_{\bar{J} } \sigma_{i}\sigma_{i+1}) \prod_{i=1}^{n} (1+ \omega_{\bar{h} } \sigma_{i})
    \end{aligned}
\end{equation}
\
Finally, the partition function can be expressed in a simpler form as follows:
\begin{align}
    Z=K\sum_{\{\sigma\}} \prod_{i=1}^{n-1} (1+\omega_{\bar{J}}\sigma_{i}\sigma_{i+1})\prod_{i=1}^{n} (1+\omega_{\bar{h}}\sigma_{i})
\end{align}
\
Where $\omega_{\bar{J}}=\tanh\bar{J}, \ \omega_{\bar{h}}=\tanh \bar{h}, \ K=(\cosh\bar{J})^{n-1}(\cosh\bar{h})^{n}$.
By extracting the leading term from each product term, the partition function can be written as follows:
\
\begin{align}
    Z=K\sum_{\{\sigma^{2}\}} \sum_{\{\sigma_{1}\}} (1+\omega_{\bar{J}}\sigma_{1}\sigma_{2})
    (1+\omega_{\bar{h}}\sigma_{1})
    \prod_{i=2}^{n-1} (1+\omega_{\bar{J}}\sigma_{i}\sigma_{i+1})\prod_{i=2}^{n} (1+\omega_{\bar{h}}\sigma_{i})
\end{align}
\
The summation is divided into two parts.  To clarify the notation, the symbol $\{\sigma^2\}$ indicates that the sum is taken over all possible spin configurations, represented by  $\{\sigma^2\} = (\sigma_{2},\sigma_{3},...,\sigma_{n})$. In contrast, the symbol $\{\sigma_1\}$ represents the sum over the possible spin configurations of $\sigma_1 = \pm 1$.  
We now proceed to evaluate the first sum:
\begin{equation}
\begin{aligned}
\sigma_{1} &={1}, \hspace{0.3cm}    Z_{+1}=K\sum_{\{\sigma^{2}\}}\left[(1+\omega_{\bar{J}}\sigma_{2})(1+\omega_{\bar{h}})\right]\prod_{i=2}^{n-1}(1+\omega_{\bar{J}}\sigma_{i}\sigma_{i+1})\prod_{i=2}^{n}(1+\omega_{\bar{h}}\sigma_{i}) \\
\sigma_{1}& ={-1}, \hspace{0.3cm}  Z_{-1}=K\sum_{\{\sigma^{2}\}}\left[(1-\omega_{\bar{J}}\sigma_{2})(1-\omega_{\bar{h}})\right]\prod_{i=2}^{n-1}(1+\omega_{\bar{J}}\sigma_{i}\sigma_{i+1})\prod_{i=2}^{n}(1+\omega_{\bar{h}}\sigma_{i})
\end{aligned}
\end{equation}
By adding these two equalities, we get the partition function, which is summed over the first spin configurations:
\begin{align}
    \begin{split}
          Z_{+1}+Z_{-1}= K\sum_{\{\sigma^{2}\}}\left[(1+\cancel{\omega_{\bar{h}}}+\cancel{\omega_{\bar{J}}\sigma_{2}}+\omega_{\bar{h}}\omega_{\bar{J}}\sigma_{2}+1-\cancel{\omega_{\bar{h}}}-\cancel{\omega_{\bar{J}}\sigma_{2}}+\omega_{\bar{J}}\omega_{\bar{h}}\sigma_{2})\right] \\
    \times \prod_{i=2}^{n-1}(1+\omega_{\bar{J}}\sigma_{i}\sigma_{i+1}) 
    \prod_{i=2}^{n}(1+\omega_{\bar{h}}\sigma_{i})
    \end{split}
\end{align}
Then, the partition function is equal to:
\begin{equation}
\begin{aligned}
    Z&=K\sum_{\{\sigma^{2}\}}\left[(1+\omega_{\bar{h}}\omega_{\bar{J}}\sigma_{2}+1+\omega_{\bar{h}}\omega_{\bar{J}}\sigma_{2})\right] 
    \times \prod_{i=2}^{n-1}(1+\omega_{\bar{J}}\sigma_{i}\sigma_{i+1}) 
    \prod_{i=2}^{n}(1+\omega_{\bar{h}}\sigma_{i}) \\
     Z&=2K \sum_{\{\sigma^{2}\}}(\alpha_{1}+\beta_{1}\sigma_{2})\prod^{n-1}_{i=2}(1+\omega_{\bar{J}}\sigma_{i}\sigma_{i+1})\prod_{i=2}^{n}(1+\omega_{\bar{h}}\sigma_{i}), \hspace{0.7cm} \text{where}\
    \alpha_{1}=1 \ \text{and} \ \beta_{1}=\omega_{\bar{h}}\omega_{\bar{J}}
\end{aligned}
\end{equation}

Subsequently, if the leading term is extracted from each product term, and the second sum is performed in a manner analogous to the preceding steps:
\begin{align}    
    Z=2K\sum_{\{\sigma^{3}\}}\sum_{\{\sigma_{2}\}}\left[(1+\omega_{\bar{h}}\omega_{\bar{J}}\sigma_{2})(1+\omega_{\bar{J}}\sigma_2\sigma_3)(1+\omega_{\bar{h}}\sigma_2)\right] \times \prod_{i=3}^{n-1}(1+\omega_{\bar{J}}\sigma_{i}\sigma_{i+1})\prod_{i=3}^{n}(1+\omega_{\bar{h}}\sigma_{i})
\end{align}
\begin{equation}
\begin{aligned}
\sigma_2&=1, \hspace{0.1cm}   Z_{+1}=2K\sum_{\{\sigma^{3}\}}\left[(1+\omega_{\bar{h}}\omega_{\bar{J}})(1+\omega_{\bar{J}}\sigma_3)(1+\omega_{\bar{h}})\right] \times \prod_{i=3}^{n-1}(1+\omega_{\bar{J}}\sigma_{i}\sigma_{i+1}) \prod_{i=3}^{n}(1+\omega_{\bar{h}}\sigma_{i})\\
\sigma_2&=-1, \hspace{0.1cm}  Z_{-1}=2K\sum_{\{\sigma^{3}\}}\left[(1-\omega_{\bar{h}}\omega_{\bar{J}})(1-\omega_{\bar{J}}\sigma_3)(1-\omega_{\bar{h}})\right] \times \prod_{i=3}^{n-1}(1+\omega_{\bar{J}}\sigma_{i}\sigma_{i+1}) \prod_{i=3}^{n}(1+\omega_{\bar{h}}\sigma_{i})
\end{aligned}
\end{equation}

By adding these two equalities, we obtain the following expression:
\begin{align}
    Z_{+1}+Z_{-1}=2^{2}K\sum_{\{\sigma^{3}\}}\left[(\alpha_{1}+\beta_{1}\omega_{\bar{h}})+(\beta_{1}\omega_{\bar{J}}+\alpha_{1}\omega_{\bar{J}}\omega_{\bar{h}})\sigma_{3}\right] 
\times \prod_{i=3}^{n-1}(1+\omega_{\bar{J}}\sigma_{i}\sigma_{i+1})\prod_{i=3}^{n}(1+\omega_{\bar{h}}\sigma_{i})
\end{align}

Thus, the partition function becomes:
\begin{equation}
    \begin{gathered}
    Z=2^{2}K\sum_{\{\sigma^{3}\}}(\alpha_{2}+\beta_{2}\sigma_{3})
    \times \prod_{i=3}^{n-1}(1+\omega_{\bar{J}}\sigma_{i}\sigma_{i+1})\prod_{i=3}^{n}(1+\omega_{\bar{h}}\sigma_{i})  \hspace{0.7cm}\\
    \text{Where}\ \alpha_{2}=\alpha_{1}+\beta_{1}\omega_{\bar{h}} \ \ \text{and} \ \ \beta_{2}=\beta_{1}\omega_{\bar{J}}+\alpha_1\omega_{\bar{J}}\omega_{\bar{h}}
    \end{gathered}
\end{equation}
\
By repeating the same process over the other spins, it becomes evident that the recursion relation holds:
\
\begin{align}\label{eq:recursion relation}
   \alpha_{i+1}=\alpha_{i}+\beta_{i}\omega_{\bar{h}}, \hspace{0.3cm}\beta_{i+1}=\beta_{i}\omega_{\bar{J}}+\alpha_{i}\omega_{\bar{J}}\omega_{\bar{h}} 
\end{align}
\
For $j \leqq n-1 $, we have:
\
\begin{align}
    Z=2^{j}K\sum_{\{\sigma^{j+1}\}}( \alpha_{j} + \beta_{j}\sigma_{j+1})
    \prod_{i=j+1}^{n-1} (1+\omega_{\bar{J}}\sigma_{i}\sigma_{i+1})\prod_{i=j+1}^{n} (1+\omega_{\bar{h}}\sigma_{i})
\end{align}
\
For $j=n-1$, we can express $Z$ as: 
\
\begin{align}
    Z=2^{n-1}K\sum_{\{\sigma^{n}\}} (\alpha_{n-1} + \beta_{n-1}\sigma_{n})
    (1+\omega_{\bar{h}}\sigma_{n})
\end{align}
Upon performing the indicated sum, the following expression is obtained:
\begin{align}
     Z=2^{n}K (\alpha_{n-1} + \beta_{n-1}\omega_{\bar{h}})
\end{align}
Then,
\begin{align}\label{eq:partition function}
    Z=2^{n}K \alpha_{n}
\end{align}
The value of  $\alpha_{n}$ can now be determined by expressing (\ref{eq:recursion relation}) as a matrix equation.
\begin{align}
    \begin{pmatrix}
        \alpha_{i+1} \\
        \beta_{i+1}  
    \end{pmatrix}
    =
    \begin{pmatrix}
        1 & \omega_{\bar{h}} \\
        \omega_{\bar{h}}\omega_{\bar{J}} & \omega_{\bar{J}} 
    \end{pmatrix}
    \begin{pmatrix}
        \alpha_{i}  \\
        \beta_{i}  
    \end{pmatrix}
    =
    \textit{A}
    \begin{pmatrix}
        \alpha_{i}  \\
        \beta_{i}  
    \end{pmatrix}
\end{align}
This system can be solved easily. First, $A$ should be diagonalized:
\begin{align}
    C^{-1}AC=\Lambda=
    \begin{pmatrix}
        \lambda_{1} & 0 \\
        0 & \lambda_{2}
    \end{pmatrix}
\end{align}
The eigenvalues of $A$ can be readily determined by solving the characteristic equation.
In order to obtain the characteristic equation, it is necessary to determine the solution of the equation which is given by the determinant of the matrix, $A-\lambda I$, being equal to zero.
\
\begin{gather}
   \begin{vmatrix}
1-\lambda & \omega_{\bar{h}}\\
\omega_{\bar{J}}\omega_{\bar{h}} & \omega_{\bar{J}}-\lambda
    \end{vmatrix}
    =0
\end{gather}

\begin{align}
    \lambda_{1,2}=\frac{1}{2}(1+\omega_{\bar{J}}) \pm \frac{1}{2}\sqrt{\Delta}
\end{align}
\
Note that under the condition $\mid \omega_{\bar{h}} \mid < 1$ eigenvalues are real, positive and different, which is always true. After finding the eigenvalues, now the matrix \textit{C}, whose columns are the eigenvectors of \textit{A}, can be found. 
\
In order to find the eigenvectors, the equation $(A-\lambda I)x=0$ should be solved.
\
\begin{equation}\label{characteristic eq}
    \begin{pmatrix}
    1-\lambda_1 & \omega_{\bar{h}}\\
    \omega_{\bar{J}}\omega_{\bar{h}} & \omega_{\bar{J}}-\lambda_1
    \end{pmatrix}
    \begin{pmatrix}
    x_1\\
    x_2
    \end{pmatrix}
    =
    \begin{pmatrix}
    0\\
    0
    \end{pmatrix}
\end{equation}
\
It is evident that the values of $x_1$ and $x_2$ can be expressed as follows:
\
\begin{align}
    \begin{aligned}
        x_1&= 1 \\
        x_2&= \frac{\lambda_1-1}{\omega_{\bar{h}}}
    \end{aligned}
\end{align}
Plugging them to (\ref{characteristic eq}):
\begin{equation}
    \begin{pmatrix}
    1-\lambda_1 & \omega_{\bar{h}}\\
    \omega_{\bar{J}}\omega_{\bar{h}} & \omega_{\bar{J}}-\lambda_1
    \end{pmatrix}
    \begin{pmatrix}
    1\\
    \frac{\lambda_1-1}{\omega_{\bar{h}}}
    \end{pmatrix}
    =
    \begin{pmatrix}
    (1-\lambda_1)+(\lambda_1-1)\\
    \frac{\omega_{\bar{h}}^2\omega_{\bar{J}}+(\omega_{\bar{J}}-\lambda_1)(\lambda_1-1)}{\omega_{\bar{h}}}
    \end{pmatrix}
    =
    \begin{pmatrix}
    0\\
    0
    \end{pmatrix}
\end{equation}
\
Similarly, the second eigenvector is:
\begin{align}
    \begin{pmatrix}
    1\\
    \frac{\lambda_2-1}{\omega_{\bar{h}}}
\end{pmatrix}
\end{align}
\
Then, the eigenvector matrix $C$ becomes:
\
\begin{align}
    C=
    \begin{pmatrix}
        1 & 1 \\
        \frac{\lambda_{1}-1}{\omega_{\bar{h}}} &
        \frac{\lambda_{2}-1}{\omega_{\bar{h}}}
    \end{pmatrix}
\end{align}
\
If we arrange the matrix equation similar to the following expression, finding the solutions becomes easier:
\
\begin{equation}
\begin{pmatrix}
\gamma_{i+1} \\
\zeta_{i+1}
\end{pmatrix}
=
\begin{pmatrix}
\chi_1 & 0\\
0 & \chi_2
\end{pmatrix}
\begin{pmatrix}
\gamma_i \\
\zeta_i
\end{pmatrix} 
\end{equation}
\
Therefore, we make the following definition to solve this system easily.
\
\begin{equation}\label{eq:41}
\begin{pmatrix}
\alpha_{i+1} \\
\beta_{i+1}
\end{pmatrix}
= C
\begin{pmatrix}
u_{i+1} \\
v_{i+1}
\end{pmatrix}
\end{equation}
\begin{equation}
\begin{pmatrix}
\alpha_{i+1} \\
\beta_{i+1}
\end{pmatrix}
=
A
\begin{pmatrix}
\alpha_{i} \\
\beta_{i}
\end{pmatrix}
\xrightarrow{} C
\begin{pmatrix}
 u_{i+1} \\
v_{i+1}
\end{pmatrix} = 
C\Lambda C^{-1}C
\begin{pmatrix}
u_i \\
v_i
\end{pmatrix}
=
C\Lambda
\begin{pmatrix}
u_i \\
v_i
\end{pmatrix}
\end{equation}
\
We want to express the system in terms of $u'$s and $v'$s. Hence,
\
\begin{equation}
\begin{pmatrix}
\alpha_{i+1} \\
\beta_{i+1}
\end{pmatrix}
=
C
\begin{pmatrix}
 u_{i+1} \\
v_{i+1}
\end{pmatrix}
=
C \Lambda
\begin{pmatrix}
u_{i} \\
v_{i}
\end{pmatrix}
\end{equation}
\
Matrix $C$ on both sides cancel each other, and then we have:
\
\begin{equation}
\begin{pmatrix}
u_{i+1} \\
v_{i+1}
\end{pmatrix}
=
\begin{pmatrix}
\lambda_1 & 0 \\
0 & \lambda_2
\end{pmatrix}
\begin{pmatrix}
u_i \\
v_i
\end{pmatrix}
\end{equation}
\
We obtain two recurrence relations,
\begin{align}
    \lambda_{1}u_{i}=u_{i+1}, \hspace{0.3cm}  \lambda_{2}v_{i}=v_{i+1}
\end{align}
\
We can express the relations as the following form by plugging $u_1$ and $v_1$:
\
\begin{align}
    u_{i+1}=(\lambda_1)^{i}u_{1},\ \text{and} \hspace{0.2cm} v_{i+1}=(\lambda_2)^{i}v_1
\end{align} 
\
Then plugging $u_{i+1}$ and $v_{i+1}$ to (\ref{eq:41}),
\
\begin{equation}
\begin{pmatrix}
\alpha_{i+1} \\
\beta_{i+1}
\end{pmatrix}
=
\begin{pmatrix}
1 & 1 \\
\frac{\lambda_{1} -1}{\omega_{\bar{h}}} &\frac{\lambda_{2}-1}{\omega_{\bar{h}}} 
\end{pmatrix}
\begin{pmatrix}
u_{i+1} \\
v_{i+1}
\end{pmatrix}
=
\begin{pmatrix}
1 & 1 \\
\frac{\lambda_{1} -1}{\omega_{\bar{h}}} &\frac{\lambda_{2}-1}{\omega_{\bar{h}}} 
\end{pmatrix}
\begin{pmatrix}
\lambda_{1}^i u_1 \\
\lambda_{2}^i v_1
\end{pmatrix}
=
\begin{pmatrix}
\lambda_{1}^i u_1+\lambda_2^i v_1 \\
\frac{\lambda_{1} -1}{\omega_{\bar{h}}}\lambda_{1}^i u_1+\frac{\lambda_{2}-1}{\omega_{\bar{h}}} \lambda_{2}^i v_1 
\end{pmatrix}
\end{equation}

\begin{align}\label{alpha}
    \alpha_{i+1}=\lambda_{1}^{i}u_1+\lambda_{2}^{i}v_1, \hspace{0.3cm} \beta_{i+1}=\left(\frac{\lambda_1-1}{\omega_{\bar{h}}}\right)\lambda_{1}^iu_{1}+\left(\frac{\lambda_2-1}{\omega_{\bar{h}}}\right)\lambda_{2}^iv_1
\end{align}
\
Since $\alpha_1=1$ and $\beta_1=\omega_{\bar{J}}\omega_{\bar{h}}$,
\
\begin{align}
    1=u_{1}+v_{1}
\end{align}
\begin{align}
    \omega_{\bar{J}}\omega_{\bar{h}}=\left(\frac{\lambda_{1}-1}{\omega_{\bar{h}}}\right) u_{1}+ \left(\frac{\lambda_{2}-1}{\omega_{\bar{h}}}\right)v_1
\end{align}
\
We can easily solve for $u_{1}$ and $v_{1}$:
\
\begin{align}\label{eq:u1 and v1}
   u_{1} = \frac{\omega _{\bar{h}^2} \omega _{\bar{j}}-\lambda _2+1}{\lambda _1-\lambda _2}, \ v_{1} = \frac{-\omega _{\bar{h}^2} \omega _{\bar{j}} + \lambda _1 -1}{\lambda _1-\lambda _2}
\end{align}
\
In (\ref{eq:partition function}), we found the partition function as:
\
\begin{equation}
    \begin{gathered}
         Z=2^nK(\alpha_{n-1}+\beta_{n-1}\omega_{\bar{h}})=2^nK\alpha_n \\
         \text{Where} \ \
        \alpha_n=\alpha_{n-1}+\beta_{n-1}\omega_{\bar{h}}, \hspace{0.3cm}
        \beta_n=\beta_{n-1}\omega_{\bar{J}}+\alpha_{n-1}\omega_{\bar{J}}\omega_{\bar{h}}
    \end{gathered}
\end{equation}
\
From (\ref{eq:recursion relation}) we know that:
\
\begin{align}
    \alpha_{i+1}=\alpha_{i}+\beta_{i}\omega_{\bar{h}},  \hspace{0.3cm}
    \beta_{i+1}=\beta_{i}\omega_{\bar{J}}+\alpha_{i}\omega_{\bar{J}}\omega_{\bar{h}}
\end{align}
\
If we set:
\
\begin{align}
   n=i+1 
\end{align}
\
Then if we plug the $\alpha_n$ from (\ref{alpha}), the partition function becomes:
\
\begin{align}
    Z=2^nK\alpha_n=2^{i+1}K\alpha_{i+1}=2^{i+1}(\cosh{\bar{J}})^{i}(\cosh{\bar{h}})^{i+1}(\lambda_{1}^iu_{1}+\lambda_{2}^iv_1)
\end{align}
\
Relabeling $i \xrightarrow{} n$ :
\
\begin{align}\label{final exp for z}
    Z=2^{n+1}(\cosh{\bar{J}})^{n}(\cosh{\bar{h}})^{n+1}(\lambda_{1}^nu_{1}+\lambda_{2}^nv_1)
\end{align}
\
Finally, when $u_1$ and $v_1$ from (\ref{eq:u1 and v1}) plugged to the (\ref{final exp for z}), the partition function is:
\
\begin{align}
    Z=2^{n+1}(\cosh{\bar{J}})^{n}(\cosh{\bar{h}})^{n+1}\left[\lambda_{1}^n\left( \frac{\omega _{\bar{h}^2}\omega _{\bar{J}}-\lambda _2+1}{\lambda _1-\lambda _2}\right)+\lambda_{2}^n\left(\frac{-\omega _{\bar{h}^2} \omega _{\bar{J}} +\lambda _1-1}{\lambda _1-\lambda _2}\right)\right]
\end{align}
\
Let us examine the case where $h=0$ to check whether the result found is correct.
\\
If $h=0$, the partition function becomes the following expression:
\begin{equation}
    \begin{aligned}
        Z=2^{n+1} (\cosh{\bar{J}})^n \left(\lambda_1^n \frac{(-\lambda_2+1)}{\lambda_1-\lambda_2}+ \lambda_2^n \frac{(\lambda_1-1)}{\lambda_1-\lambda_2}\right)
    \end{aligned}
\end{equation}
\
Similarly, as $\omega _{\bar{h}}$ is equal to $0$, the $\lambda$s become:
\begin{equation}
    \begin{aligned}
    \lambda_{1,2}= \frac{1}{2}(1+\omega_{\bar{J}})\pm\frac{1}{2}\sqrt{(\omega_{\bar{J}}-1)^{2}}
    \end{aligned}
\end{equation}
\
By labeling $\lambda$s arbitrarily:
\begin{equation}
    \begin{aligned}
        \lambda_1= \omega_{\bar{J}}, \ \ \ \lambda_2=1
    \end{aligned}
\end{equation}
\
Plugging these to the partition function, the following result is obtained:
\begin{equation}
    \begin{aligned}
        Z&=2^{n+1} (\cosh{\bar{J}})^n\left(\omega_{\bar{J}}\frac{-1+1}{\lambda_1-\lambda_2}+\frac{\omega_{\bar{J}}-1}{\omega_{\bar{J}}-1} \right) \\
        Z&=2^{n+1} (\cosh{\bar{J}})^n
    \end{aligned}
\end{equation}

\section{Appendix 2: Solving the One-Dimensional Ising Model by Mathematical Induction \textit{(Uveys Turhan)}}
\label{chapter:9}

One-dimensional Ising chain consisting of $N$ spins can be solved by using the transfer method and the result for the partition function using this method is given by:
\begin{gather}
    Z_N=\Tr(T^{N})= \lambda_1^{N} +\lambda_2^{N}
\end{gather} where $T$ is the transfer matrix and $\lambda_1$ and $\lambda_2$ are eigenvalues of $T$:
\begin{equation}
    T =\begin{pmatrix}
        e^{(\beta(J+h)} & e^{-\beta J}  \\
        e^{-\beta J} & e^{\beta(J-h)} \\
    \end{pmatrix} 
\end{equation}
In thermodynamic limit, $N \xrightarrow{} \infty$, largest eigenvalue, say $\lambda_1$ dominates the expression since $(\frac{\lambda_2}{\lambda_1})^N$ goes to zero as $N$ tends to infinity. In this case, $Z_N$ is given by:
\begin{gather}
    Z_N \approx \lambda_1^{N}=e^{N\beta J}\biggl[ \cosh{\beta h} + \sqrt{(\sinh{\beta h})^2 + e^{-4\beta J}}\biggl]^N
\end{gather}
In this section, we derive this result using mathematical induction, following the paper \cite{wang2019}. 

\subsection{Solving via Mathematical Induction}
Consider a one-dimensional, closed Ising chain with spins interacting with their closest neighbors and also with an external magnetic field. How can we construct the partition function ${Z}_{N+1}$, given the partition function $Z_N$? One can see that inserting an extra spin into the chain only affects three interactions: the interaction between $\sigma_1$ and $\sigma_N$ vanishes and new interactions between ${\sigma}_{N+1}$ and $\sigma_N$ , $\sigma_1$ are created. So, we can conclude that configurations of only the first and last spins are relevant. 
That is why we can split our partition function into four components according to the configurations of the spins at the boundaries. For example, ${Z}^{\uparrow \uparrow}_{N}$ means that the configurations of $\sigma_1$ and $\sigma_N$ are $\uparrow$, while for the rest of the spins there are $2^{N-2}$ possibilities, as usual. From this, it is clear that:

\begin{equation}\label{partition_function}
    Z_N = {Z}^{\uparrow \uparrow}_{N} + 
{Z}^{\uparrow \downarrow}_{N} + 
{Z}^{\downarrow \uparrow}_{N} +
{Z}^{\downarrow \downarrow}_{N}
\end{equation}
We can represent these four components with a "vector" partition function for convenience:
\begin{equation}\label{vector_partition_function}
   \vec z_N=(
{Z}^{\uparrow \uparrow}_{N} , 
{Z}^{\uparrow \downarrow}_{N}, 
{Z}^{\downarrow \uparrow}_{N},
{Z}^{\downarrow \downarrow}_{N})^T 
\end{equation}

In order to understand why and how to use this $\vec{z_N}$, let us try to express ${Z}^{\uparrow \uparrow}_{N+1}$ in terms of the components of $\vec{z_N}$. First, notice that ${Z}^{\uparrow \uparrow}_{N+1}$ can be expressed as a linear combination of ${Z}^{\uparrow \uparrow}_{N}$ and ${Z}^{\uparrow \downarrow}_{N}$, because we know that $\sigma_1$ should be $\uparrow$ and the newly added spin ${\sigma}_{N+1}$ should also be $\uparrow$ but $\sigma_N$ can be $\uparrow$ or $\downarrow$.

\begin{equation}\label{up up partition function}
    {Z}^{\uparrow \uparrow}_{N+1}={c_1}{Z}^{\uparrow \uparrow}_{N} + {c_2}{Z}^{\uparrow \downarrow}_{N}
\end{equation}

\begin{figure}[!]
  \centering
  \includegraphics[scale=0.5]{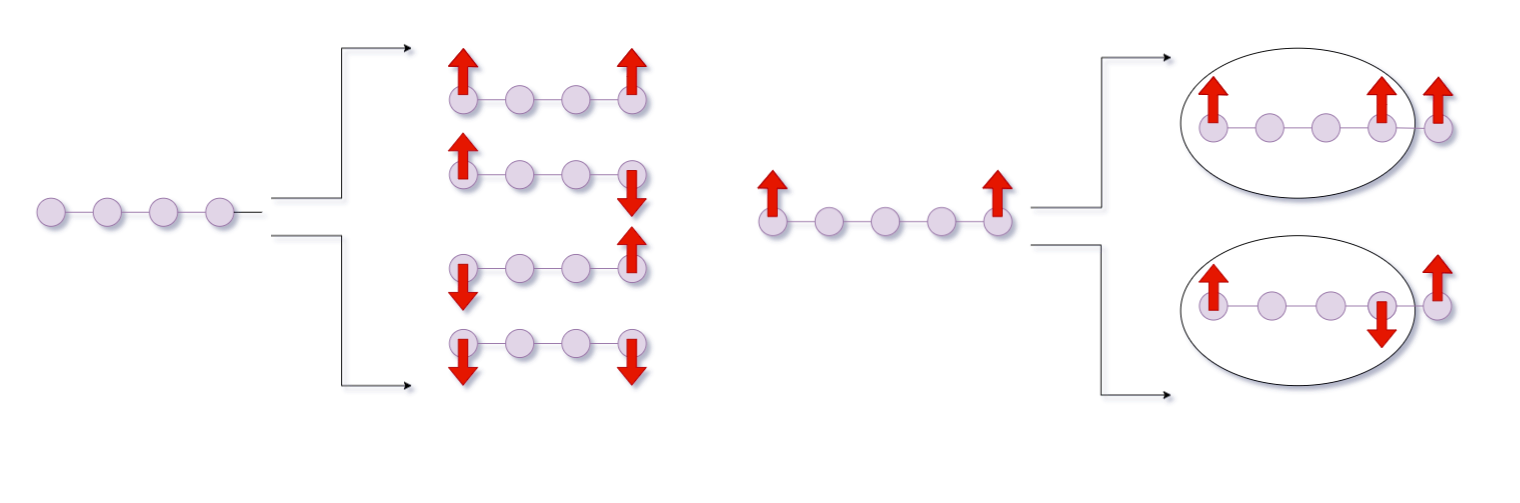}
  \caption{Illustrations of \ref{vector_partition_function} and \ref{up up partition function}.    The first panel represents possible spin configurations of the $N=4$ chain. The second panel represents possible spin configurations of $N+1=5$ spin chain with boundary spins are up spins.}
  \label{fig:induction chain}
\end{figure}

We should determine the Boltzmann weights $c_1$ and $c_2$. In the beginning, I mentioned that inserting an extra spin just affects three interactions. If $\sigma_N$ is $\uparrow$, this means:
\\
$\sigma_N-\sigma_1$ interaction, which corresponds to $+J$, vanished, and new interactions between ${\sigma}_{N+1}$ - $\sigma_1$, ${\sigma}_{N+1}$ - $\sigma_N$ are created, each corresponds to $+J$. So in total, we have $-J+J+J=+J.$ Adding the external field interaction contribution, we conclude that $c_1$= $e^{\beta(J+h)}$ and with the same argument we find $c_1 = c_2$. We actually transform the vector partition $\vec z_{N}$ into $\vec z_{N+1}$ . So that we can write:
\begin{equation}\label{recurrence_relation}
    \vec{z}_{N+1}=M{\vec{z}_{N}}
\end{equation}
We have this nice recurrence relation, where $M$ is the recurrence matrix. Now, the goal is to calculate the elements of this matrix. $M$ is a matrix $4\times4$ and we can find the elements simply by hand, just like in the example we just did.  By considering only three interactions between the spins, adding them to get the total effect, and including the contribution from the external magnetic field interaction, we can find the $M$:

\begin{equation}
    M=
    \begin{pmatrix}
        e^{\beta(J+h)} & e^{\beta(J+h)} & 0 & 0 \\
        e^{-\beta(3J+h)} & e^{\beta(J-h)} & 0 & 0 \\
        0 & 0 & e^{\beta(J+h)} & e^{-\beta(3J-h)} \\
        0 & 0 & e^{\beta(J-h)} & e^{\beta(J-h)} \\
    \end{pmatrix}
\end{equation}

The matrix $M$ is block-diagonal. To begin, we can check the base cases for $N=1$ or $N=2$ and verify the equation \ref{recurrence_relation}. Then, for the inductive step, we need to prove that equation \ref{recurrence_relation} is true for all $N$. This means that it gives the same result as the standard transfer matrix method. To achieve this, we express $\vec{z_1}$ as a system with one spin that interacts with itself and an external magnetic field. We can also express the recurrence relation as follows:

\begin{equation}
    \vec{z_1}=(e^{\beta(J+h)},0,0,e^{\beta(J-h)})^T
\end{equation}
\begin{equation}
    \vec{z}_{N}={M}^{N-1}{\vec{z}_{1}}
\end{equation}

Then we should express $M$ more compactly, by using the fact that it is a block diagonal matrix. Call the first block $M_{11}$ and the second $M_{22}$:

\begin{equation}
     \vec{z}_{n}=
    \begin{pmatrix}
        M^{N-1}_{11} & 0 \\
        0 & M^{N-1}_{22} \\
    \end{pmatrix}
    {\vec{z}_{1}}
\end{equation}

\begin{equation}
    \vec{z}_{n}=
    \begin{pmatrix}
         M^{N-1}_{11} {\vec{z}_{1}}(1:2)  \\
         M^{N-1}_{22} {\vec{z}_{1}}(3:4)\\
    \end{pmatrix}
\end{equation}

$\vec{u}(m,n)$ means the vector formed by the elements from $mth$ to $nth$ of the vector $\vec{u}$. From \ref{partition_function} we can write:
\begin{equation}\label{partition_function_sum}
    Z_N=\sum{\vec{z}_N}=\sum{\vec{z}_N}(1:2) + 
    \sum{\vec{z}_N}(3:4)
\end{equation}

The sum is over the components of $\vec{z_N}$, i.e $\sum{\vec{z}_N}={Z}^{\uparrow \uparrow}_{N} + 
{Z}^{\uparrow \downarrow}_{N} + 
{Z}^{\downarrow \uparrow}_{N} +
{Z}^{\downarrow \downarrow}_{N}$. To calculate the $Nth$ power of $M$, we will apply a similarity transform, and in some steps, we will need to express the $Z_{N}$ in terms of traces. So we use a simple identity. $\times$ sign just corresponds to ordinary matrix multiplication and $[\textit{b,b}]$ stands for $2\times2$ whose column vectors are vector \textit{b} and the sum $\Sigma$ is over the components of the vector as in \ref{partition_function_sum}. If $A$ is a $2\times2$ matrix and \textit{b} is a $2\times1$ vector, then:

\begin{equation}\label{simple_identity}
    \sum{Ab}=\Tr(A \times [b,b])
\end{equation}

Using \ref{simple_identity}, we can express the partition function as:
\begin{equation}
    Z_N=\Tr\Biggl(M^{N-1}_{11}[{\vec{z}_{1}}(1:2),{\vec{z}_{1}}(1:2)]\Biggr) + 
    \Tr\Biggl(M^{N-1}_{22}[{\vec{z}_{1}}(3:4),{\vec{z}_{1}}(3:4)]\Biggr)
\end{equation}

We can symmetrize $M$ matrices as:
\begin{equation}
     {M}_{11}=
    \begin{pmatrix}
        e^{\beta(J+h/2)} & 0 \\
        0 & e^{-\beta(J+h/2)} \\
    \end{pmatrix}
    \begin{pmatrix}
        e^{(\beta(J+h)} & e^{-\beta J}  \\
        e^{-\beta J} & e^{\beta(J-h)} \\
    \end{pmatrix}
    \begin{pmatrix}
        e^{-\beta(J+h/2)} & 0 \\
        0 & e^{\beta(J+h/2)} \\
    \end{pmatrix}
\end{equation}

\begin{equation}
     {M}_{22}=
    \begin{pmatrix}
        e^{-\beta(J-h/2)} & 0 \\
        0 & e^{\beta(J-h/2)} \\
    \end{pmatrix}
    \begin{pmatrix}
        e^{(\beta(J+h)} & e^{-\beta J}  \\
        e^{-\beta J} & e^{\beta(J-h)} \\
    \end{pmatrix}
    \begin{pmatrix}
        e^{\beta(J-h/2)} & 0 \\
        0 & e^{-\beta(J-h/2)} \\
    \end{pmatrix}
\end{equation}

We can write $M^{N-1}$ as $AT^{N-1}A^{-1}$. Using the fact that the trace is cyclic, we can compute the first and the second term:
\begin{gather}
     \Tr\Biggl(M^{N-1}_{11}[{\vec{z}_{1}}(1:2),{\vec{z}_{1}}(1:2)]\Biggr)\\=\Tr \Biggl(
    P^{N-1} 
    \begin{pmatrix}
        e^{-\beta(J+h/2)} & 0 \\
        0 & e^{\beta(J+h/2)} \\
    \end{pmatrix}
    \begin{pmatrix}
        e^{(\beta(J+h)} & e^{\beta(J+h)}  \\
        0 & 0 \\
    \end{pmatrix}
    \begin{pmatrix}
        e^{\beta(J+h/2)} & 0 \\
        0 & e^{-\beta(J+h/2)} \\
    \end{pmatrix}\Biggr)
\end{gather}

\begin{gather}
    \Tr\Biggl(M^{N-1}_{22}[{\vec{z}_{1}}(3:4),{\vec{z}_{1}}(3:4)]\Biggr)\\ = \Tr\Biggl(
    P^{N-1}
    \begin{pmatrix}
        e^{-\beta(J-h/2)} & 0 \\
        0 & e^{\beta(J-h/2)} \\
    \end{pmatrix}
    \begin{pmatrix}
        0 & 0 \\
        e^{\beta(J-h)} & e^{\beta(J-h)} \\
    \end{pmatrix}
    \begin{pmatrix}
        e^{\beta(J-h/2)} & 0 \\
        0 & e^{-\beta(J-h/2)} \\
    \end{pmatrix}\Biggr)
\end{gather}

Performing the multiplication of the last three terms in the right hand in the equations above and adding them to get the partition function we get:

\begin{equation}
    Z_N=\Tr \Biggl(
    T^{N-1}
    \begin{pmatrix}
        e^{\beta(J+h)} & e^{-\beta J}  \\
        0 & 0 \\
    \end{pmatrix}\Biggr)
    +
     \Tr\Biggl(T^{N-1}
    \begin{pmatrix}
        0 & 0 \\
        e^{-\beta J} & e^{\beta(J-h)}  \\
        
    \end{pmatrix}\Biggr) \\ = \Tr(T^{N})
\end{equation}

Finally, we have the result: $Z_N=\Tr(T^{N})$, which is identical to the result we get by using the transfer matrix method.


\end{document}